%% file: main.tex
\documentclass[11pt,a4paper]{article}
\usepackage[T1]{fontenc}
\usepackage[utf8]{inputenc}
\usepackage{mathptmx}
\usepackage[a4paper,top=2.5cm,bottom=3cm,left=2.5cm,right=2.5cm]{geometry}
\usepackage{booktabs}
\usepackage{array}
\usepackage{longtable}
\usepackage{graphicx}
\usepackage{caption}
\PassOptionsToPackage{hyphens}{url}
\usepackage[authoryear,round,semicolon]{natbib}
\usepackage[hidelinks]{hyperref}
\usepackage{microtype}
\usepackage{amsmath}
\usepackage{amssymb}
\usepackage{textcomp}
\usepackage{setspace}
\usepackage{parskip}
\usepackage{fancyhdr}
\usepackage{titlesec}
\usepackage{enumitem}
\usepackage{alltt}
\usepackage{xcolor}
\usepackage[mathlines]{lineno}

\titleformat{\section}{\large\bfseries}{\thesection.}{0.5em}{}
\titleformat{\subsection}{\normalsize\bfseries}{\thesubsection}{0.5em}{}
\titleformat{\subsubsection}{\normalsize\itshape}{\thesubsubsection}{0.5em}{}
\titlespacing*{\section}{0pt}{14pt}{8pt}
\titlespacing*{\subsection}{0pt}{10pt}{4pt}
\titlespacing*{\subsubsection}{0pt}{8pt}{3pt}

\newcounter{sixrow}
\newcommand{\srow}[1]{\leavevmode\refstepcounter{sixrow}\label{#1}\textbf{\thesixrow.}~}
\newenvironment{paperabstract}{%
  \begin{center}\textbf{Abstract}\end{center}%
  \begin{quotation}\small\noindent
}{%
  \end{quotation}\medskip
}
\begin{document}
\thispagestyle{empty}
\begin{center}
{\LARGE\bfseries Accounting for Purchased Artificial Intelligence Inference in Corporate Greenhouse Gas Inventories: A Token-Based Method}\\[12pt]
{\normalsize\bfseries Guillermo Llopis-Garcia}\\[4pt]
{\small SOMA AI, Barcelona \quad \texttt{guillermollopis@somaai.earth}}
\end{center}
\vspace{2pt}
{\footnotesize\noindent Accepted manuscript of the article published in \textit{Sustainable Production and Consumption}, \url{https://doi.org/10.1016/j.spc.2026.11.003}. \copyright{} 2026. This manuscript version is made available under the CC-BY-NC-ND 4.0 license \url{https://creativecommons.org/licenses/by-nc-nd/4.0/}.\par}
\vspace{6pt}
\rule{\linewidth}{0.8pt}
\vspace{6pt}
\begin{paperabstract}
Purchased artificial intelligence (AI) inference belongs in Scope 3 Category 1 of a corporate greenhouse gas inventory, yet practice omits it or applies a spend-based factor that gives 1.7 to 12.3 times the physical estimate and cannot tell one model or region from another. This paper sets out a token-based method and open dataset: a cradle-to-gate boundary, four tiers of activity data, an effective-token basis, a four-term factor (serving electricity, embodied hardware, data-centre construction, disclosed final training runs) and a rule resolving any model identifier and region to a factor. Inference footprints span twelve times between model classes and nine between regions. The mid-class factor, 0.039 kg of carbon dioxide equivalent (CO\textsubscript{2}e) per million billed tokens at the world-average grid, is the low end of a stated structural range: a less efficient serving configuration raises it 2.8 times, and replacing the training term with all model development, on its own boundary, gives 1.5 to 2.9 times it. Input bands alone give $-$18\% to $+$29\%. Published per-query energy estimates that can describe a mid-class model are 2.4 to 7.2 times the mid-class energy, above its structural range in four of five rows; those that can describe a frontier-class model bracket it at 0.7 to 1.6 times. Worked examples give 8.5 kg CO\textsubscript{2}e a year for a 200-person firm, and 2.1 tonnes for fifty developers using agentic coding tools on a mid-class model or 8.3 tonnes on a frontier-class one. The contribution is a traceable figure; accuracy against metered energy is untested.
\end{paperabstract}

\input{nomenclature}

\section{Introduction}\label{sec:1}

Regulation has moved faster than practice. The largest European Union (EU) companies (more than 1,000 employees and EUR 450 million net turnover after the Omnibus I Directive; \citealp{euparl2022,euparl2026b}) must disclose gross Scope 3 emissions for every significant category under the Corporate Sustainability Reporting Directive (CSRD, Directive (EU) 2022/2464), where climate change is material. The datapoint is European Sustainability Reporting Standard (ESRS) disclosure E1-6 for financial years up to 2026~\citep{eucom2023}, and E1-8 under the revised standards of Commission Delegated Regulation (EU) 2026/1563 from 2027 (or for 2026, where an undertaking opts for them under Article 2 of that Regulation), which keep both the significance test and the per-category breakdown~\citep{eucom2026a}. Those revised standards also ask an undertaking, where applicable, to report emissions from purchased cloud computing and data-centre services as a sub-category of Category 1, which this paper reads as covering purchased artificial intelligence (AI) inference, whether bought as application programming interface (API) access, as enterprise seats or embedded in software~\citep{eucom2026a}. The International Energy Agency expects electricity consumption by AI-focused data centres to more than triple between 2025 and 2030, to about 465 TWh, while total data-centre consumption roughly doubles from 485 TWh in 2025 to 950 TWh in 2030~\citep{iea2026}, and AI line items have become visible in corporate budgets. No standard, however, gives a preparer the factors and the activity-data rules needed to produce a figure from a billing statement. In current practice the AI line is either omitted or left inside the generic spend-based Category 1 calculation, priced with an environmentally extended input-output (EEIO) factor for the information and communications technology (ICT) sector as a whole. That practice is what this paper sets out to replace: on the same purchase the spend-based figure is 1.7 to 12.3 times a physical estimate on deflated spend and 2.0 to 15.1 times undeflated (Section~\ref{sec:7.1}), and it cannot respond to a change of model or of region. The method claims a traceable figure, not precision against metered energy, which no check here reaches (Section~\ref{sec:5}).

The inputs for something better already exist. Accelerator energy per token has been benchmarked by academic researchers~\citep{mlenergy2025,niu2026}. Grid carbon intensities are published by national operators and synthesised by \citet{ember2026} and the \citet{epa2025}. Server manufacturing footprints, data-centre construction emissions, model training emissions, water use effectiveness (WUE) and provider retirement policies are all published somewhere. What has been missing is the assembly: a procedure that takes the data a customer actually has, states a boundary, resolves a commercial model identifier and a region to a factor, and produces a number with a stated uncertainty that an auditor can follow.

A usable method has to do four things. It must start from what a billing export contains, which is tokens, messages, seats or spend, and never from data only the provider holds. It must meet the Category 1 minimum boundary of the Greenhouse Gas (GHG) Protocol Scope 3 Standard, which is cradle-to-gate, and the calculation-method recommendations of its Technical Guidance, rather than counting serving electricity alone~\citep{wri2011,wri2013}. It must resolve any model identifier and any region, including the closed commercial models that are most of the market and the regions that providers do not disclose. And it must stay current, because model names, prices, grid years and benchmark releases all change faster than an annual reporting cycle. Developer tools, provider disclosures and the token-based methods published while this work was in preparation each meet several of the four; what none provides, on the material consulted, is the combination that Section~\ref{sec:2} and Table~\ref{tab:requirements} set out: published defaults for messages and seats, a boundary with every exclusion bounded or named, and the resolution of any model identifier and any region under a maintenance protocol that states a cadence and a trigger for each input.

This paper sets out such a method and the dataset that parameterises it. Its components are the contributions: a \emph{scope and boundary} for a purchased inference service (Section~\ref{sec:3.1}); \emph{four tiers of activity data} matched to what a customer can obtain, including a tier for seat-licensed products that report messages or seats but no tokens (Section~\ref{sec:3.2}); a \emph{token basis} that reconciles an energy figure measured per output token with input-heavy billing (Section~\ref{sec:3.3}); a \emph{factor equation} with four terms on two stated denominators, with all model development, including the final training runs, on a separate line (Section~\ref{sec:3.4}); a \emph{resolution rule} from a model identifier and a region to a factor, with named fallbacks (Section~\ref{sec:3.5}); \emph{water} as a short extension of the same chain rather than a separate method (Sections S12.2 and S9.8); \emph{uncertainty and reporting rules} that keep three kinds of spread apart (Sections~\ref{sec:3.8} and S12.3); a \emph{maintenance protocol} naming the source, the cadence and the out-of-cycle trigger of every input (Section S12.3); and an \emph{open dataset and reference calculator}, released with the paper, implementing all of it for 18 measured open-weight model and precision variants of 16 base models, three model classes, 210 countries and economies, 27 United States (US) grid subregions and 143 public cloud regions, with 187 identifier patterns (Section~\ref{sec:4.9} and Data Availability). One further contribution is not a component of the method but a check on it: a comparison with published per-query estimates, a first-party energy decomposition, the activity defaults checked without customer data and another published method on a common workload (Section~\ref{sec:5} and S10). A worked example of a 200-person firm illustrates the calculation (Section~\ref{sec:6}), with a second at the opposite end of the volume range in S11.

Section~\ref{sec:2} reviews the literature and states the requirements, Section~\ref{sec:3} is the method with its parameterisation from dataset v3 (Section S9), Section~\ref{sec:results} prints the factor tables for three model classes and seventeen regions (Section~\ref{sec:4.9}) and the case study (Section~\ref{sec:6}), Section~\ref{sec:5} reports the checks, Section~\ref{sec:7} discusses the drivers, the limits, provider disclosure, the standards in force and water, and Section~\ref{sec:8} concludes.

\section{Literature Review}\label{sec:2}

Six bodies of work bear on this problem, and each supplies part of it.

\textbf{Measurement research.} Training energy has had sustained attention since \citet{strubell2019} and \citet{patterson2021}, and a reporting practice grew around it. Calculators and trackers turn hardware, hours, region and power usage effectiveness (PUE) into a carbon figure~\citep{lacoste2019,henderson2020,anthony2020}. One study measured a whole training run with its idle and infrastructure power~\citep{luccioni2023}, and others extend the boundary to embodied hardware~\citep{faiz2024,morrison2025}. Cloud measurements found that the region changed the emissions of one training job by 3.7 times and the hour of the day by up to 8\%~\citep{dodge2022}. Inference has had less, although it is the dominant operational workload for enterprise AI. \citet{samsi2023} and \citet{luccioni2024} measured inference energy for open-weight models on earlier accelerator generations and showed that energy per query depends on the task, the output length and the serving batch size as well as on the model, which is why this paper fixes an operating point and a token basis before any factor is stated. The ML.ENERGY Leaderboard~\citep{mlenergy2025} measures accelerator energy in joules per output token for open-weight models under one protocol, across sweeps of accelerator count and serving batch size, on both NVIDIA H100 and B200 accelerators. \citet{niu2026} extend measurement to the whole node and report the accelerator share of node energy. Neither of the two measures facility overhead. Green Algorithms~\citep{lannelongue2021} estimates the footprint of general computational workflows, but needs the hardware type, core count, memory and run time of the job, which a buyer of a hosted model does not see.

\textbf{Per-query and per-model estimates.} A second group of recent studies publishes energy, carbon or water per query for named models, mostly closed commercial ones, from outside or from inside. From outside, \citet{jegham2025} benchmark commercial models through their public APIs, with an inferred hardware configuration and company-specific multipliers for PUE, water and carbon, on an operational boundary (facility electricity, its Scope 2 carbon and on-site plus off-site water) that excludes embodied hardware. \citet{you2025} estimates a GPT-4o query in ChatGPT from an assumed parameter count, accelerator utilisation and response length, with server power and facility overhead inside the boundary and embodied hardware outside it, so the result scales with each assumption. AI Energy Score~\citep{luccioni2025} measures accelerator energy per query for mainly open-weight models on standardised tasks, and its second version reports reasoning at about 30 times the energy of models without it on average. \citet{oviedo2026} model per-query energy for models above 200B parameters on H100 nodes with a modelled rather than measured facility overhead, and add idle capacity and data-centre networking to their fleet-level estimate only. From inside, \citet{elsworth2025} report Google's median Gemini Apps text prompt from the provider's own fleet telemetry, with host, idle machines, facility overhead, embodied emissions and water in the boundary but no token count. \citet{mistral2025} publish a lifecycle assessment, made with Carbone 4 and the French Agency for Ecological Transition (ADEME), that gives one figure for a 400-token Le Chat response with no input count and excludes training from it. \citet{li2025} add the cooling water consumed on site to the water consumed off site in generating the electricity, and Section S12.2 uses their 2023 site series where nothing more specific is published. \citet{kim2026} review the published assessments and describe amortising training emissions over the inference requests expected in a model's life as one possible approach, which Section~\ref{sec:3.1} adopts per token for the final training runs. None of these is an accounting method. Each answers what one query costs, at token counts that are mostly undisclosed, not what an organisation's purchased usage costs, so they have no column in Table~\ref{tab:requirements}, and their figures are restated at the token counts behind them in Section~\ref{sec:5.1}.

\textbf{Developer tools.} EcoLogits~\citep{rince2025} and an earlier tool~\citep{lottick2019} instrument application code to attribute emissions call by call, EcoLogits by modelling each inference request and the earlier tool by metering the energy of each function on the user's own processor, both needing pipeline-level access that a sustainability manager working from a quarterly billing statement does not have. EcoLogits has taken its per-token energy from the same ML.ENERGY Leaderboard since November 2025, so the two rest on one measurement, but on its current published methodology it charges energy on output tokens only and does not model prompt caching~\citep{rince2025}, so on input-heavy billing this paper's figures are the higher of the two, by the input weighting net of the divisor $D$ (Section~\ref{sec:3.3}).

\textbf{Provider reporting.} Three tools allocate cloud emissions to customers and all three now report Scopes 1, 2 and 3~\citep{cloudtools2026}: Amazon Web Services (AWS) Sustainability, which succeeded and expanded the Customer Carbon Footprint Tool; Google Cloud's Carbon Footprint; and the Azure Emissions Impact Dashboard, which Microsoft will retire on 31 March 2027 in favour of Azure Carbon Optimization. None of them, as of October 2026, discloses AI-specific usage at model level, and S4 gives their service granularity. The EU AI Act obliges providers of general-purpose AI models to document the known or estimated energy consumption of the model for the AI Office and national competent authorities rather than for customers \citep[Article 53(1)(a) and Annex XI, Section~1, point 2(e)]{euparl2024}. This paper reads that datapoint as mainly about training and development energy, a reading the Act does not settle and the Digital Omnibus on AI of July 2026 left untouched~\citep{euparl2026a}, and S4 sets out the reading and the case against it. Nothing requires per-token inference figures to reach customers (Section~\ref{sec:7.3}), and as of this writing no major provider publishes inference energy per token. Google and Mistral publish per-prompt and per-response figures~\citep{elsworth2025,mistral2025}, which this paper uses as calibration points (Section~\ref{sec:5.1}).

\textbf{Spend-based practice.} The GHG Protocol's spend-based method takes its factors from environmentally extended input-output databases, described in the Scope 3 Standard and in its Technical Guidance~\citep{wri2011,wri2013}, and EXIOBASE 3.8.2~\citep{stadler2018} is the common open implementation. Neither text names multi-regional tables, and the Guidance lists single-region coverage among the limitations of such data, so a multi-regional table is consistent with the published method rather than provided for by it. Version 3.8.2 is used here and not updated, for the licence reason given in the reference-list entry for \citet{stadler2018}. Its ``Computer and related services'' sector carries 0.1181 kg CO\textsubscript{2}e per EUR (Austria, 2019 basic prices, retrieved via \citealp{climatiq2026}), a factor calibrated before the AI compute boom on activity spanning software development, information-technology consulting and cloud services of every kind. Applied to AI inference at API list prices it gives several times a physical estimate (Sections~\ref{sec:3.7} and \ref{sec:7.1}).

\textbf{Concurrent token-based methods.} Three have appeared since the first version of this paper (arXiv:\allowbreak 2606.10660, June 2026). \citet{bistline2026}, published in August 2026, propose a three-tier framework (spend-based, activity-based and provider-reported) with a functional unit of kg CO\textsubscript{2}e per million tokens under Category 1, with hyperscaler defaults of 0.32 and 0.96 Wh per 1,000 input and output tokens before a PUE of 1.10, 0.017 kg per million tokens of embodied hardware inside the total, and 0.1 kg per million tokens of training reported as a separate line, all proposed as defaults for feedback. The 3-to-1 ratio between their two energy intensities is a decode-to-prefill electricity ratio they take from Splitwise~\citep{patel2024} rather than a result of their framework (Section S10.2). \citet{horswill2026}, published by Greenpixie in September 2026, benchmark open-weight models on H100 and B200, separate input from output tokens, fit energy per token to model size, traffic and deployment by Bayesian regression, place closed models in size buckets sampled by Monte Carlo, and add embodied emissions, a training uplift and water. The Closed-model Latent Energy Estimation Range (CLEER) report of \citet{jegham2026}, published by the Sustainable AI Group in September 2026, estimates accelerator energy per token for individual closed models by matching each to its five nearest measured open models and inverting its published throughput and latency to an implied batch size; a companion accounting document~\citep{gamazaychikov2026} adds host, idle and facility overheads and a model-development term and bounds several exclusions. Their dashboard data~\citep{saigdata2026} give figures per session for six reference workloads rather than per-token intensities.

\textbf{Requirements.} Table~\ref{tab:requirements} states what a preparer needs from a method and how each body of work meets it. The rows are the requirements this paper's components are built to satisfy, so they state what this method set out to do rather than a neutral benchmark, and a method built for a different purpose will read ``not stated'' on rows it never addressed. The first row describes each method rather than scoring it, by the data a preparer must hold before the method can be applied at all. Measurement research and the per-query estimates have no column, because they supply inputs and calibration points rather than an accounting method.

{\scriptsize\setlength{\tabcolsep}{2.5pt}
\begin{longtable}{>{\raggedright\arraybackslash}p{1.9cm} >{\raggedright\arraybackslash}p{1.7cm} >{\raggedright\arraybackslash}p{1.4cm} >{\raggedright\arraybackslash}p{1.6cm} >{\raggedright\arraybackslash}p{2.0cm} >{\raggedright\arraybackslash}p{1.7cm} >{\raggedright\arraybackslash}p{2.0cm} >{\raggedright\arraybackslash}p{2.3cm}}
\caption{Requirements for accounting purchased AI inference, and how existing practice and the three concurrent token-based methods meet them. Cells state what the cited material says; ``not stated'' means it does not address the requirement in the published material consulted. MPL-2.0 is the Mozilla Public License 2.0; CC BY, CC BY-SA and CC BY-NC are the Creative Commons Attribution, Attribution-ShareAlike and Attribution-NonCommercial licences; PUE is power usage effectiveness; US, United States.}\label{tab:requirements}\\
\toprule
Requirement & Spend-based practice & Developer tools~\citep{rince2025,lottick2019} & Provider reporting & \citet{bistline2026} & Greenpixie \citep{horswill2026} & Sustainable AI Group~\citep{jegham2026,gamazaychikov2026} & This paper \\
\midrule
\endfirsthead
\multicolumn{8}{l}{\textit{Table~\ref{tab:requirements} (continued)}}\\
\toprule
Requirement & Spend-based practice & Developer tools~\citep{rince2025,lottick2019} & Provider reporting & \citet{bistline2026} & Greenpixie \citep{horswill2026} & Sustainable AI Group~\citep{jegham2026,gamazaychikov2026} & This paper \\
\midrule
\endhead
\midrule\multicolumn{8}{r}{\textit{continues on the next page}}\\
\endfoot
\bottomrule
\endlastfoot
Data required from the preparer & Spend by supplier sector, in the currency and price year of the factor & Provider, model, output tokens and, where available, latency of each request, captured inside the application; Lottick et al.\ instrument the user's own running code & None beyond the cloud account; the provider uses its own usage data & Spend at the lowest tier; total tokens, or input and output tokens with the model family and serving region; provider per-token figures at the highest tier & Input and output tokens by model from billing lines, with cache status and batch tier; the cloud region if declared, otherwise global averages & Uncached input, cached input and output tokens by model, or prompts, messages, sessions or active users converted to tokens; the serving region if disclosed, otherwise a provider fleet factor or the US national default & Whatever the billing export holds: input, cached and output tokens by model (Tier 2a), messages or seats (Tier 2b), deflated spend (Tier 1) or a supplier figure (Tier 3); the region if known, otherwise the global row \\
Boundary & Whole-sector economic footprint; no physical boundary & Output-token accelerator energy, non-accelerator server energy and an embodied allocation & Differs by provider; AI services appear as service lines (Google) with no token or per-request field & Serving energy at PUE 1.10, embodied, training as a separate line & Serving energy, embodied, a training uplift & Accelerator, host, idle, facility, embodied hardware, construction; development inside the footprint; three exclusions bounded, networking reported negligible & Electricity, embodied, construction, final-run training; development a separate line; every exclusion bounded or stated \\
Token basis & No tokens & Output tokens only; prompt caching not modelled & No token data published & Separate input and output intensities; their 3-to-1 ratio is a decode-to-prefill electricity ratio taken from Splitwise & Input and output separated, fitted by regression; cached input distinguished & Separate input and output intensities; cached input at 0.10 to 0.50 of an uncached input token, from provider cache-read price ratios~\citep{gamazaychikov2026} & Effective tokens: input 0.30, cached 0.05 of an output token, divided by $1+W_{\mathrm{in}}R_{\mathrm{bench}}$ to remove the prefill already inside the measured figure (defined in Section~\ref{sec:3.3}) \\
Seat products without token counts & Covered by spend & Needs pipeline access & not stated & Start at the spend tier & not stated & Non-token units converted to tokens or energy by a documented method; no defaults published & Tier 2b: messages or seats to tokens with measured, derived and assumed defaults (S9.7) \\
Closed models & No model distinction & Parameter counts estimated per model & No model-level disclosure & Defaults by hyperscaler or unknown provider; model-specific values where available & Size buckets, Monte Carlo & Matched to the five nearest measured open models & Three classes by disclosed size or vendor tier; 187 released patterns \\
Regions & Producing country of the vendor & Country mixes from the ADEME database, the world mix by default and the zone set by the user & Provider's own facilities & Serving-region factor under the Scope 2 hierarchy; US national average as fallback & Cloud-region PUE and hourly grid intensity; global averages where undeclared & Serving-region factor where disclosed; provider fleet average or US national default otherwise & 210 countries and economies, 27 grid subregions, 143 cloud regions \\
Water & No & Estimated per request & Not reported per service & Outside the boundary & Included & Not addressed & Two terms, two hydropower conventions, scarcity weighting, 17 regions, 11 with a confirmed total \\
Uncertainty & None stated & Minimum-to-maximum range from the parameter-count interval of a closed model & not stated & not stated & Bayesian posterior and Monte Carlo over size buckets & Statistical and systematic tracks, compounded at $+$62\% & Five parts kept apart (Section~\ref{sec:3.8}): central value, Monte Carlo parameter interval ($-$2\% to $+$30\% on the footprint), structural range, class range, activity interquartile range; one compounded high case at $+$64\%, per-input sensitivities, data-quality scores \\
Update protocol & Release cycle of the input-output tables & Versioned library releases & not stated & Next version and refresh promised, defaults dated June 2026; no cadence or procedure & Continuous maintenance planned; no cadence or version label & Versioned factor register, public change record and event triggers; no cadence, register not yet published & 15 input families with source, cadence and trigger, and a released script that reports changes \\
Open per-token factors & Per euro, not per token & Library code MPL-2.0; methodology CC BY-SA & No & Defaults published in the paper & Not published; absolute per-token energy values not printed & Per-session figures only; per-token intensities not published; data CC BY-NC~\citep{saigdata2026} & 3 classes and 18 measured models across 240 grid rows and 143 cloud regions, CC BY, with a calculator \\
\end{longtable}}

\textbf{The gap.} Read across the families, the approaches differ less in their numbers than in what they ask of the preparer (the first row of Table~\ref{tab:requirements}), what they assume and where they draw the boundary. On assumptions, every family rests on at least one quantity it cannot observe. Spend-based factors assume that emissions follow price and that a pre-AI sector average describes AI services, the per-query estimates assume a serving operating point, a response length or both, and every method that covers closed models infers their size, from nearest open neighbours, from size buckets or, in this paper, from the vendor's pricing tier. On boundaries, the published figures run from accelerator energy alone, through the node and the facility, to figures with embodied hardware and some share of training, and one adds data-centre networking. The whole-sector economic boundary of the spend-based factor is not a physical boundary at all. Two figures with the same unit can therefore differ by their boundary before they differ by any measurement, which is why Section~\ref{sec:5.1} restates each external figure before comparing it. What none of the published methods provides, on the material consulted, is the combination: a procedure that starts from whichever activity data a billing export holds, with published defaults for messages and seats, applies a stated cradle-to-gate boundary with every exclusion bounded or named, and resolves any model identifier and any region to a factor under a maintenance protocol that states a cadence and a trigger for each input. This paper adds that combination. It does not add a new measurement. Its accelerator energy comes from the same open-weight benchmark as EcoLogits, and its closed-model classes are an assumption of the same kind as the others'.

\section{Methods}\label{sec:3}

This section is the procedure. It holds for any model and any region: the factor values appear as symbols, with Section S9 and Table~\ref{tab:inputs} for the numbers.

\subsection{Scope and Boundary}\label{sec:3.1}

This is a method for one category of an organisational greenhouse gas inventory, not a life cycle assessment. It supplies average-data emission factors and activity-data rules for purchased AI inference in Scope 3 Category 1 of an inventory prepared under the GHG Protocol Scope 3 Standard and ESRS E1~\citep{wri2011,eucom2026a}; under International Organization for Standardization (ISO) 14064-1:2018 the same purchases are indirect emissions from products used by the organisation, its Category 4~\citep{iso2018}. It does not apply ISO 14040 and 14044 or ISO 14067: it has no product system defined by a functional unit, no life cycle impact assessment and no indicator profile. ``Cradle-to-gate'' is used in the sense of the Scope 3 Standard, whose minimum boundary for Category 1 it is (Scope 3 Standard, Table 5.4), and the four terms of the factor are the upstream emissions of the purchased service that this boundary requires. The method applies to AI services purchased by entities inside the reporting company's organisational boundary, under whichever consolidation approach the company uses for the rest of its inventory, and does not change that boundary. The provider's emissions from serving those purchases are the supplier's Scope 1 and Scope 2 emissions and appear in the reporting company's inventory only in its Category 1.

\begin{figure}[htbp]
\centering
\includegraphics[width=0.70\textwidth]{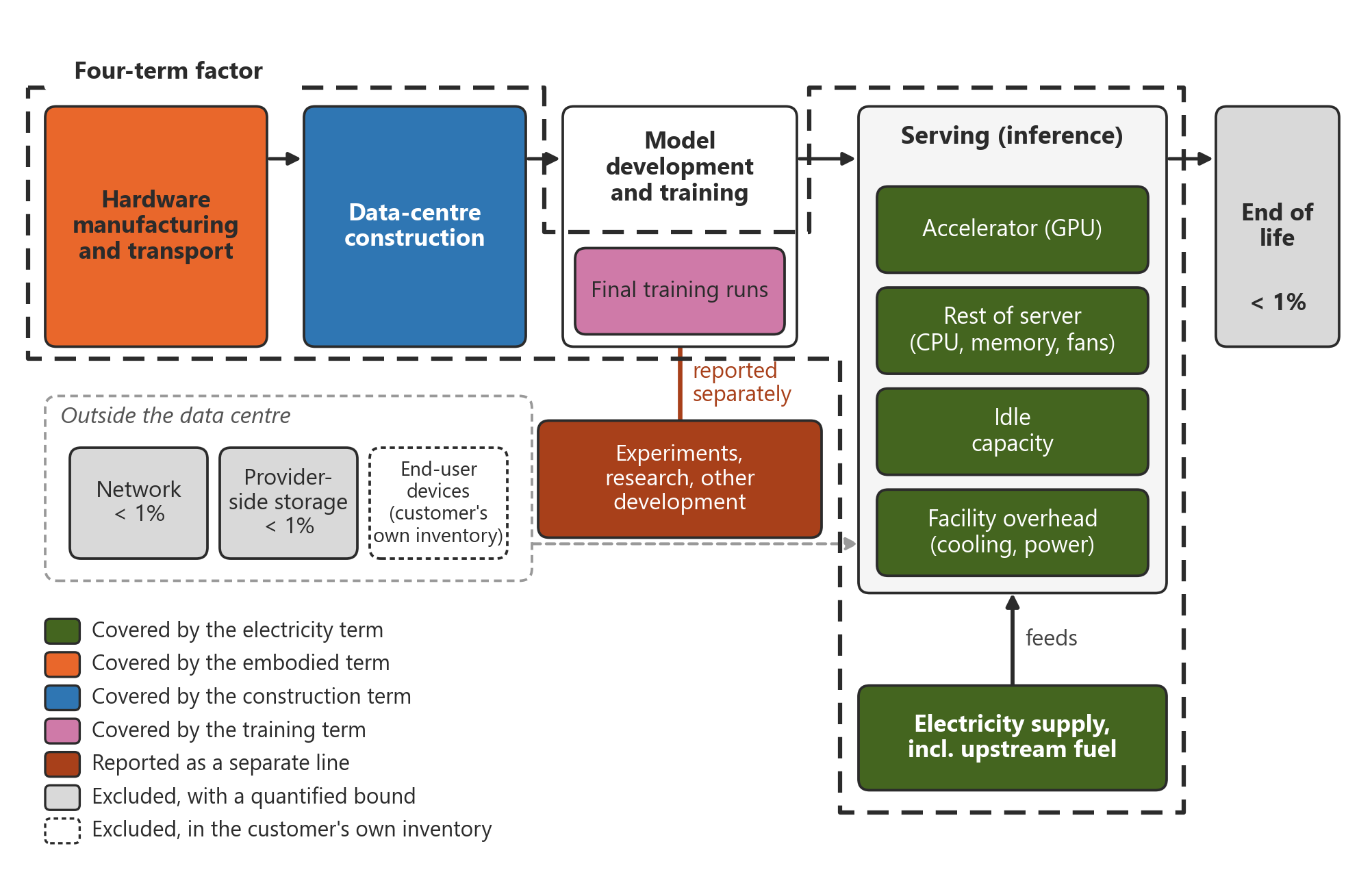}
\caption{Upstream value chain of a purchased artificial intelligence (AI) inference service, with the cradle-to-gate boundary of the four-term factor, the Scope 3 Standard's minimum boundary for Category 1, drawn on it as a dashed line. Each element is coloured by the factor term that covers it. Serving is broken into the accelerator, the rest of the server, idle provisioned capacity and facility overhead, fed by an electricity supply that includes the upstream fuel cycle. Model development sits inside the value chain but outside the factor.}
\label{fig:lifecycle}
\end{figure}

The upstream value chain of a purchased inference service runs from hardware manufacture to the delivered token, and Figure~\ref{fig:lifecycle} draws it with the boundary of the factor on it. The \emph{central factor} has four terms. \emph{Serving electricity} covers the accelerators, the rest of the server, machines provisioned but idle so that capacity stays available, the facility's cooling and power distribution, and the upstream fuel cycle of the electricity supplying all four. \emph{Embodied hardware} covers manufacturing and transport of the server, and \emph{data-centre construction} is a separate term on the server electricity. \emph{Training} covers the final training runs of the model amortised over the tokens it serves in its service life, the approach noted in Section~\ref{sec:2}. All model development, including the final training runs, is reported as a separate line, the development line, and never added to the factor (Section~\ref{sec:3.4}). Two further boundaries are named once and used throughout: the \emph{inference footprint} is electricity plus embodied plus construction, which describes serving efficiency and is the figure to use when comparing models or regions, and the \emph{development-inclusive total} replaces the training term with the development line. Table~\ref{tab:inclusion}, in S5, maps every element of that value chain to the term that carries it, or states how it is treated when no term carries it.

\textbf{Inclusion threshold.} The inclusion test is stated rather than left implicit, and it is a choice of this method, not a requirement of any standard. An element is inside the factor where it is attributable to serving a token and is worth at least 1\% of the inference footprint of the mid class at the global-average grid. Construction passes that test at 1.3\% and is inside, and embodied hardware is 2.8\%. The threshold was set after the terms were sized, at a level that separates them. It applies per element, on the location basis and at the global-average grid. No bounded excluded element exceeds 0.43\%, so the result is insensitive to any cut-off between that value and the construction share. The test is applied at central values. At its low value of 5.3 g construction is about 0.9\% of the mid-class footprint, below the threshold, so its inclusion rests on the central constant.

Seven elements sit outside the factor and four more are out of scope, each named and treated in Table~\ref{tab:inclusion}. Three carry a bound, quoted by their source as shares of operational serving emissions, which is within 4\% of the mid-class inference footprint: provider-side storage at about 0.11\%, external network transmission at 0.17 to 0.43\% and hardware end-of-life at about 0.12 to 0.17\% on a location basis and 0.38\% market-based, together 0.71\% location-based and 0.92\% market-based~\citep{gamazaychikov2026}. Data-centre internal networking is reported as negligible for a text prompt by its source and is not independently bounded here. Three are stated conventions. Grid transmission and distribution losses, about 4\% on the United States rows (Table~\ref{tab:sens}), are excluded for consistency with the Scope 2 Guidance, not as immaterial, because location-based grid factors are per kWh generated and the provider reports those losses under its own Category 3~\citep{wri2011,wri2015}, so the factor is low by about that amount. Behind-the-meter generation is carried as a named sensitivity (Section~\ref{sec:5.5}). No recycling or cascaded-use credit is taken, although \citet{schneider2025}, as reported by \citet{gamazaychikov2026}, assessed one of up to 4\% of embodied emissions. Out of scope, named but not sized, are end-user devices, which are already in the customer's own inventory, provider staff, offices and research overhead outside compute, customer fine-tuning jobs, which are a training workload billed by job and are disclosed as excluded with their spend, and embeddings and image, audio and video work, for which no measurement exists under the protocol used here (Table~\ref{tab:rules} in S5).

The terms inside this boundary follow from the standards rather than from convenience, while the inclusion threshold and the exclusion from the factor of development other than the final runs are choices of this method (S1.3). The minimum boundary of Category 1 is cradle-to-gate, allocating the provider's capital goods over a service life is not the full-year rule for a preparer's own capital goods, and there is no double counting with the preparer's own Category 2 or Category 8~\citep{wri2011}, as Section S12.1 sets out. Before the token-based figure is added, the same supplier spend is removed from the spend-based Category 1 line and any AI service line in a cloud provider's carbon report is excluded, so that the purchase is counted once (S5). The factor is not reconciled to any provider's inventory and should not be presented as the provider's own allocation.

\subsection{Activity Data: The Four Tiers}\label{sec:3.2}

\begin{figure}[tbp]
\centering
\includegraphics[width=0.62\textwidth]{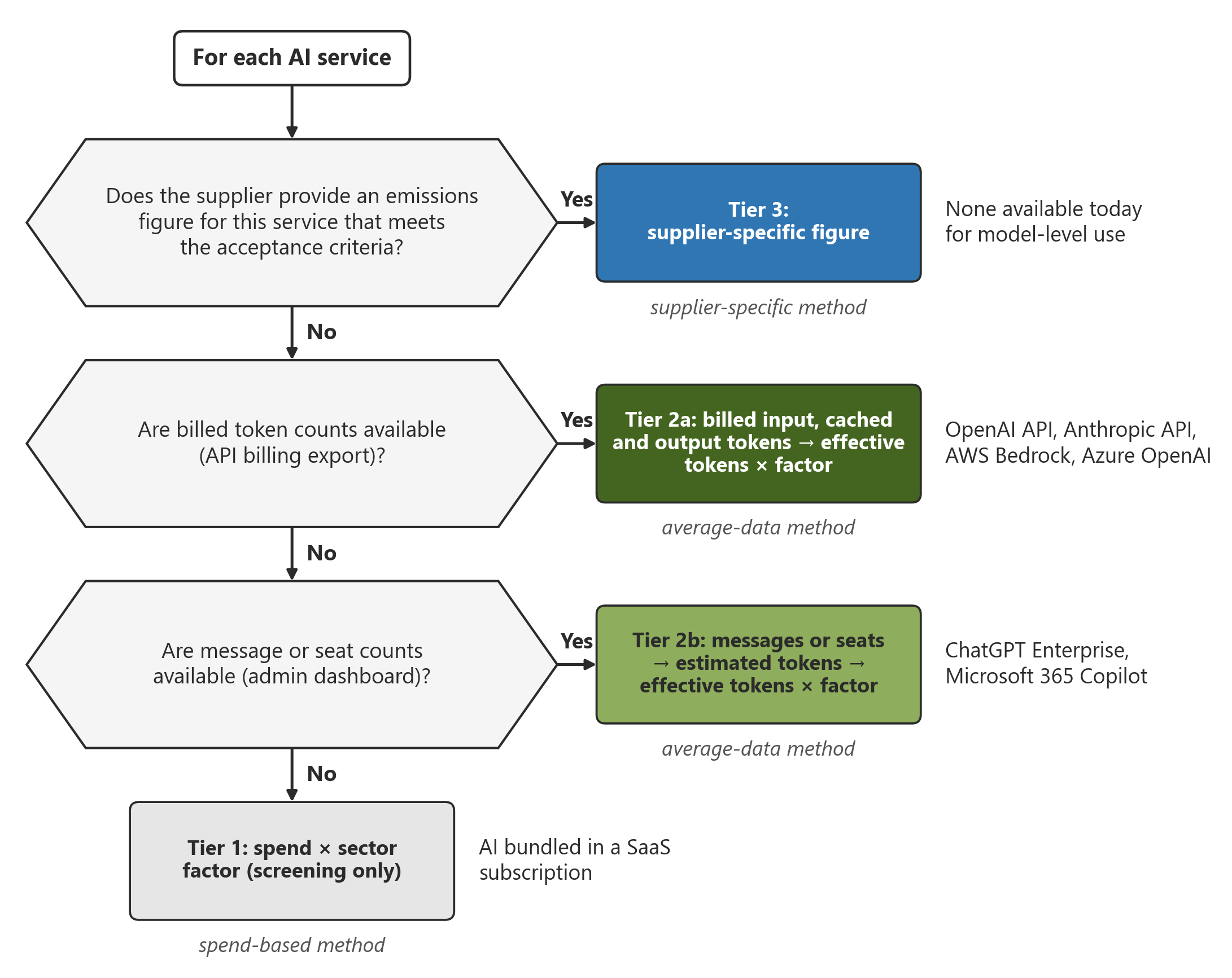}
\caption{Assigning a tier to one artificial intelligence (AI) service. Work down from the top; each outcome is labelled with the Greenhouse Gas (GHG) Protocol calculation method it corresponds to and with examples of products that reach it. Both Tier 2 branches convert billed or estimated tokens to \emph{effective} tokens before applying the factor (Section~\ref{sec:3.3}). Reserved capacity does not appear here, because it is charged on accelerator-hours (Table~\ref{tab:overview}).}
\label{fig:flowchart}
\end{figure}

The GHG Protocol does not set a strict hierarchy for purchased goods and services. Its Technical Guidance ranks the four Category 1 methods by how specific each is to the individual supplier but lets data availability, data quality and cost drive the choice~\citep{wri2013}. Applied to AI services, this gives four tiers ordered by the quality of data an organisation can realistically obtain, numbered as the Intergovernmental Panel on Climate Change numbers tiers, with Tier 1 on default factors and Tier 3 the most specific. Tier 3 is the supplier-specific method, Tiers 2a (exact tokens) and 2b (estimated tokens) the average-data method, a split introduced here, and Tier 1 the spend-based method~\citep{wri2013}. Two variants sit beside Tier 2a, the hybrid method, where a supplier gives an energy quantity rather than a CO\textsubscript{2}e figure (S5), and reserved capacity, which is charged on accelerator-hours (Section~\ref{sec:3.4}). A tier is assigned to each AI service individually, not to the company as a whole (Figure~\ref{fig:flowchart} and S12.5).

Table~\ref{tab:overview} is the quick-start summary, with the rules for an unknown serving region and an unmapped model identifier.

\begin{table}[htbp]
\centering
\setlength{\tabcolsep}{3pt}
\scriptsize
\begin{tabular}{>{\raggedright\arraybackslash}p{1.5cm} >{\raggedright\arraybackslash}p{2.2cm} >{\raggedright\arraybackslash}p{3.5cm} >{\raggedright\arraybackslash}p{2.4cm} >{\raggedright\arraybackslash}p{1.5cm} >{\raggedright\arraybackslash}p{1.1cm} >{\raggedright\arraybackslash}p{2.1cm}}
\toprule
Tier or case & Activity data needed & Calculation & Default where a value is missing & GHG Protocol method, Category 1~\citep{wri2013} & Data quality, activity / factor (author scale, 1--5; Section~\ref{sec:3.8}) & Where to look \\
\midrule
3, provider carbon report & The supplier's own cradle-to-gate CO\textsubscript{2}e for the customer's usage, with its assurance status & Use the supplier's figure as reported; do not re-derive it & None. Where the acceptance criteria are not met, drop to Tier 2 & Supplier-specific & 1--2 / 1--2 & Acceptance criteria, Table~\ref{tab:rules} \\
2a, exact tokens & Uncached input, cached input and output tokens per model per period & $T_{\mathrm{eff}}=(T_{\mathrm{out}}+W_{\mathrm{in}}T_{\mathrm{in,uncached}}+W_{\mathrm{cache}}T_{\mathrm{in,cached}})/D$, then CO\textsubscript{2}e $= T_{\mathrm{eff}}\times$ inference footprint $+\;T_{\mathrm{billed}}\times$ training term & Default input share and no cache credit, so $T_{\mathrm{eff}}=m\,T_{\mathrm{billed}}$ (Section~\ref{sec:3.3}) & Average-data, physical units (tokens) & 2 / 3 & Table~\ref{tab:carbon} (footprint and the convenience factor, defined in Section~\ref{sec:3.4}), Table~\ref{tab:regionindep} (training) \\
2a, hybrid & The supplier's kWh per million tokens for the model served, plus the preparer's own grid row & CO\textsubscript{2}e $=$ supplier kWh $\times$ grid intensity, then add the embodied, construction and training terms & None; this is the GHG Protocol hybrid method and needs the supplier's energy quantity & Hybrid & 2 / 2--3 & Grid column of Table~\ref{tab:carbon}, Table~\ref{tab:regionindep}, Table~\ref{tab:rules} \\
2b, estimated tokens & Message count, or active seat count, per product & Tokens $=$ messages $\times$ the input and output defaults, kept \emph{separate}, the output alone multiplied by the reasoning uplift $1 + s_r(k-1)$ (2.195 by default; Section~\ref{sec:3.2}), then the Tier 2a equation on that split. Do not also apply $m$ & 30 messages per 30-day-active seat month (a seat used at least once in the past 30 days), or 24 per licensed seat month, used or not, where only licences are known; class from the vendor's documented default model & Average-data, proxy activity data & 4 / 3 & Table~\ref{tab:tier2b}, Table~\ref{tab:classmap} \\
2a, reserved capacity & Reserved or provisioned accelerator-hours & CO\textsubscript{2}e $=$ hours $\times$ kg per reserved accelerator-hour. No idle uplift and no utilisation divisor apply, because the hours are charged whether used or not; the training term follows the tokens actually served & None; the hours are on the contract & Average-data, accelerator-hours & 1 / 3 & Section~\ref{sec:3.4}, Table~\ref{tab:reserved} \\
1, spend-based & Annual spend, excluding value-added tax (VAT) & CO\textsubscript{2}e $=$ spend $\times$ consumer price index (CPI) ratio (CPI\textsubscript{factor year}/CPI\textsubscript{spend year}) $\times$ the EEIO factor of the vendor's producing country & The example factor and deflator of Section~\ref{sec:4.9} & Spend-based & 5 / 5 & Sections~\ref{sec:3.7} and \ref{sec:4.9}; worked line in Section~\ref{sec:6.1} \\
\midrule
Rule: serving region unknown & --- & Use the global-average row. For a customer-pinned region use its own row; for an EU data zone or a contractual EU-only processing commitment use the EU-27 row & --- & --- & --- & Table~\ref{tab:carbon} \\
Rule: model identifier unmapped & --- & Assign the highest-intensity class, the frontier class, flag the service in the inventory, and disclose the share of activity that fell back & --- & --- & --- & Section~\ref{sec:3.5}, Table~\ref{tab:classmap} \\
\bottomrule
\end{tabular}
\caption{Quick start: the four tiers and the two rules most preparers need. The symbols are defined in Sections~\ref{sec:3.3} and \ref{sec:3.4}. The token weights used throughout are $W_{\mathrm{in}} = 0.30$ and $W_{\mathrm{cache}} = 0.05$, the divisor is $D = 1.339$ and at the \emph{default mix}, an input share $s_{\mathrm{in}} = 0.85$ with no cache credit, the effective-to-billed multiplier is $m = 0.302$ (Section S9.6). Factors are kg CO\textsubscript{2}e, per million effective tokens for the inference footprint and per million billed tokens for the training term. Data-quality scores are on the author scale of Section~\ref{sec:3.8}. GHG is greenhouse gas; EEIO, environmentally extended input-output.}
\label{tab:overview}
\end{table}

\textbf{Which products sit where.} A supplier figure is accepted at Tier 3 only where it meets the criteria of Table~\ref{tab:rules}, in short a location-based service- or model-level CO\textsubscript{2}e figure for the customer's own usage, on a boundary covering at least the four terms of Section~\ref{sec:3.4}, with the allocation method, a named methodology standard and the assurance status stated (where assurance is obtained, at least limited assurance in the sense of ISO 14064-3:2019~\citep{iso2019}), an unassured figure scoring 2 rather than 1, and for a reporting period aligned with the inventory. Otherwise the service drops to Tier 2. No major AI provider yet offers such a figure, so Tier 3 is at present empty, and the energy documentation required by the EU AI Act does not fill it (Section~\ref{sec:7.3}). Tier 2a covers API access with token billing and any product whose export reports tokens by model, Tier 2b covers seat-licensed products that report activity but no tokens, and Tier 1 covers subscriptions in which the vendor, not the customer, holds the AI API relationship, so token-level data never surfaces in the customer's billing portal. The rule is to \emph{read the product's own message count first, and use the per-seat default only where it cannot be retrieved}. Token counts at every tier are billed tokens, so the input column includes the system prompt, any retrieval or tool context and the conversation history re-sent with every turn (Section S10.1). Which admin export each product offers today is in S5 and S12.5.

Four further defaults set the class split, the class mapping, the band and the reasoning treatment at Tier 2b, and their values are in Table~\ref{tab:inputs} and S1.4. Weight each class by the days the vendor's documented default model was in force during the reporting year, or read the split off the product's own export where it reports messages by model family~\citep{openai2026a}. Map that model, or its absence, by the rule of Section~\ref{sec:3.5}, Microsoft 365 Copilot being the usual case of absence (S12.5). Add an asymmetric band on the estimated tokens, taken from the interquartile ranges of the measured workload distributions (Section S9.7).

\textbf{Reasoning modes.} This method models reasoning as a change in the tokens per interaction, not in the energy per token or the class, so one reasoning multiplier $k$ serves all three classes. The rise of energy per token with context length is not modelled (Section~\ref{sec:7.2}). Tier 2a billed output counts already include the reasoning tokens and must not be multiplied. At Tier 2b a product's estimated output tokens, but not its input tokens, are multiplied by $1 + s_r(k - 1)$, because hidden reasoning is text the model writes rather than context it reads. Here $k$ is 25 centrally within a range of 10 to 50, and $s_r$ is the share of conversations that use a reasoning mode, applied per message as an assumption, read from the product where it reports it and otherwise the measured 4.98\% of Section S9.7. That share was measured on Claude.ai in February and March 2025, before reasoning models became the default route on major chat products, so it is likely low for later years, and the worked example carries the later user shares of \citet{altman2025} as two sensitivities (Section~\ref{sec:6.1} and Table~\ref{tab:casestudy}).

\subsection{Token Basis}\label{sec:3.3}

The energy anchor is measured per \emph{output} token on one benchmark workload, while a customer is billed for input and output tokens on a different workload, and the two are reconciled before any factor is applied. Three quantities do the work: $W_{\mathrm{in}}$, the energy of an uncached input token relative to an output token; $W_{\mathrm{cache}}$, the same for a cached input token; and $R_{\mathrm{bench}}$, the input-to-output token ratio of the reference mix at which the energy figures are stated. Because the published benchmark figure is the whole accelerator energy of a request divided by that request's output tokens, the energy of processing the benchmark's own input is already inside it. Each benchmark run has its own input-to-output ratio and its own share of input served from the engine's prefix cache, both read from the benchmark's run files, and each measurement is first restated from that mix to the reference mix (Section S9.6). Writing $e_d$ for the decode energy per output token, the restated figure is $e_d(1 + W_{\mathrm{in}} R_{\mathrm{bench}})$, so applying $W_{\mathrm{in}}$ on top of it without rescaling would charge that prefill twice. The correction divides by
\[
D = 1 + W_{\mathrm{in}} R_{\mathrm{bench}} ,
\qquad
T_{\mathrm{eff}} = \frac{T_{\mathrm{out}} + W_{\mathrm{in}} T_{\mathrm{in,uncached}} + W_{\mathrm{cache}} T_{\mathrm{in,cached}}}{D} .
\]
$T_{\mathrm{eff}}$ is the quantity called \emph{effective tokens} throughout: the number of output tokens, at the reference mix, that would consume the same accelerator energy as the billed mix. At that mix $T_{\mathrm{eff}} = T_{\mathrm{out}}$ by construction, which is why a measurement made per output token can be applied to effective tokens unchanged. The reference mix, and with it $D$, is a convention: $D$ cancels in every quantity per billed token, and the prefix caching measured in each benchmark run enters through that run's own divisor (Section S9.6), so no further cache correction is applied to the class energy. The adopted $W_{\mathrm{in}}$ is the reading of the prefill--decode measurement that matches the batch sizes at which the energy is measured, not a value below it.

Emissions are $T_{\mathrm{eff}}$ times the electricity, embodied and construction terms plus $T_{\mathrm{billed}}$ times the training term, where $T_{\mathrm{billed}} = T_{\mathrm{in}} + T_{\mathrm{out}}$ including cached tokens. Training is amortised over the tokens a model sells, which are the billed ones, and water and the development line both use $T_{\mathrm{eff}}$. The embodied term is allocated by server \emph{time}, not by energy, so charging it on effective tokens slightly overstates it, by 0.6\% to 0.8\% at the default mix and by at most 1.5\% on the most input-heavy workload tabulated here, and one weight set is kept for all three compute-proportional terms (S1.3).

Where only the billed total is known, multiply it by $m = \left((1 - s_{\mathrm{in}}) + W_{\mathrm{in}}\,s_{\mathrm{in}}\right)/D$, where $s_{\mathrm{in}}$ is the assumed input share, and claim no cache credit. Claiming none is the conservative choice. It stops an organisation claiming, on activity it cannot evidence, a reduction of up to about four times, the effect of full cache credit on the tabulated agentic request (Table~\ref{tab:workload}). Where the billing export reports the split, use it, and at Tier 2b apply the weights to the split defaults without $m$ (Table~\ref{tab:overview}). The weights lower the result for every workload, and by most for the input-heavy ones.

\subsection{Factor Equation}\label{sec:3.4}

Facility electricity per 1,000 tokens is accelerator energy times three uplifts,
\[
E_{\mathrm{fac}} = E_{\mathrm{GPU}} \times H \times I \times \mathrm{PUE},
\]
numerically equal to kWh per million effective tokens, and the server value $E_{\mathrm{srv}} = E_{\mathrm{GPU}} \times H \times I$ is the basis the construction term uses. The four terms are
\[
\mathrm{EF_{el}} = g\,E_{\mathrm{fac}}, \quad
\mathrm{EF_{emb}} = \frac{10^{6}\,(M_{\mathrm{mfg}} + M_{\mathrm{tr}})}{L \times 8{,}760 \times 3{,}600 \times U \times T_8}, \quad
\mathrm{EF_{con}} = k_c\,E_{\mathrm{srv}}, \quad
\mathrm{EF_{tr}} = \frac{10^{6}\,E_{\mathrm{train}}}{(S/12) \times 365 \times V},
\]
where $g$ is the grid intensity of the serving region on a life-cycle basis (combustion plus upstream fuel cycle), $M_{\mathrm{mfg}}$ and $M_{\mathrm{tr}}$ the manufacturing and transport footprints of an eight-accelerator server, $L$ its service life, $U$ the utilisation of the serving fleet, $T_8$ the measured output throughput of that server at the operating point (the serving configuration at which the class energy is taken, Section~\ref{sec:3.5}), $k_c$ the construction emissions per kWh of server electricity, $E_{\mathrm{train}}$ the disclosed emissions of the final training runs, $S$ the model's service life in months and $V$ the tokens it serves per day. All four terms are in kg CO\textsubscript{2}e per million tokens when $g$ and $k_c$ are per kWh, energies per million tokens, footprints in kg, $T_8$ in output tokens per second per eight accelerators and $V$ in tokens per day. $I$ and $U$ describe one fleet and are tied by $I = 1 + p\,(1-U)/U$, where $p$ is the power an idle accelerator draws as a fraction of its serving power, so the pair must move together (Section S9.2).

The first three terms are per million \emph{effective} tokens and their sum is the \emph{inference footprint}; the fourth is per million \emph{billed} tokens. They are never summed into one unlabelled total, because they have different denominators. A reported quantity of tokens is converted by
\[
\mathrm{CO_2e} = T_{\mathrm{eff}} \left( \mathrm{EF_{el}} + \mathrm{EF_{emb}} + \mathrm{EF_{con}} \right) + T_{\mathrm{billed}}\, \mathrm{EF_{tr}} ,
\]
with both token counts in millions. For the common case in which only a billed total is known, the two collapse into one \emph{convenience factor} per million billed tokens at the default mix, $m \times \mathrm{inference\ footprint} + \mathrm{EF_{tr}}$, which Table~\ref{tab:carbon} prints so that a preparer never has to add quantities on two denominators by hand. \emph{Do not multiply a billed token count by an inference footprint}: at the default mix that overstates the factor, training included, by about a factor of 3.1, and on an agentic workload by about a factor of 12; on the inference footprint alone, which is what Table~\ref{tab:workload} tabulates, the ratios are 2.9 and 17.

\textbf{The development line.} All model development, including the final training runs, is reported as $\mathrm{EF_{dev}} = d \times E_{\mathrm{GPU}} \times H \times \mathrm{PUE} \times g$, where $d$ is the energy of that development as a multiple of serving energy, charged on serving energy without the idle uplift, because development workloads are scheduled for throughput rather than to a latency target and no capacity is held idle for them. The \emph{development-inclusive total} is the inference footprint plus that line, with the training term dropped because the line already includes the final training runs. The line is never added to the central factor, because $d$ is a sector spend ratio spanning a factor of nearly four rather than a disclosure about the model used (Section S9.5). The training term and the line also differ in attribution, in the direction of their bias and in their dependence on the serving volume $V$ (S1.3). The training term is biased low in its numerator, final runs being 9.6\% to 22.6\% of research and development compute~\citep{denain2026}, and the development-inclusive total does not use $V$, so it does not inherit the bias of that denominator. It is therefore printed beside the factor by class and region, as a different boundary rather than as the top of an interval around the factor (Table~\ref{tab:structrange}). Wherever the line is reported, two properties are stated with it. It carries no embodied or construction term, so it is about 4\% short of the boundary the inference footprint uses, and its class split, which scales with serving energy, understates the frontier class's development and overstates the small class's.

\textbf{Reserved capacity.} Where accelerator-hours are contracted rather than tokens billed, the activity quantity is the reserved hour, charged at the serving operating point whether used or not, with no idle uplift and no utilisation divisor in the embodied term, because the hours themselves are the provisioned capacity. Electricity per accelerator-hour is the measured average accelerator power at the class operating point times $H$, the PUE and $g$, the embodied term is $(M_{\mathrm{mfg}} + M_{\mathrm{tr}})/(8 L \times 8{,}760)$, construction is $k_c$ times the server electricity of that hour, and the training term follows the tokens actually served. Reserved hours cover the reserved portion and tokens the on-demand overflow, never both for the same hours. A contract containing a lease of identified hardware is reported in the lessee's own operations (ESRS 1, paragraphs 71 and 73; \citealp{eucom2026a}). The full statement, the values and the consistency check are in S1.3.

\begin{figure}[htbp]
\centering
\includegraphics[width=0.95\textwidth]{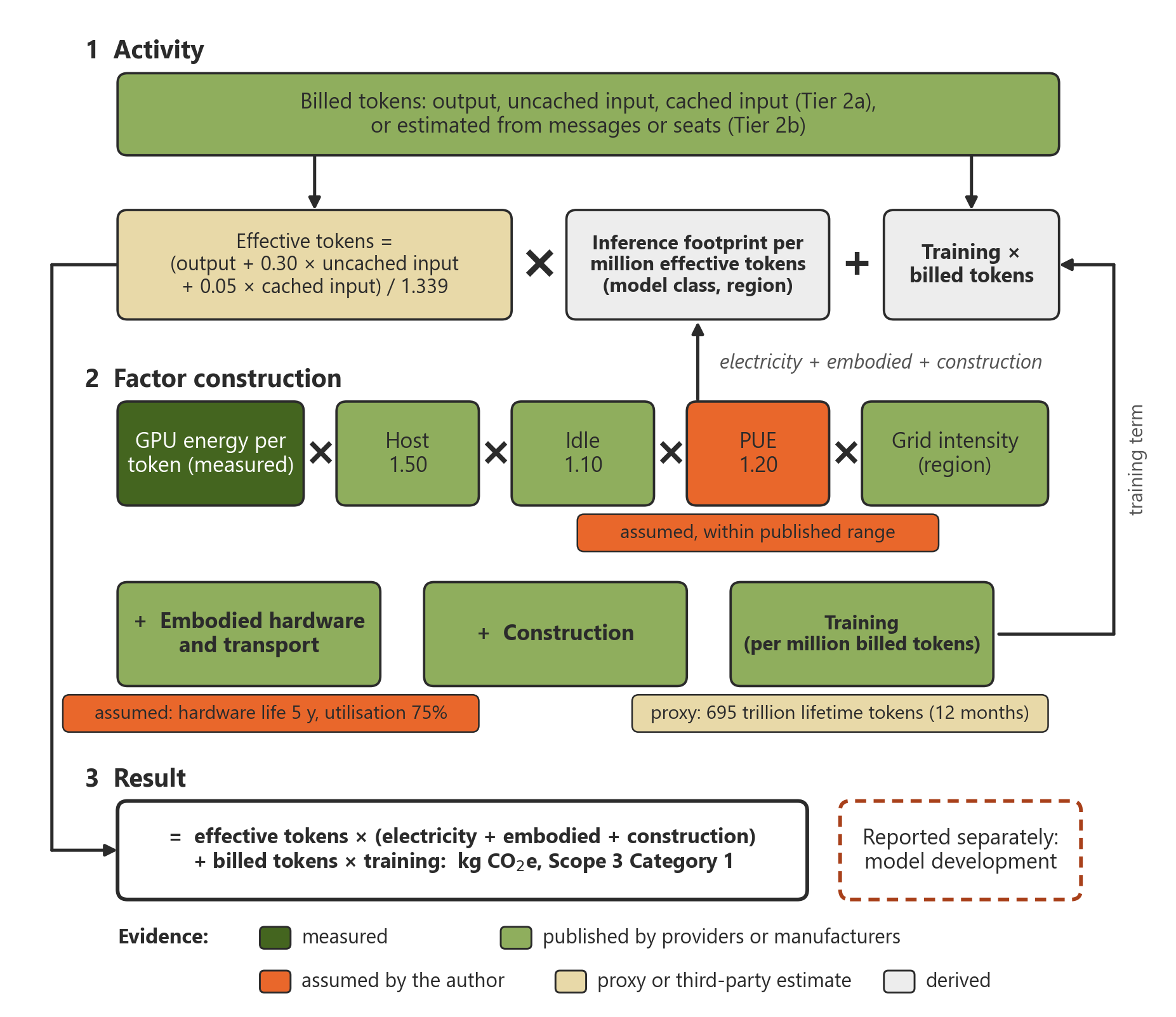}
\caption{The calculation chain, from activity data to a Scope~3 Category~1 figure in kg CO\textsubscript{2}e. Tokens enter from the tier logic of Section~\ref{sec:3.2}. The compute-proportional terms are charged on effective tokens ($D = 1.339$) and the training term on billed tokens, so the two branches rejoin only at the result (Sections~\ref{sec:3.3} and \ref{sec:3.4}). Shading marks the evidence class of each input: measured, published by a provider or manufacturer, assumed by the author, a proxy or third-party estimate, or derived. All model development, including the final training runs, is a separate line, never inside the factor.}
\label{fig:chain}
\end{figure}

Figure~\ref{fig:chain} draws the whole chain from activity data to the reported figure, with the evidence class of each input.

\subsection{Resolution: From a Model Identifier and a Region to a Factor}\label{sec:3.5}

\textbf{Model.} A model with an energy measurement under the benchmark protocol takes its own factor, with the class training term (Section S9.5). Where no measurement exists, which is every closed commercial model, the model is assigned to a size class and the class value is used. \emph{The class value is the arithmetic mean of the measured models of that class at a common operating point}, defined in Section S9.1 as the lowest-energy configuration of the measured sweep that still streams at interactive speed.

\textbf{Class assignment.} Class follows the disclosed total parameter count where the provider discloses it, and the vendor's pricing tier where it does not. For dense models the small class is up to 15B parameters, the mid class above 15B up to and including 70B, and the frontier class above 70B; for mixture-of-experts models, counted by total rather than active parameters because all experts must be resident in accelerator memory, the small class is below 50B, the mid class from 50B to below 200B, and the frontier class at 200B and above. Without a disclosed size, the efficiency tier (\emph{mini}, \emph{nano}, \emph{flash}, \emph{lite}, \emph{haiku} and similar) is the small class, the standard tier (\emph{sonnet}, \emph{pro}, \emph{plus}, an unsuffixed mid-tier model) the mid class, and the frontier tier (\emph{opus}, \emph{ultra}, \emph{max}, the top reasoning or flagship model of a generation) the frontier class. Three classes are used for a model without a disclosed size because a finer bucket would claim a precision the evidence does not support; a measured model takes its own factor. The tier is read from the vendor's price list for the generation (Table~\ref{tab:classmap} in S6, with worked examples in S12.4), and \emph{a disclosed total parameter count always overrides the tier word}.

\textbf{Matching and fallback.} Identifiers are normalised and matched against the 187 identifier patterns of the deposited \texttt{model\_\allowbreak class\_\allowbreak mapping.csv}, the longest match winning (S6). A seat product that exposes no identifier takes the class of the vendor's current default general-purpose model as the vendor documents it at the reporting date. Where the vendor documents none, or an identifier matches no pattern, take the highest-intensity class, flag the service and disclose the share of activity that fell back.

\textbf{Region.} The electricity term needs one number, the grid intensity $g$, location-based and on a life-cycle basis, which is not the combustion-only factor a provider uses for its own Scope 2. The region rule is in Table~\ref{tab:overview}. Global routing and a cross-region profile count as an unknown region, and a marketing statement about data residency for storage is not an EU-only processing commitment. A region resolves at three levels: a country or economy, a United States grid subregion, or a named public cloud region mapped to one of those, which outside the United States is its national average (S12.4).

\textbf{What the resolution does not do.} It does not produce a model-specific factor for a closed model (Section~\ref{sec:7.2}), and it does not distinguish self-hosting from a provider API; self-hosted inference is in any case the preparer's own Scope 2 and Category 2 rather than Category 1, and only the energy chain transfers (S12.4).

\textbf{Water.} The same activity data and energy chain also give a water-use inventory, with no change to the carbon factor and no water term ever added to a carbon figure. Two terms are reported per 1,000 effective tokens, on-site cooling water, from a published WUE applied to the energy basis that WUE is stated on (facility or IT energy), and off-site water consumed in generating the electricity, on facility energy. The preparer declares the basis, the hydropower convention and the scarcity weighting, the results are never combined, and a region without a value is reported as not available. The method is in Section S12.2 and its inputs in Section S9.8.

\subsection{Spend-Based Fallback}\label{sec:3.7}

Where no usage data can be obtained, apply the EEIO factor for ``Computer and related services'' of the vendor's producing country~\citep{stadler2018} to spend excluding VAT, deflated from the spend year to the factor's price basis with the harmonised index of consumer prices (HICP) of the factor's own economy, or the euro-area HICP where that is unavailable. Where no factor for the producing country can be retrieved, use a retrievable country factor, disclose the substitution and give a second country's factor beside it as a sensitivity; the example uses Austria for a United States vendor, with Germany beside it. \emph{The deflated figure is the method's own result}; an undeflated figure is reported only for comparison, and every ratio to a physical estimate in this paper is stated on the deflated basis first, with the undeflated figure beside it. The example factor, the deflator and their sources are at the end of Section~\ref{sec:4.9}, and the worked line is in Section~\ref{sec:6.1}. Two limitations govern its use. The factor is a boundary-inclusive screening figure, several times the central physical estimate, and never an upper bound (Section~\ref{sec:7.1}). And it covers the entire subscription, not only its AI component: unless the AI share can be evidenced, for example by a separately priced AI add-on, leave the subscription in the spend-based calculation as a subscription screening line and disclose that its AI share is not separately estimable. Do not relabel the whole subscription as AI.

\subsection{Uncertainty and Data Quality}\label{sec:3.8}

Three different things are commonly merged under one error bar, and the method keeps them apart. The first is the \emph{class range}. Where a model is assigned to a class rather than measured, the factor carries the spread of the measured models of that class at the same operating point, the lowest and the highest member relative to the class mean, with the number of members. It is a data-driven spread, not a confidence interval, and \emph{it is removed entirely where the model has its own factor}. The second is the \emph{activity band}. At Tier 2b the tokens are estimated from messages or seats, and the band is the interquartile range of the measured, right-skewed per-message distributions, a per-message spread carried as a scenario on the activity quantity, not an interval on an annual total and not a band on the factor. At Tier 2a there is none, because the tokens are counted. The third is \emph{systematic uncertainty of the factor}, handled by scenarios rather than by a band. One \emph{central} value is reported per formula. One \emph{conservative high case} recomputes the electricity, construction and embodied terms with the upper value of the host uplift, the low-utilisation idle pair and the high PUE together, and holds training central, so it moves the weakest-evidenced link, PUE, as well as the two best-evidenced ones. It is never inside the formulas but is reported beside the central value wherever a result is given in the body or a table, and Section~\ref{sec:5.5} shows what each of its three moves contributes. Beyond it, each input is varied between its low and high value one at a time, and a hardware scenario on a newer accelerator generation is a sensitivity, not a reporting bound. No low case is reported.

\textbf{Two intervals, printed side by side and never combined.} Every central value carries two further statements. The \emph{parameter interval} is the 5th to 95th percentile of a Monte Carlo propagation of the input bands (Section~\ref{sec:5.5}). The \emph{structural range} has two typed axes. On the \emph{serving-regime} axis the energy per token is taken at the most efficient configuration that meets interactive latency (Section S9.1), and the mean over all configurations meeting the same limit, called \emph{the mean rule} from here on, is printed as the upper end, at 2.0 to 2.8 times the central factor per million billed tokens across the classes and the three regions of Table~\ref{tab:structrange} (2.8 times for the mid class at the global average). It is a choice of serving regime rather than an input with a band, so it enters neither the high case nor the parameter interval. The \emph{boundary} axis is the development-inclusive total of Section~\ref{sec:3.4}, 1.3 to 3.0 times the central factor over $d = 0.67$ to 2.5 across the classes and regions of Table~\ref{tab:structrange} (1.5 to 2.9 times for the mid class at the global average), a different quantity printed beside the factor and never as the top of an interval around it.

In the terms of the Scope 3 Standard~\citep{wri2011}, the input bands propagated in Section~\ref{sec:5.5} and the Tier 2b activity band are \emph{parameter} uncertainty; the operating-point rule, the training allocation and the token-weight convention are \emph{scenario} uncertainty, arising from methodological choices; and the class value standing in for a closed model and the use of benchmark energy as a fleet average are \emph{model} uncertainty, which is reported as the class range and the adjacent-class values and is not quantified probabilistically. A preparer therefore reports five parts separately and never combines them into one interval: the central value, the parameter interval propagated from the input bands, the structural range on its two axes, the class range, and at Tier 2b the activity band.

\textbf{Data quality.} Each tier carries two scores, one for activity data and one for the emission factor, each a composite of the five data-quality indicators of the Scope 3 Standard: technological, temporal and geographical representativeness, completeness and reliability~\citep{wri2011}. Scores 1 to 4 are the Standard's example levels very good, good, fair and poor. Score 5, below poor, is the author's and is used only for spend-based data, whose technological and temporal representativeness is that of a whole economic sector in the factor's base year. The Standard prescribes no aggregation, so the combining rule is the author's and is computed, not judged (Table~\ref{tab:dq}). For the Scope 3 Standard's metric, the percentage of emissions calculated using data obtained from suppliers~\citep{wri2011}, which the current ESRS E1 also asks for (application requirement (AR)~46(g); \citealp{eucom2023}), Tier 2a and 2b emissions count in full under this method's convention, because their activity data come from the supplier's own export, reconcilable to an invoice. In Tier 2b only the message count is supplier data, the tokens per message and the reasoning share being defaults, so a preparer also states the share with Tier 2a alone, about 63\% in the worked example (5.4 of 8.5 kg, Table~\ref{tab:casestudy}). None of Tiers 1, 2a or 2b uses supplier-specific emission factors, which only Tier 3 provides, so the share of emissions calculated with supplier-specific emission factors is 0\%; a preparer states it beside the activity-data share. In the Tier 2a hybrid case the supplier's energy quantity is also supplier data and only the grid factor is secondary.

The reporting rules make the figure traceable by an auditor, and they are a proposal of this method that goes beyond what ESRS E1 or the GHG Protocol prescribe. For each AI service a preparer discloses the tier, the boundary with its exclusions, the dataset version and grid year, the class or own-model factor with the share of activity that fell back, the region rule, the token weights and cache treatment, the five uncertainty parts above with each structural end labelled by its axis, the data-quality scores and both supplier-data shares, and its reliance on proxies, as ESRS 2 asks of value-chain metrics~\citep{eucom2026a}. The figure is reported gross, one dataset version is pinned per reporting year, and a change of version or of grid year is a methodology change under the base-year recalculation policy. The full rules, the records to retain for assurance and the maintenance protocol are in Section S12.3, and a worked disclosure that meets them is in Section S11.2.

\subsection{Case Study Design}\label{sec:casedesign}
The method is applied to one illustrative organisation, Verdant Solutions, a hypothetical 200-person European professional-services firm preparing a Scope 3 Category 1 inventory for fiscal year 2025. It buys three AI services, chosen so that each falls in a different tier: an OpenAI model accessed through the API and billed in tokens (Tier 2a), ChatGPT Enterprise seats, whose export reports messages but no tokens (Tier 2b), and a Notion subscription that bundles AI without pricing it separately, so that the spend-based fallback applies to the whole subscription (Tier 1). The export quantities are invented, and every other input, the serving region included (Section~\ref{sec:3.5}), is a default of the method applied as a preparer without further information would apply it. The example illustrates the calculation and does not validate it (Section~\ref{sec:5}). The results are in Section~\ref{sec:6}, and a second example at the other end of the volume range, fifty developers using agentic coding tools, is in S11.1.

\subsection{Parameterisation: Dataset v3}\label{sec:4}
Every input takes its value from a measured or published source wherever one exists (Table~\ref{tab:inputs}), and four rules govern the choices. First, accelerator energy comes from one open benchmark, the ML.ENERGY Leaderboard v3~\citep{mlenergy2025}, at the operating point and with the class statistic of Section~\ref{sec:3.5} (Section S9.1). Second, where the evidence gives a range but no single value, the central value is set at or towards the conservative end and the choice is named: the PUE sits above the three disclosed hyperscaler fleet values, the hardware life is shorter than the six years Alphabet files, and the construction constant is the upper end of its range. Third, low and high values are taken from the sources rather than set as symmetric bands, and a bound that rests on no source is stated as the author's sensitivity. Fourth, each input carries the direction in which its central value is more likely wrong, stated in its subsection of S9. Most are biased low, chiefly because the operating point is a best-case serving configuration applied as a fleet average. The grid intensities are biased high, most for the European rows, because they are for calendar year 2023, while leaving out transmission and distribution losses biases the electricity term low by a few per cent (Section~\ref{sec:3.1}).

Two pairs of inputs are tied rather than chosen separately. The idle uplift and the utilisation of the embodied term describe one fleet and move together through $I = 1 + p\,(1-U)/U$, with an assumed idle-power fraction $p$. The lifetime-token basis of the training term combines the minimum service life published by two large hosting platforms with the one per-model serving volume any provider has disclosed. The water inputs follow the precedence of S9.8. The register of assumptions in S7 gives the reason for each source and its sensitivity, Table~\ref{tab:master} in S1 the bands of the central factor, and S5 the refresh schedule.

\begingroup\small\setlength{\tabcolsep}{3pt}
\begin{longtable}{>{\raggedright\arraybackslash}p{3.1cm} >{\raggedright\arraybackslash}p{2.9cm} >{\raggedright\arraybackslash}p{2.7cm} >{\raggedright\arraybackslash}p{1.9cm} >{\raggedright\arraybackslash}p{2.9cm} >{\raggedright\arraybackslash}p{0.95cm}}
\caption{Inputs of the factor, the token basis, the Tier 2b defaults and the water extension, with the subsection of the Supplementary Material where each is derived. Evidence classes are the five of Figure~\ref{fig:chain}, detailed in Figure~\ref{fig:evidence}, with \emph{derived} for a value computed from other inputs. A, B and C are the small, mid and frontier classes. United States grid rows use the Emissions \& Generation Resource Integrated Database (eGRID2023) scaled to the life-cycle basis of Ember. AWARE is the Available Water Remaining characterisation factor. The resulting terms are in Table~\ref{tab:regionindep}.}\label{tab:inputs}\\
\toprule
Input & Central value & Range & Evidence & Source & Where \\
\midrule
\endfirsthead
\multicolumn{6}{l}{\textit{Table~\ref{tab:inputs} (continued)}}\\
\toprule
Input & Central value & Range & Evidence & Source & Where \\
\midrule
\endhead
\midrule\multicolumn{6}{r}{\textit{continues on the next page}}\\
\endfoot
\endlastfoot
Accelerator energy per output token, H100, class means A / B / C & 0.1309 / 0.4285 / 1.5958 J & Class range $-$40/$+$47\%, $-$38/$+$24\%, $-$67/$+$55\% & Measured; proxy for a closed model & \citet{mlenergy2025} & S9.1 \\
Host uplift $H$ & 1.50 & 1.43 to 1.88 & Measured (central and low); bound (high) & \citet{latif2025}; \citet{elsworth2025}; \citet{schneider2025} & S9.2 \\
Idle uplift $I$ and utilisation $U$ & $I$ 1.10 with $U$ 0.75 & 1.05 with $U$ 0.857; 1.20 with $U$ 0.60 & Published ($I$ central); author sensitivity ($I$ low); derived ($U$, from an assumed $p$) & \citet{elsworth2025}; \citet{oviedo2026} & S9.2 \\ Idle-power fraction $p$, tying $I$ to $U$ & 0.30 & --- & Author choice & None published & S9.2 \\
PUE & 1.20 & 1.09 to 1.45 & Author choice & \citet{google2026a}; \citet{shehabi2024} & S9.2 \\
Grid intensity $g$, life-cycle basis, calendar year 2023 & By region (Table~\ref{tab:carbon}); global average 0.4832 kg CO\textsubscript{2}e/kWh & --- & Published & \citet{ember2026}; eGRID2023~\citep{epa2025} & S9.3 \\
Server manufacturing $M_{\mathrm{mfg}}$ & 3,300 kg CO\textsubscript{2}e & 2,872 to 4,241 kg & Author choice in a published range & \citet{dell2025} & S9.4 \\
Server transport $M_{\mathrm{tr}}$ & 672 kg CO\textsubscript{2}e & 427 to 712 kg & Published & \citet{dell2025} & S9.4 \\
Hardware life $L$ & 5 years & 4 to 6 years & Author choice & \citet{secfilings2026}; \citet{schneider2025} & S9.4 \\ Server throughput $T_8$, class harmonic means A / B / C & 37,865 / 10,044 / 2,615 output tokens/s & --- & Measured; proxy for a closed model & \citet{mlenergy2025} & S9.4 \\
Construction $k_c$ & 7.66 g CO\textsubscript{2}e per kWh of server electricity & Low 5.3 g; no high value & Published & \citet{schneider2025}; \citet{carbontrust2026} & S9.4 \\
Final-run training $E_{\mathrm{train}}$, A / B / C & 420 / 2,040 / 8,930 tCO\textsubscript{2}e & --- & Proxy & \citet{meta2026} & S9.5 \\
Model service life $S$ & 12 months & 6 to 18 months & Proxy & \citet{aws2026a}; \citet{microsoft2026b} & S9.5 \\
Tokens per day $V$ and lifetime tokens $N$ & $V = 1.905 \times 10^{12}$; $N = (S/12) \times 365 \times V = 6.95 \times 10^{14}$ & --- & Proxy & \citet{openai2025a} & S9.5 \\
Development ratio $d$ (separate line only) & 1.5 & 0.67 to 2.5 & Proxy & \citet{gamazaychikov2026}; \citet{denain2026} & S9.5 \\
Token weights $W_{\mathrm{in}}$ and $W_{\mathrm{cache}}$ & 0.30 and 0.05 & 0.10 to 0.40; 0.02 to 0.10 & Proxy & \citet{chung2025}; provider prices & S9.6 \\ Default input share $s_{\mathrm{in}}$, giving $m$ & 0.85, so $m = 0.302$ & 76.7\% (input share of the short-chat default) to 99.5\% (agentic default) & Author choice within the range of the workload defaults & \citet{azure2024}; \citet{openrouter2026} & S9.6 \\
Reference mix $R_{\mathrm{bench}}$ and divisor $D$ & 1.13 and 1.339 & Runs 0.47 to 2.01; 1.08 to 1.45 & Author choice; run mixes measured & \citet{mlenergy2025}; \citet{mlenergy2026} & S9.6 \\
Tier 2b tokens per message, short chat & 928 input, 282 output & $-$50\%/$+$130\% on estimated tokens & Measured medians from two corpora, combined & \citet{azure2024}; \citet{zhao2024} & S9.7 \\
Tier 2b messages per seat month & 30 per 30-day-active seat; 325 heavy use & 20 to 65 & Derived & \citet{ukgov2025}; \citet{openai2026a}; \citet{adminreports2026} & S9.7 \\
Reasoning share and multiplier & 4.98\% of conversations; 25 times & unweighted interquartile range across the 601 clusters, 2.92 to 6.44\%; 10 to 50 times & Measured (share); author choice (multiplier) & \citet{anthropic2025} & S9.7 \\
On-site WUE & By region (Table~\ref{tab:water}) & 0.02 to 1.57 L/kWh & Published for 8 of 17 regions and derived from operator site data for 4; proxy or aggregate for five & \citet{aws2026b}; \citet{microsoft2026a} & S9.8 \\
Withdrawal to consumption & 1.5 & As published, as a sensitivity & Derived & \citet{aws2026b} & S9.8 \\
Off-site water intensity & Consumption factors by region under two co-equal conventions, all reservoir evaporation charged to hydropower and a hydropower-excluded variant reconstructed from App.~4--6 & --- & Published & \citet{reig2020} & S9.8 \\
Scarcity weighting, on-site term only & AWARE 2.0, non-agricultural, annual & 0.64 (Japan) to 13.56 (Texas) & Published & \citet{seitfudem2025} & S9.8 \\
\bottomrule
\end{longtable}
\endgroup

\section{Results}\label{sec:results}
\subsection{The Factor Tables}\label{sec:4.9}

\begin{figure}[htbp]
\centering
\includegraphics[width=0.95\textwidth]{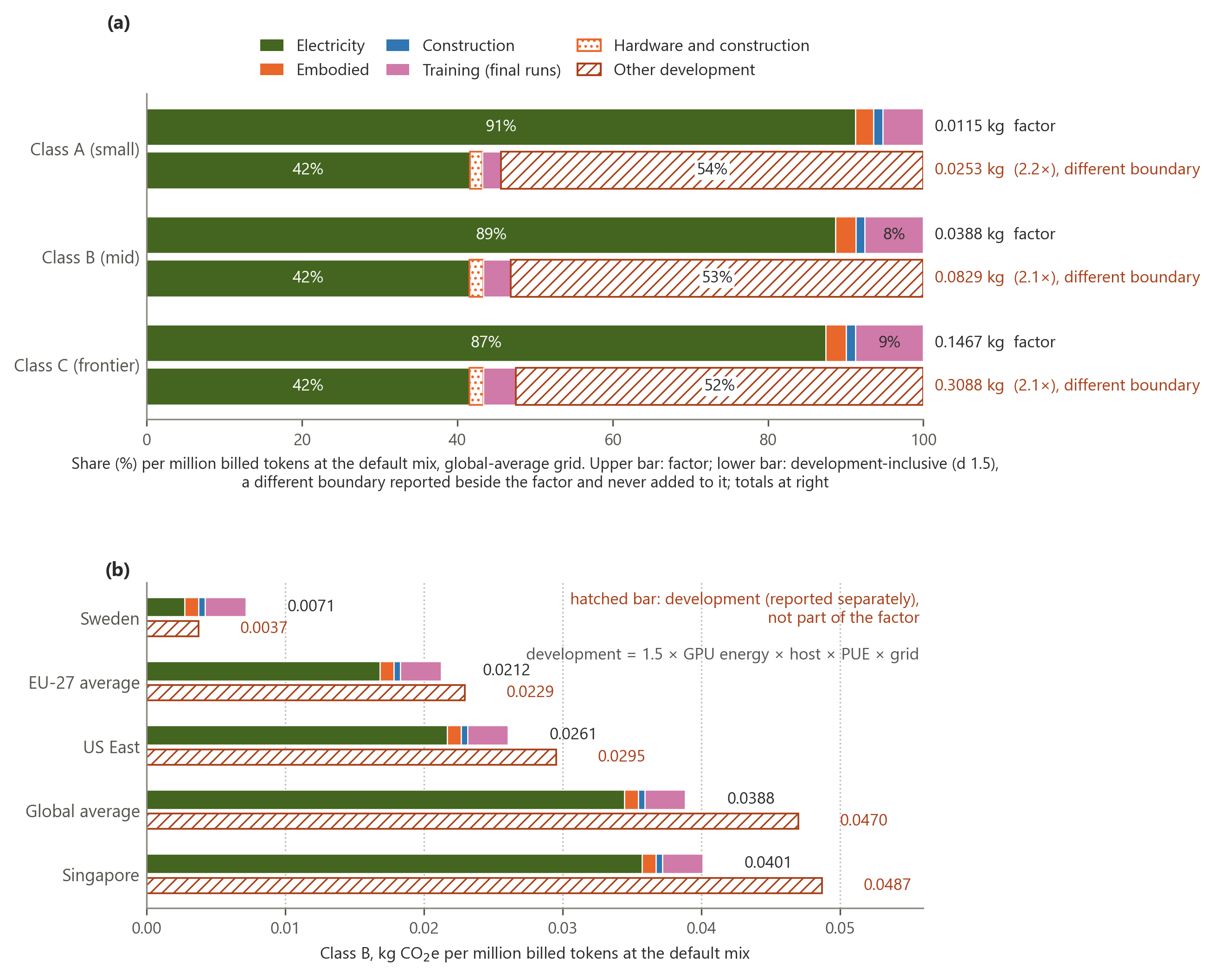}
\caption{Composition of the factor per million \emph{billed} tokens at the default mix, in kg CO\textsubscript{2}e. Panel (a) gives two bars for each class at the global-average grid. The first is the serving factor in its four terms, with the factor printed at the right. The three compute-proportional terms enter at the default effective-to-billed multiplier and training at full weight, so training is a larger share here than of the inference footprint. Electricity is 87\% to 91\% of the factor and training 5\% to 9\%. The second bar is the development-inclusive total at $d = 1.5$ (Table~\ref{tab:structrange}), split into electricity, hardware and construction, final runs and other development: for the mid class 41.6\%, 1.8\%, 3.5\% and 53.1\%, and within two points of that for the other two classes. This split implies that final runs are about 6\% of development, below the 9.6\% to 22.6\% of \citet{denain2026}, because the training term and the development line come from different sources, Meta's disclosure over a serving denominator against a sector ratio on serving energy (Section~\ref{sec:3.4}). Panel (b) gives the mid-class factor by region, with the development line at $d = 1.5$ hatched beneath each bar on the same per-billed basis, the Table~\ref{tab:devline} value times the 0.302 multiplier. Both development bars carry the caveats of Section~\ref{sec:3.4}.}
\label{fig:composition}
\end{figure}

Table~\ref{tab:energy} gives the energy chain per class, Table~\ref{tab:regionindep} the three region-independent terms, Table~\ref{tab:carbon} the carbon factors by class and region, Table~\ref{tab:water} the water factors and Figure~\ref{fig:composition} the composition of the factor by term and by region. The three classes and seventeen regions are the form of the method that can be checked by hand, not its limit. The deposit, listed under Data Availability~\citep{llopis2026}, gives each of the 18 measured open-weight models its own factor on every accelerator type measured and resolves every public AWS, Azure and Google Cloud region to a grid row with a stated mapping confidence. Regions outside the printed rows need its grid and cloud-region files, and identifiers outside the extract of Table~\ref{tab:classmap} its 187 model patterns. Closed commercial models stay at the class (Section~\ref{sec:3.5}), and water is not extended beyond Table~\ref{tab:water} (S1.5). The conservative high case at any printed region is $g$ times the high-case facility energy of Table~\ref{tab:energy}, plus the high-case embodied term of Table~\ref{tab:regionindep} and a high-case construction term of 0.63, 2.06 and 7.66 g per million effective tokens. The B200 case takes the B200 facility energy and construction term with the H100 embodied term.

\begin{table}[htbp]
\centering
\setlength{\tabcolsep}{3pt}
\footnotesize
\begin{tabular}{>{\raggedright\arraybackslash}p{1.35cm} >{\raggedright\arraybackslash}p{1.95cm} >{\raggedleft\arraybackslash}p{1.0cm} >{\raggedleft\arraybackslash}p{1.25cm} >{\raggedleft\arraybackslash}p{1.1cm} >{\raggedleft\arraybackslash}p{1.15cm} >{\raggedleft\arraybackslash}p{1.25cm} >{\raggedright\arraybackslash}p{1.9cm}}
\toprule
Class & Scenario & Members $n$ & J / output token & GPU Wh/1k & Server Wh/1k & Facility Wh/1k & Class range (lowest / highest member) \\
\midrule
A --- Small & H100, central & 7 & 0.1309 & 0.0364 & 0.0600 & \textbf{0.0720} & $-$40\% / $+$47\% \\
A --- Small & B200, hardware sensitivity & 7 & 0.1133 & 0.0315 & 0.0519 & 0.0623 & $-$34\% / $+$49\% \\
B --- Mid & H100, central & 5 & 0.4285 & 0.1190 & 0.1964 & \textbf{0.2357} & $-$38\% / $+$24\% \\
B --- Mid & B200, hardware sensitivity & 4 & 0.2853 & 0.0792 & 0.1308 & 0.1569 & $-$38\% / $+$33\% \\
C --- Frontier & H100, central & 3 & 1.5958 & 0.4433 & 0.7314 & \textbf{0.8777} & $-$67\% / $+$55\% \\
C --- Frontier & B200, hardware sensitivity & 3 & 1.6473 & 0.4576 & 0.7550 & 0.9060 & $-$31\% / $+$36\% \\
\bottomrule
\end{tabular}
\caption{Energy per 1,000 \emph{effective} tokens by model class, in Wh. Each class value is the arithmetic mean of its measured members at the 50 ms operating point, the most efficient configuration within 50 ms per output token (ML.ENERGY v3); server $=$ graphics processing unit (GPU) $\times$ 1.50 host $\times$ 1.10 idle; facility $=$ server $\times$ power usage effectiveness (PUE) 1.20.}
\label{tab:energy}

\begin{minipage}{\textwidth}\footnotesize \textit{Facility Wh per 1,000 effective tokens is numerically kWh per million effective tokens. The class range is the lowest and the highest member relative to the class mean (members in Table~\ref{tab:classmembers}, definition in Section~\ref{sec:3.8}). The frontier class mixes bfloat16 and FP8 members (16-bit and 8-bit floating-point formats). B200 to H100 facility ratios are 0.87, 0.67 and 1.03, but the mid and frontier classes differ in membership and precision between the two platforms (Table~\ref{tab:classmembers}), so the ratios are not a pure hardware effect. For Qwen3-235B, the one frontier member measured in bfloat16 on both, the ratio is 0.63. The conservative high case ($H$ 1.88, $I$ 1.20, PUE 1.45) gives 0.1189, 0.3894 and 1.4500 Wh. At the \emph{central} uplifts on H100, the minimum of the smallest accelerator count with no latency limit gives 0.0803, 0.2786 and 0.7322 Wh, and the mean rule 0.1510, 0.6788 and 2.1325 Wh, the upper end of the serving-regime range of Table~\ref{tab:structrange}.}\end{minipage}
\end{table}

\begin{table}[htbp]
\centering
\setlength{\tabcolsep}{3pt}
\footnotesize
\begin{tabular}{>{\raggedright\arraybackslash}p{1.45cm} >{\raggedleft\arraybackslash}p{1.25cm} >{\raggedleft\arraybackslash}p{1.0cm} >{\raggedleft\arraybackslash}p{1.0cm} >{\raggedleft\arraybackslash}p{1.6cm} >{\raggedleft\arraybackslash}p{1.5cm} >{\raggedleft\arraybackslash}p{1.35cm} >{\raggedleft\arraybackslash}p{1.0cm} >{\raggedleft\arraybackslash}p{1.45cm} >{\raggedleft\arraybackslash}p{1.35cm}}
\toprule
Class & Manufac\-turing & Trans\-port & Embod\-ied & Embodied, server-footprint and transport range & Embodied at $U$ 0.60 (high case) & Construc\-tion (H100) & Train\-ing & Training, $S$ $=$ 18 to 6 months & Training, class volumes \\
\midrule
A --- Small & 0.74 & 0.15 & \textbf{0.89} & 0.74--1.11 & 1.11 & \textbf{0.46} & \textbf{0.60} & 0.40--1.21 & 0.06 \\
B --- Mid & 2.78 & 0.57 & \textbf{3.34} & 2.78--4.17 & 4.18 & \textbf{1.50} & \textbf{2.93} & 1.96--5.87 & 1.47 \\
C --- Frontier & 10.67 & 2.17 & \textbf{12.84} & 10.67--16.01 & 16.05 & \textbf{5.60} & \textbf{12.84} & 8.56--25.69 & 12.84 \\
\bottomrule
\end{tabular}
\caption{The three region-independent terms by model class, in g CO\textsubscript{2}e: embodied hardware and construction per million \emph{effective} tokens, training per million \emph{billed} tokens.}
\label{tab:regionindep}

\begin{minipage}{\textwidth}\footnotesize \textit{Divide by 1,000 for the kg units of Table~\ref{tab:carbon}: for the mid class, embodied plus construction is 0.00485 kg per million effective tokens and training 0.00293 kg per million billed tokens, and the two are not added. Each class value is the mean of its members' own terms. Construction is charged on \emph{server} electricity, so it alone differs between the H100 and B200 scenarios, where it is 0.40, 1.00 and 5.78 g. Training is the Meta Llama 3.1 disclosure over 695 trillion lifetime billed tokens (Table~\ref{tab:inputs}), with Llama 3.1 405B as the frontier-class value, and the class volumes are 10, 2 and 1 times that denominator. Training is held at its H100 value in the B200 and high cases, and embodied is held in the B200 case but recomputed at $U = 0.60$ in the high case.}\end{minipage}
\end{table}

\begin{table}[htbp]
\centering
\setlength{\tabcolsep}{3pt}
\footnotesize
\begin{tabular}{>{\raggedright\arraybackslash}p{2.75cm} >{\raggedright\arraybackslash}p{2.5cm} >{\raggedleft\arraybackslash}p{0.95cm} >{\raggedleft\arraybackslash}p{1.0cm} >{\raggedleft\arraybackslash}p{1.0cm} >{\raggedleft\arraybackslash}p{1.0cm} >{\raggedleft\arraybackslash}p{1.0cm} >{\raggedleft\arraybackslash}p{1.0cm} >{\raggedleft\arraybackslash}p{1.0cm}}
\toprule
 & & Grid & \multicolumn{3}{c}{Inference footprint,} & \multicolumn{3}{c}{Convenience factor,} \\
 & & & \multicolumn{3}{c}{per M \emph{effective} tokens} & \multicolumn{3}{c}{per M \emph{billed} tokens} \\
\cmidrule(lr){4-6}\cmidrule(lr){7-9}
Region & Cloud example & kg/kWh & A & B & C & A & B & C \\
\midrule
EU North --- Sweden & AWS eu-north-1 & 0.0384 & 0.0041 & 0.0139 & 0.0522 & 0.0018 & 0.0071 & 0.0286 \\
EU West --- France & AWS eu-west-3 & 0.0533 & 0.0052 & 0.0174 & 0.0652 & 0.0022 & 0.0082 & 0.0326 \\
EU North --- Finland & Google europe-north1 & 0.0813 & 0.0072 & 0.0240 & 0.0898 & 0.0028 & 0.0102 & 0.0400 \\
EU West --- Belgium & Google europe-west1 & 0.1347 & 0.0110 & 0.0366 & 0.1366 & 0.0039 & 0.0140 & 0.0542 \\
EU West --- UK (London) & Google europe-west2 & 0.2356 & 0.0183 & 0.0604 & 0.2252 & 0.0061 & 0.0212 & 0.0810 \\
EU-27 average & EU data zone & 0.2360 & 0.0183 & 0.0605 & 0.2256 & 0.0061 & 0.0212 & 0.0811 \\
EU West --- Netherlands & Google europe-west4 & 0.2682 & 0.0206 & 0.0681 & 0.2538 & 0.0068 & 0.0235 & 0.0896 \\
EU West --- Ireland & AWS eu-west-1 & 0.2825 & 0.0217 & 0.0714 & 0.2664 & 0.0072 & 0.0245 & 0.0934 \\
EU Central --- Germany & AWS eu-central-1 & 0.3636 & 0.0275 & 0.0905 & 0.3376 & 0.0089 & 0.0303 & 0.1149 \\
US East --- N. Virginia & AWS us-east-1 (SRVC) & 0.3038 & 0.0232 & 0.0765 & 0.2851 & 0.0076 & 0.0261 & 0.0991 \\
US West --- Oregon & AWS us-west-2 (NWPP) & 0.3237 & 0.0246 & 0.0811 & 0.3025 & 0.0081 & 0.0275 & 0.1043 \\
US South Central --- Texas & Google us-south1 (ERCT) & 0.3753 & 0.0284 & 0.0933 & 0.3478 & 0.0092 & 0.0312 & 0.1180 \\
US South --- Southeast & None identified (SRSO) & 0.4310 & 0.0324 & 0.1064 & 0.3967 & 0.0104 & 0.0351 & 0.1328 \\
US Central --- Iowa & Google us-central1 (MROW) & 0.4721 & 0.0353 & 0.1161 & 0.4328 & 0.0113 & 0.0381 & 0.1437 \\
Asia Pacific --- Japan & AWS ap-northeast-1 & 0.4926 & 0.0368 & 0.1210 & 0.4508 & 0.0117 & 0.0395 & 0.1492 \\
Asia Pacific --- Singapore & AWS ap-southeast-1 & 0.5009 & 0.0374 & 0.1229 & 0.4580 & 0.0119 & 0.0401 & 0.1514 \\
\textbf{Global average} & Unknown region & \textbf{0.4832} & \textbf{0.0361} & \textbf{0.1187} & \textbf{0.4425} & \textbf{0.0115} & \textbf{0.0388} & \textbf{0.1467} \\
\bottomrule
\end{tabular}
\caption{Carbon factors by model class and region, in kg CO\textsubscript{2}e, H100 central, 2023 grid data. The inference footprint (electricity $+$ embodied $+$ construction) is \textbf{per million effective tokens}; the convenience factor is \textbf{per million billed tokens} at the default mix. The region-independent training term, which is per million billed tokens, is in Table~\ref{tab:regionindep} and is already inside the convenience factor. Region names follow the cloud providers (AWS, Amazon Web Services; EU, European Union; UK, United Kingdom); codes in parentheses are eGRID subregions (Section S9.3).}
\label{tab:carbon}

\begin{minipage}{\textwidth}\footnotesize \textit{Inference footprint $=$ grid intensity $\times$ facility kWh per million effective tokens (Table~\ref{tab:energy}, H100) $+$ the embodied and construction terms of Table~\ref{tab:regionindep}. Convenience factor $=$ inference footprint $\times$ 0.302 $+$ the training term. For any other mix apply the equation of Section~\ref{sec:3.4} instead. Hand check for US East, mid class: $596.3/2{,}204.6 = 0.2705$ kg/kWh $\times$ 1.1232 $=$ 0.3038, $\times$ 0.2357 $=$ 0.0716 electricity, $+\,0.0033 + 0.0015 = 0.0765$ footprint, and $0.302 \times 0.0765 + 0.0029 = 0.0261$ per million billed on the unrounded terms. The United States Southeast is a grid subregion with no public hyperscale AI region, and the EU-27 and global rows are averages. The B200 and conservative high values are in Table~\ref{tab:b200high}, the development line in Table~\ref{tab:devline} and the structural range of the convenience factor in Table~\ref{tab:structrange}.}\end{minipage}
\end{table}

\begin{table}[htbp]
\addtocounter{table}{-1}\renewcommand{\thetable}{\arabic{table}b}\renewcommand{\theHtable}{\arabic{table}b}
\centering
\setlength{\tabcolsep}{3pt}
\footnotesize
\begin{tabular}{>{\raggedright\arraybackslash}p{1.5cm} >{\raggedright\arraybackslash}p{1.9cm} >{\raggedleft\arraybackslash}p{1.2cm} >{\raggedleft\arraybackslash}p{2.0cm} >{\raggedleft\arraybackslash}p{2.0cm} >{\raggedleft\arraybackslash}p{2.0cm} >{\raggedleft\arraybackslash}p{2.0cm}}
\toprule
 & & & Serving regime (same quantity) & \multicolumn{3}{c}{Boundary (different quantity): development-inclusive} \\
\cmidrule(lr){4-4}\cmidrule(lr){5-7}
Class & Region & Central & Mean rule & $d = 0.67$ & $d = 1.5$ & $d = 2.5$ \\
\midrule
A --- Small & Global average & \textbf{0.0115} & 0.0237 (2.05$\times$) & 0.0173 (1.50$\times$) & 0.0253 (2.19$\times$) & 0.0348 (3.02$\times$) \\
A --- Small & EU-27 & \textbf{0.0061} & 0.0124 (2.01$\times$) & 0.0087 (1.41$\times$) & 0.0126 (2.04$\times$) & 0.0172 (2.80$\times$) \\
A --- Small & US East & \textbf{0.0076} & 0.0155 (2.03$\times$) & 0.0110 (1.45$\times$) & 0.0160 (2.10$\times$) & 0.0221 (2.89$\times$) \\
B --- Mid & Global average & \textbf{0.0388} & 0.1072 (2.76$\times$) & 0.0569 (1.46$\times$) & 0.0829 (2.13$\times$) & 0.1142 (2.94$\times$) \\
B --- Mid & EU-27 & \textbf{0.0212} & 0.0564 (2.66$\times$) & 0.0285 (1.34$\times$) & 0.0412 (1.94$\times$) & 0.0565 (2.66$\times$) \\
B --- Mid & US East & \textbf{0.0261} & 0.0703 (2.70$\times$) & 0.0363 (1.39$\times$) & 0.0527 (2.02$\times$) & 0.0723 (2.78$\times$) \\
C --- Frontier & Global average & \textbf{0.1467} & 0.3407 (2.32$\times$) & 0.2120 (1.45$\times$) & 0.3088 (2.10$\times$) & 0.4254 (2.90$\times$) \\
C --- Frontier & EU-27 & \textbf{0.0811} & 0.1813 (2.24$\times$) & 0.1064 (1.31$\times$) & 0.1537 (1.90$\times$) & 0.2106 (2.60$\times$) \\
C --- Frontier & US East & \textbf{0.0991} & 0.2250 (2.27$\times$) & 0.1353 (1.37$\times$) & 0.1962 (1.98$\times$) & 0.2695 (2.72$\times$) \\
\bottomrule
\end{tabular}
\caption{The central factor with its structural range, in kg CO\textsubscript{2}e per million \emph{billed} tokens at the default mix, H100, with each value's ratio to the central factor. The two axes of Section~\ref{sec:3.8} are typed and never combined. The \emph{serving regime} is the same quantity under the mean rule at the 50 ms latency limit (Section S9.1), with electricity, construction and embodied recomputed and training unchanged. The \emph{development-inclusive} columns are a \emph{different boundary}, not the top of an interval for the factor: the inference footprint plus the development line of Section~\ref{sec:3.4}, times the default multiplier 0.302, with the final-run training term dropped.}
\label{tab:structrange}

\begin{minipage}{\textwidth}\footnotesize \textit{The caveats of the development line are those of Section~\ref{sec:3.4} (S9.5). The development line alone, mid class, global average, is 0.0210, 0.0470 and 0.0783 per million billed tokens at the three values of $d$. The parameter interval of the central factor is in Section~\ref{sec:5.5}.}\end{minipage}
\end{table}

\begin{table}[htbp]
\centering
\setlength{\tabcolsep}{3pt}
\footnotesize
\begin{tabular}{>{\raggedright\arraybackslash}p{3.1cm} >{\raggedright\arraybackslash}p{0.9cm} >{\raggedleft\arraybackslash}p{1.1cm} >{\raggedleft\arraybackslash}p{1.2cm} >{\raggedleft\arraybackslash}p{1.9cm} >{\raggedleft\arraybackslash}p{1.9cm} >{\raggedleft\arraybackslash}p{1.1cm} >{\raggedleft\arraybackslash}p{1.9cm}}
\toprule
Region & Conf. & WUE & On-site & Total, hydro allocated & Total, hydro excluded & AWARE & On-site scarcity-weighted \\
\midrule
EU North --- Sweden & P & 0.020 & 0.003 & \textbf{1.421} & \textbf{0.329} & 2.17 & 0.006 \\
EU West --- France* & C & 0.050 & 0.007 & 0.872 & 0.492 & 4.99 & 0.033 \\
EU North --- Finland* & S & 0.010 & 0.002 & 1.072 & 0.405 & 2.17 & 0.004 \\
EU West --- Belgium* & C & 0.050 & 0.007 & 0.622 & 0.524 & 3.52 & 0.023 \\
EU West --- UK (London) & P & 0.140 & 0.018 & \textbf{0.572} & \textbf{0.454} & 2.29 & 0.042 \\
EU-27 average* & C & 0.050 & 0.007 & --- & --- & --- & --- \\
EU West --- Netherlands & D & 0.036 & 0.006 & \textbf{0.818} & \textbf{0.818} & 1.89 & 0.011 \\
EU West --- Ireland & P & 0.020 & 0.003 & \textbf{0.351} & \textbf{0.270} & 1.17 & 0.003 \\
EU Central --- Germany & P & 0.170 & 0.022 & \textbf{0.477} & \textbf{0.418} & 2.09 & 0.047 \\
US East --- N. Virginia & P & 0.060 & 0.008 & \textbf{0.570} & \textbf{0.475} & 1.62 & 0.013 \\
US West --- Oregon & P & 0.120 & 0.016 & \textbf{2.255} & \textbf{0.152} & 3.91 & 0.061 \\
US South Central --- Texas & D & 0.292 & 0.046 & \textbf{0.349} & \textbf{0.332} & 13.56 & 0.621 \\
US South --- Southeast & D & 0.123 & 0.019 & \textbf{0.564} & \textbf{0.430} & 1.78 & 0.034 \\
US Central --- Iowa & D & 0.123 & 0.019 & \textbf{0.751} & \textbf{0.415} & 3.41 & 0.066 \\
Asia Pacific --- Japan & P & 1.220 & 0.160 & \textbf{0.704} & \textbf{0.512} & 0.64 & 0.102 \\
Asia Pacific --- Singapore$\dagger$ & P & 1.570 & 0.206 & --- & --- & 0.87 & 0.178 \\
Global average* & S & 0.550 & 0.108 & 0.849 & 0.478 & --- & --- \\
\bottomrule
\end{tabular}
\caption{Water by region for the mid class, H100 central, in mL per 1,000 \emph{effective} tokens: on-site on a consumption basis, totals under both hydropower conventions, and an on-site water scarcity footprint in mL world-equivalent. For the small and frontier classes, scale each value by the class's facility energy relative to the mid class (Table~\ref{tab:energy}). The two off-site components behind the totals are in Table~\ref{tab:wateroffsite}. WUE is water use effectiveness; AWARE, the Available Water Remaining scarcity factor.}
\label{tab:water}

\begin{minipage}{\textwidth}\footnotesize \textit{WUE is in L/kWh. Confidence is P for a published named-region value~\citep{aws2026b} and D for a ratio derived from a provider's published site water and site electricity~\citep{microsoft2026a,meta2025}. It is S for the 2023 Microsoft site series~\citep{li2025} and C for a provider's continental aggregate (the AWS series again). Only P and D rows count as confirmed, and the eleven confirmed totals are in bold. Rows marked * are not confirmed, $\dagger$ marks a region with no off-site source, and the two rows that are not locations carry no scarcity factor. The two totals are two conventions, not an uncertainty range, and the scarcity-weighted column is never added to either. Off-site water is 69\% to 99.8\% of every volumetric total, so the scarcity weighting reaches little of the volume. Denominators, the consumption conversion and a worked row are in S1.5, the rankings and sensitivities in S3.}\end{minipage}
\end{table}

\textbf{Spend-based inputs.} The Tier 1 rule of Section~\ref{sec:3.7} needs two numbers. The factor is the EXIOBASE 3.8.2 ``Computer and related services'' coefficient for Austria, 2019 data, 0.1181 kg CO\textsubscript{2}e per EUR, read from the distributor's record~\citep{stadler2018,climatiq2026}; the same record gives 0.1333 for Germany. The deflator is the euro-area HICP all-items index, 104.80 for 2019 and 128.75 for 2025, so spend in 2025 euro is multiplied by 0.8140~\citep{eurostat2026}. \emph{Evidence}: published statistical series. A United States factor for a United States vendor could not be retrieved, and the Austrian coefficient against the euro-area rather than the Austrian index is a stated inconsistency worth about 6\% on the deflator: the Austrian series, 106.98 and 139.01, would give 0.7696. Two further mismatches are not sized. List-price spend is at purchaser prices while the coefficient is at basic prices, and the all-items index does not follow computer-services prices. \emph{Bias}: unknown in level; the gap to a physical estimate is reported in Section~\ref{sec:7.1}.

\subsection{Comparison and Consistency Checks}\label{sec:5}
Three checks are reported here: a comparison with published per-query estimates, the consistency of the serving chain with a first-party energy decomposition, and a sensitivity analysis. Three further checks are in S10. Against per-query research the comparison is a consistency check, not a contest: the question is whether the published figures fall within the structural range of Table~\ref{tab:structrange}. The gap to current practice, the spend-based calculation on the same purchase, is in Section~\ref{sec:7.1}. None of the checks compares predicted with metered energy for any commercial or self-hosted service, tests the method on a real organisation's billing data or calibrates the propagated interval against an observed error. Each is a consistency or plausibility check that reaches only part of the chain.

\subsubsection{Comparison with Published Per-Query Estimates}\label{sec:5.1}

Every external figure is published per prompt, per query or per response, so the comparison is made per query, each figure restated at the token counts it discloses (Table~\ref{tab:comparison}). One chat exchange at the Tier 2b short-chat default, 928 input and 282 output tokens, is 418.5 effective tokens, at which the three classes draw 0.0301, 0.0986 and 0.3673 Wh of facility electricity (0.0498, 0.1630 and 0.6069 Wh in the conservative high case) and emit 0.0158, 0.0532 and 0.2007 g CO\textsubscript{2}e at the global-average grid.

\begin{table}[htbp]
\centering
\setlength{\tabcolsep}{4pt}
\footnotesize
\begin{tabular}{>{\raggedright\arraybackslash}p{4.3cm} >{\raggedright\arraybackslash}p{1.5cm} >{\raggedright\arraybackslash}p{1.9cm} >{\raggedright\arraybackslash}p{1.9cm} >{\raggedleft\arraybackslash}p{0.95cm} >{\raggedleft\arraybackslash}p{0.95cm} >{\raggedleft\arraybackslash}p{0.95cm} >{\raggedleft\arraybackslash}p{1.25cm}}
\toprule
Reference figure, per query, and the class it describes & Value & Tokens behind the figure & Token basis used here & $\times$ A & $\times$ B & $\times$ C & $\times$ C high case \\
\midrule
Jegham et al.\ \citeyearpar{jegham2025}, GPT-4o query; mid class by pricing tier & 0.42 Wh & 100 in / 300 out & matched & 23.68 & 7.23 & 1.94 & 1.18 \\
Oviedo et al.\ \citeyearpar{oviedo2026}, per query; frontier class, above 200B & 0.31 Wh & 500 in (fixed) / 300 out (median) & matched & 12.82 & 3.91 & 1.05 & 0.64 \\
Google, median Gemini Apps text prompt~\citep{elsworth2025}; class not stated & 0.24 Wh & not stated & 928 in / 282 out & 7.97 & 2.43 & 0.65 & 0.40 \\
 & & & 100 in / 300 out & 13.53 & 4.13 & 1.11 & 0.67 \\
Average ChatGPT query, a provider chief executive's figure in a 2025 blog post with no published method~\citep{altman2025b}, noted as such by \citet{horswill2026}; class not stated & 0.34 Wh & not stated & 928 in / 282 out & 11.29 & 3.45 & 0.93 & 0.56 \\
 & & & 100 in / 300 out & 19.17 & 5.85 & 1.57 & 0.95 \\
Mistral Large 2, per response~\citep{mistral2025}; frontier class & 1.14 g CO\textsubscript{2}e & 400 out; input not stated & 928 in / 400 out & 59.67 & 17.80 & 4.73 & 2.97 \\
 & & & 0 in / 400 out & 103.33 & 31.11 & 8.30 & 5.15 \\
\bottomrule
\end{tabular}
\caption{The five published per-query figures against this method's classes, each at the token counts behind it, in facility Wh per query for the energy rows and g CO\textsubscript{2}e per query for the Mistral row. Ratios above 1 mean the external figure is higher than this method's. A figure that discloses no input count is shown at the chat exchange of Section S9.7 and at a short 100-in/300-out query as a bracket.}
\label{tab:comparison}

\begin{minipage}{\textwidth}\footnotesize \textit{The Mistral row is at the global-average grid. Mistral's figure excludes training and is computed for serving on a French grid, about nine times cleaner than the global average, so its ratio is not a like-for-like regional comparison. Boundaries differ: this method is accelerator $\times$ host $\times$ idle $\times$ PUE; Google is accelerator, host, idle and facility overhead at a disclosed fleet PUE; Oviedo et al.'s 0.31 Wh is node energy at a sampled PUE with no idle or networking term; their factor of 1.31 for networking, diurnal load and node redundancy scales only their fleet total, so this method's idle uplift of 1.10 has no counterpart in that row; and Jegham et al.\ benchmark commercial API models from outside. Google's disclosure is also the source of this method's idle uplift and of the low ends of its host uplift and PUE (Section S9.2), so that row is a partially dependent comparison rather than an independent one. The high-case column also takes its idle uplift of 1.20 from Oviedo et al., so that cell of the Oviedo row is not independent either. The comparison is class-matched. The $\times$C columns of the Jegham et al.\ row are shown for completeness and do not enter the frontier-class bracket, because this paper assigns gpt-4o to the mid class (S6). The $\times$B column of the Oviedo et al.\ row likewise does not enter the mid-class range, because that figure describes models above 200B parameters.}\end{minipage}
\end{table}

\textbf{Where the method sits, plainly.} For the mid class the per-query values lie \emph{below} every external figure that can describe it, by 2.4 to 7.2 times on energy (Jegham et al., Google and the ChatGPT average), and no published figure describes a small-class model, so the small-class columns are shown only for scale; for the frontier class the three energy figures that can describe it (Oviedo et al., Google and the ChatGPT average, the last two at two token counts) \emph{bracket} the central value, from 0.65 to 1.57 times, and under the conservative high case they are 0.40 to 0.95 times the frontier-class value, all below it (Table~\ref{tab:comparison}). The comparison is class-matched. Jegham et al.'s GPT-4o figure is read against the mid class only, because the mapping assigns gpt-4o to it (S6), and Oviedo et al.'s against the frontier class only, because it describes models above 200B parameters.

Read against the serving-regime range of Table~\ref{tab:structrange}, the two tested classes differ. The mid-class estimates sit at or above the \emph{top} of that range, they are 0.84 to 2.51 times the value under the mean rule and above it in four of the five energy rows, the exception being Google's figure at the chat exchange. The frontier-class estimates bracket the central value and are 0.27 to 0.65 times the mean-rule value. For the mid class the published estimates therefore lie above the central value and mostly beyond the upper end of the range (S9.1). The mean rule was adopted as that upper end after this comparison was made, so the comparison does not test it. No published figure describes a small-class model, so Class A is untested here, and no check in this paper tests the accelerator-energy term against an independent measurement of the same models at a production operating point, the method's largest untested link (S10.4).

\subsubsection{Consistency with a First-Party Decomposition}\label{sec:5.2}

Google reports the median Gemini Apps text prompt in May 2025 as 0.24 Wh across active accelerators, host, idle machines and facility overhead, a scaling of 1.72 on accelerator energy~\citep{elsworth2025}. Rebuilding that ratio from the same four components is an identity rather than a test. What it establishes is that this method's chain has the same four links as the only serving stack disclosed component by component.

The one comparison of values is between the two sources of the host uplift. The node and accelerator power metered by \citet{latif2025}, the source of the central 1.50, imply a host ratio of 1.51 to 1.52, 2 to 3\% above the top of the 1.38 to 1.48 that the two-decimal rounding of Google's table allows. The chain used here is $1.50 \times 1.10 \times 1.20 = 1.98$, 15\% above Google's, of which the PUE accounts for 10\% and the host uplift for 5\% (Table~\ref{tab:chains}).

\begin{table}[htbp]
\centering
\setlength{\tabcolsep}{3pt}
\footnotesize
\begin{tabular}{>{\raggedright\arraybackslash}p{4.3cm} >{\raggedleft\arraybackslash}p{1.2cm} >{\raggedleft\arraybackslash}p{1.2cm} >{\raggedleft\arraybackslash}p{1.2cm} >{\raggedleft\arraybackslash}p{1.5cm} >{\raggedleft\arraybackslash}p{2.0cm} >{\raggedleft\arraybackslash}p{2.4cm}}
\toprule
Serving chain & Host & Idle & PUE & Product & Relative to this method & Mid-class facility Wh/1k \\
\midrule
Google (\citealp{elsworth2025}) & 1.43 & 1.10 & 1.09 & 1.7146 & 0.87 & 0.2041 \\
This method & 1.50 & 1.10 & 1.20 & 1.9800 & 1.00 & 0.2357 \\
Sustainable AI Group~\citep{gamazaychikov2026} & 1.50 & 1.10 & 1.45 & 2.3925 & 1.21 & 0.2848 \\
\bottomrule
\end{tabular}
\caption{The three published serving chains on accelerator energy, and the mid-class facility energy per 1,000 effective tokens each produces from the same measured accelerator energy. All three take the idle uplift from the same Google disclosure, which therefore does not test it, and the second and third share a host uplift, so the second and third rows differ only in power usage effectiveness (PUE), and the first also in its host uplift.}
\label{tab:chains}
\end{table}

\subsubsection{Sensitivity Analysis}\label{sec:5.5}

Table~\ref{tab:sens} gives the inputs and structural choices that move the result by more than a per cent, on both denominators, together with the hardware block, which does not. The full one-at-a-time table, covering every input of Table~\ref{tab:master}, is Table~\ref{tab:sensfull} in S2.

\begin{table}[htbp]
\centering
\setlength{\tabcolsep}{3pt}
\footnotesize
\begin{tabular}{>{\raggedright\arraybackslash}p{5.6cm} >{\raggedleft\arraybackslash}p{1.25cm} >{\raggedleft\arraybackslash}p{1.25cm} >{\raggedleft\arraybackslash}p{1.25cm} >{\raggedleft\arraybackslash}p{1.35cm} >{\raggedleft\arraybackslash}p{1.35cm} >{\raggedleft\arraybackslash}p{1.35cm}}
\toprule
 & \multicolumn{4}{c}{Inference footprint, per M effective} & \multicolumn{2}{c}{Factor, per M billed} \\
\cmidrule(lr){2-5}\cmidrule(lr){6-7}
Sensitivity & A & B & C & B vs central & B & B vs central \\
\midrule
Central & 0.0361 & 0.1187 & 0.4425 & --- & 0.0388 & --- \\
\multicolumn{7}{l}{\textit{Structural choices}} \\
\textbf{Operating point: mean over all configurations meeting the latency limit} & \textbf{0.0763} & \textbf{0.3446} & \textbf{1.0840} & \textbf{$+$190.3\%} & \textbf{0.1072} & \textbf{$+$175.9\%} \\
Training default 0.1 kg per M tokens~\citep{bistline2026} & 0.0361 & 0.1187 & 0.4425 & $+$0.0\% & 0.1359 & $+$249.9\% \\
\textbf{Conservative high case} ($H$ 1.88, $I$ 1.20 with $U$ 0.60, PUE 1.45) & \textbf{0.0592} & \textbf{0.1944} & \textbf{0.7243} & \textbf{$+$63.7\%} & \textbf{0.0617} & \textbf{$+$58.9\%} \\
\multicolumn{7}{l}{\textit{Chain uplifts}} \\
Host uplift $H = 1.88$ & 0.0450 & 0.1480 & 0.5514 & $+$24.6\% & 0.0477 & $+$22.8\% \\
PUE 1.45 & 0.0434 & 0.1425 & 0.5309 & $+$20.0\% & 0.0460 & $+$18.5\% \\
$I = 1.20$ with $U = 0.60$ (tied pair) & 0.0395 & 0.1301 & 0.4848 & $+$9.5\% & 0.0423 & $+$8.8\% \\
\multicolumn{7}{l}{\textit{Class statistics}} \\
Lowest-footprint member of each class & 0.0215 & 0.0730 & 0.1477 & $-$38.5\% & 0.0250 & $-$35.6\% \\
Highest-footprint member of each class & 0.0533 & 0.1463 & 0.6839 & $+$23.2\% & 0.0472 & $+$21.4\% \\
\multicolumn{7}{l}{\textit{Token and training basis}} \\
$W_{\mathrm{in}} = 0.40$ / 0.10, the band around the within-cell 0.30 & 0.0371 / 0.0341 & 0.1229 / 0.1092 & 0.4572 / 0.4093 & $+$3.5 / $-$8.1\% & 0.0444 / 0.0260 & $+$14.3 / $-$33.1\% \\
No restatement to the reference mix (previous treatment) & 0.0337 & 0.1085 & 0.4067 & $-$8.6\% & 0.0358 & $-$8.0\% \\
Training-high: service life $S = 6$ months $\times$ usage shape 0.4 & 0.0361 & 0.1187 & 0.4425 & $+$0.0\% & 0.0506 & $+$30.2\% \\
Service life $S = 6$ months & 0.0361 & 0.1187 & 0.4425 & $+$0.0\% & 0.0418 & $+$7.6\% \\
Hardware block: server footprint, transport, life 4--6 y and construction 5.3 g, moved together & 0.0357 / 0.0366 & 0.1172 / 0.1206 & 0.4368 / 0.4497 & $-$1.3 / $+$1.6\% & 0.0384 / 0.0394 & $-$1.2 / $+$1.5\% \\
\bottomrule
\end{tabular}
\caption{The sensitivities that move the result, H100, global-average grid unless stated. Inference-footprint columns are kg CO\textsubscript{2}e per million \emph{effective} tokens; factor columns are kg CO\textsubscript{2}e per million \emph{billed} tokens at the default mix, with the effective-to-billed multiplier re-derived for each row's weights. Rows are grouped and ordered by the size of the move within each group. Two stated grid conventions are not inputs and are therefore not rows: charging the United States (US) gross grid loss raises the United States rows by 3.9\% to 4.0\%, and overriding the grid intensity with a reported on-site gas intensity of 0.64 kg/kWh raises the mid-class footprint by 31\% at the global average and 104\% at US East.}
\label{tab:sens}

\begin{minipage}{\textwidth}\footnotesize \textit{Each row changes one input or one structural choice and holds the rest central, except the hardware block, which moves four inputs together to their low and high ends. $W_{\mathrm{cache}}$ claims no cache credit at the default mix, so there it acts only through the restatement of each benchmark run (Section S9.6), and its band on the customer's tokens is in Table~\ref{tab:workload}. Every other one-at-a-time row, including the class-statistic alternatives and the unscaled eGRID basis, is in Table~\ref{tab:sensfull}.}\end{minipage}
\end{table}

The two largest movers are choices, not inputs. The operating point moves the result by more than any input and is printed as the upper end of the serving-regime range (Table~\ref{tab:structrange}). The training denominator is the second lever and accounts for a third of the gap to \citet{bistline2026}. The training-high row, a service life of six months with a usage shape of 0.4, raises the training term five times and the factor by 30.2\%, and it is a sensitivity rather than part of the conservative high case, which moves only the three facility uplifts. On the per-billed denominator the token basis matters more than any input of the embodied or construction terms, because $W_{\mathrm{in}}$ sits at a within-cell value inside a band that extends further below it than above (S10.4).

\textbf{Propagated uncertainty.} The one-at-a-time rows do not say how the bands combine, so the bands were also propagated together, in 20,000 Monte Carlo draws (Table~\ref{tab:mc} in S2 gives the method and every value). For the mid class at the global-average grid, the inference footprint has a median 12\% above the central value (0.1326 kg) and a 5th to 95th percentile interval of $-$2\% to $+$30\% (0.1158 to 0.1549 kg). The factor per million billed tokens, which adds the service life behind the training term and the two token weights, has a median 5\% above the central value (0.0409 kg) and an interval of $-$18\% to $+$29\% (0.0319 to 0.0502 kg). On the chain the central value thus sits near the lower end of what its own input bands allow, because those bands were set skewed upwards (Section S9). On the token axis the interval is two-sided because of the $W_{\mathrm{in}}$ band, which is why it reaches $-$18\% while the footprint interval stops at $-$2\%. The compounded high case, at $+$64\%, lies beyond the 97.5th percentile of the footprint, at $+$34\%, as it must when three independently drawn inputs are set at their tops together, so it is a scenario and not a likely value. The interval excludes the class range, the operating-point rule, the serving region and the activity estimate, which Section~\ref{sec:3.8} keeps apart.

\subsection{Case Study Results}\label{sec:6}

\subsubsection{A Professional-Services Firm}\label{sec:6.1}

Verdant Solutions (Section~\ref{sec:casedesign}) reports for fiscal year 2025 (FY2025). It is outside the CSRD scope under the Omnibus I Directive~\citep{euparl2026b} and was not in a reporting wave for FY2025~\citep{euparl2025}. From financial year 2027 the Omnibus also lets a value-chain undertaking of no more than 1,000 employees decline any request made for the purpose of CSRD reporting that goes beyond the value-chain cap, the closed list of datapoints in Annex II to Delegated Regulation (EU) 2026/1560, which does not include Scope 3, a statutory right that takes effect through national law due by 19 March 2027~\citep{euparl2026b,eucom2026b}. For FY2025 no rule obliged the firm to answer such a request either, so it is free to decline a customer request for this inventory.

\textbf{Tier 2a, OpenAI API.} The model is gpt-4o, the mid class by vendor pricing tier: it was priced at the standard tier of its generation, below o1 and GPT-4.5, so the tier rule gives the mid class, although it was also the default flagship and the energy reported by \citet{jegham2025} would place it at or above the frontier class, which makes this a low-side assignment (S6). OpenAI does not disclose the region, so the global average applies. Annual usage is 120 million billed tokens, about USD 525, which is EUR 465 at the 2025 annual average rate~\citep{ecb2026}; the export separates input from output at 90 and 30 million and reports no cache reads, so no cache credit is taken.

\begin{alltt}\small
T\(\sb{\mathrm{eff}}\) = (30 + 0.30 \(\times\) 90) / 1.339 = 42.6 M effective tokens  (Section \ref{sec:3.3})
Inference footprint = 0.1139 electricity + 0.00334 embodied
                      + 0.00150 construction
                    = 0.1187 kg CO\textsubscript{2}e/M effective   (Table \ref{tab:carbon})
Training            = 0.00293 kg CO\textsubscript{2}e/M billed      (Table \ref{tab:regionindep})
CO\textsubscript{2}e (central) = 42.6 \(\times\) 0.1187 + 120 \(\times\) 0.00293 = 5.4 kg CO\textsubscript{2}e
\end{alltt}

The US East variant gives 3.6 kg, and the B200 and conservative high cases are in Table~\ref{tab:casestudy}. The Tier 1 EEIO factor applied to the same spend gives $465 \times 0.1181 \times 0.8140 = 44.7$ kg on the deflated basis the method requires, about 8.3 times the Tier 2a figure (54.9 kg and 10.2 times undeflated). Data quality is 2 for activity and 3 for the factor.

\textbf{Tier 2b, ChatGPT Enterprise.} Here the activity is read from the firm's billing export rather than assumed: the workspace analytics user export~\citep{openai2026a} gives, for this illustrative firm, 18,000 messages across 50 active seats for FY2025, which is 30 per active seat per month, set for illustration equal to the Tier 2b default. The export's by-model breakdown does not cover the whole reporting year, so the class split is time-weighted from the documented default-model dates: GPT-4o (mid class) was the default from 1 January to 11 August 2025 and GPT-5 (frontier class) from 12 August, when it reached ChatGPT Enterprise workspaces five days after its 7 August launch~\citep{openai2025b,openai2025d}, giving weights $223/365 = 0.6110$ and $142/365 = 0.3890$. The export reports no reasoning share, so the measured 4.98\% at the central multiplier of 25 applies (Section~\ref{sec:3.2}, S9.7) and scales the output tokens by 2.195, which at this exchange mix raises the billed tokens per exchange 1.279 times, from 1,210 to 1,547; input tokens are not scaled. That share was measured on Claude.ai and predates the routing of default traffic to reasoning models that came with GPT-5. The only later public figure is a statement by OpenAI's chief executive that after GPT-5 the share of daily users on reasoning models rose to 24\% on the Plus tier and 7\% on the free tier~\citep{altman2025}. It is a share of users rather than of messages and does not cover Enterprise, so it is used only as two named sensitivities.

\begin{alltt}\small
Reasoning uplift = 1 + 0.0498 \(\times\) (25 \(-\) 1) = 2.195, on output tokens  (Section \ref{sec:3.2})
Output tokens    = 282 \(\times\) 2.195 = 619.0 per exchange
Billed tokens    = 928 + 619.0 = 1,547 per exchange; 18,000 \(\times\) 1,547 = 27.8 M
Effective tokens = (619.0 + 0.30 \(\times\) 928) / 1.339 = 670.2 per exchange  (Section \ref{sec:3.3})
                 = 18,000 \(\times\) 670.2 = 12.1 M
Inference footprint = 0.6110 \(\times\) 0.1187 + 0.3890 \(\times\) 0.4425
                    = 0.2447 kg CO\textsubscript{2}e/M effective
Training            = 0.6110 \(\times\) 0.00293 + 0.3890 \(\times\) 0.01284
                    = 0.00679 kg CO\textsubscript{2}e/M billed
CO\textsubscript{2}e (central) = 12.1 \(\times\) 0.2447 + 27.8 \(\times\) 0.00679 = 3.1 kg CO\textsubscript{2}e
\end{alltt}

The US East variant gives 2.1 kg. Without the reasoning uplift the line is 2.0 kg, and at the ends of the 10 to 50 multiplier range the token-based total is 7.8 and 9.7 kg. At the user shares of the later statement, held at the central multiplier, the line is 7.5 kg at 24\% and 3.6 kg at 7\%, and the token-based total 12.9 kg ($+$51\%) and 9.0 kg ($+$5\%). On the heavy-use default of 325 messages per active seat per month the line is 34.0 kg. With $-$50\%/$+$130\% on the token conversion the central value spans 1.6 to 7.2 kg. That band is retained although the message count is exported, because the tokens per message remain an estimate. Data quality is 4 for activity and 3 for the factor.

\textbf{Tier 1, Notion subscription (AI share not estimable).} Annual spend is EUR 8,000, about 34 Notion Business seats at EUR 19.50 per member per month billed annually, the plan that bundles Notion AI~\citep{notion2026}. Notion is a United States vendor and takes the Austrian factor of Section~\ref{sec:4.9}. The German factor of the same record, 0.1333 kg CO\textsubscript{2}e per EUR, is applied the same way as the second-country sensitivity of Section~\ref{sec:3.7}: 868.0 kg deflated and 1,066.4 kg undeflated (German-factor rows of Table~\ref{tab:casestudy}).

\begin{alltt}\small
CO\textsubscript{2}e (method, deflated) = 8,000 \(\times\) (104.80 / 128.75) \(\times\) 0.1181 = 769.0 kg CO\textsubscript{2}e
CO\textsubscript{2}e (undeflated)           = 8,000 \(\times\) 0.1181 = 944.8 kg CO\textsubscript{2}e
\end{alltt}

This is a screening line for the whole subscription: the Business plan does not price AI separately, so no AI figure is derived from it. Data quality is 5 for activity and 5 for the factor.

\begin{table}[htbp]
\centering
\setlength{\tabcolsep}{4pt}
\small
\begin{tabular}{>{\raggedright\arraybackslash}p{6.4cm} >{\raggedright\arraybackslash}p{0.9cm} >{\raggedleft\arraybackslash}p{1.8cm} >{\raggedleft\arraybackslash}p{2.4cm} >{\raggedleft\arraybackslash}p{2.4cm}}
\toprule
Line (kg CO\textsubscript{2}e) & Tier & Central & B200 (hardware sensitivity) & Conservative high \\
\midrule
OpenAI API & 2a & 5.4 & 3.8 & 8.6 \\
ChatGPT Enterprise (time-weighted GPT-4o / GPT-5, with the reasoning uplift) & 2b & 3.1 & 2.9 & 5.0 \\
\textbf{Token-based AI, Tiers 2a + 2b} & & \textbf{8.5} & \textbf{6.7} & \textbf{13.6} \\
\quad Parameter interval, 5th to 95th percentile ($-$18\% / $+$29\% on the mid-class factor, Table~\ref{tab:mc}, applied to the total) & & 7.0--11.1 & --- & --- \\
\quad Class range on the energy term $-$38\%/$+$24\% mid, $-$67\%/$+$55\% frontier (Table~\ref{tab:energy}); every model one class higher / lower & & 26.1 / 2.5 & --- & --- \\
\quad Tier 2b tokens $-$50\% / $+$130\% (per-message scenario, not an interval on the total) & & 7.0--12.6 & 5.2--10.5 & 11.1--20.2 \\
\quad Heavy-use seats (325 messages/seat/month) & & 39.4 & 35.4 & 63.0 \\
\quad US East variant (sensitivity; region unconfirmed) & & 5.7 & 4.5 & 8.9 \\
\quad Serving-region range over the seventeen printed regions, Sweden to Singapore (central case only) & & 1.5--8.8 & --- & --- \\
\quad No reasoning uplift, for comparison & & 7.4 & 5.6 & 11.8 \\
\quad Reasoning share 24\% / 7\% of users (sensitivity, \citealp{altman2025}) & & 12.9 / 9.0 & --- & --- \\
\textbf{Development-inclusive, $d = 1.5$ (0.67 to 2.5); a different boundary} & & \textbf{18.5} (12.7--25.5) & --- & --- \\
\quad US East variant & & 11.7 (8.1--16.1) & --- & --- \\
\midrule
Notion subscription screening line, deflated & 1 & 769.0 & 769.0 & 769.0 \\
\quad undeflated, for comparison & 1 & 944.8 & 944.8 & 944.8 \\
\quad German factor 0.1333 (second-country sensitivity, Section~\ref{sec:3.7}), deflated & 1 & 868.0 & 868.0 & 868.0 \\
\quad German factor, undeflated, for comparison & 1 & 1,066.4 & 1,066.4 & 1,066.4 \\
\bottomrule
\end{tabular}
\caption{Worked example, fiscal year 2025 (FY2025), in kg CO\textsubscript{2}e (global-average grid except where a row states otherwise, inference footprint on effective tokens plus training on billed tokens, central token weights). Totals are sums of the unrounded lines and can differ from the sum of the printed lines in the last digit. The development-inclusive line is computed line by line, each line's effective tokens times the inference footprint plus the development line, with the training term dropped (Tier 2a 11.7 kg, Tier 2b 6.8 kg at $d = 1.5$). The reasoning-share rows hold the multiplier at 25.}
\label{tab:casestudy}
\end{table}

Token-based AI emissions are 8.5 kg CO\textsubscript{2}e central (Table~\ref{tab:casestudy}), and neither the Notion screening line nor the development-inclusive figure, 18.5 kg at $d = 1.5$ (12.7 to 25.5 kg), is added to them, each being a different boundary. The serving region, which neither provider discloses, moves the central figure between 1.5 kg at the lowest-intensity printed region (EU North, Sweden) and 8.8 kg at the highest (Asia Pacific, Singapore), and the global-average row sits near the top of that range. The mean rule, the upper end of the serving-regime range (Section~\ref{sec:3.8}), would raise this firm's token-based total by about 170\% on its own model and token mix. The class of a closed model is itself an assignment: with every model one class higher than assigned (the frontier class staying frontier) the token-based total would be 26.1 kg, and one class lower 2.5 kg, and that range is what a preparer should disclose where the class rests on the tier name alone. The method matters where token volumes are large, and this firm's are not: for a firm like this the AI inventory is a documentation and methodology task rather than a magnitude one.

\textbf{Water for the same firm.} The 54.6 million effective tokens behind those figures carry 32.3 litres of water with hydropower reservoir evaporation excluded and 57.2 litres with it allocated to the electricity, of which 7.3 litres is on-site cooling water. These figures use the global-average row of Table~\ref{tab:water}, whose on-site value is not confirmed, as an illustrative proxy and are stated as such. Water has no unknown-region fallback (S12.2), so in its disclosure the firm would report water for an undisclosed region as not available. At US East, the one confirmed region shown, the same usage is 32.0 and 38.4 litres volumetric with 0.5 litres on-site, and the on-site water scarcity footprint is 0.9 litres world-equivalent.

\section{Discussion}\label{sec:7}

\subsection{What Drives the Result}\label{sec:7.1}

\textbf{Model class and region are of the same order, and which dominates depends on the boundary.} On the electricity term alone the small-to-frontier ratio is 12.2 and the ratio between the cleanest and the most carbon-intensive region printed here, Sweden and Singapore, is 13.0. The grid-independent terms compress the regional spread to 8.8 times on the mid-class inference footprint while the class spread stays at 12.3 times, so a preparer needs both: either alone leaves about an order of magnitude open, and the two together span a factor of about a hundred (S2).

\textbf{Where the central value sits.} Per million billed tokens, the mid-class factor at the world-average grid is 0.0388 kg CO\textsubscript{2}e, the low end of a stated structural range rather than a best guess with symmetric error (Table~\ref{tab:structrange}). On the serving-regime axis the same quantity at the mean rule is 2.8 times it for this class and region. On the boundary axis, a different quantity, the development-inclusive total is 1.5 to 2.9 times it over $d = 0.67$ to 2.5, the development line alone being 0.0210 to 0.0783 kg. The parameter interval, $-$18\% to $+$29\%, is narrower than either axis and is never combined with them. The published per-query estimates for the mid class lie above the central value and mostly beyond the upper end of the range (Section~\ref{sec:5.1}). Current practice sits further off. On the same purchase the spend-based figure is 1.7 to 12.3 times the physical estimate on deflated spend (2.0 to 15.1 times undeflated), from gpt-4o-mini against the small class at the global-average grid to gpt-4o against the mid class in US East. None of these ratios measures an error in either figure (S2).

\textbf{The orderings hold across the range, and clean grids have a floor.} The three classes keep their order at the central operating point, in the conservative high case and at the mean rule (Table~\ref{tab:energy}), in every column of Table~\ref{tab:structrange} and in every region of Table~\ref{tab:b200high}, and the training-high row cannot reverse it, because the training term it raises is itself ordered by class (Table~\ref{tab:regionindep}). The regions keep their order because a region enters only through its grid intensity, which scales the electricity term and the development line, while the embodied, construction and training terms are the same in every region and so move every row by the same amount. Embodied hardware and construction add 4.3\% to the mid-class electricity term at the global average and 53.5\% in Sweden (Table~\ref{tab:nonelec}), and with the training term they are the larger part of the Swedish factor per million billed tokens in all three classes, whereas at the world average the factor is 87\% to 91\% electricity (Figure~\ref{fig:composition}). Sweden still stays the lowest row in the central, B200 and conservative high columns, and as grids decarbonise the factor falls towards a floor set by embodied hardware, construction and training rather than towards zero. What changes along the range is the size of a gap, not its direction. The EU-27 mid-class factor is 0.0212 against 0.0388 at the global average centrally, and 0.0565 against 0.1142 at $d = 2.5$. A choice between two classes or two regions made on the central factor is therefore the same choice at any point of the range. The level itself is known only to within a factor of about three on each structural axis, before the class range and the class assignment are added.

\textbf{What reporting development separately means in practice.} Keeping model development outside the central factor has four consequences for a preparer. First, the figure reported in Category 1 for purchased inference is the central value, and the development line is disclosed beside it as a separately labelled amount on its own boundary, never summed into the Scope 3 total. In the worked example the token-based figure is 8.5 kg (13.6 kg in the conservative high case) and the development-inclusive total is 18.5 kg, or 12.7 to 25.5 kg over the development ratios (Table~\ref{tab:casestudy}). Second, a figure from another source that includes development, such as a study that charges research compute to inference, can be 1.5 to 2.9 times the central value for the mid class at the global average from the boundary difference alone (Table~\ref{tab:structrange}), without either figure being wrong, so any comparison is made on the matching boundary. Third, reductions are tracked on the central value. The development line scales with a sector ratio, $d$, that no customer controls, so a change in $d$ between reporting years reflects the sector rather than the preparer and is handled as a methodology change. Its class split also understates the development of frontier models (Section~\ref{sec:3.4}), so it should not guide a choice between models. Fourth, where the inventory serves a materiality assessment or a target meant to cover the full upstream footprint, the development-inclusive total is the level compared with the threshold or the target, while year-on-year progress is still read from the central value, and the preparer states which of the two boundaries the target uses. The method takes no position on whether a future standard will place development inside Category 1. If one does, the development-inclusive total is already computed on that boundary and can replace the central value without new inputs, with the two properties that Section~\ref{sec:3.4} states for it.

\subsection{Limitations, Largest First}\label{sec:7.2}

The first is the pair of choices behind the low-end placement of the central value. The next two are ordered by their effect on a reported carbon figure (Table~\ref{tab:sens}), and the last two are of a different kind. Three lower-severity limitations, on the water conventions, reasoning models and the location basis, are in S2.

\textbf{The operating point and the training denominator.} The mean rule instead of the minimum raises the mid-class factor per million billed tokens at the global average by 176\%, the 2.8 times of Table~\ref{tab:structrange}, and the inference footprint per effective token by 190\% (S9.1). The direction of the rule chosen is stated, and the mid-class per-query estimates lie beyond even the mean rule in four of the five energy rows (Section~\ref{sec:5.1}), but no external evidence fixes which serving regime a commercial provider actually runs. The training denominator rests on one disclosed per-model volume whose own token basis the source does not state. Training is 8\% of the mid-class factor per million billed tokens, so a tenfold larger denominator would lower that factor by 7\% and a tenfold smaller one would raise it by 68\%. The development-inclusive total does not use that denominator at all and is printed beside the factor as a different boundary. A third choice of the same kind is smaller. The idle uplift rests on one provider disclosure, and its low-utilisation pair adds 8.8\% to the factor.

\textbf{The class is the unit of estimation for closed models.} No closed commercial model from OpenAI, Anthropic, Google, xAI or Amazon discloses a size or an energy figure, so every closed commercial model is class-only. Google's open-weight Gemma models do, and carry their own factors in the dataset, as do the other sixteen measured models. One class of misclassification is worth 3.3 times on the electricity term from small to mid and 3.7 times from mid to frontier, and two classes 12.2 times. A model-specific factor for a closed model would have to be inferred from its price, its observed throughput or its vendor tier and would carry at least the uncertainty of the class range, so none is published. The frontier-class mean rests on three open-weight models whose range is $-$67\%/$+$55\%, and that range is the limit of the evidence for closed frontier models.

\textit{Reading a class-based estimate.} Where the provider of a commercial model discloses neither its size nor its energy use, the factor is the mean of the measured open-weight models of the class to which the vendor's pricing tier is assigned, at the stated operating point, not a figure for the model itself, and a preparer should read and present it that way. Four practical rules follow. First, the class and the basis of the assignment are stated with the figure, and the values one class higher and one class lower are reported beside the central value. In the worked example they move the token-based total from 8.5 kg to 26.1 kg and 2.5 kg (Table~\ref{tab:casestudy}). Second, a difference between the services of two vendors that arises only from their class assignments is not evidence of a difference in emissions, whereas a switch between pricing tiers of the same vendor is a change of model that the method does show: its direction holds because the classes keep their order across the whole range (Section~\ref{sec:7.1}), while its size is only as certain as the class step. Third, a change of class assignment that follows a vendor's disclosure of a size or a revision of the mapping is a methodology change under the base-year recalculation policy and not a reduction (Section S12.3). Fourth, where no class can be determined, the highest-intensity class of Section~\ref{sec:3.5} applies, so the figure leans high and is flagged. Read in this way, a class-based figure supports screening, materiality assessment and the choice between model tiers, while its absolute level carries the class range and the adjacent-class values in addition to the parameter interval and the structural range of Section~\ref{sec:3.8}. A disclosed size fixes the class and so removes the adjacent-class values, but the class range remains until the model has an energy figure of its own.

\textbf{The systematic high case and the token basis.} The compounded high case is $+$64\% on the footprint and $+$59\% on the convenience factor. It compounds the high ends of three facility uplifts at the central operating point, so it is neither a bound on the chain nor a confidence interval. $W_{\mathrm{cache}}$ has no energy measurement behind it, and it enters both the customer's token count and the restatement of each benchmark run, whose own token mix and cache use are measured. Energy per output token also rises with the length of the context each decode step attends over, and a reasoning trace lengthens that context, so a per-token factor measured at the benchmark's prompt lengths understates a long reasoning interaction by an amount not quantified here. The rest of the token basis is in S2.

\textbf{No validation against metered service-level energy.} No check in Section~\ref{sec:5} compares predicted with metered energy, one of them partly shares inputs with the method, and the worked examples are illustrations. The test that would close this is a metered deployment, one open model served on H100 nodes with per-request token logs and node power readings, against which the predicted energy is compared. Such a study has been proposed to a supercomputing centre and is not part of this paper.

\textbf{Data vintages.} The grid intensities are for calendar year 2023, the energy measurements were published in February 2026, and the worked examples are for fiscal year 2025. A reporter should take the grid year that matches the reporting year where one is published (Section S12.3). Recomputing every printed region with the 2024 grid intensities, with eGRID 2024 not yet published so that the US rows are scaled uniformly by Ember's national change, lowers every factor (Table~\ref{tab:gridyear} in S2): the inference footprint by 0.4\% (Singapore) to 17.3\% (France), the factor per million billed tokens by 0.4\% to 11.1\%, and the global-average billed factor by 2.2\%, the size of change to expect from one grid year to the next. The spread between model classes does not depend on the grid year, and the spread between regions does. The intensities are also annual averages. An hourly or a marginal intensity, which moved a cloud training job's emissions by up to 8\% with the hour of the day in the measurements of \citet{dodge2022}, is not used here, so the regional spread of the tables is a spread between annual averages and a provider's hour-by-hour placement of load is invisible to the method (Section~\ref{sec:7.4}).

\subsection{Which Provider Disclosure Would Replace Which Input}\label{sec:7.3}

The binding constraint on precision is not the energy benchmarks or the grid intensities but what providers publish: token counts per model per billing period, the infrastructure region serving an account, energy per million tokens for the model served, a per-model lifetime serving volume and the share of inference served from renewable energy would each retire a named input or fallback of this method, from the Tier 2b activity band to the global-average region and the single datapoint behind the training denominator. The EU AI Act creates indirect pressure towards the third of these through the energy documentation of Annex XI, which reaches the authorities rather than customers (Section~\ref{sec:2}; \citealp{euparl2024}), and on 7 April 2026 the European Commission opened a targeted consultation towards a framework for measuring the energy consumption and emissions of AI models and systems and a possible AI energy and emission label~\citep{eucom2026c}. Either route ends in an energy figure per model, not a carbon figure for a customer's usage, so it would replace an input and not the method. Where it covers serving it would replace the accelerator energy and, if stated at the facility, the host, idle and PUE uplifts, and where it covers development, as Section~\ref{sec:2} reads the Annex XI datapoint, it would inform the training term. The activity data, the token basis, the region rule, the remaining terms and the reporting rules would all still be needed, so such a disclosure would feed the hybrid case of Tier 2a and, once a supplier converts it into a figure for the customer's usage, Tier 3. Supplementary S4 gives the roadmap, disclosure by disclosure and tier by tier.

\subsection{Relation to Existing Standards}\label{sec:7.4}

Five points place the method among existing standards and models. It fits the GHG Protocol as an application of the four Category 1 calculation methods of the Technical Guidance to one kind of purchase, each tier corresponding to one method (Section~\ref{sec:3.2}), and no standard supplies a preparer's factors and activity data for AI inference. The Software Carbon Intensity specifications and International Telecommunication Union Telecommunication Standardization Sector (ITU-T) Recommendation L.1801 address AI assessment without supplying either (S4).

Under ESRS E1 the result is a gross Scope 3 Category 1 figure, and the reporting rules of Section~\ref{sec:3.8} are written for the CSRD and the ESRS. The same factors and activity-data rules would serve an International Financial Reporting Standards (IFRS) S2 or a California disclosure, but the reporting rules of Section S12.3 are not mapped to them.

The factor is an annual average for a region, which is the basis the inventory asks for. An organisational inventory under the GHG Protocol is kept for a year and accounts electricity on a location basis, with the market-based figure of Scope 2 dual reporting beside it. How much one additional query would emit at a given place and hour is a different question, of dynamic routing and marginal grid response, and answering it needs the provider's hardware and utilisation data tied to hourly grid conditions, which the preparer does not hold. \citet{bistline2026} leave hourly matching, dynamic load balancing and the choice of a long-run marginal rate as unresolved questions, and recommend location- and market-based figures with a marginal rate at most as an optional addition. The method leaves the marginal and hourly question out of scope, and the location basis it does use is a stated limitation (S2).

Under a partnership with ISO announced in September 2025, the GHG Protocol is consolidating its standards into a single Corporate Standard, Version 3.0, expected to be co-branded with ISO, with a draft for consultation estimated for the second quarter of 2027 and publication estimated for the fourth quarter of 2028. The existing standards stay in effect until the GHG Protocol says otherwise~\citep{ghgp2026}, and ISO 14064-1:2018 is itself listed for revision~\citep{iso2018}. The allocation, market-based and data-quality rules of Sections~\ref{sec:3.1} and \ref{sec:3.8} would need revisiting against the consolidated text. The Scope 3 technical working group's draft meeting papers, marked not for citation, discuss digital products under Category 11, give energy per AI token or per floating-point operation as examples of computation energy, and note that AI workloads may require separate and more detailed methods~\citep{ghgp2026}, which is the question this method is a candidate answer to.

Read as a model rather than against a standard, the method is parametric in the sense of parametric life cycle assessment, in structure only: a preparer or provider substitutes its own parameters, such as a metered PUE or a contracted region, and obtains a factor on the same boundary. S2 sets out that reading, the extensions it allows and what it does not supply, which is evidence for the parameter values of closed services.

\subsection{Water as an Outlook}\label{sec:7.5}

The same chain yields water factors for seventeen regions, eleven of them with a confirmed total (S3). They are an outlook rather than a disclosure for one reason. The revised ESRS rename E3 ``Water'' and ask for six water metrics expressly ``for its own operations''~\citep{eucom2026a}, so a purchased-inference water figure has no datapoint in ESRS E3 and could appear only as an entity-specific value-chain metric (ESRS 1, paragraph 11 and AR 38). It serves a materiality assessment, a customer data request or an internal target, and not a reporting line.

\section{Conclusions}\label{sec:8}

Purchased AI inference is an ordinary purchased service, and a preparer should be able to account for it from a billing export with the same discipline as any other Category 1 line. The method set out here allows that with the components listed in Section~\ref{sec:1}, from a cradle-to-gate boundary and four tiers of activity data to a four-term factor, a resolution rule with named fallbacks, rules that keep three kinds of uncertainty apart and a maintenance protocol for every input. Model development stays outside the factor as a labelled line on its own boundary. For the mid class at the global average that line alone is 0.54 to 2.0 times the factor per million billed tokens over the development ratios available, and the development-inclusive total is 1.5 to 2.9 times it.

The checks available are consistency and plausibility checks, one of them an identity rather than a test (Section~\ref{sec:5.2}), and none reaches the accelerator-energy term. Of four published per-query energy figures (Google's at two token counts), the three that can describe a frontier-class model bracket its value and the three that can describe a mid-class model are 2.4 to 7.2 times the mid-class one, above the top of the serving-regime range in four of five rows, so for the mid class the central value is more likely low than the range alone suggests. The unresolved terms of the chain lean low, as the bias column of Table~\ref{tab:master} records, and the alternative operating-point rule, the largest structural choice, gives a value well above the central one, so the central figure is a lower-leaning estimate rather than a best guess with symmetric error. Several input bands and structural alternatives reach below it, by up to 33\% on the mid-class factor per million billed tokens for the input-token weight and 36\% for the lowest-footprint member of a class (Table~\ref{tab:sens}), and the input-token weight moves the factor per million billed tokens furthest down of any input.

None of these checks is a validation. The factors have not been compared with metered energy for a real inference workload (Section~\ref{sec:7.2}), so they are internally consistent estimates built from published measurements and stated assumptions, and their uncertainty statements describe the spread of those inputs and choices rather than a demonstrated error against observed energy. Until such a comparison exists, the method is suited to estimating and disclosing purchased inference with the five uncertainty parts of Section~\ref{sec:3.8} stated, to materiality assessment and to choosing between model tiers and regions. Its results should be presented as estimates, never as measured values. A metered deployment of one open model on H100 nodes, with per-request token logs and node power readings, is the next step. It would test the accelerator-energy term that no present check reaches, though not the serving regime of any commercial provider.

What the method cannot do is replace provider disclosure. Every closed commercial model is estimated from a class, the training denominator rests on one disclosed number, and the serving region is usually unknown. Token counts and a region in a billing export would move most seat-licensed products from an estimated activity quantity to a counted one and retire the global-average fallback, and energy per token would retire the longest chain of proxies in the method. Until then, the factors, the evidence files and a reference calculator are released with this paper so that any preparer, auditor or researcher can reproduce, criticise or replace any factor and table, and the activity defaults can be re-derived from the named third-party sources with analysis scripts available from the author.

\section*{CRediT authorship contribution statement}

\textbf{Guillermo Llopis-Garcia:} Conceptualization, Methodology, Software, Validation, Formal analysis, Investigation, Data curation, Writing -- original draft, Writing -- review \& editing, Visualization.

\section*{Declaration of competing interest}

The author is a founder of SOMA AI, which develops sustainability-reporting software that applies this methodology, and is the author of the dataset~\citep{llopis2026}. The three concurrent methods compared in Table~\ref{tab:requirements} are published by organisations that compete commercially with SOMA AI; every cell about them is drawn from their published material on the access dates given in the references, and the requirements against which they are scored are this paper's own.

\section*{Funding}

This research did not receive any specific grant from funding agencies in the public, commercial, or not-for-profit sectors.

\section*{Declaration of generative AI and AI-assisted technologies in the manuscript preparation process}

During the preparation of this work the author used Claude (Anthropic) in order to draft and edit prose, to write and check the analysis and verification scripts, and to retrieve and cross-check published sources. After using this tool, the author reviewed and edited the content as needed, verified every number, source attribution and claim against the primary material, and takes full responsibility for the content of the published article.

\section*{References}
\small
\sloppy
\raggedright

\input{references_main}
\fussy

\section*{Data Availability}
\small
\begin{sloppypar}
All emission and water factors reported in this paper are released with it as an open dataset under CC BY 4.0~\citep{llopis2026}, deposited on Zenodo as version 3.0.0 at \href{https://doi.org/10.5281/zenodo.23371447}{10.5281/zenodo.23371447}, under the concept DOI \href{https://doi.org/10.5281/zenodo.20443585}{10.5281/zenodo.20443585}, which always resolves to the latest version; the record at 10.5281/zenodo.22767475 remains for the superseded values. The version is also identified by its dataset date and by the model-class mapping's assignment months, and the reporting rule of Section S12.3 is applied on the version DOI. The file list below follows \texttt{MANIFEST.md}, whose machine-readable form is \texttt{deposit\_\allowbreak manifest.csv}; in the file names, ``lifecycle'' denotes the four-term boundary of Section~\ref{sec:3.1}.

\emph{Factor tables.} \texttt{factors\_\allowbreak v3\_\allowbreak lifecycle.csv} (102 rows: one per model class, accelerator type and printed region, with every term named by its denominator, the inference footprint, the convenience factor, the conservative high case, the development line and about thirty water columns); \texttt{model\_\allowbreak factors.csv} (32 rows: one per measured open-weight model and accelerator type, 18 distinct models, the only models with their own factor); \texttt{grid\_\allowbreak intensity\_\allowbreak all.csv} (240 rows: 210 countries and economies, 27 eGRID subregions, one provincial row for Quebec and the world and EU-27 aggregates); \texttt{cloud\_\allowbreak region\_\allowbreak factors.csv} (143 rows: every public AWS, Azure and Google Cloud region, each resolved to a grid row with a mapping confidence and a source URL); \texttt{term\_\allowbreak constants.csv} (the grid-independent terms per class); \texttt{reserved\_\allowbreak capacity\_\allowbreak factors.csv} (the factor per reserved accelerator-hour per class); \texttt{lifecycle\_\allowbreak sensitivities.csv} (30 rows: ten one-at-a-time treatments per class at the global-average grid).

\emph{Lookup tables.} \texttt{model\_\allowbreak class\_\allowbreak mapping.csv} (187 identifier patterns to a class or an out-of-scope marker, with the basis, the own-factor model where there is one, the provider page each was read from and the month assigned; the last row is the \texttt{*} fallback that makes the unmapped-identifier rule data rather than prose); \texttt{token\_\allowbreak basis\_\allowbreak constants.csv} (the weights, $R_{\mathrm{bench}}$, $D$, the default input share, the default multiplier and the default class for an unmapped identifier); \texttt{region\_\allowbreak grid\_\allowbreak crosswalk.csv} (joins the printed region identifiers to the grid identifiers and names the unknown-region and EU-only defaults); \texttt{region\_\allowbreak water\_\allowbreak factors.csv} (the selected on-site value, its denominator and published basis, both off-site factors, the high-hydropower sensitivity and the scarcity factor, per printed region).

\emph{Scripts.} \texttt{recompute\_\allowbreak v3\_\allowbreak lifecycle.py} is the single definition of every factor term and writes \texttt{factors\_\allowbreak v3\_\allowbreak lifecycle.csv}, \texttt{tables\_\allowbreak v3\_\allowbreak lifecycle.md}, \texttt{headline\_\allowbreak v3\_\allowbreak lifecycle.md}, \texttt{numbers\_\allowbreak v3\_\allowbreak lifecycle.json} and \texttt{water\_\allowbreak v3\_\allowbreak lifecycle.md}. \texttt{recompute\_\allowbreak v3\_\allowbreak final.py} holds the superseded definitions, is retained for diffing and is imported by the first for its unchanged pieces. \texttt{build\_\allowbreak detailed.py} writes every detailed and lookup table except the model mapping, importing the first so that no constant is duplicated, and runs 558 cross-checks. \texttt{ai\_\allowbreak inference\_\allowbreak emissions.py} is the reference calculator and library: dependency-free, standard library only, reading only the published CSV files, taking a model identifier, a region and a token count and returning kg CO\textsubscript{2}e with the file, line and column behind every number; \texttt{-{}-selftest} runs 447 internal checks. \texttt{tests\_\allowbreak end\_\allowbreak to\_\allowbreak end.py} runs 589 end-to-end checks. \texttt{verify\_\allowbreak printed\_\allowbreak numbers.py} checks every number printed in this paper against the generated dataset, table cell by table cell, and checks the figures restated in the abstract, introduction, discussion, conclusion and suggested disclosure text against the table values they come from. \texttt{refresh\_\allowbreak check.py} re-downloads the machine-readable inputs named in the register and reports what has changed, per row, with the recompute action each change triggers. \texttt{recompute.py} is the superseded v2 script, retained for the version history. The model mapping has no generator: it is maintained by hand from provider model lists, and \texttt{tests\_\allowbreak end\_\allowbreak to\_\allowbreak end.py} fails if any identifier in the deposited service-life list or in the OpenRouter top 150 by tokens falls to the Class C default.

\emph{Generator inputs}, without which the dataset cannot be rebuilt: \texttt{mlenergy\_\allowbreak v3\_\allowbreak lm-arena-chat\_\allowbreak 2026-09-29.json} (the ML.ENERGY Leaderboard v3 \texttt{lm-arena-chat} snapshot, data of 16 February 2026, downloaded 29 September 2026; the source of every joule per output token); \texttt{ember\_\allowbreak yearly\_\allowbreak full\_\allowbreak release\_\allowbreak long\_\allowbreak format.csv} (Ember yearly electricity data, full release); \texttt{egrid2023\_\allowbreak subregions.csv} (transcribed from the EPA eGRID2023 summary tables, Table 1); \texttt{cloud\_\allowbreak regions.csv} (compiled from the published region lists of the three providers, each row mapped by hand with a confidence and a source URL); \texttt{factors\_\allowbreak v3\_\allowbreak final.csv} and \texttt{factors\_\allowbreak v3\_\allowbreak candidate.csv} (read only by the back-compatibility check and by the superseded script); and in \texttt{v4\_inputs/}, \texttt{water\_\allowbreak onsite\_\allowbreak selected.csv}, \texttt{aware\_\allowbreak factors.csv}, \texttt{tier2b\_\allowbreak defaults.csv}, \texttt{crosscheck\_\allowbreak cost\_\allowbreak to\_\allowbreak tokens.csv} and \texttt{model\_\allowbreak service\_\allowbreak life.csv}.

\emph{Evidence files}, in \texttt{v4\_inputs/}: \texttt{overhead\_\allowbreak evidence.md} (the uplifts, PUE and the construction constant, with every candidate value and why the central was chosen); \texttt{service\_\allowbreak life\_\allowbreak summary.md} and \texttt{VERIFICATION\_\allowbreak NOTES\_\allowbreak lifetime.md}; \texttt{token\_\allowbreak weighting\_\allowbreak evidence.md}; \texttt{tier2b\_\allowbreak evidence.md}; \texttt{ewif\_\allowbreak sources.md}, \texttt{hydro\_\allowbreak scenarios.md} and \texttt{VERIFICATION\_\allowbreak NOTES.md}; \texttt{water\_\allowbreak onsite\_\allowbreak sources.csv}; \texttt{mlenergy\_\allowbreak token\_\allowbreak basis.csv} and \texttt{mlenergy\_\allowbreak token\_\allowbreak basis.txt}; the measured per-request token distributions and reasoning share, with the raw snapshots not deposited and the analysis scripts available from the author; and at the deposit root \texttt{assumptions\_\allowbreak A.md}, \texttt{assumptions\_\allowbreak B.md} and \texttt{assumptions\_\allowbreak C.md} (the per-assumption source audit) and the three independent reviews of an intermediate draft, \texttt{review\_\allowbreak lca.md}, \texttt{review\_\allowbreak ghg.md} and \texttt{review\_\allowbreak editorial.md}.

\emph{Raw corpora.} \texttt{v4\_inputs/raw/}, which is not deposited, holds the material the analysis scripts read. It contains the Azure LLM inference traces~\citep{azure2024} and the WildChat-1M and LMArena samples~\citep{zhao2024,mlenergy2026}. It also contains the Anthropic Economic Index extract~\citep{anthropic2025} and the OpenRouter rankings snapshots~\citep{openrouter2026}, the last of which is also the usage list the model mapping is tested against. These are needed only to re-run the analysis scripts, not the two generators, and each keeps its own licence and attribution requirement; they are therefore not deposited, and \texttt{THIRD\_\allowbreak PARTY\_\allowbreak DATA.md} names each source, its licence, the attribution it requires, the exact file or interface and the date retrieved.

\emph{Generated reports and maintenance.} \texttt{tables\_\allowbreak v3\_\allowbreak lifecycle.md} (the generated form of every table here), \texttt{numbers\_\allowbreak v3\_\allowbreak lifecycle.json} (every number printed here, each with its superseded value and the rounded-as-printed form the figures use), \texttt{headline\_\allowbreak v3\_\allowbreak lifecycle.md}, \texttt{water\_\allowbreak v3\_\allowbreak lifecycle.md} and \texttt{refresh\_\allowbreak report.md}, with the four superseded reports \texttt{tables\_\allowbreak v3.md}, \texttt{tables\_\allowbreak v3\_\allowbreak final.md}, \texttt{headline\_\allowbreak v3.md} and \texttt{headline\_\allowbreak v3\_\allowbreak final.md} retained for diffing; \texttt{input\_\allowbreak register.csv} (70 rows, one per individual input, grouped into the fifteen families of Table~\ref{tab:refresh}; 55 checked by script and 15 manual), \texttt{REFRESH.md} and \texttt{refresh\_\allowbreak state.json}. \emph{Documentation}: \texttt{README.md}, \texttt{MANIFEST.md}, \texttt{THIRD\_\allowbreak PARTY\_\allowbreak DATA.md}, \texttt{DETAILED\_\allowbreak TABLES\_\allowbreak README.md} (the data dictionary: every published file and column, with units, denominators, the identifier normalisation rule and the fallbacks) and \texttt{V3\_\allowbreak LIFECYCLE\_\allowbreak SPEC.md} (the computation spec, whose later sections override earlier ones).
\end{sloppypar}
\normalsize
\newpage
\normalsize
\setcounter{section}{0}
\setcounter{table}{0}
\setcounter{figure}{0}
\renewcommand{\thesection}{S\arabic{section}}
\renewcommand{\thesubsection}{S\arabic{section}.\arabic{subsection}}
\renewcommand{\thetable}{S\arabic{table}}
\renewcommand{\thefigure}{S\arabic{figure}}
\clearpage
\pagenumbering{arabic}
\renewcommand{\thepage}{S\arabic{page}}
\resetlinenumber
\makeatletter
\def\@extra@b@citeb{@S}
\let\NAT@wrout@main\NAT@wrout
\def\NAT@wrout#1#2#3#4#5{\NAT@wrout@main{#1}{#2}{#3}{#4}{#5\@extra@b@citeb}}
\makeatother
\begin{center}
{\Large\bfseries Supplementary Material for: Accounting for Purchased Artificial Intelligence Inference in Corporate Greenhouse Gas Inventories: A Token-Based Method}\\[12pt]
{\normalsize\bfseries Guillermo Llopis-Garcia}\\[4pt]
{\small SOMA AI, Barcelona \quad \texttt{guillermollopis@somaai.earth}}
\end{center}
\vspace{6pt}
Section, table and figure numbers prefixed S refer to this document; unprefixed ones refer to the main text.
\medskip

\section*{\large Supplementary Material}
\rule{\linewidth}{0.8pt}
\medskip

\textit{These sections support the main paper: technical derivations (S1), the full sensitivity table and the limitations (S2), the water rankings and their sensitivities (S3), the provider transparency roadmap (S4), the decision rules and the refresh schedule (S5), the model class mapping (S6), the register of assumptions (S7), the version history (S8), the parameterisation of every input (S9), further consistency checks (S10), a second worked example with suggested disclosure text (S11) and the boundary justification, water method and reporting rules (S12). References cited in these sections are listed at the end of this document. Version labels are as follows. v1 is arXiv:2606.10660v1 (June 2026) with dataset v1. v2 is the September 2026 journal-submission manuscript with dataset v2 (10.5281/zenodo.22767475), not posted to arXiv. v3 denotes the factors of the present version.}

\subsection*{S1. Factor Derivation Details}

\subsubsection*{S1.1 Accelerator Energy to Facility Energy}

Accelerator energy measurements from ML.ENERGY Leaderboard v3 (joules per output token) are converted to facility Wh per 1,000 effective tokens as follows.

\begin{alltt}\small
GPU Wh/1k tokens      = (GPU J/token \(\times\) 1{,}000) / 3{,}600
Server Wh/1k tokens   = GPU Wh/1k \(\times\) H \(\times\) I,   H = 1.50, I = 1.10
Facility Wh/1k tokens = Server Wh/1k \(\times\) PUE, PUE = 1.20 (both platforms)
\end{alltt}

Every joule-per-token value used is a direct measurement from the leaderboard (\texttt{lm-arena-chat} task, vLLM, bfloat16 unless marked FP8, data of 16 February 2026), restated to the reference mix by $D/D_{\mathrm{run}}$ (Section S9.6), and none is extrapolated. ML.ENERGY reports energy per \emph{output} token at each model's own response length on the same prompts, and response lengths differ by model family, so cross-family comparisons mix verbosity with model size.

\textbf{What the published quantity contains.} The benchmark defines energy per token as the whole accelerator energy of a request divided by the number of output tokens of that request, so the energy of processing the input is inside the numerator while only output tokens are in the denominator~\citep{chung2025}. The identity holds exactly on the export used here: dividing \texttt{energy\_per\_request\_joules} by \texttt{avg\_output\_len} reproduces \texttt{energy\_per\_token\_joules} for all 328 configurations, with a maximum absolute deviation of zero. The leaderboard export carries no input-length field, but the benchmark's raw run files do: each run has a result file with the client-side token counts of every request and a dump of the serving engine's metrics, which gives the server-side prompt tokens, generated tokens, prefix-cache queries and prefix-cache hits. The task builds the prompt of each turn by concatenating all earlier user prompts and model responses~\citep{mlenergy2026} and sends each turn as its own request, with 1,024 distinct prompts per run. Across the 328 configurations the server-side ratio of prompt to generated tokens is 0.47 to 2.01 (median 1.06, mean 1.22), and it differs by model because output length does. An earlier draft reconstructed one ratio, 1.13, by replaying the task on a sample of the source dataset (mean input 727 tokens against a mean output of 641); that value is kept as the reference mix and is no longer evidence about any run.

\begin{table}[htbp]
\centering
\setlength{\tabcolsep}{3pt}
\tiny
\begin{tabular}{>{\raggedright\arraybackslash}p{0.95cm} >{\raggedright\arraybackslash}p{2.3cm} >{\raggedright\arraybackslash}p{1.9cm} >{\raggedright\arraybackslash}p{1.0cm} >{\raggedright\arraybackslash}p{1.0cm} >{\raggedright\arraybackslash}p{3.0cm} >{\raggedright\arraybackslash}p{3.4cm} >{\raggedright\arraybackslash}p{0.6cm}}
\toprule
Symbol & Meaning & Central (A / B / C) & Low & High & Source & Bias in central & Sec. \\
\midrule
$E_{\mathrm{GPU}}$ & Accelerator J per output token at the reference mix, which equals J per effective token, H100, class mean & 0.1309 / 0.4285 / 1.5958 & $-$40 / $-$38 / $-$67\% & +47 / +24 / +55\% & ML.ENERGY v3~\citep{mlenergy2025} & Low (accelerator only; a best-case serving point applied as a fleet average) & S9.1 \\
$H$ & Host uplift on accelerator energy & 1.50 & 1.43 & 1.88 & Google~\citep{elsworth2025} (low); \citet{latif2025}, node over measured accelerator power (central); \citet{schneider2025} Table 1 (high) & Unknown & S9.2 \\
$I$ & Idle provisioned-capacity uplift & 1.10 & 1.05 & 1.20 & Google~\citep{elsworth2025} (central; 1.11 to 1.13 from Google's printed percentage shares); \citet{oviedo2026}, point estimate $\approx$1.19 (high); author sensitivity (low) & Low & S9.2 \\
PUE & Facility overhead & 1.20 & 1.09 & 1.45 & Author choice; above the disclosed fleet values of \citet{google2026a}, AWS~\citep{amazon2024} and \citet{microsoft2026a} and below the Uptime industry average~\citep{uptime2024}; 1.45 is the top of the LBNL 2024 band~\citep{shehabi2024} & Unknown; low if any serving is outside hyperscale facilities & S9.2 \\
$g$ & Grid intensity, kg/kWh, lifecycle & Table~\ref{tab:carbon}; global 0.4832 & --- & --- & \citet{ember2026}; eGRID2023~\citep{epa2025} & High for EU rows (2023 year) & S9.3 \\
$M_{\mathrm{mfg}}$ & Server manufacturing, kg & 3,300 & 2,872 & 4,241 & \citet{dell2025} PCFs; \citet{nvidia2025a} & Low & S9.4 \\
$M_{\mathrm{tr}}$ & Server transport, kg & 672 & 427 & 712 & \citet{dell2025} PCFs; \citet{schneider2025} & Slightly low & S9.4 \\
$L$ & Hardware life, years & 5 & 4 & 6 & Operator 10-K filings~\citep{secfilings2026};~\citep{schneider2025};~\citep{dell2025};~\citep{gamazaychikov2026} & Unknown; a shorter life raises the embodied term (+25\% at 4 years, $-$17\% at 6). Six years is Alphabet's general life and the upper end of the Amazon and Microsoft classes & S9.4 \\
$U$ & Utilisation, tied to $I$ by $I = 1 + 0.30(1-U)/U$ & 0.75 & 0.60 & 0.857 & Derived from $I$~\citep{elsworth2025} at an assumed idle power of 0.30 of serving power & Unknown; lower utilisation raises the embodied term (+25\% at 0.60) & S9.2, S9.4 \\
$T_8$ & Output tokens/s per 8-accelerator server, class harmonic mean of the members' measured rates, restated to the reference mix and scaled to eight accelerators & 37,865 / 10,044 / 2,615 & --- & --- & \citet{mlenergy2025}, the member points of Table~\ref{tab:classmembers} & Measured, but the scaling to eight accelerators assumes independent replicas & S9.4 \\
$k_c$ & Construction, g/kWh of \emph{server} (IT) electricity & 7.66 & 5.3 & --- & \citet{schneider2025} Table 1 restated per machine kWh; low from Carbon Trust assumptions~\citep{carbontrust2026} & Unknown; 7.66 is the top of the five-machine range, 7.58--7.66 & S9.4 \\
$E_{\mathrm{train}}$ & Final-run training, tCO\textsubscript{2}e & 420 / 2,040 / 8,930 & --- & --- & Meta Llama 3.1~\citep{meta2026} & Low (final runs only) & S9.5 \\
$S$ & Model service life, months & 12 & 6 & 18 & \citet{anthropic2026a}; \citet{aws2026a}; \citet{microsoft2026b} & Unknown; a shorter life raises the training term (+7.6\% at 6 months, $-$2.5\% at 18 on the mid-class factor per M billed) & S9.5 \\
$V$ & Tokens served per day, per model & $1.905 \times 10^{12}$ & --- & --- & GPT-5-Codex disclosure~\citep{openai2025a} & Low on the training term, the rate itself being high (an early-life rate, and the token definition is not stated in the source). Shape factor 0.4 as a sensitivity & S9.5 \\
$d$ & All development, final runs included, to serving energy & 1.5 & 0.67 & 2.5 & \citet{gamazaychikov2026}; \citet{patterson2022}; Epoch AI~\citep{denain2026} & n/a (separate line; a spend ratio used as an energy ratio) & S9.5 \\
$W_{\mathrm{in}}$ & Uncached input token per output token & 0.30 & 0.10 & 0.40 & ML.ENERGY~\citep{chung2025}; provider prices~\citep{openai2026b,anthropic2026b,google2026b}; Greenpixie~\citep{horswill2026}; \citet{vellaisamy2026} & Central: set at the reading matched to this paper's batch sizes, 0.295--0.304; the band reaches further below than above & S9.6 \\
$W_{\mathrm{cache}}$ & Cached input token per output token & 0.05 & 0.02 & 0.10 & Provider cache prices~\citep{openai2026b,anthropic2026b,google2026b} (lower bound); ML.ENERGY~\citep{chung2025} (upper bound) & Low for cache-heavy workloads; no energy measurement exists & S9.6 \\
$R_{\mathrm{bench}}$ & Reference input per output token & 1.13 & --- & --- & Convention; the run values (0.47 to 2.01) are measured per configuration from the benchmark's run files~\citep{mlenergy2025} & None on results computed through effective tokens & S9.6 \\
$D$ & De-biasing divisor of the reference mix, $1 + W_{\mathrm{in}} R_{\mathrm{bench}}$ & 1.339 & 1.113 & 1.452 & Follows from the definition of the measured quantity~\citep{chung2025} & Each run is restated by $D/D_{\mathrm{run}}$; $D_{\mathrm{run}}$ is 1.08 to 1.45 at the operating points, measured; $D$ cancels per billed token; no restatement is $-$8.6\% on the mid class & S9.6 \\
\bottomrule
\end{tabular}
\caption{Inputs of the central factor and of the token basis, with their bands, sources and the direction of bias in the central value (all quantities in the units given in the Meaning column).}
\label{tab:master}
\end{table}

\footnotesize
\textit{Low and high are the numerically lower and higher values used in the one-at-a-time sensitivities of Table~\ref{tab:sensfull}, not a confidence interval; the bias column states which direction each moves the result. For $E_{\mathrm{GPU}}$ the low and high columns are the class range, the lowest and highest measured member relative to the class mean. $I$ and $U$ describe one fleet and are tied, so the pair moves together ($I$ 1.05 with $U$ 0.857; $I$ 1.20 with $U$ 0.60) as well as separately. A lower $W_{\mathrm{in}}$ lowers the result on both denominators once each run is restated (Table~\ref{tab:sens}), and the low and high values of $D$ follow from the band of $W_{\mathrm{in}}$ rather than being a separate input. The conservative high case uses $H = 1.88$, the $I = 1.20$ / $U = 0.60$ pair and PUE 1.45 together, and holds every other input central. Figure~\ref{fig:evidence} grades the same inputs by the kind of evidence each rests on.}
\normalsize

\textbf{Class members and their operating points.} Table~\ref{tab:classmembers} lists every measured model that the rule of Section~\ref{sec:3.5} assigns to each class, with the configuration the operating-point rule selects, the measured values at that point and the member's ratio to the class mean. One entry per distinct model: the bfloat16 measurement where one exists at the operating point on that accelerator type, and the FP8 measurement only where none does. The same rows, each with its own factors, are published as \texttt{model\_\allowbreak factors.csv}.

\begin{table}[htbp]
\centering
\setlength{\tabcolsep}{3pt}
\scriptsize
\begin{tabular}{>{\raggedright\arraybackslash}p{0.5cm} >{\raggedright\arraybackslash}p{0.65cm} >{\raggedright\arraybackslash}p{3.3cm} >{\raggedright\arraybackslash}p{0.95cm} >{\raggedleft\arraybackslash}p{1.1cm} >{\raggedleft\arraybackslash}p{0.85cm} >{\raggedleft\arraybackslash}p{0.75cm} >{\raggedleft\arraybackslash}p{1.1cm} >{\raggedleft\arraybackslash}p{0.75cm} >{\raggedleft\arraybackslash}p{1.0cm} >{\raggedleft\arraybackslash}p{0.9cm} >{\raggedleft\arraybackslash}p{1.25cm}}
\toprule
Class & Accel. & Model & Precision & GPUs $\times$ batch & ITL (ms) & GPU W & Output tok/s & $D_{\mathrm{run}}$ & J / token (ref.\ mix) & Ratio to mean & Embodied g per M eff. \\
\midrule
A & H100 & Llama-3.1-8B-Instruct & bfloat16 & 1 $\times$ 512 & 47.4 & 688 & 9,806 & 1.202 & 0.0782 & 0.60 & 0.48 \\
A & H100 & Qwen3-8B & bfloat16 & 1 $\times$ 512 & 47.2 & 616 & 8,535 & 1.158* & 0.0834 & 0.64 & 0.57 \\
A & H100 & NVIDIA-Nemotron-Nano-9B-v2 & bfloat16 & 1 $\times$ 256 & 43.4 & 693 & 4,465 & 1.409\textsuperscript{n} & 0.1475 & 1.13 & 0.89 \\
A & H100 & gemma-3-12b-it & bfloat16 & 1 $\times$ 128 & 34.7 & 589 & 3,546 & 1.154 & 0.1927 & 1.47 & 1.37 \\
A & H100 & NVIDIA-Nemotron-Nano-12B-v2 & bfloat16 & 1 $\times$ 256 & 46.0 & 694 & 3,847 & 1.452\textsuperscript{n} & 0.1663 & 1.27 & 1.01 \\
A & H100 & Qwen3-14B & bfloat16 & 1 $\times$ 256 & 38.4 & 645 & 5,760 & 1.169* & 0.1282 & 0.98 & 0.84 \\
A & H100 & Qwen3-30B-A3B-Instruct-2507 & bfloat16 & 2 $\times$ 512 & 47.4 & 953 & 9,650 & 1.105 & 0.1197 & 0.91 & 1.05 \\
A & B200 & Llama-3.1-8B-Instruct & bfloat16 & 1 $\times$ 256 & 31.7 & 540 & 7,664 & 1.171 & 0.0806 & 0.71 & --- \\
A & B200 & Qwen3-8B & bfloat16 & 1 $\times$ 512 & 34.6 & 752 & 12,164 & 1.104 & 0.0750 & 0.66 & --- \\
A & B200 & NVIDIA-Nemotron-Nano-9B-v2 & bfloat16 & 1 $\times$ 256 & 40.7 & 800 & 5,425 & 1.168\textsuperscript{n} & 0.1690 & 1.49 & --- \\
A & B200 & gemma-3-12b-it & bfloat16 & 1 $\times$ 512 & 43.5 & 928 & 9,733 & 1.151* & 0.1109 & 0.98 & --- \\
A & B200 & NVIDIA-Nemotron-Nano-12B-v2 & bfloat16 & 1 $\times$ 256 & 33.9 & 885 & 6,233 & 1.142\textsuperscript{n} & 0.1664 & 1.47 & --- \\
A & B200 & Qwen3-14B & bfloat16 & 1 $\times$ 512 & 39.1 & 913 & 11,858 & 1.115 & 0.0925 & 0.82 & --- \\
A & B200 & Qwen3-30B-A3B-Instruct-2507 & bfloat16 & 1 $\times$ 512 & 38.0 & 703 & 8,840 & 1.076 & 0.0990 & 0.87 & --- \\
B & H100 & Llama-3.1-70B-Instruct & bfloat16 & 4 $\times$ 256 & 39.8 & 2,354 & 4,915 & 1.285 & 0.4992 & 1.16 & 3.56 \\
B & H100 & Llama-3.3-70B-Instruct & bfloat16 & 4 $\times$ 256 & 39.1 & 2,519 & 5,130 & 1.241 & 0.5301 & 1.24 & 3.53 \\
B & H100 & Qwen3-32B & bfloat16 & 2 $\times$ 256 & 44.5 & 1,178 & 5,155 & 1.158 & 0.2642 & 0.62 & 1.88 \\
B & H100 & gemma-3-27b-it & bfloat16 & 2 $\times$ 192 & 39.8 & 1,210 & 4,181 & 1.159* & 0.3346 & 0.78 & 2.32 \\
B & H100 & Llama-4-Scout-17B-16E-Instruct & bfloat16 & 8 $\times$ 384 & 47.8 & 3,189 & 6,783 & 1.223 & 0.5146 & 1.20 & 5.42 \\
B & B200 & Llama-3.1-70B-Instruct & bfloat16 & 2 $\times$ 256 & 35.8 & 1,942 & 5,372 & 1.275 & 0.3795 & 1.33 & --- \\
B & B200 & Llama-3.3-70B-Instruct & bfloat16 & 2 $\times$ 256 & 36.7 & 1,942 & 5,515 & 1.242 & 0.3796 & 1.33 & --- \\
B & B200 & Qwen3-32B & bfloat16 & 1 $\times$ 256 & 36.0 & 971 & 6,444 & 1.144 & 0.1764 & 0.62 & --- \\
B & B200 & gemma-3-27b-it & bfloat16 & 1 $\times$ 256 & 39.6 & 974 & 5,510 & 1.151* & 0.2057 & 0.72 & --- \\
C & H100 & Qwen3-235B-A22B-Instruct-2507 & bfloat16 & 8 $\times$ 128 & 46.6 & 3,903 & 2,626 & 1.115 & 1.7851 & 1.12 & 15.36 \\
C & H100 & Llama-3.1-405B-Instruct-FP8 & fp8 & 8 $\times$ 128 & 48.0 & 4,606 & 1,896 & 1.316 & 2.4728 & 1.55 & 18.03 \\
C & H100 & Llama-4-Maverick-17B-128E-Instruct-FP8 & fp8 & 8 $\times$ 384 & 47.9 & 3,462 & 7,149 & 1.225 & 0.5294 & 0.33 & 5.14 \\
C & B200 & Qwen3-235B-A22B-Instruct-2507 & bfloat16 & 8 $\times$ 256 & 43.5 & 5,078 & 5,403 & 1.112 & 1.1319 & 0.69 & --- \\
C & B200 & Llama-3.1-405B-Instruct & bfloat16 & 8 $\times$ 256 & 49.3 & 7,643 & 3,587 & 1.272 & 2.2436 & 1.36 & --- \\
C & B200 & DeepSeek-V3.1 & fp8 & 8 $\times$ 256 & 48.6 & 5,935 & 4,534 & 1.119 & 1.5664 & 0.95 & --- \\
\bottomrule
\end{tabular}
\caption{Measured class members at the 50 ms operating point (ML.ENERGY v3), with the configuration selected, the measured values at it and the member's ratio to the class mean. Output tok/s and GPU W, the graphics processing unit (GPU) power in watts, are measured, and their quotient is the measured J per output token. $D_{\mathrm{run}}$ is the divisor of the run at its own token mix and cache-hit share, and J / token is the measured value times $D/D_{\mathrm{run}}$, that is at the reference mix (Section S9.6). An asterisk marks a run whose cache counter shows recomputation, for which the hit share of the clean runs of the same model and accelerator type is used, and \textsuperscript{n} a run with prefix caching off. Batch is the configured maximum number of concurrent sequences and GPU W the measured average power summed over the accelerators of the point.}
\label{tab:classmembers}
\end{table}

\textbf{Example (mid class, H100).} The class mean is 0.4285 J per output token over the five members above.
\begin{alltt}\small
GPU Wh/1k      = (0.4285 \(\times\) 1{,}000) / 3{,}600 = 0.1190 Wh/1k tokens
Server Wh/1k   = 0.1190 \(\times\) 1.50 \(\times\) 1.10 = 0.1964 Wh/1k tokens
Facility Wh/1k = 0.1964 \(\times\) 1.20 PUE = 0.2357 Wh/1k tokens (= kWh per M tokens)
\end{alltt}

\begin{figure}[tbp]
\centering
\includegraphics[width=1.0\textwidth]{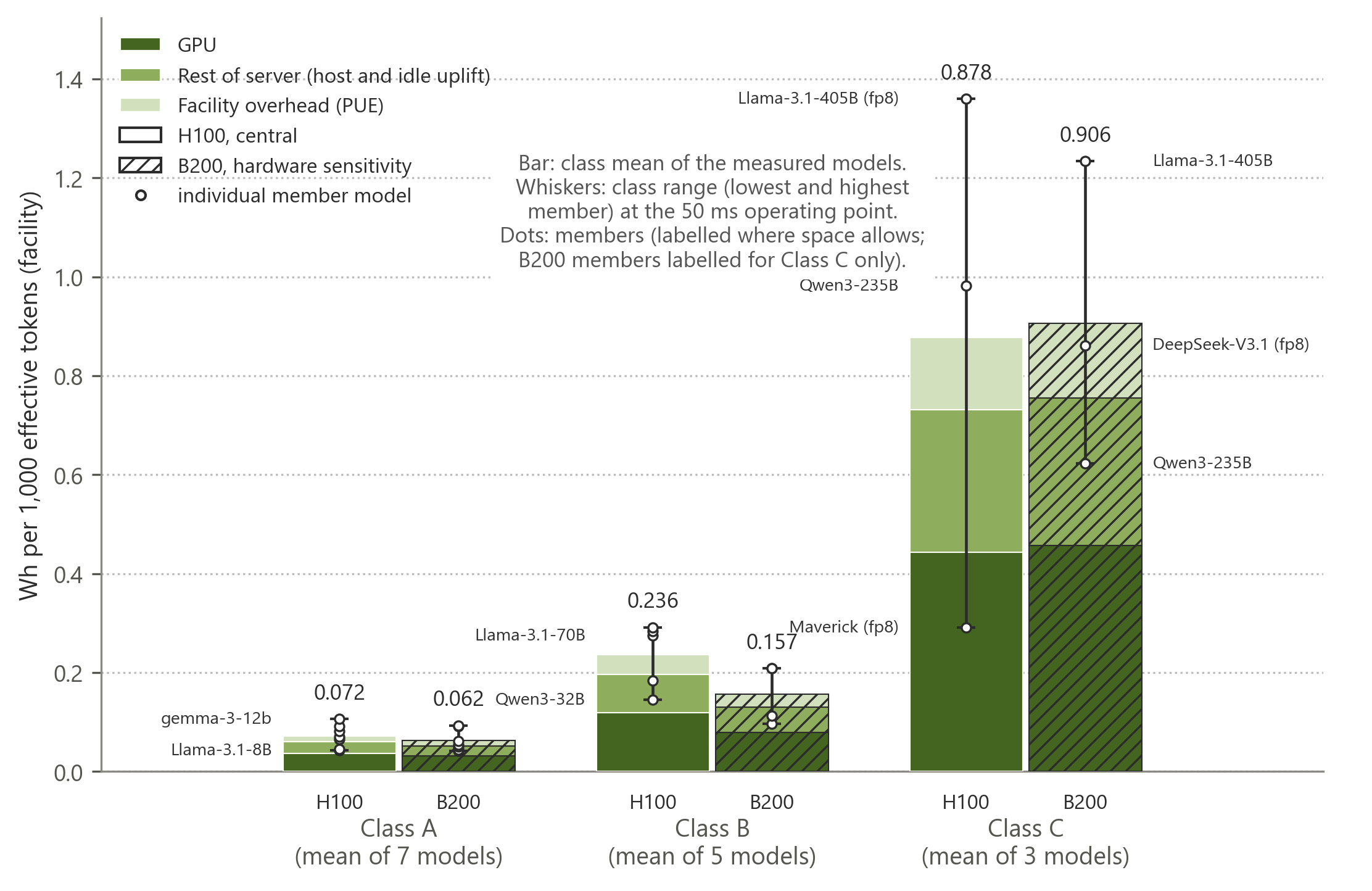}
\caption{Facility energy by model class, in Wh per 1,000 \emph{effective} tokens, H100 central against the B200 hardware sensitivity, both at the 50 ms operating point. Each bar is the class mean of the measured members, stacked into the measured accelerator energy, the rest of the server (host and idle uplift) and facility overhead (power usage effectiveness, PUE), with the total printed above it. Whiskers are the class range, the lowest and the highest member at the same operating point on the same platform, and the dots are the individual members, labelled where space allows. The range is a data-driven spread and not a confidence interval, and the frontier class mixes bfloat16 and eight-bit floating-point (FP8) members.}
\label{fig:energy}
\end{figure}

Figure~\ref{fig:chain} draws the whole calculation chain, Figure~\ref{fig:composition} the composition of the factor per million billed tokens and Figure~\ref{fig:energy} the class means with their ranges and their members.

\textbf{Statistics of each class, and the alternatives.} At the 50 ms rule the class means on H100 are 0.1309 J (n $=$ 7), 0.4285 J (n $=$ 5) and 1.5958 J (n $=$ 3), with geometric means of 0.1249, 0.4131 and 1.3270 and medians of 0.1282, 0.4992 and 1.7851. On B200 the means are 0.1133 (n $=$ 7), 0.2853 (n $=$ 4) and 1.6473 (n $=$ 3). The small-to-frontier ratio in accelerator energy is 12.2 on H100 and 14.5 on B200 with the means, 10.6 and 14.7 with geometric means, and 21.2 and 13.3 with the previous named proxies; the corresponding ratios on the inference footprint are 12.3 and 21.4, and on the convenience factor 12.7. The class means at the minimum of the smallest accelerator count with no latency limit, kept as a sensitivity, are 0.1459, 0.5066 and 1.3313 on H100 and 0.1002, 0.2096 and 0.7065 on B200; the means over all configurations meeting the latency limit are 0.2746, 1.2343 and 3.8773 on H100 and 0.2850, 1.0163 and 6.0084 on B200.

\textbf{Why the unconstrained minimum was rejected as the operating point.} The alternative configuration sensitivity is the lowest-energy point at each member's \emph{smallest} accelerator count with no latency limit. In most of the six cells that point is the last point of the sweep, and in several the average batch is below the configured maximum, a sign that the key-value cache is full, so the point records how far each sweep was run rather than a serving regime. B200 sweeps also run deeper than H100 sweeps, so at that point the B200-to-H100 facility ratios mix hardware with differences in batch depth and latency. It is an orthogonal configuration choice and does not bound the operating-point rule from below: nothing in this paper does.

\textbf{PUE comparisons.} The high value of 1.45 is the top of the Lawrence Berkeley National Laboratory 2024 projection band for that year rather than a United States average: that series reaches about 1.43 in 2023, and its 2023 aggregates by space type are about 1.22 for hyperscale and 1.14 for AI-specialised facilities~\citep{shehabi2024}, so 1.45 is a scenario high for a forecast year applied to a facility class the same source puts well below it. Oviedo et al.\ model PUE as a log-normal distribution with a 5th-to-95th-percentile range of 1.05 to 1.40, whose geometric centre of 1.21 is about 1 per cent above the value used here~\citep{oviedo2026}.

\textbf{Precision, and the frontier class.} Two of the three frontier-class H100 members are FP8 measurements, used because no bfloat16 point exists for those models at the operating point on that accelerator type. Serving the whole class in FP8 where a member has such a measurement lowers the frontier-class footprint by 10\%, and the single-model FP8 readings for Qwen3 235B A22B give $\times$1.27 on 4 accelerators and $\times$0.81 on 8 against the class mean. On B200, DeepSeek V3.1 alone as the frontier class would give 0.95 of the class mean.

\textbf{Note on the host uplift and the TokenPowerBench ceiling.} The host uplift of 1.50 is taken from the metered node measurements of Section S9.2, not from TokenPowerBench. The split of 1.50 into about 1.10 for processor and memory and 1.36 for the remainder comes from the Sustainable AI Group. Only the 1.10 leg is their own B300 node telemetry (summary statistics in their Annex B, configuration data on request), and the 1.36 is their midpoint of Latif et al.'s 1.41 to 1.50 after removing the processor and memory term at H100 rated power (1.07)~\citep{gamazaychikov2026}. Latif et al.\ publish no component breakdown and name one as future work, so only the product is cited to them. TokenPowerBench~\citep{niu2026} states only that accelerators ``typically account for over 60\% of total energy use'', which is a floor on the accelerator share and therefore a ceiling of 1.667 on the host uplift; its Figure 3 compares total and accelerator energy per token across models on logarithmic axes and prints no values, but read off it, GPU energy is about 0.8 of node energy for all ten models (total over GPU about 1.2 to 1.3), a digitised reading below the low host uplift of 1.43; its hardware is an eight-node cluster of four-accelerator nodes. That ceiling bounds the host uplift alone, because a node measurement contains no idle machines. The central host uplift of 1.50 sits 10\% below it and the high value of 1.88 sits 13\% above it, so the ceiling is retained only as a consistency check and the reason the high value exceeds it is given below. The high value of 1.88 is the mean of the machine-to-accelerator power ratios of Google's five TPU generations, recomputed here from Schneider et al.'s Table 1 as machine power divided by chip count times per-chip power excluding host: 1.97 (v4i), 2.22 (v5e), 1.78 (v6e), 1.78 (v4) and 1.64 (v5p), mean 1.878~\citep{schneider2025}. Those are all-workload machine ratios rather than serving-specific ones, and the TPU notion of a machine includes the host tray, so the value is a high reading for the host uplift rather than a measurement of an accelerator serving node. The high value therefore exceeds the 1.667 ceiling because it comes from a wider machine boundary, an all-workload TPU machine against a GPU node benchmarked on inference, while the digitised TokenPowerBench reading of about 1.2 to 1.3 is a low-side reading below the low value of 1.43. With readings on both sides of the band, the direction of the bias in $H$ is unknown.

\subsubsection*{S1.2 Grid Carbon Intensity Sources}

\textbf{United States regions (\citealp{epa2025}, eGRID2023 summary tables rev.~2, Table 1).} These are annual CO\textsubscript{2}e total-output emission rates at subregion level, converted from lb/MWh to kg/kWh by dividing by 2{,}204.6 and scaled by 1.1232 to Ember's lifecycle basis; that constant was set from an Ember vintage giving a United States 2023 intensity of 392.755 g/kWh, and the current release gives 392.9 g/kWh and a factor of 1.1236. Subregions approximate the power pool a data centre draws from, which state boundaries do not, and the EPA recommends subregion rates for purchased electricity for that reason. Ember's state series ignores imports and is not used. Rates are generation-based and exclude the 4.2\% grid gross loss (Section~\ref{sec:5.5}). Where that loss is charged it is applied as a multiplier on delivered energy, 1.042, an approximation of $1/(1 - 0.042) = 1.044$. Unscaled value, scaled value, and the mid-class electricity term on the unscaled value as a sensitivity:
\begin{itemize}[topsep=2pt,itemsep=1pt,parsep=0pt]
\item SRVC (N. Virginia; AWS us-east-1): 596.3 lb/MWh $=$ 0.2705, scaled 0.3038 kg/kWh; mid-class unscaled 0.0638 kg per million effective tokens. The Ashburn--Loudoun cluster is in Dominion Energy territory, which eGRID assigns to SRVC.
\item NWPP (Oregon; AWS us-west-2): 635.3 lb/MWh $=$ 0.2882, scaled 0.3237; 0.0679. NWPP spans the whole Northwest power pool, including coal-heavy Utah and Wyoming.
\item ERCT (Texas; Google us-south1): 736.6 lb/MWh $=$ 0.3341, scaled 0.3753; 0.0787.
\item SRSO (Southeast; no public hyperscale AI region identified): 846.0 lb/MWh $=$ 0.3837, scaled 0.4310; 0.0904.
\item MROW (Iowa; Google us-central1): 926.6 lb/MWh $=$ 0.4203, scaled 0.4721; 0.0991.
\end{itemize}

\begin{figure}[htbp]
\centering
\includegraphics[width=0.92\textwidth]{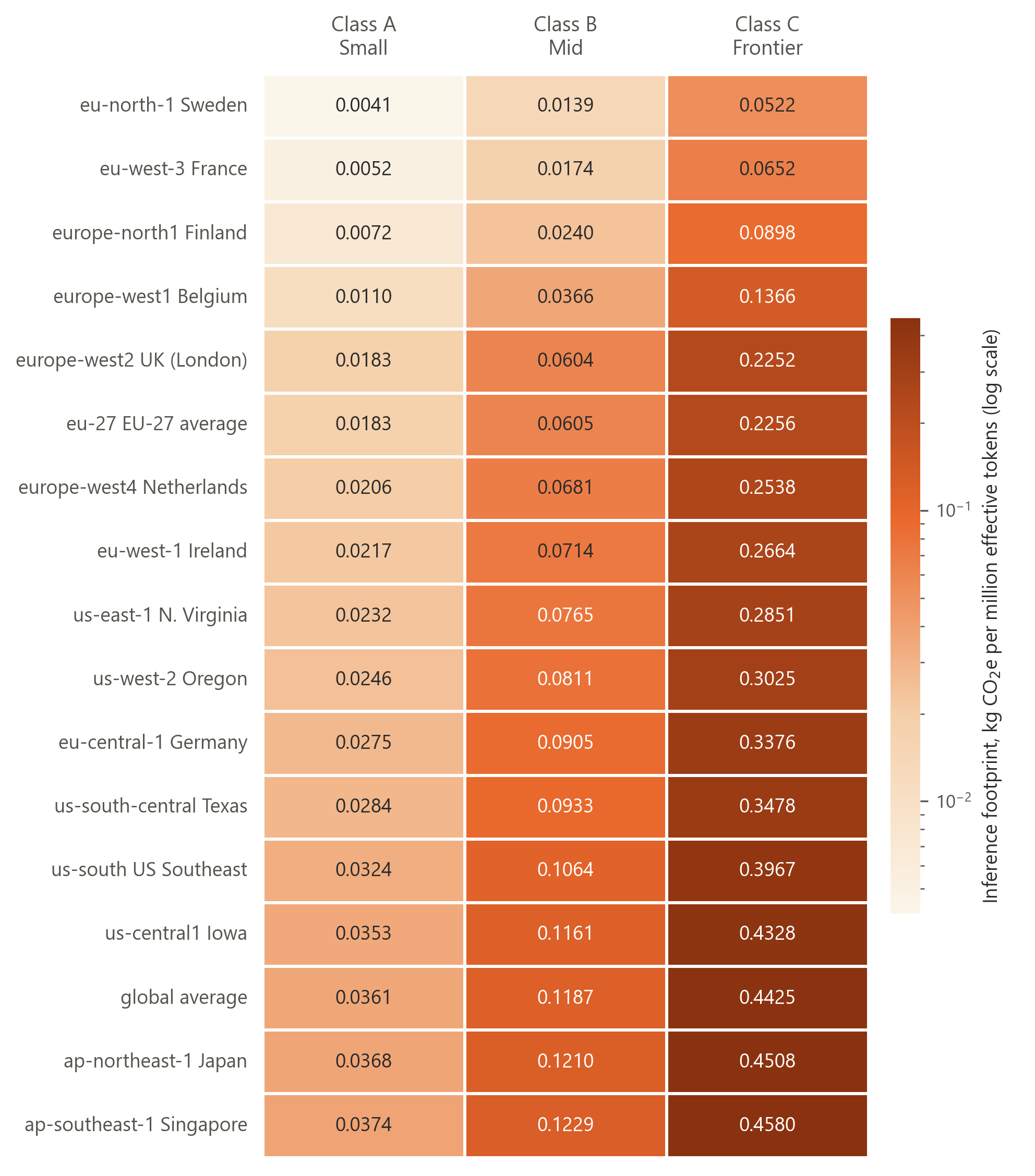}
\caption{Inference footprint by model class and region: electricity plus embodied hardware plus data-centre construction, in kg~CO\textsubscript{2}e per million \emph{effective} tokens (H100 central, class means, 2023 grid data), on a logarithmic colour scale, ordered by grid carbon intensity, cleanest first. The training term, which is per million \emph{billed} tokens, is not in this figure; the values are the inference-footprint columns of Table~\ref{tab:carbon}.}
\label{fig:carbon}
\end{figure}

Figure~\ref{fig:carbon} plots the resulting inference footprints by class and region.

\textbf{EU and Asia-Pacific (\citealp{ember2026}).} Annual national generation-based averages, calendar year 2023, lifecycle CO\textsubscript{2}e (IPCC AR5 factors), in g/kWh: Sweden 38.42, France 53.26, Finland 81.33, Belgium 134.67, UK 235.60, EU-27 236.0, Netherlands 268.17, Ireland 282.50, Germany 363.59, Japan 492.6, Singapore 500.87, and the world average 483.19 for the global row. The Singapore Energy Market Authority grid emission factor for 2023 is 0.412 kg CO\textsubscript{2}/kWh~\citep{ema2025}, a system-average figure of combustion CO\textsubscript{2} over all grid-connected generation, and Ember's lifecycle basis explains most of the gap. Cross-border trade is not reflected, which matters most for net importers such as Finland, Ireland and Belgium. Japan uses the national average, although the Tokyo area served by AWS ap-northeast-1 differs from it. Beyond the printed rows, the dataset carries 210 countries and economies, 27 eGRID subregions, a Quebec provincial row and the two aggregates, and resolves the 143 public AWS, Azure and Google Cloud regions listed in October 2026 to those rows, 20 of them resting on a location the provider states only as a state or country.

\begin{table}[htbp]
\centering
\setlength{\tabcolsep}{4pt}
\footnotesize
\begin{tabular}{>{\raggedright\arraybackslash}p{4.2cm} >{\raggedleft\arraybackslash}p{1.4cm} >{\raggedleft\arraybackslash}p{1.4cm} >{\raggedleft\arraybackslash}p{1.4cm} >{\raggedleft\arraybackslash}p{1.4cm} >{\raggedleft\arraybackslash}p{1.4cm} >{\raggedleft\arraybackslash}p{1.4cm}}
\toprule
 & \multicolumn{3}{c}{B200 inference footprint} & \multicolumn{3}{c}{Conservative high inference footprint} \\
Region & A & B & C & A & B & C \\
\midrule
EU North --- Sweden & 0.0037 & 0.0104 & 0.0534 & 0.0063 & 0.0212 & 0.0794 \\
EU West --- France & 0.0046 & 0.0127 & 0.0669 & 0.0081 & 0.0270 & 0.1009 \\
EU North --- Finland & 0.0064 & 0.0171 & 0.0923 & 0.0114 & 0.0379 & 0.1416 \\
EU West --- Belgium & 0.0097 & 0.0255 & 0.1406 & 0.0178 & 0.0587 & 0.2190 \\
EU West --- UK (London) & 0.0160 & 0.0413 & 0.2321 & 0.0298 & 0.0980 & 0.3653 \\
EU-27 average & 0.0160 & 0.0414 & 0.2324 & 0.0298 & 0.0981 & 0.3659 \\
EU West --- Netherlands & 0.0180 & 0.0464 & 0.2616 & 0.0336 & 0.1107 & 0.4126 \\
EU West --- Ireland & 0.0189 & 0.0487 & 0.2746 & 0.0353 & 0.1162 & 0.4333 \\
EU Central --- Germany & 0.0239 & 0.0614 & 0.3480 & 0.0450 & 0.1478 & 0.5509 \\
US East --- N. Virginia & 0.0202 & 0.0520 & 0.2939 & 0.0379 & 0.1245 & 0.4642 \\
US West --- Oregon & 0.0215 & 0.0551 & 0.3119 & 0.0402 & 0.1323 & 0.4930 \\
US South Central --- Texas & 0.0247 & 0.0632 & 0.3586 & 0.0464 & 0.1524 & 0.5679 \\
US South --- Southeast & 0.0282 & 0.0720 & 0.4091 & 0.0530 & 0.1741 & 0.6487 \\
US Central --- Iowa & 0.0307 & 0.0784 & 0.4463 & 0.0579 & 0.1901 & 0.7082 \\
Asia Pacific --- Japan & 0.0320 & 0.0816 & 0.4649 & 0.0603 & 0.1981 & 0.7380 \\
Asia Pacific --- Singapore & 0.0325 & 0.0829 & 0.4724 & 0.0613 & 0.2013 & 0.7500 \\
\textbf{Global average} & \textbf{0.0314} & \textbf{0.0802} & \textbf{0.4564} & \textbf{0.0592} & \textbf{0.1944} & \textbf{0.7243} \\
\bottomrule
\end{tabular}
\caption{B200 hardware sensitivity and conservative high case: inference footprint in kg CO\textsubscript{2}e per million \emph{effective} tokens. B200 $=$ B200 electricity and construction $+$ the H100 embodied term; high $=$ electricity, construction and embodied at $H = 1.88$, the $I = 1.20$ / $U = 0.60$ pair and power usage effectiveness (PUE) 1.45. The training term, which is per million \emph{billed} tokens, is central in both cases.}
\label{tab:b200high}
\end{table}

\begin{table}[htbp]
\centering
\setlength{\tabcolsep}{3pt}
\footnotesize
\begin{tabular}{>{\raggedright\arraybackslash}p{2.5cm} >{\raggedleft\arraybackslash}p{1.1cm} >{\raggedleft\arraybackslash}p{1.0cm} >{\raggedleft\arraybackslash}p{1.15cm} >{\raggedleft\arraybackslash}p{1.25cm} >{\raggedleft\arraybackslash}p{1.3cm} >{\raggedleft\arraybackslash}p{1.25cm} >{\raggedleft\arraybackslash}p{1.1cm} >{\raggedleft\arraybackslash}p{1.1cm} >{\raggedleft\arraybackslash}p{1.2cm}}
\toprule
Region & Electri\-city & Embod\-ied & Construc\-tion & Infer\-ence foot\-print & Non-electri\-city uplift & Train\-ing (per M billed) & Factor (per M billed) & B200 foot\-print & High-case foot\-print \\
\midrule
US East --- N. Virginia & 0.0716 & 0.0033 & 0.0015 & 0.0765 & +6.8\% & 0.0029 & 0.0261 & 0.0520 & 0.1245 \\
EU North --- Sweden & 0.0091 & 0.0033 & 0.0015 & 0.0139 & +53.5\% & 0.0029 & 0.0071 & 0.0104 & 0.0212 \\
EU West --- Ireland & 0.0666 & 0.0033 & 0.0015 & 0.0714 & +7.3\% & 0.0029 & 0.0245 & 0.0487 & 0.1162 \\
EU-27 average & 0.0556 & 0.0033 & 0.0015 & 0.0605 & +8.7\% & 0.0029 & 0.0212 & 0.0414 & 0.0981 \\
Global average & 0.1139 & 0.0033 & 0.0015 & 0.1187 & +4.3\% & 0.0029 & 0.0388 & 0.0802 & 0.1944 \\
Asia Pacific --- Singapore & 0.1181 & 0.0033 & 0.0015 & 0.1229 & +4.1\% & 0.0029 & 0.0401 & 0.0829 & 0.2013 \\
Asia Pacific --- Japan & 0.1161 & 0.0033 & 0.0015 & 0.1210 & +4.2\% & 0.0029 & 0.0395 & 0.0816 & 0.1981 \\
\bottomrule
\end{tabular}
\caption{Effect of the grid-independent terms on the mid-class factor by region, in kg CO\textsubscript{2}e. Electricity, embodied, construction and the inference footprint are per million \emph{effective} tokens; the training term and the convenience factor are per million \emph{billed} tokens.}
\label{tab:nonelec}

\begin{minipage}{\textwidth}\footnotesize \textit{Footprints and uplifts are computed from unrounded values (embodied 0.00334 and construction 0.00150 kg per million effective tokens, training 0.00293 kg per million billed tokens). ``Non-electricity uplift'' is embodied plus construction as a percentage of the electricity term, not a share of the footprint. Across all seventeen rows it ranges from about 4\% (Singapore) to about 54\% (Sweden).}\end{minipage}
\end{table}

\begin{table}[htbp]
\centering
\setlength{\tabcolsep}{3pt}
\footnotesize
\begin{tabular}{>{\raggedright\arraybackslash}p{1.5cm} >{\raggedleft\arraybackslash}p{1.4cm} >{\raggedleft\arraybackslash}p{1.4cm} >{\raggedleft\arraybackslash}p{1.4cm} >{\raggedleft\arraybackslash}p{1.7cm} >{\raggedleft\arraybackslash}p{1.9cm} >{\raggedleft\arraybackslash}p{1.4cm} >{\raggedleft\arraybackslash}p{1.4cm} >{\raggedleft\arraybackslash}p{1.5cm}}
\toprule
 & \multicolumn{3}{c}{Development line, global average} & & Development-inclusive & \multicolumn{3}{c}{Development line at $d = 1.5$} \\
\cmidrule(lr){2-4}\cmidrule(lr){7-9}
Class & $d = 0.67$ & $d = 1.5$ & $d = 2.5$ & Inference footprint & total at $d = 1.5$ & Sweden & US East & Singapore \\
\midrule
A --- Small & 0.0212 & 0.0474 & 0.0790 & 0.0361 & 0.0835 & 0.0038 & 0.0298 & 0.0492 \\
B --- Mid & 0.0694 & 0.1553 & 0.2588 & 0.1187 & 0.2740 & 0.0123 & 0.0976 & 0.1610 \\
C --- Frontier & 0.2583 & 0.5783 & 0.9638 & 0.4425 & 1.0208 & 0.0460 & 0.3636 & 0.5995 \\
\bottomrule
\end{tabular}
\caption{The development line and the development-inclusive total, in kg CO\textsubscript{2}e per million \emph{effective} tokens. $\mathrm{EF_{dev}} = d \times$ accelerator energy $\times H \times$ power usage effectiveness (PUE) $\times g$, with no idle uplift; the development-inclusive total is the inference footprint plus that line, with the training term dropped. The line is never added to the inference footprint plus training.}
\label{tab:devline}
\end{table}

\subsubsection*{S1.3 Embodied, Construction and Training: Supporting Detail}

\emph{Which Dell footprint carries which value.} The manufacturing stage is 2,872 kg CO\textsubscript{2}e for the XE9680, 3,065 for the XE9680L, 4,009 for the XE7745 and 4,241 for the XE9685L, and the transport stage 712, 632, 427 and 530 kg respectively~\citep{dell2025}. The transport values are printed on page 2 of each report. The manufacturing values are not: page 2 prints a value for each of End of Life, Transport, Use and Total but leaves the Manufacturing cell empty, so the four figures above are sums of the nine printed component lines, of which the mainboard is 88 to 92\%. Subtracting the other three stages from the printed Total instead gives 2,871, 3,064, 4,011 and 4,240 kg, so the printed Total is rounded and disagrees with the component sum by one or two kilograms; the component sum is used and the Total is not quoted as corroboration. The high end of the footprint-plus-transport band, 4,953 kg, pairs the XE9685L manufacturing value with the XE9680 transport value and so describes no one server. A bottom-up cross-check adds NVIDIA's published HGX H100 baseboard footprint of 1,312 kg~\citep{nvidia2025a} to a host server and gives about 2,800 to 3,800 kg, which brackets the central value; the implied host-server footprint of 1,488 to 2,488 kg is an author range rather than a published figure, so that check bounds the central value rather than corroborating it, and the Sustainable AI Group's own stack of about 5,325 kg for baseboard, host tray and distribution~\citep{gamazaychikov2026} sits above it. The XE7745 and XE9685L are 2025 platforms whose accelerator generation is not declared, which is why the central value is weighted to the two XE9680 variants. \textbf{GPU Quantity and GPU Memory are blank on all four reports}, so Dell discloses nothing about accelerator content and the presence of eight accelerators is the inference from the processor count (10 $=$ 2 CPUs $+$ 8 accelerators), the system mass, the installed memory and the size of the mainboard line; the two June 2025 reports therefore cannot be assumed to be eight-accelerator platforms either. The two April 2025 reports state PAIA model version 1.4.6 and the two June 2025 reports version 1.4.7, so the four are not on one tool revision. All four assume a four-year product lifetime and state that the use stage assumes a PUE of 1.0~\citep{dell2025}; \citet{schneider2025} use a six-year machine life and the Sustainable AI Group three~\citep{gamazaychikov2026}.

\emph{Hardware life in operator filings.} The other three large operators file longer or wider classes for the same asset type: six years generally for Alphabet, two to six years for Microsoft and 5.5 years for most of Meta's servers and network assets, the last extended with effect from 1 January 2025, the same month in which Amazon shortened its own; the Meta filing states the new life but not the previous one~\citep{secfilings2026}.

\emph{Utilisation evidence.} Dell's footprint implicitly assumes continuous operation and \citet{luccioni2023} use 85\% for a high-performance computing allocation, both above the 75\% used here and well above the 60\% of the high-case pair.

\emph{Embodied band arithmetic.} Taking the extremes of footprint plus transport, hardware life and the tied utilisation pairs together gives a low end of $(3{,}299/3{,}972) \times (5/6) \times (0.75/0.857) = 0.61$ and a high end of $(4{,}953/3{,}972) \times (5/4) \times (0.75/0.60) = 1.95$, that is $-$39\%/$+$95\%. Google's disclosed Scope 1 and embodied Scope 3 emissions per median Gemini prompt, 0.010 g, restated on this paper's effective-token basis at 418.5 effective tokens per chat exchange, are about 24 g per million effective tokens~\citep{elsworth2025}, so $23.9/3.34 = 7.1$ times the mid-class term and above the band, which leaves its upper tail open.

\emph{Alternative embodied allocations.} Server power is reported in the dataset but does not enter the time allocation. The time allocation at the smallest-accelerator-count point gives lower values for the mid and frontier classes, because larger batches serve more tokens per second. The embodied term is allocated by server time while the weights of Section~\ref{sec:3.3} are energy weights; \citet{vellaisamy2026} report smoothed GPU power for a prefill-dominated first-token window at 2.05 to 3.07 times that of a complete 128-token request for the two dense models it quotes (H200, batch 4, context 2,048), so the time cost of an uncached input token is $W_{\mathrm{in}}$ divided by that, 0.098 to 0.146 rather than 0.30. Re-deriving $D$ at those time weights, charging the embodied term on effective tokens overstates the embodied charge by 16.7\% to 24.7\% on the chat exchange, 28.5\% to 44.1\% at the default token mix, 37.9\% to 61.3\% on the document-heavy workload, 75.9\% to 147.7\% on inline completion and 6.0\% to 8.5\% on an agentic request. Because embodied is 2.8\% of the mid-class inference footprint, the effect on the reported mid-class figure is 0.38\% to 0.52\% on the chat exchange, 0.58\% to 0.80\% at the default mix and 1.10\% to 1.51\% on inline completion, the most input-heavy workload tabulated here.

\emph{Construction constant.} \citet{schneider2025} give construction emissions per TPU in their Table 1, on page 2 (59, 59, 109, 117 and 218 kg CO\textsubscript{2}e for TPU v4i, v5e, v6e, v4 and v5p) over a six-year machine life, and publish no per-kWh figure. Dividing each value by the matching machine electricity over the same life gives 7.58, 7.66, 7.63, 7.62 and 7.62 g CO\textsubscript{2}e per kWh of machine (server) electricity, hence $k_c = 7.66$ g, the top of a range whose width is 1\%. Dividing again by Google's 2024 fleet PUE of 1.10, the PUE used in that paper's own chain and stated in its Appendix A.3, would give 6.89 to 6.97 g per kWh of \emph{facility} electricity; the two are the same number on different bases, and applying the IT-basis value after a PUE of 1.20 would overstate construction by 20\%. An intermediate version of this paper used the facility-basis restatement, 6.9 g, against \emph{server} electricity, which understated the term by 10\%. The source amortises the data-centre construction footprint over twenty years and allocates it pro rata to machine energy (Appendices A.5.1 and A.5.2), and calls its own estimate of this stage rough. Two independent closures confirm the denominator: the location-based operational row of Table 1 is a near-constant 403 g per kWh of machine energy across all five machines, and 403.1 divided by 1.10 is 366.5 g, matching that paper's own statement of Google's 2023 location-based factor.

The low value, 5.3 g per kWh of server electricity, is the Carbon Trust low scenario taken directly on that basis: its materiality table gives 117 kg CO\textsubscript{2}e per square metre per year for building and services embodied carbon over a 60-year life, and dividing by 5 kW per square metre of information technology load at 50\% utilisation gives 5.34 g per kWh of that load~\citep{carbontrust2026}. An intermediate version of this paper used 5.8 g, which restated the same figure at a PUE of 1.10 as if it had been on a facility basis; the correction moves the mid-class inference footprint by 0.1\%. Two qualifications travel with the low value. It is the lowest-density of three scenarios and gives the highest value per kWh of the three. The medium and high rack densities give 1.71 and 0.76 g per kWh, so density and utilisation rather than building life separate them, and an AI serving hall sits at the dense end. And two of the three authors of \citet{gamazaychikov2026} sat on that report's expert working group, so it is not an independent check on it.

\emph{Scale checks on the training denominator.} The 695 trillion-token denominator can be placed against two public provider totals. Google reported processing over 3.2 quadrillion tokens a month across its surfaces in May 2026~\citep{pichai2026}, so 695 trillion is 22\% of one month of that total, although the source does not break the total down by model and much of it is likely to be served by small models. The same keynote puts Google's model APIs at roughly 19 billion tokens a minute, which is about $8.3\times10^{14}$ a month on our own conversion rather than a figure the source states, and of which 695 trillion is 0.8 months. OpenAI's API processed over 6 billion tokens per minute in October 2025~\citep{openai2025a}, which is at least $2.6\times10^{14}$ per month, so 695 trillion is at most about 2.6 to 2.7 months of that API's whole throughput across all models. That total comes from the same announcement as $V$, so Google's totals are the only independent scale check. Both source figures are floors, so the comparison is a lower bound on throughput and an upper bound on the number of months.

\emph{What the disclosed volume counts.} The one disclosed per-model volume does not state its own token basis. The training term is charged on billed tokens including cached ones, so the matching reading is input plus output with cache reads inside; any other reading makes the denominator too small and the term too high. Because training is 8\% of the mid-class factor per million billed tokens, a tenfold larger denominator would lower that factor by 7\% and a tenfold smaller one would raise it by 68\%.

\emph{Model service lives, by kind of observation.} The 80-model first-party series is not a sample of completed lifetimes and should not be read as one. It holds 45 completed retirements, 21 announced future dates and 14 Anthropic published dates that sit at exactly release plus 12.0 months; a further 10 Google generally available models with no announced shutdown date are right-censored and excluded. The Anthropic regularity is a pattern inferred from the retirement dates the vendor lists, not a commitment it states: the words ``12 month'', ``twelve month'' and ``one year'' do not appear on the page, no published date has been brought forward, most completed retirements have run longer than twelve months, and one, Claude Sonnet 3.7, ran 360 days against its announcement date, five days short of it; the deposited series anchors on the snapshot date in the model identifier and records 12.0 months for the same model~\citep{anthropic2026a}. The medians are 12.8 months over all 80 (interquartile range 11.9 to 17.0), 12.2 over the completed retirements alone and 14.6 with the floors excluded. Per provider the all-status medians are 12.0 months for Anthropic (n $=$ 25), 16.1 for OpenAI (n $=$ 40) and 9.0 for the Gemini API (n $=$ 15); on third-party platforms Azure OpenAI gives 18.0 (n $=$ 45). The boundary used is first-party, the date the vendor itself stops serving the model, which is both the point at which the serving fleet is released and the conservative choice for the training term.

\emph{Usage shape factor.} Usage over a model's life ramps, peaks and decays, so a lifetime-average serving rate is below a peak rate. The shape factor of 0.4 is the median ratio of the mean active week to the peak week across 14 provider accounts on a public routing platform. It is carried as a sensitivity and not as a central constant, because the series is per provider rather than per model and is confounded by 21-fold growth of the platform over the window.

\emph{Reconstructions behind the development ratio.} \citet{patterson2022} report that across three years about three fifths of Google's machine-learning energy was inference and two fifths training, covering research, development, testing and production, which gives $d = 0.67$ and is the only energy measurement in the set. The Sustainable AI Group's $d = 1.5$~\citep{gamazaychikov2026} rests on Epoch AI's split of Anthropic's 2025 compute \emph{spending} (USD 4.10 billion of development compute against USD 2.70 billion of serving compute, giving 1.52) and on a round 50/10/40 split of frontier-lab accelerator \emph{capacity} stated in an interview and reported in \citet{gamazaychikov2026}, which gives exactly 1.50 because the numbers are round; their agreement to 1.2\% is an artefact of rounding rather than corroboration. In Epoch AI's own words the USD 2.70 billion is not a disclosed serving figure but its inference from a reported 40\% gross margin on USD 4.5 billion of revenue, and both figures are themselves second-hand~\citep{denain2026}. Epoch AI's reconstruction of OpenAI's 2024 spending, at about USD 5 billion of research and development compute against about USD 2 billion of inference compute, gives $d = 2.5$~\citep{denain2026}. The one energy measurement in the set, Patterson et al.'s three fifths against two fifths, rests on one week of April in each of 2019, 2020 and 2021~\citep{patterson2022}.

\emph{Reserved and provisioned capacity.} The rule is in Section~\ref{sec:3.4} and Table~\ref{tab:reserved} gives the values. At the global-average grid a reserved H100 accelerator-hour is 0.5653, 0.5068 and 0.4509 kg CO\textsubscript{2}e for the small, mid and frontier classes, falling with class because the larger models draw less power per accelerator at their operating points; a reserved hour therefore costs less for a frontier model but serves far fewer tokens, about 1.2 million against 4.5 million for the mid class. As a reconciliation of the two calculation routes, a mid-class reserved accelerator-hour run at the benchmark operating point would serve about 4.5 million output tokens, from the class harmonic-mean throughput $T_8$ of Table~\ref{tab:master} divided by eight accelerators, so 0.5068 kg per accelerator-hour corresponds to about 0.11 kg per million effective tokens against the 0.1187 of Table~\ref{tab:carbon}. Most of the difference is the idle uplift and the utilisation divisor that a reserved hour does not carry, and the rest comes from averaging, since the reserved hour divides mean power by harmonic-mean throughput while the per-token factor averages each member's joules per token.

\begin{table}[htbp]
\centering
\setlength{\tabcolsep}{3pt}
\footnotesize
\begin{tabular}{>{\raggedright\arraybackslash}p{1.5cm} >{\raggedleft\arraybackslash}p{1.1cm} >{\raggedleft\arraybackslash}p{1.5cm} >{\raggedleft\arraybackslash}p{1.6cm} >{\raggedleft\arraybackslash}p{1.6cm} >{\raggedleft\arraybackslash}p{1.7cm} >{\raggedleft\arraybackslash}p{1.8cm} >{\raggedleft\arraybackslash}p{1.3cm} >{\raggedleft\arraybackslash}p{1.2cm}}
\toprule
Class & Members $n$ & Avg GPU W per GPU & Server kWh per GPU-h & Facility kWh per GPU-h & Embodied kg per GPU-h & Construc\-tion kg per GPU-h & Global average & EU-27 \\
\midrule
A & 7 & 628.7 & 0.9430 & 1.1316 & 0.0113 & 0.0072 & \textbf{0.5653} & 0.2856 \\
B & 5 & 562.2 & 0.8434 & 1.0120 & 0.0113 & 0.0065 & \textbf{0.5068} & 0.2566 \\
C & 3 & 498.8 & 0.7482 & 0.8978 & 0.0113 & 0.0057 & \textbf{0.4509} & 0.2290 \\
\bottomrule
\end{tabular}
\caption{Reserved (provisioned) capacity: kg CO\textsubscript{2}e per reserved H100 accelerator-hour, by model class and grid, with the energy and the grid-independent terms behind it.}
\label{tab:reserved}

\begin{minipage}{\textwidth}\footnotesize \textit{The accelerator power is the mean over the class members of the measured average power per accelerator at each member's own operating point. The last two columns are the total kg CO\textsubscript{2}e per reserved accelerator-hour at two grids; at US East they are 0.3623, 0.3252 and 0.2898, and in the conservative high case at the global average 0.8485, 0.7600 and 0.6755. The training term is not per hour and is allocated on the tokens actually served (Table~\ref{tab:regionindep}).}\end{minipage}
\end{table}

\textbf{Development line and reserved capacity: full statements.} The two paragraphs of Section~\ref{sec:3.4} in full, with the arguments that the main text summarises:

\textbf{The development line.} All model development, including the final training runs, is reported as $\mathrm{EF_{dev}} = d \times E_{\mathrm{GPU}} \times H \times \mathrm{PUE} \times g$, where $d$ is the energy of that development as a multiple of serving energy, charged on serving energy without the idle uplift, because development workloads are scheduled for throughput rather than to a latency target and no capacity is held idle for them. The \emph{development-inclusive total} is the inference footprint plus that line, with the training term dropped because the line already includes the final training runs. The line is never added to the central factor, for three reasons given in Section S9.5: $d$ spans a factor of nearly four, it is a sector ratio rather than a disclosure about the model a customer is using, and it is a spend ratio applied to energy. Three arguments keep the training term and the line apart. \emph{Attribution}: the final training runs are a disclosed quantity attributable to one released model, whereas other development is a same-period flow of sector compute between research and serving that no single model owns. \emph{Direction}: the training term is biased low in its numerator, because final runs are 9.6\% to 22.6\% of research and development compute in the published reconstructions~\citep{denain2026}, and the serving volume $V$ is an early-life rate that biases its denominator high, although for the small class a larger true volume would make the term high (row~\ref{r:lifetok}). \emph{Independence}: the development-inclusive total does not use $V$ at all, since the line is charged on serving energy, so it does not depend on the true serving volume. It is therefore printed beside the factor by class and region, as a different boundary rather than as the top of an interval around the factor (Table~\ref{tab:structrange}). Two properties travel with it wherever it is reported, and both are stated in Section S9.5: it carries no embodied or construction term, so it is about 4\% short of the boundary the inference footprint uses, and its class split, which scales with serving energy, understates the frontier class's development and overstates the small class's.

\textbf{Reserved capacity.} Where accelerator-hours are contracted rather than tokens billed, the activity quantity is the reserved hour and two rules of the chain change: there is no idle uplift, because the hours themselves are the provisioned capacity, and there is no utilisation divisor in the embodied term, because the hours are contracted rather than served. Electricity per accelerator-hour is the measured average accelerator power at the class operating point times the host uplift, the PUE and the grid intensity; the embodied term is $(M_{\mathrm{mfg}} + M_{\mathrm{tr}})/(8 L \times 8{,}760)$; construction is $k_c$ times the server electricity of that hour; and the training term follows the tokens actually served. Every contracted hour is charged at the serving operating point, so an hour the customer does not use is charged at serving power rather than at the idle fraction $p$, which is the conservative choice. The values and the consistency check against the per-token factors are in S1.3. A preparer with both reserved capacity and on-demand billing on the same service uses reserved hours for the reserved portion and tokens for the on-demand overflow, and does not add the two for the same hours. Under the revised ESRS, where a reserved-capacity contract contains a lease of identified hardware, the emissions from its use are reported in the lessee's own operations rather than in Category 1 or Category 8 (ESRS 1, paragraphs 71 and 73; \citealp{eucom2026a}); a contract that identifies no asset stays in Category 1.

\subsubsection*{S1.4 Token Basis and Activity Defaults: Supporting Detail}

{\footnotesize\setlength{\tabcolsep}{4pt}
\begin{longtable}{>{\raggedright\arraybackslash}p{3.1cm} >{\raggedleft\arraybackslash}p{1.6cm} >{\raggedleft\arraybackslash}p{1.3cm} >{\raggedleft\arraybackslash}p{1.5cm} >{\raggedright\arraybackslash}p{7.0cm}}
\caption{Tier 2b default activity values, with the evidence behind each (billed tokens per exchange, or per model request for the agentic row; messages or actions per seat per month, the reasoning share and the reasoning multiplier for the lower block).}\label{tab:tier2b}\\
\toprule
Workload & Input & Output & Band on the total & Source \\
\midrule
\endfirsthead
\multicolumn{5}{l}{\textit{Table~\ref{tab:tier2b} (continued)}}\\
\toprule
Workload & Input & Output & Band on the total & Source \\
\midrule
\endhead
\midrule\multicolumn{5}{r}{\textit{continues on the next page}}\\
\endfoot
\bottomrule
\endlastfoot
Short chat & 928 & 282 & 600--2,600 & Input: Azure production inference trace, conversation service, median over all 27.3 M requests of the public one-week 2024 trace (12 to 18 May 2024)~\citep{azure2024}. Output: WildChat-1M, 19,234 exchanges~\citep{zhao2024} \\
Document-heavy & 3,293 & 383 & 2,282--8,536 & WildChat-1M conversations opening with at least 1,000 input tokens (496 conversations); median total 3,676~\citep{zhao2024} \\
Coding, inline completion & 1,944 & 8 & 960--3,372 & Azure production inference trace, coding service, median over all 16.8 M requests of the public one-week 2024 trace (10 to 16 May 2024); median total 1,952~\citep{azure2024} \\
Agentic coding, per model request & 7,614 uncached $+$ 76,986 cached & 400 & 71,179--119,346 & OpenRouter month view by application (Claude Code 96,884 tokens per request over 258.7 M requests; Codex 85,269; Cline 119,346)~\citep{openrouter2026}; the band runs from the documented Claude Code session (71,179) to Cline (119,346). The central 85,000 is the author's choice between the listed applications, close to Codex's 85,269; Claude Code's 96,884 carries more request volume and would raise the row by about 14\%. Cache-read share 0.91 (0.91--0.95) from a separate documented example on the same page, which is not stated to describe the same session as the token breakdown~\citep{anthropic2026c}, so it is illustrative rather than measured \\
\midrule
Messages (actions) per active enterprise seat per month & \multicolumn{2}{l}{30 (20--65)} & & UK Department for Business and Trade telemetry, 1,000 Microsoft 365 Copilot licences, 1 October to 30 December 2024: 1.14 actions per user per working day, reported per user and, being unadjusted for leave or part-time staff, read here as a mean over all licences rather than over active users, that is 24 actions per licensed seat per month (72 actions over 63 working days)~\citep{ukgov2025}. Converted to 30 per 30-day-active seat at the 80\% thirty-day-active rate reported for the 20,000-seat Government Digital Service experiment~\citep{ukgov2025}, of which the department was one participating organisation. This applies the whole experiment's rate to one department's counts and is an assumption rather than a measurement, and the department's own weekly-active rate was 64\%. Low end 20.6 per month is our conversion of that department's diary sub-study, in which 300 consenting licence holders logged 1,424 use cases over one week in November 2024, that is 4.75 a week~\citep{ukgov2025}. High end is Microsoft's power-user threshold, an average of 15 or more Copilot actions a week with use in at least 9 of the past 12 weeks, classified over a rolling 12-week window~\citep{adminreports2026} \\
Heavy-use seat & \multicolumn{2}{l}{325} & & OpenAI's power-user definition for ChatGPT Enterprise: the top 20\% of message senders, with 75 or more messages a week and three or more different tools a month~\citep{openai2026a}. The monthly figure is our restatement, $75 \times 52/12 = 325$. It is a threshold at the edge of the top quintile, not a mean of heavy users, and OpenAI elsewhere describes its heaviest users by percentile instead, placing ``frontier workers'' at the 95th percentile of adoption intensity and sending six times the median worker's messages~\citep{openai2025c} \\
Messages (actions) per licensed enterprise seat per month, used or not & \multicolumn{2}{l}{24 (no band)} & & Same telemetry as the active-seat row: 72 actions per licensed user over the pilot's three months, before the 80\% thirty-day-active rate; one measurement on one population, so no band. Used with licence counts, and the active-seat row with active-seat counts, never both \\
Reasoning share of conversations & \multicolumn{3}{>{\raggedright\arraybackslash}p{4.9cm}}{4.98\% (unweighted interquartile range across the 601 clusters, 2.92--6.44\%)} & Anthropic Economic Index open data, 601 clusters covering 96\% of records~\citep{anthropic2025} \\
Reasoning token multiplier $k$ & \multicolumn{3}{>{\raggedright\arraybackslash}p{4.9cm}}{25 (10--50)} & Author assumption: no public source gives a hidden-to-visible token ratio (S7). Applied as $1 + s_r(k-1)$ to the estimated \emph{output} tokens of a Tier 2b exchange only; input tokens are not multiplied, and Tier 2a billed counts never are \\
\end{longtable}}

\begin{table}[htbp]
\centering
\setlength{\tabcolsep}{4pt}
\footnotesize
\begin{tabular}{>{\raggedright\arraybackslash}p{3.6cm} >{\raggedleft\arraybackslash}p{1.5cm} >{\raggedleft\arraybackslash}p{1.6cm} >{\raggedleft\arraybackslash}p{1.6cm} >{\raggedleft\arraybackslash}p{1.6cm} >{\raggedleft\arraybackslash}p{1.5cm} >{\raggedleft\arraybackslash}p{1.6cm}}
\toprule
Workload & Billed tokens & Uncached input & Cached input & Output & $T_{\mathrm{eff}}$ & $T_{\mathrm{eff}}$ / billed \\
\midrule
Short chat & 1,210 & 928 & 0 & 282 & 418.5 & 0.346 \\
Document-heavy & 3,676 & 3,293 & 0 & 383 & 1,023.8 & 0.279 \\
Coding, inline completion & 1,952 & 1,944 & 0 & 8 & 441.5 & 0.226 \\
Agentic coding (per model request) & 85,000 & 7,614 & 76,986 & 400 & 4,879.4 & 0.057 \\
\bottomrule
\end{tabular}
\caption{Effective tokens for the four Tier 2b workloads at the central weights (tokens per exchange, or per model request for the agentic row).}
\label{tab:workload}

\begin{minipage}{\textwidth}\footnotesize \textit{Token counts are the Tier 2b defaults of Table~\ref{tab:tier2b}, before the reasoning uplift of Section~\ref{sec:3.2}, which multiplies the output column only (the short-chat exchange becomes 619.0 output, 1,547 billed and 670.2 effective tokens in the worked example). For the document-heavy and inline rows the uncached input is the median billed total minus the median output, because medians of input, output and total do not sum. At the low and high weights the ratio to billed tokens is 0.278 to 0.372 for short chat, 0.174 to 0.319 for document-heavy, 0.093 to 0.277 for inline completion and 0.029 to 0.090 for an agentic request, whose effective-token count therefore spans a factor of 3.2 across both weights and 2.5 across the cached weight alone. Only the agentic row moves with $W_{\mathrm{cache}}$, because it is the only one with cached input.}\end{minipage}
\end{table}

\emph{Readings behind $W_{\mathrm{in}} = 0.30$.} There is one energy-measurement anchor, read two ways. Differencing the published energies of ML.ENERGY's prefill--decode disaggregation experiment to remove fixed overheads gives 0.272, 0.172 and 0.171 on the three Llama 3.1 8B deployments (the 70B row is excluded as a deployment artefact rather than an energy ratio), and reading the balanced 512-in/512-out column of the same table directly gives 0.295, 0.304 and 0.299~\citep{chung2025}. The two readings are not independent: they are the same four rows. The within-cell reading is confounded only by idle power booked to an under-utilised prefill or decode pool, and it is the reading adopted, rounded to 0.30, as the author's choice: the table does not state the batch size or the hardware of its runs, so neither reading can be matched to this paper's batch sizes. Beside it sit two things that are not energy measurements of this quantity. Provider list prices put uncached input at 0.12 to 0.25 times output across OpenAI, Anthropic and Google current-generation models, clustering on 0.20~\citep{openai2026b,anthropic2026b,google2026b}; prices reflect commercial strategy. Greenpixie~\citep{horswill2026} state that the exact prefill-to-decode factor cannot be inferred from their figures, which read generously supports 0.03 to 0.3. Two further values are not used, one below the band and one inside it: the affine fits of \citet{vellaisamy2026} (Table III), a fixed energy per batch window plus a per-token decode energy on batches of at most 16 without continuous batching, imply 0.009 to 0.024 on this paper's own derivation, which spreads the fixed energy over the batch's input tokens and divides by the per-token step energy, a lower bound for production serving; and \citet{bistline2026} use input and output intensities of 0.32 and 0.96 Wh per 1,000 tokens, an implied 0.333 that rests on a 3-to-1 decode-to-prefill electricity ratio they take from Splitwise~\citep{patel2024} rather than measure.

\emph{Bounds behind $W_{\mathrm{cache}} = 0.05$.} The lower bound is commercial: all three large providers price a cache read at 0.10 times uncached input or less on current models (OpenAI 0.05 on one flagship), and Anthropic at 0.025 to 0.05 on Claude Fable 5.1, Mythos 5.1, Opus 5.5 and Sonnet 5.5, which gives $W_{\mathrm{cache}}$ about 0.02 of an output token~\citep{openai2026b,anthropic2026b,google2026b}. That is a lower bound on energy, because the price reflects the compute the provider avoids and not the decode-side attention it still pays. The upper bound is physical: decomposing the same disaggregation table into its prefill and decode halves shows that, on the two Llama 3.1 8B deployments whose prefill and decode pools stay balanced at long input (2P2D and 3P1D), 48.5\% and 49.6\% of the marginal energy of an extra uncached input token, at 512 output tokens, is extra decode work rather than prefill. A cache hit skips only the prefill part, so the decode part of that marginal energy, divided by the marginal energy of an extra output token on the same rows, is 0.084 and 0.085 of an output token, rounded up to an upper bound of 0.10~\citep{chung2025}. The 1P3D deployment gives 65.0\% and 0.18 because its three decode instances wait on a single prefill instance at long input: its decode energy rises to 140~J against 84 and 89~J on the balanced rows, while prefill energy is 68 to 70~J on all three. The 70B row, excluded above, gives 46.6\%. The same decode share applied to the adopted within-cell $W_{\mathrm{in}}$ of 0.30 would give 0.15, one reason the weight is biased low for cache-heavy workloads. The adopted band is exactly those two bounds and nothing wider; an earlier version used 0.01 to 0.20, which was wider than either derived bound in both directions. A third argument, physical but not a measurement, scales the ratio with output length: the decode-attention share of a cached token is $O/(512+O)$ of the differenced $W_{\mathrm{in}}$ of the balanced rows, 0.17, which at 282, 383 and 400 output tokens gives 0.061 to 0.076 of an output token. The adopted 0.05 sits below all three readings, between them and the price-implied value. Restated on the energy basis of this paper, the cache-read price ratio of 0.10 times $W_{\mathrm{in}} = 0.30$ gives a price-implied 0.03, whereas the printed lower bound of 0.02 takes the ratio against an input price of about 0.20 of output. Setting the central value below every physical reading is a named departure from the rule that sets a central value at or towards the conservative end of its evidence, the second rule of Section~\ref{sec:4}, so the weight is biased low on cache-heavy workloads. The band of 0.02 to 0.10 spans the two derived bounds and the readings of 0.061 to 0.085, while the 0.15 of the decode share applied to $W_{\mathrm{in}}$ lies above it.

\emph{Benchmark prefix caching.} An earlier draft could not establish whether the serving engine's automatic prefix caching was on in the leaderboard runs and carried $D = 1$ as a bounding case. The benchmark's raw run files settle it. Of the 388 \texttt{lm-arena-chat} runs, the leaderboard lists 328; the engine reports prefix caching on in 300 of them and off in the 28 of the two NVIDIA Nemotron models. Where it was on and the counters are clean (279 configurations), the cache served 52\% to 93\% of the prompt tokens, with a median of 62\%. It matters because the task re-sends the whole conversation history on every turn, which is the pattern prefix caching removes, so the prefill energy inside a published figure is below $W_{\mathrm{in}} R_{\mathrm{run}}$. Each measurement is therefore divided by its own $D_{\mathrm{run}}$ and multiplied by $D$, which states it at the reference mix; throughput is restated inversely, so the embodied term follows. The operating point is still selected on the measured values. In 21 of the 328 configurations the cache-query counter exceeds the prompt tokens, by up to 10.6 times. The benchmark's authors attribute this to preemption and recomputation when the configured batch is too large for the memory, describe such runs as unstable, and discard those in which the number of running requests sits more than about 15\% below the configured maximum (J.-W.~Chung, personal communication, 5 October 2026); the same communication gives the serving engine as vLLM 0.11.1 with fixes later merged upstream. For the 21 that remain listed the run's own counter is not a hit share of the prompt, and the median of the clean runs of the same model and accelerator type is used. Five of the 29 operating points are affected: Qwen3-8B and Qwen3-14B on H100, gemma-3-12b on B200 and gemma-3-27b on both. They are kept, and since recomputation spends energy without producing tokens their values lean high. Two limits remain. The counters are totals over a whole run while the energy is that of the steady-state window, so the steady-state hit share is approximated. And the restatement inherits $W_{\mathrm{in}}$ and $W_{\mathrm{cache}}$, the second of which has no energy measurement behind it. Taking the measured values as they are, the previous treatment, lowers the mid-class inference footprint by 8.6\% and the factor per million billed tokens by 8.0\% (Table~\ref{tab:sensfull}).

\subsubsection*{S1.5 Water Factor Derivation}

Water per 1,000 effective tokens for the mid class is computed as

\begin{alltt}\small
On-site  = denominator energy (Wh/1k) \(\times\) WUE
Off-site = Facility\_Wh\_per\_1K \(\times\) EWIF
Total    = On-site + Off-site
\end{alltt}

with WUE and the energy water intensity factor (EWIF) in L/kWh, numerically equal to mL/Wh, so the result is in mL per 1,000 effective tokens. The denominator energy is the server value, 0.1964 Wh per 1,000 effective tokens, for a WUE whose own denominator is IT load, and the facility value, 0.2357 Wh, for a site ratio derived from total site electricity. Off-site uses facility energy, because an EWIF is defined per kWh drawn from the grid. On-site values published on a withdrawal basis are then divided by 1.5 to reach consumption. The scarcity-weighted result multiplies the \emph{on-site} consumption, and only that, by the region's AWARE 2.0 non-agricultural annual factor~\citep{seitfudem2025}. That method is defined at watershed and monthly resolution, and its developers supply country and annual aggregates only where the watershed and the month are unknown, recording that these carry very large uncertainty; classing a data centre in the non-agricultural aggregation, which is the one built for domestic and industrial consumption, is the inference, since they nowhere name the sector. Water uses the same uniform PUE of 1.20 as the carbon chain; the site PUEs of 1.14 to 1.28 reported by Li et al.\ would change off-site water by $-$5\% to $+$7\%.

\textbf{Worked example (Ireland, mid class).} On-site as published is $0.1964 \times 0.020 = 0.0039$ mL, which on a consumption basis is $0.0039/1.5 = 0.0026$ mL; off-site is $0.2357 \times 1.476 = 0.3479$ mL, and the total is 0.351 mL, which matches Table~\ref{tab:water}. The on-site water scarcity footprint is $0.0026 \times 1.17 = 0.0031$ mL world-equivalent, printed as 0.003. For a derived-site row the denominator differs: Iowa's on-site term is $0.2357 \times 0.1227 = 0.0289$ mL as published, on facility rather than server energy, and 0.0193 mL on a consumption basis.

\textbf{Caveats on the derived site ratios.} Three attach to the four rows that rest on a provider's own published site water and site electricity. Their numerator is all water withdrawn at the site and their denominator total site electricity, so they are site water intensities applied to facility rather than server energy. Some are partly circular, because Microsoft estimates withdrawal for sites without primary data from a method that itself considers electricity consumption, and its withdrawal definition was informed by Global Reporting Initiative disclosure 303-3~\citep{microsoft2026a}. And operator values in the same metropolitan area can differ several-fold: Microsoft's two northern-Virginia sites, Ashburn and Manassas, give 0.288 and 0.275 L/kWh against the AWS region value of 0.06 used here, so that row should be read as the low end of its range.

\textbf{Hydropower-excluded factors.} Reig et al.\ publish one consumption factor per grid. The excluded variant rebuilds it from the source's own per-technology factors and generation mix (Appendices 4 and 6), reads off the share of the \emph{reconstructed water} that is hydropower, which for Sweden is 77\% against a 41.6\% share of generation, and multiplies the published factor by one minus that share: $6.019 \times 0.230 = 1.386$ L/kWh. The reconstruction reproduces the published factor within a factor of 0.60 to 1.20 for fourteen of the fifteen rows; the exception is the Netherlands, at 0.27, whose reconstructed hydropower share is zero and whose excluded factor therefore equals the published one.

\textbf{Cross-checks on the off-site term.} The off-site term is the one to scrutinise, and it is checkable for the United States, where three routes agree on the order of magnitude, although the methods behind the LBNL and Meta figures are not stated here. \citet{reig2020} give 3.14 L/kWh of consumptive water per kWh of United States grid electricity, the LBNL assessment gives 4.35 L/kWh for the national average and 4.52 L/kWh weighted to the grid mix of data-centre locations~\citep{shehabi2024}, and Meta's own disclosure of water embedded in purchased electricity, divided by its data-centre electricity, gives 4.00 L/kWh for 2024 and 3.70 L/kWh for 2023~\citep{meta2025}. The factors used here are therefore about 30\% below the two United States averages, which biases the United States rows low. No comparable cross-check exists for the European or Asian rows.

\textbf{What is still not available.} No source gives an EWIF for Singapore. No operating Finnish site publishes water with electricity, so Finland rests on a projection for a site then under construction. No hydropower water factor exists for Sweden. The published fleet water use effectiveness figures of the five operators that publish one span an order of magnitude and do not share a basis, at 0.12 (AWS, withdrawal over IT load, 2025), 0.19 (Meta, withdrawal over IT load, 2024), 0.27 (Microsoft, cooling and humidification water over IT load, FY25), 0.91 (Equinix, cooling water per IT kWh with neither withdrawal nor consumption specified, 2025, \citealp{equinix2025}) and 1.15 L/kWh (Google, consumption over IT load, published only by \citealp{elsworth2025}), so the choice of a global row is a judgement that has to be stated. Google's own reports carry no water use effectiveness value: its 2026 Environmental Report refers in an endnote to a 2024 fleetwide figure without giving one~\citep{google2026c}. Oracle publishes no such figure in any disclosure checked~\citep{equinix2025}, and no row here rests on an Equinix or Oracle value. For the same reason water is not extended in the dataset beyond the seventeen printed regions: no operator publishes a water use effectiveness figure for the additional regions, whereas the off-site source covers 47 countries and all eGRID subregions.

\subsection*{S2. Full Sensitivity Table, Data Quality and Further Limitations}

\footnotesize
\setlength{\tabcolsep}{3pt}
\begin{longtable}{>{\raggedright\arraybackslash}p{6.4cm} >{\raggedleft\arraybackslash}p{1.3cm} >{\raggedleft\arraybackslash}p{1.3cm} >{\raggedleft\arraybackslash}p{1.3cm} >{\raggedleft\arraybackslash}p{1.5cm} >{\raggedleft\arraybackslash}p{1.3cm} >{\raggedleft\arraybackslash}p{1.5cm}}
\caption{One-at-a-time sensitivities, H100, global-average grid unless stated. Inference-footprint columns are kg CO\textsubscript{2}e per million \emph{effective} tokens; factor columns are kg CO\textsubscript{2}e per million \emph{billed} tokens at the default mix, with the effective-to-billed multiplier re-derived for each row's weights.}\label{tab:sensfull}\\
\toprule
Sensitivity & A footprint & B footprint & C footprint & B vs central & B factor & B factor change \\
\midrule
\endfirsthead
\multicolumn{7}{l}{\emph{Table~\thetable\ (continued)}} \\
\toprule
Sensitivity & A footprint & B footprint & C footprint & B vs central & B factor & B factor change \\
\midrule
\endhead
\bottomrule
\endfoot
\bottomrule
\endlastfoot
Central & 0.0361 & 0.1187 & 0.4425 & --- & 0.0388 & --- \\
Host uplift $H = 1.43$ & 0.0345 & 0.1133 & 0.4225 & $-$4.5\% & 0.0372 & $-$4.2\% \\
Host uplift $H = 1.88$ & 0.0450 & 0.1480 & 0.5514 & $+$24.6\% & 0.0477 & $+$22.8\% \\
\quad $H = 1.667$ (TokenPowerBench ceiling check) & 0.0400 & 0.1316 & 0.4904 & $+$10.8\% & 0.0427 & $+$10.0\% \\
Idle uplift $I = 1.05$, $U$ unchanged & 0.0345 & 0.1135 & 0.4230 & $-$4.4\% & 0.0373 & $-$4.1\% \\
Idle uplift $I = 1.20$, $U$ unchanged & 0.0393 & 0.1292 & 0.4816 & $+$8.8\% & 0.0420 & $+$8.2\% \\
$I = 1.05$ with $U = 0.857$ (tied pair) & 0.0344 & 0.1131 & 0.4214 & $-$4.8\% & 0.0371 & $-$4.4\% \\
$I = 1.20$ with $U = 0.60$ (tied pair) & 0.0395 & 0.1301 & 0.4848 & $+$9.5\% & 0.0423 & $+$8.8\% \\
PUE 1.09 & 0.0329 & 0.1083 & 0.4037 & $-$8.8\% & 0.0357 & $-$8.1\% \\
PUE 1.45 & 0.0434 & 0.1425 & 0.5309 & $+$20.0\% & 0.0460 & $+$18.5\% \\
\textbf{Conservative high case} ($H$ 1.88, $I$ 1.20 with $U$ 0.60, PUE 1.45) & \textbf{0.0592} & \textbf{0.1944} & \textbf{0.7243} & \textbf{$+$63.7\%} & \textbf{0.0617} & \textbf{$+$58.9\%} \\
Hardware life 4 years & 0.0363 & 0.1196 & 0.4457 & $+$0.7\% & 0.0391 & $+$0.7\% \\
Hardware life 6 years & 0.0360 & 0.1182 & 0.4404 & $-$0.5\% & 0.0387 & $-$0.4\% \\
Server footprint 2,872 kg & 0.0360 & 0.1184 & 0.4411 & $-$0.3\% & 0.0387 & $-$0.3\% \\
Server footprint 4,241 kg & 0.0363 & 0.1195 & 0.4456 & $+$0.7\% & 0.0391 & $+$0.6\% \\
Transport 427 kg & 0.0361 & 0.1185 & 0.4417 & $-$0.2\% & 0.0388 & $-$0.2\% \\
Transport 712 kg & 0.0361 & 0.1188 & 0.4427 & $+$0.0\% & 0.0389 & $+$0.0\% \\
Construction 5.3 g per server kWh & 0.0360 & 0.1183 & 0.4408 & $-$0.4\% & 0.0387 & $-$0.4\% \\
Hardware block combined low (server 2,872 kg, transport 427 kg, life 6 y, construction 5.3 g) & 0.0357 & 0.1172 & 0.4368 & $-$1.3\% & 0.0384 & $-$1.2\% \\
Hardware block combined high (server 4,241 kg, transport 712 kg, life 4 y, construction central) & 0.0366 & 0.1206 & 0.4497 & $+$1.6\% & 0.0394 & $+$1.5\% \\
Service life $S = 6$ months & 0.0361 & 0.1187 & 0.4425 & $+$0.0\% & 0.0418 & $+$7.6\% \\
Service life $S = 18$ months & 0.0361 & 0.1187 & 0.4425 & $+$0.0\% & 0.0379 & $-$2.5\% \\
Usage shape factor 0.4 & 0.0361 & 0.1187 & 0.4425 & $+$0.0\% & 0.0432 & $+$11.3\% \\
Lifetime tokens 500 trillion (superseded) & 0.0361 & 0.1187 & 0.4425 & $+$0.0\% & 0.0400 & $+$2.9\% \\
Class-specific training volumes & 0.0361 & 0.1187 & 0.4425 & $+$0.0\% & 0.0374 & $-$3.8\% \\
Training default 0.1 kg per M tokens~\citep{bistline2026} & 0.0361 & 0.1187 & 0.4425 & $+$0.0\% & 0.1359 & $+$249.9\% \\
$W_{\mathrm{in}} = 0.10$ & 0.0341 & 0.1092 & 0.4093 & $-$8.1\% & 0.0260 & $-$33.1\% \\
$W_{\mathrm{in}} = 0.40$ & 0.0371 & 0.1229 & 0.4572 & $+$3.5\% & 0.0444 & $+$14.3\% \\
No restatement to the reference mix (previous treatment) & 0.0337 & 0.1085 & 0.4067 & $-$8.6\% & 0.0358 & $-$8.0\% \\
Minimum at the smallest accelerator count, no latency limit & 0.0403 & 0.1401 & 0.3688 & $+$18.0\% & 0.0453 & $+$16.6\% \\
Geometric mean of the class members & 0.0345 & 0.1144 & 0.3686 & $-$3.7\% & 0.0375 & $-$3.4\% \\
Median of the class members & 0.0354 & 0.1379 & 0.4960 & $+$16.2\% & 0.0447 & $+$15.0\% \\
Named proxy model (previous rule) & 0.0193 & 0.1324 & 0.4130 & $+$11.5\% & 0.0430 & $+$10.6\% \\
Mean over all configurations meeting the latency limit & 0.0763 & 0.3446 & 1.0840 & $+$190.3\% & 0.1072 & $+$175.9\% \\
Lowest-footprint member of each class & 0.0215 & 0.0730 & 0.1477 & $-$38.5\% & 0.0250 & $-$35.6\% \\
Highest-footprint member of each class & 0.0533 & 0.1463 & 0.6839 & $+$23.2\% & 0.0472 & $+$21.4\% \\
Frontier class with the FP8 measurement of every member that has one & 0.0361 & 0.1187 & 0.4005 & $+$0.0\% & 0.0388 & $+$0.0\% \\
FP8 for the frontier class (4 accelerators, 50 ms) & 0.0361 & 0.1187 & 0.5699 & $+$0.0\% & 0.0388 & $+$0.0\% \\
FP8 for the frontier class (8 accelerators, 50 ms) & 0.0361 & 0.1187 & 0.3714 & $+$0.0\% & 0.0388 & $+$0.0\% \\
DeepSeek V3.1 as the frontier class on B200 & 0.0314 & 0.0802 & 0.4346 & $-$32.5\% & 0.0272 & $-$30.0\% \\
US East, scaled eGRID (reference row for the next) & 0.0232 & 0.0765 & 0.2851 & $-$35.6\% & 0.0261 & $-$32.9\% \\
US East, unscaled eGRID & 0.0208 & 0.0686 & 0.2558 & $-$42.2\% & 0.0237 & $-$39.0\% \\
\end{longtable}
\normalsize

\textit{Each row changes one input or one structural choice and holds the rest central, except the two hardware-block rows, which move four inputs together. $I$ and $U$ describe one fleet, so the two $I$ rows move the electricity term only while the two tied-pair rows move the embodied term with it. The hardware-life, footprint and transport rows move the embodied term only, and their effect is larger on a clean grid, where the grid-independent terms are a large share of the factor (Table~\ref{tab:nonelec}). The operating-point rows move the energy and the embodied term together, the latter recomputed at the throughput of the same configurations. $W_{\mathrm{cache}}$ claims no cache credit at the default mix, so there it acts only through the restatement of each benchmark run (Section S9.6), and its band on the customer's tokens is in Table~\ref{tab:workload}. The US East rows are not at the global-average grid and their percentages are against the global central. Fifteen rows vary quantities that are not inputs of Table~\ref{tab:master}: the superseded lifetime-token denominator, class-specific training volumes, an external training default, the two tied pairs, the two hardware-block combinations, four operating-point and class-statistic alternatives, the lowest and highest member of each class, FP8 precision for the frontier class, and the unscaled eGRID basis.}

\begin{table}[htbp]
\centering
\footnotesize
\begin{tabular}{>{\raggedright\arraybackslash}p{3.2cm} l r r r r r r r}
\toprule
Quantity & Class & Central & Median & 5th pct & 95th pct & 5th / 95th vs central & 2.5th pct & 97.5th pct \\
\midrule
Inference footprint, kg per M effective tokens & A --- Small & 0.0361 & 0.0404 & 0.0352 & 0.0471 & $-$2.5\% / $+$30.5\% & 0.0345 & 0.0484 \\
 & B --- Mid & 0.1187 & 0.1326 & 0.1158 & 0.1549 & $-$2.4\% / $+$30.4\% & 0.1134 & 0.1589 \\
 & C --- Frontier & 0.4425 & 0.4944 & 0.4317 & 0.5772 & $-$2.4\% / $+$30.4\% & 0.4227 & 0.5923 \\
Factor per M billed tokens, default mix & A --- Small & 0.0115 & 0.0122 & 0.0096 & 0.0150 & $-$16.8\% / $+$29.8\% & 0.0092 & 0.0155 \\
 & B --- Mid & 0.0388 & 0.0409 & 0.0319 & 0.0502 & $-$17.8\% / $+$29.3\% & 0.0306 & 0.0520 \\
 & C --- Frontier & 0.1467 & 0.1545 & 0.1215 & 0.1892 & $-$17.2\% / $+$29.0\% & 0.1163 & 0.1959 \\
\bottomrule
\end{tabular}
\caption{Monte Carlo propagation of the input bands: 20,000 draws, H100, global-average grid. Each input is drawn independently from a triangular distribution over its band with the mode at the central value. The inference footprint varies the host uplift, the idle uplift with its tied utilisation, PUE, the server footprint, transport, the hardware life and the construction constant; the factor per million billed tokens varies in addition the service life behind the training term and the two token weights, with the benchmark runs restated at the drawn weights. Not drawn: the accelerator energy, whose class range is a spread between models, the grid intensity, the activity data, the usage shape factor and the development ratio. $H$, $I$ and PUE are drawn independently although their low and central values share the Google source, $W_{\mathrm{in}}$ is not drawn on the inference footprint, and the idle-power fraction $p$ is held fixed (S2).}
\label{tab:mc}
\end{table}

\begin{table}[htbp]
\centering
\footnotesize
\setlength{\tabcolsep}{3pt}
\begin{tabular}{>{\raggedright\arraybackslash}p{4.0cm} r r r r r r r}
\toprule
Region & Grid 2023 & Grid 2024 & Grid & Footprint A & Footprint B & Footprint C & Billed factor B \\
\midrule
EU North --- Sweden & 0.0384 & 0.0349 & $-$9.1 & $-$6.1 & $-$6.0 & $-$5.9 & $-$3.5 \\
EU West --- France & 0.0533 & 0.0405 & $-$24.0 & $-$17.7 & $-$17.3 & $-$17.2 & $-$11.1 \\
EU North --- Finland & 0.0813 & 0.0668 & $-$17.9 & $-$14.6 & $-$14.3 & $-$14.2 & $-$10.2 \\
EU West --- Belgium & 0.1347 & 0.1273 & $-$5.5 & $-$4.8 & $-$4.8 & $-$4.7 & $-$3.8 \\
EU-27 average & 0.2360 & 0.2115 & $-$10.4 & $-$9.6 & $-$9.5 & $-$9.5 & $-$8.2 \\
EU West --- UK (London) & 0.2356 & 0.2165 & $-$8.1 & $-$7.5 & $-$7.5 & $-$7.4 & $-$6.4 \\
EU West --- Netherlands & 0.2682 & 0.2507 & $-$6.5 & $-$6.1 & $-$6.0 & $-$6.0 & $-$5.3 \\
EU West --- Ireland & 0.2825 & 0.2709 & $-$4.1 & $-$3.8 & $-$3.8 & $-$3.8 & $-$3.4 \\
EU Central --- Germany & 0.3636 & 0.3371 & $-$7.3 & $-$6.9 & $-$6.9 & $-$6.9 & $-$6.2 \\
US East --- N. Virginia (proxy) & 0.3038 & 0.2968 & $-$2.3 & $-$2.2 & $-$2.2 & $-$2.2 & $-$1.9 \\
US West --- Oregon (proxy) & 0.3237 & 0.3162 & $-$2.3 & $-$2.2 & $-$2.2 & $-$2.2 & $-$1.9 \\
US South Central --- Texas (proxy) & 0.3753 & 0.3666 & $-$2.3 & $-$2.2 & $-$2.2 & $-$2.2 & $-$2.0 \\
US South --- Southeast (proxy) & 0.4310 & 0.4210 & $-$2.3 & $-$2.2 & $-$2.2 & $-$2.2 & $-$2.0 \\
US Central --- Iowa (proxy) & 0.4721 & 0.4612 & $-$2.3 & $-$2.2 & $-$2.2 & $-$2.2 & $-$2.0 \\
Asia Pacific --- Singapore & 0.5009 & 0.4987 & $-$0.4 & $-$0.4 & $-$0.4 & $-$0.4 & $-$0.4 \\
Asia Pacific --- Japan & 0.4926 & 0.4836 & $-$1.8 & $-$1.8 & $-$1.8 & $-$1.8 & $-$1.6 \\
Global average & 0.4832 & 0.4715 & $-$2.4 & $-$2.3 & $-$2.3 & $-$2.3 & $-$2.2 \\
\bottomrule
\end{tabular}
\caption{Grid-year sensitivity: every printed region recomputed with the calendar-year 2024 grid intensity in place of 2023, every other input unchanged. Grid intensities in kg CO\textsubscript{2}e per kWh; the other columns are percentage changes against the central 2023 value, for the inference footprint of each model class (A, B, C) on the H100 central case and for the Class B factor per million billed tokens. Non-US rows, the EU-27 and the global average use Ember's 2024 values from the same release as the 2023 ones. The United States Environmental Protection Agency has not published eGRID 2024, so the US rows (marked proxy) scale the 2023 subregion values uniformly by Ember's United States change, 2024 over 2023, a factor of 0.9769 (383.81 over 392.9 g per kWh), and carry no subregional information.}
\label{tab:gridyear}
\end{table}

\begin{table}[htbp]
\centering
\setlength{\tabcolsep}{4pt}
\small
\begin{tabular}{>{\raggedright\arraybackslash}p{2.6cm} >{\raggedleft\arraybackslash}p{1.8cm} >{\raggedleft\arraybackslash}p{1.6cm} >{\raggedleft\arraybackslash}p{2.0cm} >{\raggedleft\arraybackslash}p{1.9cm} >{\raggedleft\arraybackslash}p{1.7cm} >{\raggedright\arraybackslash}p{2.0cm}}
\toprule
Tier (quantity scored) & Technologi\-cal & Tempo\-ral & Geographi\-cal & Complete\-ness & Reliabil\-ity & Composite \\
\midrule
3, supplier figure & 1 & 1--2 & 1 & 2 & 1--3 & 1--2 \\
2a, activity data & 1 & 1 & n/a & 2 & 2 & 2 \\
2a, emission factor & 3 & 4 & 4 & 2 & 2 & 3 \\
2b, activity data & 4 & 3 & 4 & 4 & 4 & 4 \\
2b, emission factor & 3 & 4 & 4 & 2 & 2 & 3 \\
1, spend-based & 5 & 5 & 4 & 1 & 3 & 5 \\
\bottomrule
\end{tabular}
\caption{Data-quality indicators behind the two composite scores of Table~\ref{tab:overview}. The five indicators are those of the Scope 3 Standard, Chapter 7~\citep{wri2011}; the one-to-five numbering is ours, 1 being the Standard's ``very good'' and 5 a level below its ``poor'' (dimensionless scores). Each composite is the mean of the indicators scored, rounded half up, or 5 where any indicator is 5 (Section~\ref{sec:3.8}); for the Tier 3 ranges the rule is applied at each end.}
\label{tab:dq}

\begin{minipage}{\textwidth}\footnotesize \textit{The temporal and geographical indicators are the weakest for both token-based emission-factor rows: the grid data are calendar year 2023 applied to FY2025 activity, the benchmark data are February 2026, and the serving region is usually not disclosed, so a national or global average stands in for a facility. Completeness is good for Tier 1 because spend captures the whole purchase, and poor for Tier 2b because estimated tokens omit whatever the seat default does not describe. Reliability for Tier 3 depends on the supplier's assurance status. Each composite is the arithmetic mean of the indicators scored, rounded half up, that is towards the worse score, except that any indicator at level 5 sets the composite to 5, because a level below the Standard's own scale is not offset by good scores on the others, and only spend-based data reach it. Where an indicator is a range, as at Tier 3, the rule is applied to the low and the high ends. The rule mapping indicators to composites (Section~\ref{sec:3.8}) is the author's, and so is the numbering: the Scope 3 Standard gives the five indicators and illustrates a four-level rating of very good, good, fair and poor as an example only, leaves companies to determine how the indicators are applied, and prescribes no aggregation~\citep{wri2011}. The Technical Guidance has no data-quality chapter and cross-refers to the Standard for them~\citep{wri2013}.}\end{minipage}
\end{table}

\begin{figure}[htbp]
\centering
\includegraphics[width=0.95\textwidth]{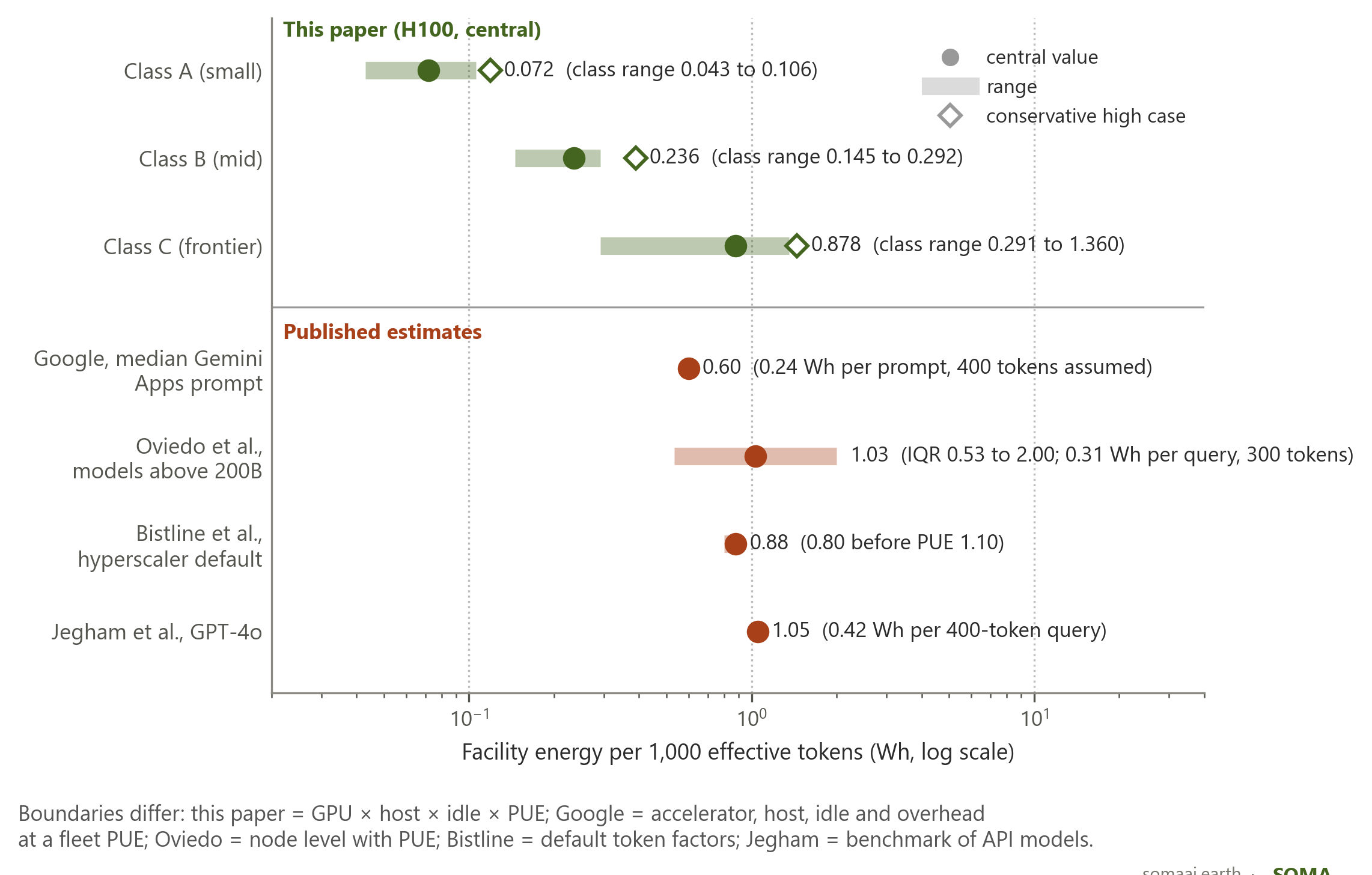}
\caption{Facility energy per 1,000 \emph{effective} tokens for the three classes, with the class range and the conservative high case, against four published estimates restated on the same basis, on a logarithmic scale. The published figures are per prompt or per query at their own assumed prompt lengths, which the panel states beside each, so the restatement carries that assumption; the boundaries also differ, as the note under the panel records. It restates the published figures per token at the token counts the panel names, so its rows and token bases differ from the per-exchange comparison of Table~\ref{tab:comparison}.}
\label{fig:cmpenergy}
\end{figure}

Figure~\ref{fig:cmpenergy} supports Section~\ref{sec:5.1} by restating the per-query comparison on a per-1,000-token basis.

\begin{figure}[htbp]
\centering
\includegraphics[width=1.0\textwidth]{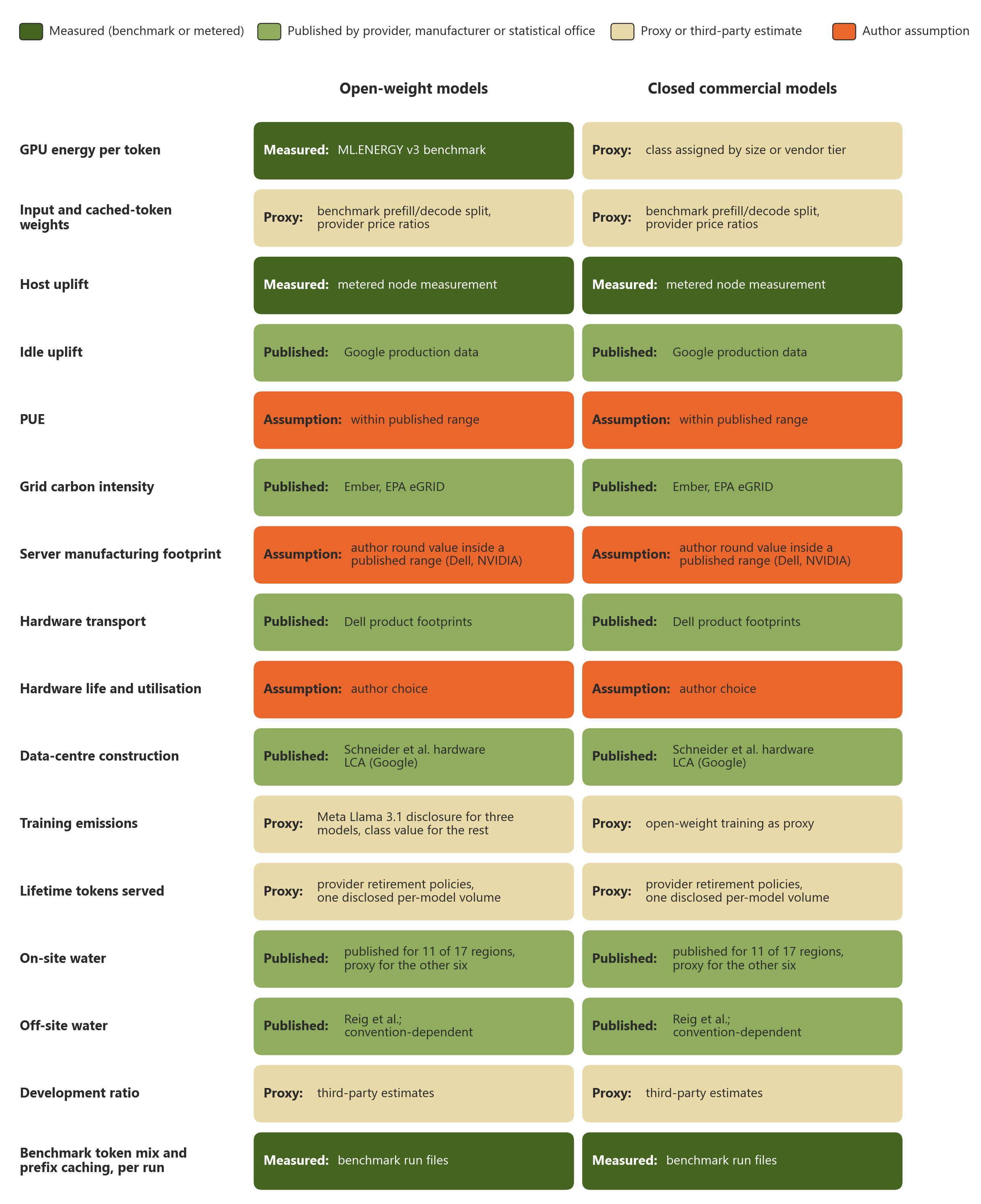}
\caption{Evidence class of each input of the carbon and water factors, for open-weight and for closed commercial models. Four classes are distinguished: measured on a benchmark or metered node; published by a provider, manufacturer or statistical office; a proxy or third-party estimate; and an author assumption. The two columns differ in one row only, the accelerator energy, which is a measurement for an open-weight model and a class value assigned by size or vendor tier for a closed one. Three rows are graded below their appearance elsewhere in this paper and the grading is the honest one: the server manufacturing footprint is an author-chosen round value inside a published range, the training emissions are a Meta disclosure for three models and a class value for the other fifteen, and the on-site water is published or derived from operator site data for twelve of seventeen regions and a proxy or aggregate for the other five.}
\label{fig:evidence}
\end{figure}

\textbf{Limitations of the token basis.} $W_{\mathrm{in}}$ is also a function of serving concurrency, which a customer cannot observe, and the 70B reading of about 0.54 lies above the band. The agentic workload sits at about 85,000 tokens of context, far beyond the 4,096-token range over which the disaggregation evidence was measured, so its effective-token figure is an extrapolation and the weakest-evidenced quantity in the token basis. The mix of each benchmark run is measured, but its cache-hit share is a whole-run total applied to a steady-state energy figure, and the restatement to the reference mix carries the same two weights as the customer's token count.

\textbf{Newer model generations.} The frontier class is calibrated on three measured open-weight models, which are low-side proxies for closed frontier models. Closed commercial models released after the benchmark data are not individually measured. Generation-to-generation changes in per-token energy are not bounded by the class range, since reasoning settings alone change energy per query several-fold within one model (\citealp{jegham2025}; see Section~\ref{sec:7.2}). The benchmark software (vLLM, February 2026) is newer than FY2025 serving stacks, which biases the factors low for FY2025 activity. Factor tables should be updated at least annually.

\textbf{Water data coverage.} Twelve of the seventeen regions have a confirmed on-site value and eleven a confirmed total, Singapore having no off-site source. France, Belgium and the EU-27 rest on a continental aggregate, Finland on a projection for a site that was under construction, and the global row on a United States average. Singapore and the EU-27 have no off-site source at all. Every value represents hyperscale operation by one of the three operators that publish enough to compute one, so none describes a non-hyperscale or hot-climate facility: AWS publishes 2.85 L/kWh for its Jakarta region and 1.57 for Singapore against 0.02 for Stockholm~\citep{aws2026b}, and the series of \citet{li2025} puts an Arizona site at 1.63 L/kWh, so a facility outside the regions tabulated here can be two orders of magnitude above the lowest row. The off-site term is the larger of the two in every region and rests on one source whose generation mixes date from 2018.

\textbf{Water conventions.} Three results are reported rather than one, because two of the conventions are unsettled in the literature and the third, the scarcity weighting, can only be applied to the smaller of the two terms. The hydropower-excluded factor is a reconstruction accurate only to a factor of 0.60 to 1.20, and every generation mix behind the off-site term dates from 2018 while the grid intensities are 2023, so the carbon and water results for a region do not rest on the same year's mix. Only eleven of the seventeen regions rest on a published on-site value, the scarcity factors are at country or state resolution, and the scarcity weighting reaches only the on-site term, so comparisons of locations on scarcity are indicative.

\textbf{Reasoning models.} Models that generate internal chain-of-thought tokens consume energy on both visible and hidden tokens, so the per-token factors are applied unchanged, which understates long reasoning interactions (Section~\ref{sec:7.2}), and the output token count per interaction can be an order of magnitude or more above standard generation. No provider publishes hidden-to-visible token ratios, so the 10 to 50 times range is an author assumption used for Tier 2b message-based estimates only; Tier 2a billing counts already include hidden tokens and must never be multiplied. That range is a multiplier on \emph{output tokens}, whereas published figures such as the 2.33 Wh and 17.15 Wh that \citet{jegham2025} report for GPT-5 at minimal and high reasoning effort on the same query are \emph{energy} ratios at a fixed interaction, and the two must never be substituted for one another.

\textbf{Location-based basis and no reconciliation.} All grid intensities are location-based annual averages, so the resulting figures do not reflect providers' renewable procurement, only supplier-specific data meeting the Tier 3 criteria can lower a Category 1 figure, and the factor is not reconciled to any provider's published inventory (Section~\ref{sec:3.1}).

\textbf{Scope of the propagation.} The Monte Carlo draws of Table~\ref{tab:mc} treat $H$, $I$ and PUE as independent, although their low and central values share the Google source (Section S9.2), so the joint spread of the three may be understated. $W_{\mathrm{in}}$ is drawn on the factor per million billed tokens but not on the inference footprint, which it also moves through the restatement of each benchmark run (Table~\ref{tab:sens}). The idle-power fraction $p$ is held at its central value in every draw.

\textbf{Temporal validity.} Preparers should use the latest grid year available at the time of reporting, disclose it, and treat a grid-year change as a factor-version change. Two inputs are known to have moved since the dataset was assembled. The Lawrence Berkeley National Laboratory issued a 2025 Update to the assessment cited here as \citet{shehabi2024} after the dataset was built, and the five figures that rest on it, the two United States water intensities, the direct-to-indirect water ratio and the 1.43 and 1.45 PUE values, are a pending refresh item rather than a correction. And the AWARE 2.0 and EXIOBASE open items remain as recorded in the deposited changelog.

\textbf{What drives the result: further discussion.} The discussion of Section~\ref{sec:7.1} in full:

\textbf{Model class and region are of the same order, and which dominates depends on the boundary.} On the electricity term alone the small-to-frontier ratio is 12.2 and the ratio between the cleanest and the most carbon-intensive region printed here, Sweden and Singapore, is 13.0. Adding the two grid-independent terms changes that: they hold flat across regions, so they compress the regional spread to 8.8 times on the mid-class inference footprint (9.1 and 8.8 for the small and frontier classes) while leaving the class spread at 12.3 times globally and 12.7 times on the Swedish grid. For the mid class those terms are 0.0048 kg per million effective tokens everywhere, which is 4\% of the electricity term at the global average and 54\% of it in Sweden. The practical reading is that a preparer needs both, since either alone leaves about an order of magnitude open and the two together span a factor of about a hundred. These ratios hold for the class definitions, the H100 measurements and the 2023 grid data used here.

\textbf{Development is of the same order as the factor, and larger at the central ratio.} Per million billed tokens at the default mix, the development line for the mid class at the global average is 0.0210 to 0.0783 kg CO\textsubscript{2}e, 0.54 to 2.0 times the central factor and 1.2 times it at the central ratio, and a development-inclusive total is 1.5 to 2.9 times the factor (Table~\ref{tab:structrange}). It is the largest quantity in the value chain that the factor does not carry, and it is not carried for the three reasons given in Section~\ref{sec:3.4}.

\textbf{Spend-based and physical figures compared.} On the same purchase, the spend-based figure is 1.7 to 12.3 times the physical estimate on the deflated spend the Tier 1 rule of Section~\ref{sec:3.7} requires, and 2.0 to 15.1 times before deflation. The lower end is the efficiency tier, gpt-4o-mini, against the small class at the global-average grid, and the upper end the standard tier, gpt-4o, against the mid class in US East, both at list price per million billed tokens of the 928-in/282-out chat exchange (Table~\ref{tab:register}). On the Swedish grid the same comparisons are 10.7 and 46.7 times deflated, because the physical estimate falls while the price does not. The ratio is not a measure of error in either figure, since it mixes a boundary difference, the whole economic footprint of the supplying sector per euro, with a price effect, but a preparer who moves a purchase from Tier 1 to Tier 2 should expect a change of this size, larger than every sensitivity in Table~\ref{tab:sens} except the operating point and the external training default of \citet{bistline2026}.

\textbf{A parametric reading of the method.} Parametric life cycle assessment expresses environmental impacts as explicit functions of process parameters, so that scenarios are explored by changing parameters and not by rebuilding the model, as in recent parametric models of industrial processes~\citep{spreafico2025}. The method here is parametric in that sense, in structure only, since it performs no impact assessment and is not a life cycle assessment. Three extensions follow directly. A preparer or a provider can substitute its own parameters, for example a metered PUE or a contracted region, and obtain a factor on the same boundary; trade-offs between indicators can be explored on one parameter set, as the carbon and water rankings of S3 do for location; and another modality needs a new accelerator-energy term and reuses the rest of the chain. What a parametric model does not supply is evidence for the parameter values of closed services, and that stays the binding limit.

\subsection*{S3. Water: Rankings, Figures and Sensitivities}

The water factors are in Table~\ref{tab:water} and their derivation in S1.5. This section gives what follows from them.

\textbf{What the three results are called.} ISO 14046 uses the unqualified term \emph{water footprint} only for the result of a comprehensive impact assessment, requires a qualifier where water-related impacts have not been comprehensively assessed, and gives the result of the inventory analysis a separate defined term of its own~\citep{iso2014}. The two volumetric results here are therefore \emph{water-use inventory results}, a coinage of ours that is consistent with the standard's defined term \emph{water use} without being one of its terms, and only the weighted on-site result is called a \emph{water scarcity footprint}, which is the standard's own leading example of such a qualified term.

\begin{figure}[htbp]
\centering
\includegraphics[width=0.86\textwidth]{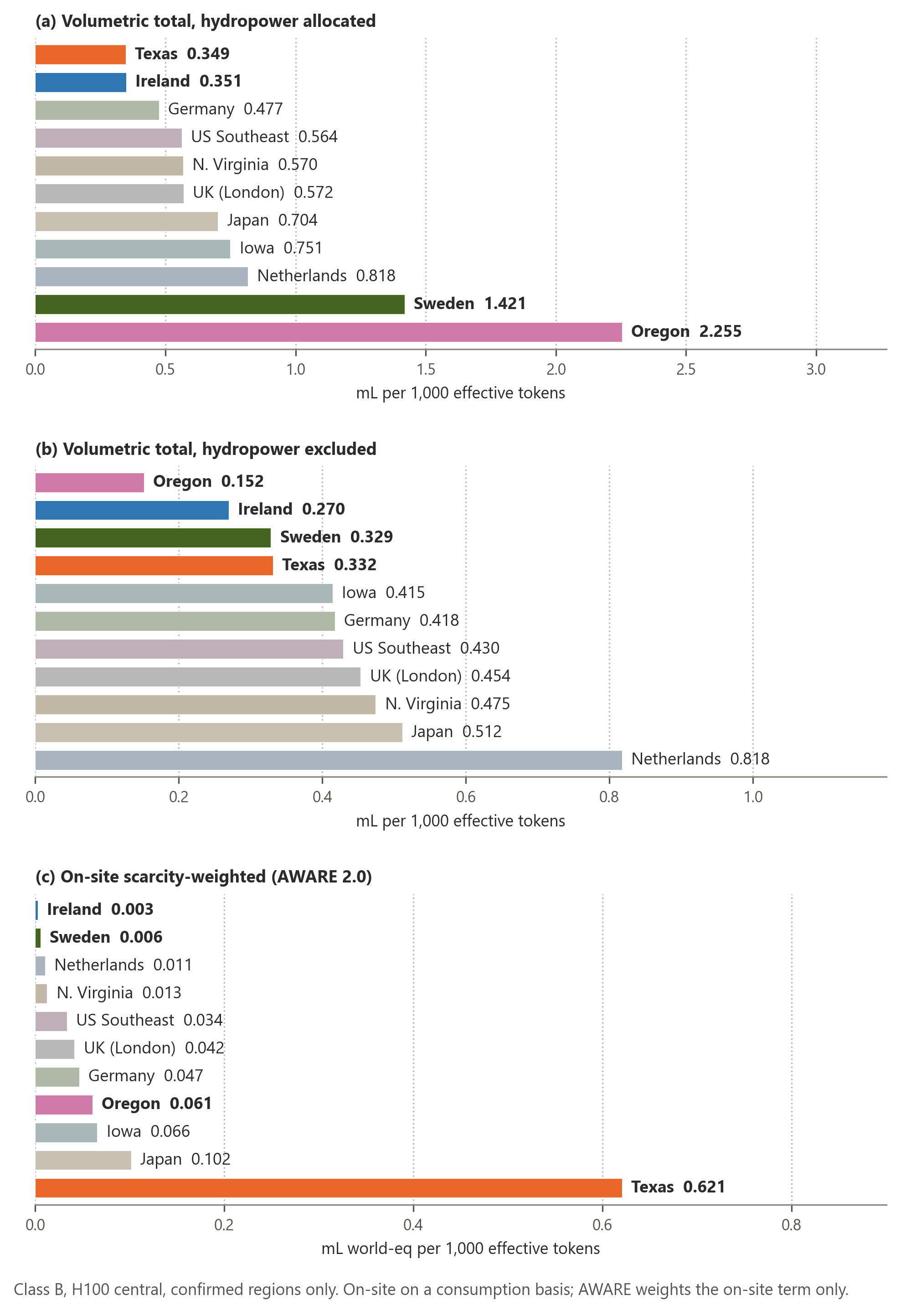}
\caption{The three rankings of the eleven confirmed regions, in mL per 1,000 \emph{effective} tokens (mid class, H100 central, on-site on a consumption basis). Panels (a) and (b) are the volumetric water-use inventory result with hydropower reservoir evaporation allocated to the electricity and with it excluded. Panel (c) is the on-site water scarcity footprint, in mL world-equivalent per 1,000 effective tokens, the on-site consumption multiplied by the region's Available Water Remaining (AWARE) 2.0 factor; it is not a total, because the basin in which the off-site water is consumed is unknown. Regions are ordered from lowest to highest within each panel and the three panels have different horizontal scales. Oregon, Sweden, Texas and Ireland are highlighted because they move most between panels or anchor an end of a ranking.}
\label{fig:wrank}
\end{figure}

\textbf{Three rankings, not one.} The ranking of the eleven confirmed regions depends on the hydropower convention and on whether the result is weighted for scarcity, and the three orderings do not agree (Figure~\ref{fig:wrank}). Both ends move. The lowest region is Texas on the volumetric total with hydropower allocated, Oregon with it excluded, and Ireland on the on-site scarcity-weighted figure; the highest is Oregon, the Netherlands and Texas respectively. Oregon spans ranks 1 to 11 and Texas 1 to 11; Sweden runs from tenth to second.

Three statements survive all three rankings. Ireland is in the lowest five in each and ranks first or second in each. Japan is in the highest five in each. And off-site water dominates every volumetric total. Two statements that earlier versions of this paper made do \emph{not} survive: the Netherlands is in the highest five on the two volumetric rankings but third-lowest on the scarcity-weighted one, because its on-site water is small and its basin is not stressed, and no single region is the lowest or the highest throughout. The practical consequence is that a region cannot be called water-efficient without naming the convention, and a preparer who reports a single water figure should report the volumetric total under both hydropower conventions and, separately and labelled, the scarcity-weighted on-site figure (Section S12.2). A scarcity-weighted figure is never reported as a volume.

\begin{figure}[htbp]
\centering
\includegraphics[width=0.95\textwidth]{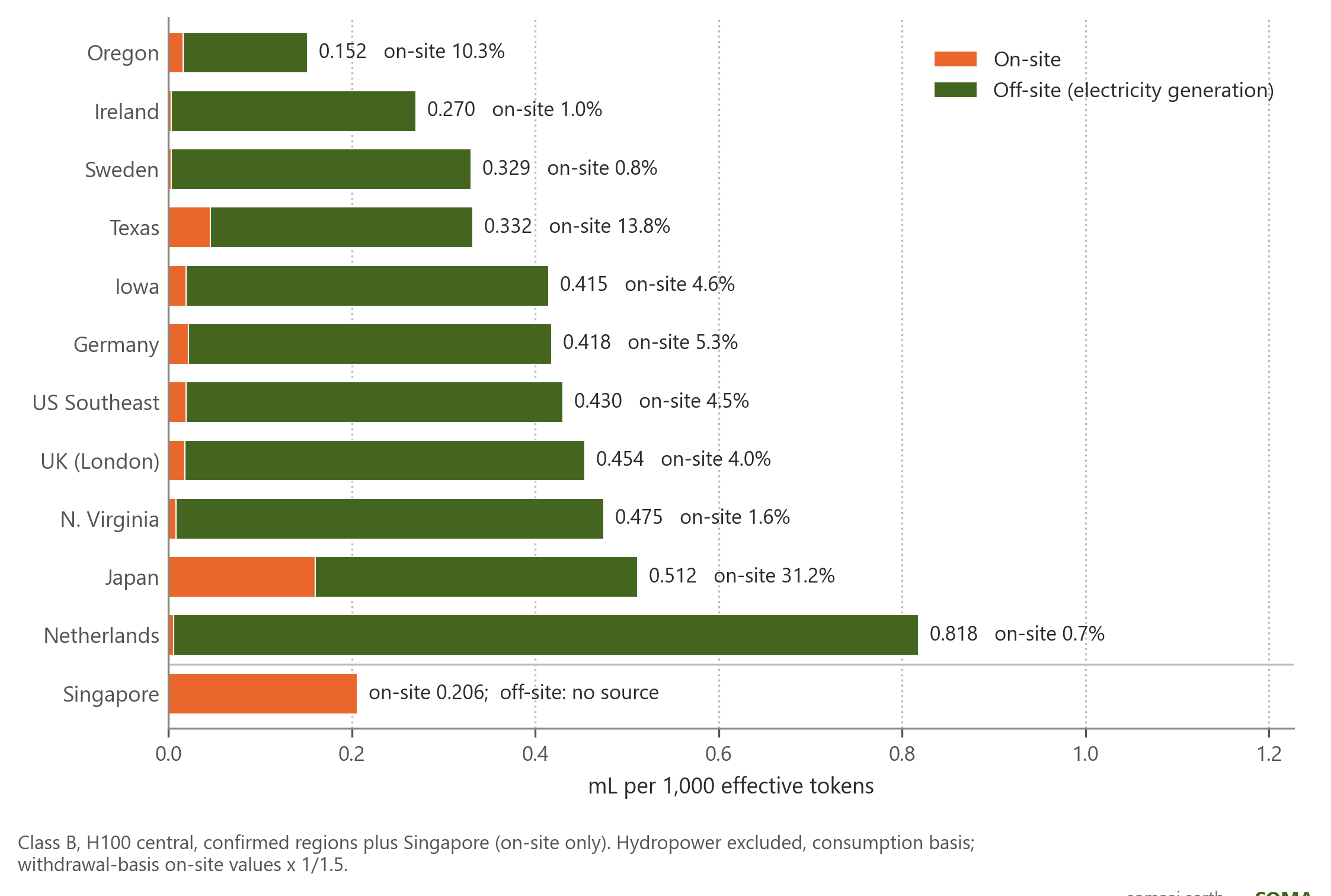}
\caption{On-site and off-site water in mL per 1,000 \emph{effective} tokens with hydropower excluded, on-site on a consumption basis (mid class, H100 central), for the eleven confirmed regions ordered from lowest total, with the total and the on-site share printed beside each bar. Singapore is shown below the rule because it has an on-site value only, with no source covering the water embedded in its electricity; it is outside the rank statistics for that reason, as are the unconfirmed rows of Table~\ref{tab:water}, which are not plotted.}
\label{fig:scope}
\end{figure}

\textbf{On-site and off-site.} Off-site water is between 69\% and 99.8\% of the volumetric total in every confirmed region (Figure~\ref{fig:scope}). The on-site share runs from 0.2\% in Sweden to 22.7\% in Japan with hydropower allocated and from 0.7\% in the Netherlands to 31.2\% in Japan with it excluded; across all seventeen rows the lowest is Finland at 0.2\%, an unconfirmed row. The 2024 Lawrence Berkeley National Laboratory assessment supports that ordering for the United States, reporting 66 billion litres of direct water consumption against nearly 800 billion litres of indirect water in 2023, a ratio of about twelve~\citep{shehabi2024}; it also records that its indirect method ignores power purchase agreements and behind-the-meter generation. \citet{siddik2021}, aggregating plant-level data for 2018, put roughly three-fourths of the operational water footprint of United States data centres in indirect dependencies, 3.83 of 5.13 $\times 10^{8}$ m\textsuperscript{3}. Their direct estimate, 1.30 $\times 10^{8}$ m\textsuperscript{3}, is about twice the Lawrence Berkeley National Laboratory figure for 2023 and their indirect estimate about half of it, which is why their indirect share is lower.

\begin{figure}[htbp]
\centering
\includegraphics[width=0.92\textwidth]{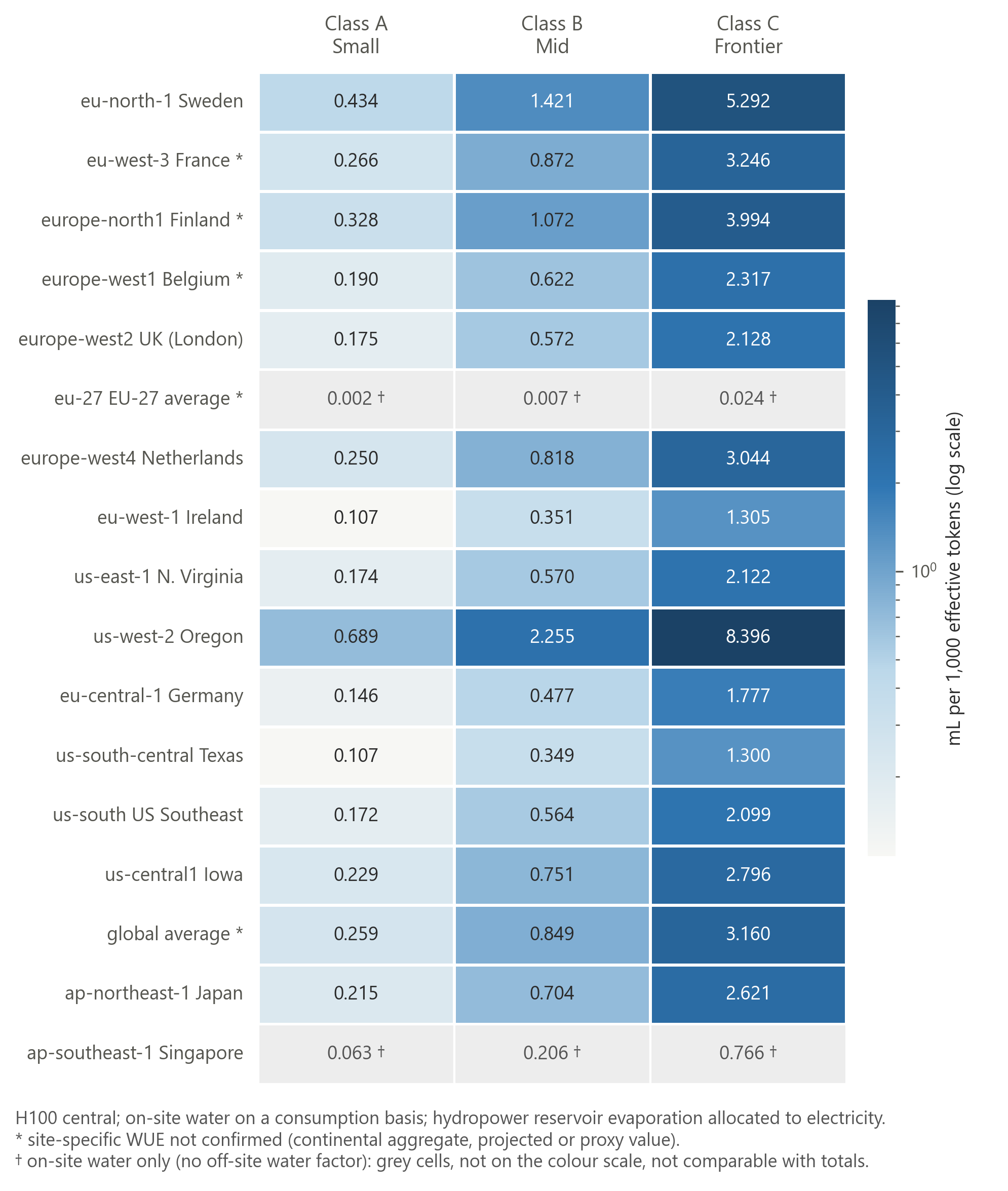}
\caption{Volumetric water-use inventory result by model class and region, in mL per 1,000 \emph{effective} tokens, on-site plus off-site with on-site on a consumption basis (H100 central, class means, power usage effectiveness (PUE) 1.20, hydropower reservoir evaporation allocated to the electricity), on a logarithmic colour scale and in the same order as Figure~\ref{fig:carbon}. Rows marked * rest on a continental aggregate, a projected site or a proxy rather than a region-specific published value. Rows marked $\dagger$, Singapore and the 27-member European Union (EU-27), carry an on-site value only, because no source gives the water embedded in their electricity; those cells are grey, are off the colour scale and are not comparable with the totals.}
\label{fig:water}
\end{figure}

Figure~\ref{fig:water} extends the volumetric result with hydropower allocated to all three model classes.

\begin{table}[htbp]
\centering
\setlength{\tabcolsep}{4pt}
\footnotesize
\begin{tabular}{>{\raggedright\arraybackslash}p{3.6cm} >{\raggedright\arraybackslash}p{0.9cm} >{\raggedleft\arraybackslash}p{1.9cm} >{\raggedleft\arraybackslash}p{1.9cm} >{\raggedleft\arraybackslash}p{2.2cm} >{\raggedleft\arraybackslash}p{2.2cm}}
\toprule
Region & Basis & EWIF, hydro allocated & Off-site, hydro allocated & EWIF, hydro excluded & Off-site, hydro excluded \\
\midrule
EU North --- Sweden & W & 6.019 & 1.419 & 1.386 & 0.327 \\
EU West --- France & W & 3.671 & 0.865 & 2.061 & 0.486 \\
EU North --- Finland & C & 4.542 & 1.071 & 1.712 & 0.403 \\
EU West --- Belgium & W & 2.612 & 0.616 & 2.195 & 0.517 \\
EU West --- UK (London) & W & 2.347 & 0.553 & 1.850 & 0.436 \\
EU West --- Netherlands & W & 3.445 & 0.812 & 3.445 & 0.812 \\
EU West --- Ireland & W & 1.476 & 0.348 & 1.132 & 0.267 \\
EU Central --- Germany & W & 1.930 & 0.455 & 1.681 & 0.396 \\
US East --- N. Virginia & W & 2.385 & 0.562 & 1.981 & 0.467 \\
US West --- Oregon & W & 9.500 & 2.239 & 0.579 & 0.136 \\
US South Central --- Texas & W & 1.287 & 0.303 & 1.213 & 0.286 \\
US South --- Southeast & W & 2.309 & 0.544 & 1.743 & 0.411 \\
US Central --- Iowa & W & 3.104 & 0.732 & 1.680 & 0.396 \\
Asia Pacific --- Japan & W & 2.309 & 0.544 & 1.494 & 0.352 \\
Global average & C & 3.142 & 0.741 & 1.572 & 0.370 \\
\bottomrule
\end{tabular}
\caption{The off-site components behind the totals of Table~\ref{tab:water}: the energy water intensity factor (EWIF) in L/kWh under each hydropower convention, and the off-site water it produces in mL per 1,000 \emph{effective} tokens for the mid class on H100 central. ``Basis'' is the basis on which that row's on-site value was published, W withdrawal and C consumption. Singapore and the 27-member European Union (EU-27) are absent because no source gives an EWIF for either.}
\label{tab:wateroffsite}
\end{table}

\textbf{Three sensitivities.} Leaving the withdrawal-basis on-site values as published instead of converting them to consumption raises totals by 0.1\% to 11.3\% with hydropower allocated and by 0.4\% to 15.6\% with it excluded, the largest moves being Japan and Texas, the two regions where on-site water is a material share (Table~\ref{tab:waterbasis}). It also swaps the two lowest regions on the hydropower-allocated ranking, Texas and Ireland, and moves Japan and Iowa past one another; the hydropower-excluded ranking does not change at all. Texas and Ireland are within 0.4\% of each other on the consumption basis, so their order is unresolved there; as published they are 5.6\% apart and Ireland is clearly the lower. Japan and Iowa are 7\% apart on the consumption basis and 3\% as published, so that swap is a real reordering rather than noise. The high-hydropower case replaces the excluded-hydropower factor with one that charges reservoir evaporation at the most-cited global average, and it moves the hydro-heavy grids by two orders of magnitude, raising Sweden by a factor of 74 and Oregon by a factor of 180 over their hydropower-excluded totals (Table~\ref{tab:waterhydro}). Table~\ref{tab:waterchange} gives the move against the superseded water dataset.

\begin{table}[htbp]
\centering
\setlength{\tabcolsep}{4pt}
\footnotesize
\begin{tabular}{>{\raggedright\arraybackslash}p{3.35cm} >{\raggedright\arraybackslash}p{0.8cm} >{\raggedleft\arraybackslash}p{1.4cm} >{\raggedleft\arraybackslash}p{1.8cm} >{\raggedleft\arraybackslash}p{1.2cm} >{\raggedleft\arraybackslash}p{1.5cm} >{\raggedleft\arraybackslash}p{1.8cm} >{\raggedleft\arraybackslash}p{1.2cm}}
\toprule
Region & Basis & Total, main & Total, as published & Change & Total excl.\ hydro, main & Excl.\ hydro, as published & Change \\
\midrule
EU North --- Sweden & W & 1.421 & 1.423 & +0.1\% & 0.329 & 0.331 & +0.4\% \\
EU West --- France & W & 0.872 & 0.875 & +0.4\% & 0.492 & 0.496 & +0.7\% \\
EU North --- Finland & C & 1.072 & 1.072 & +0.0\% & 0.405 & 0.405 & +0.0\% \\
EU West --- Belgium & W & 0.622 & 0.625 & +0.5\% & 0.524 & 0.527 & +0.6\% \\
EU West --- UK (London) & W & 0.572 & 0.581 & +1.6\% & 0.454 & 0.464 & +2.0\% \\
EU West --- Netherlands & W & 0.818 & 0.820 & +0.3\% & 0.818 & 0.820 & +0.3\% \\
EU West --- Ireland & W & 0.351 & 0.352 & +0.4\% & 0.270 & 0.271 & +0.5\% \\
EU Central --- Germany & W & 0.477 & 0.488 & +2.3\% & 0.418 & 0.429 & +2.7\% \\
US East --- N. Virginia & W & 0.570 & 0.574 & +0.7\% & 0.475 & 0.479 & +0.8\% \\
US West --- Oregon & W & 2.255 & 2.263 & +0.3\% & 0.152 & 0.160 & +5.2\% \\
US South Central --- Texas & W & 0.349 & 0.372 & +6.6\% & 0.332 & 0.355 & +6.9\% \\
US South --- Southeast & W & 0.564 & 0.573 & +1.7\% & 0.430 & 0.440 & +2.3\% \\
US Central --- Iowa & W & 0.751 & 0.761 & +1.3\% & 0.415 & 0.425 & +2.3\% \\
Asia Pacific --- Japan & W & 0.704 & 0.784 & +11.3\% & 0.512 & 0.592 & +15.6\% \\
Global average & C & 0.849 & 0.849 & +0.0\% & 0.478 & 0.478 & +0.0\% \\
\bottomrule
\end{tabular}
\caption{Sensitivity: on-site water left as published (withdrawal for most rows) instead of converted to consumption, in mL per 1,000 effective tokens, mid class, H100 central.}
\label{tab:waterbasis}
\end{table}

\textit{In the Basis column, W is withdrawal and C consumption, as published. The main basis converts every withdrawal row to consumption using the relation implied by the AWS page's own industry-average footnote, 0.84 litres of withdrawal per 0.56 of consumption at three cycles of concentration, that is withdrawal equal to 1.5 times consumption~\citep{aws2026b}. Microsoft's company-wide ratio (0.616)~\citep{microsoft2026a}, Meta's data-centre ratio (0.717)~\citep{meta2025} and Google's 2025 data-centre ratio (0.776, from its 2026 Environmental Report)~\citep{google2026c} bracket it. The as-published totals are higher because the off-site term is already consumption and does not move.}

\begin{table}[htbp]
\centering
\setlength{\tabcolsep}{4pt}
\footnotesize
\begin{tabular}{>{\raggedright\arraybackslash}p{3.6cm} >{\raggedleft\arraybackslash}p{1.8cm} >{\raggedleft\arraybackslash}p{1.8cm} >{\raggedleft\arraybackslash}p{1.6cm} >{\raggedleft\arraybackslash}p{1.8cm} >{\raggedleft\arraybackslash}p{2.2cm}}
\toprule
Region & Hydro share of generation & EWIF excl.\ hydro & EWIF high & Total, high case & Ratio to the excluded total \\
\midrule
EU North --- Sweden & 41.6\% & 1.386 & 103.3 & 24.351 & 74.0 \\
EU West --- France & 12.3\% & 2.061 & 32.2 & 7.595 & 15.4 \\
EU North --- Finland & 19.8\% & 1.712 & 50.2 & 11.839 & 29.2 \\
EU West --- Belgium & 2.1\% & 2.195 & 7.3 & 1.736 & 3.3 \\
EU West --- UK (London) & 2.6\% & 1.850 & 8.2 & 1.956 & 4.3 \\
EU West --- Netherlands & 0.1\% & 3.445 & 3.7 & 0.875 & 1.1 \\
EU West --- Ireland & 3.8\% & 1.132 & 10.4 & 2.464 & 9.1 \\
EU Central --- Germany & 4.1\% & 1.681 & 11.7 & 2.786 & 6.7 \\
US East --- N. Virginia & 1.5\% & 1.981 & 5.7 & 1.341 & 2.8 \\
US West --- Oregon & 47.2\% & 0.579 & 116.2 & 27.408 & 180.2 \\
US South Central --- Texas & 0.3\% & 1.213 & 1.9 & 0.505 & 1.5 \\
US South --- Southeast & 2.0\% & 1.743 & 6.6 & 1.585 & 3.7 \\
US Central --- Iowa & 5.0\% & 1.680 & 13.9 & 3.303 & 8.0 \\
Asia Pacific --- Japan & 8.4\% & 1.494 & 22.1 & 5.362 & 10.5 \\
Global average & 6.5\% & 1.572 & 17.5 & 4.232 & 8.8 \\
\bottomrule
\end{tabular}
\caption{Sensitivity: high-hydropower case (energy water intensity factors in L/kWh, totals in mL per 1,000 effective tokens, mid class, H100 central, on-site on a consumption basis).}
\label{tab:waterhydro}
\end{table}

\textit{The high-hydropower factor adds the hydropower share of generation times 245 L/kWh to the hydropower-excluded factor. That constant is 68 m\textsuperscript{3} per GJ converted at 1 kWh $=$ 0.0036 GJ, the generation-weighted aggregate blue water footprint of the 35 plants studied by \citet{mekonnen2012}, which covers 8\% of world hydroelectric capacity. It is an outlier at the top of the published distribution rather than a consensus value: the compiled range for gross reservoir evaporation runs from 5.4 to 234 m\textsuperscript{3}/MWh for most studies~\citep{bakken2017}, and the hydropower factors that \citet{reig2020} apply, 10.8 L/kWh for Sweden and 22.7 L/kWh for the United States, sit 23 and 11 times below 245. For the Nordic rows the global average is the weakest possible proxy: the only Nordic-specific row in the compilation of \citet{bakken2017} is two Norwegian systems whose gross average is 37.8 L/kWh, 6.5 times below 245, and whose net average is 0.17 L/kWh. At 37.8 L/kWh Sweden's high-case total would be 4.04 mL rather than 24.35, a ratio of 12.3 rather than 74.0 over its hydropower-excluded total. This case is a bound on the convention, not a central estimate.}

\textbf{Why both conventions are reported as co-equal.} The published evidence does not support a single hydropower water factor. Net evaporation, which subtracts the evapotranspiration that occurred before the reservoir existed, is reported at 45\% to 60\% of gross in some studies but is often below 40\% and sometimes about 1\%, and it can be negative; the Water Footprint Assessment Manual prescribes gross, as \citet{bakken2017} record; and the review that compiles the literature declines to recommend either, calling instead for scientific agreement to be reached~\citep{bakken2017}. Allocation between the purposes of a multipurpose reservoir is equally unsettled, with no accepted method and tested allocation factors for hydropower of 38\% to 76\%. The share at stake is large: aggregating the plant-level data behind a published United States factor of this family gives hydropower 6.8\% of generation but 70.9\% of the consumptive water embedded in United States electricity, and the same authors' no-allocation-to-hydropower scenario is 55\% below their primary-purpose baseline (supplementary dataset)~\citep{siddik2021}. The two percentages differ because the 55\% is of the data-centre footprint, which includes on-site water unaffected by the hydropower convention, while the 70.9\% is of the water embedded in electricity alone.

\begin{table}[htbp]
\centering
\setlength{\tabcolsep}{4pt}
\footnotesize
\begin{tabular}{>{\raggedright\arraybackslash}p{3.4cm} >{\raggedleft\arraybackslash}p{1.5cm} >{\raggedleft\arraybackslash}p{1.4cm} >{\raggedleft\arraybackslash}p{1.5cm} >{\raggedleft\arraybackslash}p{1.2cm} >{\raggedleft\arraybackslash}p{1.2cm} >{\raggedleft\arraybackslash}p{1.5cm} >{\raggedleft\arraybackslash}p{1.2cm}}
\toprule
Region & Previous WUE & New WUE & Previous total & New total & Change & Previous, excl.\ hydro & New, excl.\ hydro \\
\midrule
EU North --- Sweden & 0.090 & 0.020 & 0.973 & 1.421 & +40.5\% & 0.233 & 0.329 \\
EU West --- France & 0.060 & 0.050 & 0.594 & 0.872 & +41.1\% & 0.337 & 0.492 \\
EU North --- Finland & 0.010 & 0.010 & 0.727 & 1.072 & +41.8\% & 0.275 & 0.405 \\
EU West --- Belgium & 0.060 & 0.050 & 0.425 & 0.622 & +40.7\% & 0.358 & 0.524 \\
EU West --- UK (London) & 0.050 & 0.140 & 0.381 & 0.572 & +44.4\% & 0.302 & 0.454 \\
EU West --- Netherlands & 0.060 & 0.036 & 0.558 & 0.818 & +40.9\% & 0.558 & 0.818 \\
EU West --- Ireland & 0.020 & 0.020 & 0.238 & 0.351 & +41.6\% & 0.183 & 0.270 \\
EU Central --- Germany & 0.080 & 0.170 & 0.319 & 0.477 & +43.9\% & 0.279 & 0.418 \\
US East --- N. Virginia & 0.140 & 0.060 & 0.399 & 0.570 & +37.3\% & 0.335 & 0.475 \\
US West --- Oregon & 0.950 & 0.120 & 1.643 & 2.255 & +32.0\% & 0.219 & 0.152 \\
US South Central --- Texas & 0.250 & 0.292 & 0.239 & 0.349 & +40.6\% & 0.227 & 0.332 \\
US South --- Southeast & 0.060 & 0.123 & 0.377 & 0.564 & +43.8\% & 0.286 & 0.430 \\
US Central --- Iowa & 0.140 & 0.123 & 0.514 & 0.751 & +40.5\% & 0.287 & 0.415 \\
Asia Pacific --- Japan & 0.200 & 1.220 & 0.395 & 0.704 & +71.4\% & 0.265 & 0.512 \\
Asia Pacific --- Singapore & --- & 1.570 & --- & --- & --- & --- & --- \\
Global average & 0.550 & 0.550 & 0.575 & 0.849 & +34.9\% & 0.324 & 0.478 \\
\bottomrule
\end{tabular}
\caption{Water against the previous dataset version (water use effectiveness, WUE, in L/kWh; totals in mL per 1,000 effective tokens, hydropower allocated unless stated; on-site on a consumption basis in the new columns and as published in the previous ones).}
\label{tab:waterchange}
\end{table}

\textit{Three changes act together. The host and idle uplifts raise server and facility energy by a factor of 1.65, which raises both terms; the class value moves from one named proxy model to the mean of the measured members, which lowers the mid-class energy by 11\%; and the on-site values move to published region and site figures on a consumption basis. The net energy change alone accounts for about $+$42\%, so a row near $+$42\% has an unchanged water use effectiveness, a row above it a higher new one and a row below it a lower one. Oregon is the exception that changes sign: its previous on-site value was a neighbouring-site proxy eight times the published region figure, and with hydropower excluded the on-site term is a large enough share for that correction to dominate. Japan moves most, at $+$71\%, because its published region value is 6.1 times the author estimate it replaces.}

\subsection*{S4. Provider Transparency Roadmap}

Table~\ref{tab:disclosures} gives the minimum disclosure that would move a service up each tier boundary.

\begin{table}[htbp]
\centering
\setlength{\tabcolsep}{4pt}
\small
\begin{tabular}{>{\raggedright\arraybackslash}p{1.7cm} >{\raggedright\arraybackslash}p{3.1cm} >{\raggedright\arraybackslash}p{5.5cm} >{\raggedright\arraybackslash}p{3.4cm}}
\toprule
Current tier & Missing data & Required provider disclosure & Unlocks \\
\midrule
Tier 1 & No usage data & Token counts per model per billing period & Tier 2a \\
Tier 2b & No token counts & Token counts from an admin dashboard export, as Anthropic's Enterprise Analytics API already provides for Claude seats~\citep{adminreports2026} & Tier 2a \\
Tier 2a & No region confirmation & Infrastructure region per account & Tier 2a with a confirmed grid \\
Tier 2a & No energy data & kWh per million tokens per model & The hybrid method, then Tier 3 \\
Tier 2a & No serving-volume data & Lifetime tokens served per model, or tokens served to date with the model's retirement date & Replaces $V$, the single disclosed datapoint behind the training term \\
Tier 2a & No renewable-energy data & Share of inference served from renewable energy, market-based, with energy attribute certificate documentation & Tier 3 \\
\bottomrule
\end{tabular}
\caption{Minimum provider disclosures per tier upgrade.}
\label{tab:disclosures}
\end{table}

Token counts move a seat-licensed product from Tier 2b to Tier 2a and remove the activity band entirely, and a confirmed region removes the global-average fallback and with it the regional uncertainty, a factor of about six on the factor per million billed tokens and nine on the inference footprint. A supplier's energy figure replaces the accelerator-energy, host, idle and PUE chain with a measured quantity and moves the service to the hybrid method, although the token basis of Section~\ref{sec:3.3} still applies unless the supplier splits input from output; a CO\textsubscript{2}e figure meeting the Tier 3 criteria replaces the factor altogether. A renewable-energy share with certificate documentation is the one disclosure that could move the figure off a location basis. The EU AI Act Article 53~\citep{euparl2024} creates indirect pressure towards energy disclosure, but its Annex XI documentation is not a public disclosure: it goes to the AI Office and national authorities on request, is not per token, and exempts open-source models without systemic risk (Article 53(2)), with a transition to 2 August 2027 for models already on the market (Article 111(3)). Much of this data is likely to exist in provider metering used for capacity planning and billing, although per-model energy may be regarded as commercially sensitive; aggregated disclosure, per model family and region and per reporting year, would meet the accounting need.

\textbf{Cloud carbon tools and the Annex XI datapoint: further detail.} The three cloud carbon tools of Section~\ref{sec:2} differ in service granularity. AWS resolves only its EC2 compute, S3 storage and CloudFront content-delivery services in its console and groups the rest as ``Other'', Azure Carbon Optimization resolves to individual resources while the Emissions Impact Dashboard stays at subscription level, and Google reports Vertex AI and the Gemini API as separate products~\citep{cloudtools2026}. The energy datapoint of Annex XI~\citep[Article 53(1)(a) and Annex XI, Section~1, point 2(e)]{euparl2024} is read in this paper as mainly about training and development energy, because it sits in a point headed ``relevant information of the process for the development'' and immediately after the point on training compute and training time, and because the fallback clause allows it to be derived from the computational resources used. Against that reading, the text says ``energy consumption of the model'' unqualified, and where the legislator meant training alone it said so, at Annex XIII, point (c). The Act does not settle it, and the Commission may specify how the datapoint is measured under Article 53(5)~\citep{euparl2024}. The Digital Omnibus on AI amended the Act in July 2026, postponing the high-risk application dates of Article 113 while leaving Article 53, Annex XI and the general-purpose AI application date of Article 113, point (b), untouched~\citep{euparl2026a}.

\textbf{Relation to existing standards: further detail.} The discussion of Section~\ref{sec:7.4} in full:

This paper is written for the CSRD and the ESRS; the same factors and activity-data rules would serve an International Financial Reporting Standards (IFRS) S2 or a California disclosure, but the reporting rules of Section S12.3 are not mapped to them.

The Software Carbon Intensity (SCI) specification from the Green Software Foundation provides a rate-based metric, g CO\textsubscript{2}e per functional unit of software, that maps onto Tier 2a accounting; tokens are a natural functional unit. It combines an operational term and an embodied term amortised over the hardware's expected life, over one functional unit, whereas the factor here carries four terms on two denominators, so the two are compatible without being identical. Version 1.0 was adopted by ISO and the International Electrotechnical Commission (IEC) as ISO/IEC 21031:2024; the Foundation's current text is version 1.1 and is not the ISO standard~\citep{gsf2024}. The SCI for AI specification, which the Foundation announced as ratified in December 2025, makes one token its \emph{suggested} functional unit for large language models in the consumer boundary, while its provider boundary must report per floating-point operation (FLOP), per training token or per billion parameters and must count failed runs, hyperparameter sweeps and discarded checkpoints inside training~\citep{gsf2025}: the factor of Section~\ref{sec:3.4} is compatible with the first, and the second corroborates the content of this paper's development line (Section S9.5). Recommendation L.1801 of the International Telecommunication Union Telecommunication Standardization Sector (ITU-T), approved in February 2026 and in force, gives a life-cycle method for assessing the environmental impact of AI systems and is the nearest thing to an in-force international standard here~\citep{itu2026}, but it supplies no emission factors and no per-token values, its data-collection rules require life-cycle inventory data on the system itself, across every stage from inception to use, which a customer buying tokens does not hold, and it expressly excludes aggregation to a national or regional level. The claim made in this paper is therefore that no standard supplies a preparer's factors and activity data, not that none addresses AI assessment at all.

Under a partnership with ISO announced in September 2025, the GHG Protocol is consolidating the Corporate Standard, the Scope 2 Guidance, the Scope 3 Standard and the Actions and Market Instruments work into a single Corporate Standard, Version 3.0, expected to be co-branded with ISO, with a draft for consultation estimated for the second quarter of 2027 and publication estimated for the fourth quarter of 2028; the existing standards stay in effect until the GHG Protocol says otherwise~\citep{ghgp2026}, and ISO 14064-1:2018 is itself listed for revision~\citep{iso2018}. The rules of Sections~\ref{sec:3.1} and \ref{sec:3.8}, and the allocation, market-based and data-quality rules in particular, would need revisiting against the consolidated text. No published draft addresses cloud, information technology or AI services, but the Scope 3 technical working group has them under open discussion, and its draft meeting papers, marked not for citation, discuss digital products under Category 11, give energy per AI token or per FLOP as examples of computation energy, and note that AI workloads may require separate and more detailed methods~\citep{ghgp2026}. That is the question this paper's method is a candidate answer to.

\subsection*{S5. Decision Rules and the Refresh Schedule}

\textbf{The boundary, element by element.} Table~\ref{tab:inclusion} maps every lifecycle element of a purchased inference service to the factor term that carries it, or states how it is treated when no term carries it. It is the auditable form of Section~\ref{sec:3.1} and of Figure~\ref{fig:lifecycle}.

\begin{table}[htbp]
\centering
\setlength{\tabcolsep}{4pt}
\scriptsize
\begin{tabular}{>{\raggedright\arraybackslash}p{4.4cm} >{\raggedright\arraybackslash}p{8.6cm} >{\raggedright\arraybackslash}p{1.1cm}}
\toprule
Lifecycle element & Term that carries it, or how it is treated & Sec. \\
\midrule
\multicolumn{3}{l}{\textit{Inside the central factor}} \\
Accelerator energy while serving & Electricity, measured accelerator energy per token & \ref{sec:3.4} \\
Rest of the server (CPU, memory, fans, power supplies, network interface, local storage) & Electricity, host uplift $H$ & \ref{sec:3.4} \\
Provisioned capacity held idle & Electricity, idle uplift $I$ & \ref{sec:3.4} \\
Facility cooling and power distribution & Electricity, PUE & \ref{sec:3.4} \\
Upstream fuel cycle of the electricity & Electricity, lifecycle grid intensity & \ref{sec:3.4} \\
Server manufacturing and transport & Embodied hardware. The manufacturer footprints used are cradle-to-gate, so raw material acquisition and component fabrication are inside them; they carry no end-of-life stage & \ref{sec:3.4} \\
Data-centre construction and site infrastructure & Construction, on server electricity & \ref{sec:3.4} \\
Final training runs of the served model & Training, per billed token & \ref{sec:3.4} \\
\midrule
\multicolumn{3}{l}{\textit{Reported separately, never added to the factor}} \\
Model development other than the final training runs: research, experiments, synthetic data generation and other data curation, evaluation, unreleased runs, post-training and retraining & Separate line, the development line, which covers all development including the final runs and is worth 0.6 to 2.2 times the inference footprint for the mid class at the global-average grid: the largest item outside the factor & \ref{sec:3.4} \\
Cooling water and water embedded in generation & Water factor, two volumetric conventions and a scarcity-weighted on-site result & \ref{sec:3.5} \\
\midrule
\multicolumn{3}{l}{\textit{Outside, with a bound (share of operational serving emissions, from \citealp{gamazaychikov2026})}} \\
Provider-side storage & About 0.11\% & \ref{sec:3.1} \\
External network transmission & 0.17 to 0.43\% & \ref{sec:3.1} \\
Hardware end-of-life, including waste treatment & About 0.12 to 0.17\% on a location basis, and 0.38\% market-based. Every figure here is location-based, so the location-based bound applies & \ref{sec:3.1} \\
\midrule
\multicolumn{3}{l}{\textit{Outside, reported as negligible by the source, not independently bounded}} \\
Data-centre internal networking & Reported as negligible for a text prompt (Google, \citealp{elsworth2025}; as used by \citealp{gamazaychikov2026}); not independently computed here & \ref{sec:3.1} \\
\midrule
\multicolumn{3}{l}{\textit{Outside, as a stated convention: two carried as sensitivities, one as a credit not applied}} \\
Grid transmission and distribution losses & Excluded for consistency with the Scope 2 Guidance, not as immaterial: location-based grid intensities are per kWh generated and exclude these losses, which the supplier reports under its own Category 3~\citep{wri2015}. The losses exceed the 1\% threshold, so the exclusion is disclosed. Charging the United States gross grid loss raises the electricity term of the United States rows by 4.2\% & \ref{sec:5.5} \\
On-site (behind-the-meter) generation & The factor uses a grid intensity. Overriding it with a reported on-site gas intensity raises the mid-class footprint by about a third at the global average and doubles it at US East & \ref{sec:5.5} \\
Recycling, reuse and cascaded-use credit & No credit is taken: the embodied term is cradle-to-gate plus transport, and a credit for end-of-life material recovery of up to 4\% of embodied emissions, assessed by \citet{schneider2025} and reported by \citet{gamazaychikov2026}, was available and not applied & \ref{sec:3.4} \\
\midrule
\multicolumn{3}{l}{\textit{Out of scope}} \\
End-user devices & Already in the customer's own Scope 2 and Category 1 or 2 (by capitalisation), or optionally Category 7 (teleworking) where used at home~\citep{wri2011}; not bounded here & \ref{sec:3.1} \\
Provider staff, offices and research overhead outside compute & Not quantified & none \\
Customer fine-tuning jobs & A training workload billed by job, under the customer's operational control. Disclose as excluded with the spend involved & S5 \\
Embeddings, and image, audio and video work & No measurement exists under the protocol used here. The boundary, the three uplifts and the embodied, construction and grid terms all transfer; only the accelerator energy factor is modality-specific & S5 \\
\bottomrule
\end{tabular}
\caption{Lifecycle elements of a purchased artificial intelligence (AI) inference service: the term that carries each, the two reported separately, the bounded exclusions, the three stated conventions, and what is out of scope. Bounds are relative to the quantity named, not absolute kg CO\textsubscript{2}e.}
\label{tab:inclusion}
\end{table}

Table~\ref{tab:rules} covers the cases the tier logic does not settle on its own; the serving-region and unmapped-identifier rules, which a preparer meets most often, are given as rows of Table~\ref{tab:overview} instead.

\begin{table}[htbp]
\centering
\setlength{\tabcolsep}{4pt}
\footnotesize
\begin{tabular}{>{\raggedright\arraybackslash}p{3.6cm} >{\raggedright\arraybackslash}p{11.4cm}}
\toprule
Case & Rule \\
\midrule
AI bought through the organisation's own cloud account (Azure OpenAI, Bedrock, Vertex AI) & Use the cloud provider's carbon tool where it reports the AI service as its own line and meets the Tier 3 criteria below; otherwise apply Tier 2a to the billed tokens and remove that service line from the cloud-tool figure, recording the boundary and the Scope 2 basis of what was replaced. Where the tool reports market-based Scope 2 the substitution changes the basis as well as the boundary, and both are disclosed. Where the tool cannot isolate the AI line, do not subtract: either keep the tool's figure and disclose the AI service as not separately estimated, or replace the whole cloud line with Tier 2a. Keep Tier 2a as a cross-check either way. \\
Dedicated or provisioned capacity (reserved throughput units, provisioned throughput, reserved accelerator capacity) & Category 1, on a reserved-capacity activity basis: emissions follow reserved accelerator-hours, not tokens. Apply the per-accelerator-hour factor of Table~\ref{tab:reserved} and allocate the training term on the tokens actually served. Score 1 for activity data, because reserved hours are the activity that drives the factor directly with no conversion step, and 3 for the grid factor. Category 2 applies only where the arrangement is capitalised, and Category 8 only where identified hardware is leased~\citep{wri2011}. \\
Batch API and cached input tokens & Same per-token factors as interactive calls. Cached input tokens are weighted at $W_{\mathrm{cache}}$ where the export reports them separately, and at the uncached weight where it does not. Batch serving is not latency-bound and may run at larger batches than the operating point used here, so the factor is likely high for it, although no unconstrained minimum is reported here. \\
Embeddings, image, audio and video work, and fine-tuning jobs & Out of scope: no measurement exists under the protocol used here, and fine-tuning is a training workload billed by job. The boundary, the uplifts and the embodied, construction and grid terms would all transfer; only the accelerator energy factor is modality-specific. Disclose as excluded with the spend involved. \\
Which admin export each seat product offers & Microsoft 365 Copilot reports prompts submitted and active days per user and, for metered agents alone, credits rather than tokens. Gemini for Workspace reports, outside the Gemini app, only usage levels, active days and days at limit, so it borders on Tier 1. ChatGPT Enterprise sits at the Tier 2a boundary, because its workspace analytics export carries a message count by model family~\citep{openai2026a}. Claude Enterprise returns uncached input, cached input and output tokens by user and model on usage-based plans, and usage credits only on seat-based ones~\citep{adminreports2026}. \\
Comparing models or regions & Use the inference footprint per million effective tokens, not the convenience factor and not a sum across denominators. The training term describes the provider, not serving efficiency, and belongs only in the inventory total. \\
Tier 3 acceptance & A service- or model-level CO\textsubscript{2}e figure for the customer's own usage; location-based, market-based additionally with each basis stated; a boundary covering at least the four terms of Section~\ref{sec:3.4}, with the supplier stating whether research and development compute is inside it; the allocation method stated, including whether customer allocations reconcile to the supplier's reported total; a named methodology standard; the assurance status stated, and where assurance is obtained, at least at the limited level in the sense of ISO 14064-3:2019, whose Annex A sets the requirements for a limited-assurance engagement and which defines limited assurance as designed to give a reduced, rather than a high but not absolute, level of assurance on historical data and information~\citep{iso2019}; an unassured figure carries the supplier's own uncertainty statement and scores 2 rather than 1; a reporting period aligned with the inventory. Limited assurance is also the level the CSRD requires of a sustainability statement itself, although the European limited-assurance standards are not due until 1 July 2027 (Directive (EU) 2026/470, recital (4); \citealp{euparl2026b}). \\
Supplier energy data rather than a supplier CO\textsubscript{2}e figure & The GHG Protocol \emph{hybrid} method, not the supplier-specific method~\citep{wri2013}: combine the supplier's kWh per million tokens with the preparer's own grid row and add the embodied, construction and training terms. Score 2 for activity data, 2 for the energy input and 3 for the grid factor. It does not reach Tier 3. \\
\bottomrule
\end{tabular}
\caption{Treatment of cases outside the basic tier logic (author rules unless a standard is cited).}
\label{tab:rules}
\end{table}

\textbf{How the Category 1 boundary is assembled in practice.} The cradle-to-gate minimum of Section~\ref{sec:3.1} is a definition of the boundary, not a data-collection recipe. In practice a preparer assembles it from the supplier's product-allocated Scope 1 and Scope 2 emissions plus all other upstream emissions of the purchased product~\citep{wri2013}, which is a data-collection construct rather than the definition, and is why the allocation rules of the Standard's Chapter 8 apply~\citep{wri2011}.

\textbf{Four further reporting rules.} These are referred to from Section S12.3 and set out here.

\emph{Market-based claims.} All grid intensities are location-based, so three rules follow. The reporting company's own power purchase agreements, renewable energy certificates or guarantees of origin may not be applied to a supplier's emissions inside Category 1, on this paper's reading of the Scope 2 Guidance's quality criteria, which tie such instruments to the reporting entity's own electricity-consuming operations~\citep{wri2015}. A provider's public claim of renewable or carbon-neutral operation may not reduce a Category 1 figure either, unless it arrives as supplier-specific data meeting the Tier 3 criteria. And where a provider supplies a market-based figure, the location-based figure is reported alongside it; that is this paper's own convention, not a standard's, because dual reporting under the Scope 2 Guidance~\citep{wri2015} governs a reporter's own Scope 2 and the Scope 3 Standard sets no market-based convention for Category 1.

\emph{Global warming potentials.} The sources do not share one set: the grid intensities are lifecycle values over a 100-year horizon that \citet{ember2026} take from the IPCC Fifth Assessment Report (AR5), Working Group III Annex III, and for coal and gas from UNECE and Jordaan et al., and eGRID2023 states AR5 values without climate--carbon feedbacks (CH\textsubscript{4} 28, N\textsubscript{2}O 265)~\citep{epa2025}. Schneider et al.\ state AR5 100-year values, the set the grid data use, and the Dell and NVIDIA disclosures follow their own set, or do not state one~\citep{schneider2025,dell2025,nvidia2025a}. The training disclosure~\citep{meta2026} reports tonnes of CO\textsubscript{2}e without naming a set. The differences fall on the non-CO\textsubscript{2} fraction only, so the effect on any factor here is well under 1\%. State one set for the inventory, AR5 as used by the grid data source, and disclose that the hardware and training inputs follow their own set or state none. AR5 rather than the most recent IPCC set, from the Sixth Assessment Report (AR6), is used because the grid sources publish on AR5 and no AR6 series exists for them, and for CO\textsubscript{2}-dominated grid factors the difference falls on the small non-CO\textsubscript{2} fraction only. The revised ESRS allow factors on older values where they are the most suitable available, provided the inventory states which values it is based on (ESRS E1, AR~20(c), with ESRS 2, GDR-M, paragraph 49; \citealp{eucom2026a}), and this paragraph is that explanation. The factors will move to AR6 values when the grid sources do.

\emph{Factor version, grid year, base year and back-casting.} Pin one dataset version per reporting year, identified by the dataset version or, until a version identifier exists, by its dataset date. A change of dataset version or of grid year is a methodology change and is disclosed as such, and the discovery of a material error, in the dataset or in the preparer's own activity data, is treated in the same way; a dataset erratum is issued as a new version. Recalculate the base year only where the resulting change in the AI line exceeds the company's own disclosed significance threshold applied to total Scope 3. Where no base-year token data exist, back-cast on spend only after adjusting for the change in the vendor's list price per million tokens for the model in use, because price per token has fallen by roughly an order of magnitude per model generation and an unadjusted spend ratio overstates base-year tokens and flatters any reduction; where the price history cannot be retrieved, do not back-cast but take the first year with token data as the base year and disclose the change.

\emph{Completeness, de minimis and integration.} Build the service list from accounts-payable vendors and general-ledger codes, the software-management export, cloud-bill service lines and expense-claim keywords, and allocate usage by usage date from the usage export, not by invoice date, prorating billing periods that straddle the year end by days. Apply the company's own Scope 3 significance threshold as the de minimis cut-off for an individual AI service and disclose the aggregate spend of everything cut. Free-tier, trial and out-of-process use through personal or reimbursed accounts is a known completeness gap: disclose it, and where no data exist give an order-of-magnitude estimate from headcount times the Tier 2b seat default. Before adding Tier 2a, 2b or 1 results, remove the corresponding supplier spend from the spend-based Category 1 calculation, and exclude any AI spend already covered by a cloud provider's customer carbon report, or the same purchase is counted twice.

\begin{table}[htbp]
\centering
\setlength{\tabcolsep}{4pt}
\footnotesize
\begin{tabular}{>{\raggedright\arraybackslash}p{3.2cm} >{\raggedright\arraybackslash}p{5.0cm} >{\raggedright\arraybackslash}p{2.0cm} >{\raggedright\arraybackslash}p{4.6cm}}
\toprule
Input & Source & Cadence & Out-of-cycle trigger \\
\midrule
Accelerator energy per token & ML.ENERGY Leaderboard~\citep{mlenergy2025} & Each release & A new leaderboard release, or a new accelerator generation entering the sweep \\
Grid carbon intensity & Ember Electricity Data Explorer~\citep{ember2026}; eGRID~\citep{epa2025} & Annual & Publication of a new complete calendar year \\
Host and idle uplifts & Google serving disclosure~\citep{elsworth2025};~\citep{gamazaychikov2026}; \citet{latif2025}; \citet{oviedo2026} & Annual & A first-party measurement of a full serving stack on GPU hardware \\
PUE & \citet{google2026a}; AWS~\citep{amazon2024}; Microsoft~\citep{microsoft2026a}; \citet{uptime2024} & Annual & A provider restating its fleet value \\
Server and transport footprints & Dell PCFs~\citep{dell2025}; NVIDIA PCFs~\citep{nvidia2025a,nvidia2025b} & Annual & Publication of a footprint that names the accelerator \\
Hardware life & Operator Form 10-K filings~\citep{secfilings2026} & Annual & A change in estimate disclosed by any large operator \\
Model service life & Anthropic model deprecations~\citep{anthropic2026a}; AWS Bedrock model lifecycle~\citep{aws2026a}; Microsoft Foundry Models lifecycle and support policy~\citep{microsoft2026b} & Quarterly & A provider changing its retirement policy or withdrawing a published minimum \\
Per-model token volume & Provider token-volume disclosures, as tabulated by Epoch AI~\citep{openai2025a} & Quarterly & Any provider disclosing a per-model serving volume \\
Development ratio &~\citep{gamazaychikov2026}; \citet{patterson2022}; Epoch AI~\citep{denain2026} & Annual & A first-party disclosure of development against serving energy \\
On-site water & AWS regional PUE and WUE series~\citep{aws2026b}; Microsoft environmental data~\citep{microsoft2026a}; Meta environmental data index~\citep{meta2025}; \citet{li2025} & Annual & A provider publishing a water use effectiveness figure for a region that currently rests on an aggregate or a proxy \\
Off-site water and scarcity factors & \citet{reig2020}; AWARE 2.0~\citep{seitfudem2025} & On release & A replacement grid-average water factor set, or a hydropower-excluded variant published directly~\citep{siddik2024} \\
Token weights & ML.ENERGY benchmark~\citep{chung2025}; provider cache and output prices~\citep{openai2026b,anthropic2026b,google2026b} & Annual & Any published energy measurement of a cached-input read, or of prefill against decode at production batch sizes \\
Tier 2b activity defaults & Azure inference traces~\citep{azure2024}; WildChat~\citep{zhao2024}; government Copilot evaluations~\citep{ukgov2025}; vendor thresholds~\citep{adminreports2026,openai2026a}; Anthropic Economic Index~\citep{anthropic2025} & Annual; quarterly for the reasoning share & A vendor publishing an absolute per-seat message or token rate; a new release of the index that reports thinking use \\
Model class mapping, prices and rates & Provider model and pricing pages~\citep{openai2026b,openai2025b}; \citet{ecb2026}; \citet{eurostat2026} & Quarterly for the mapping; annual for prices and rates & A new model name in a provider model list, in a provider pricing page or in the OpenRouter top 150 by tokens; a vendor changing its documented default model or pricing tier \\
Cloud region list and grid mapping & Published AWS, Azure and Google Cloud region lists & Semi-annual & A new region, or a region whose stated location changes \\
\bottomrule
\end{tabular}
\caption{Inputs, sources, refresh cadence and the events that trigger an out-of-cycle revision (Section S12.3).}
\label{tab:refresh}
\end{table}

\subsection*{S6. Model Class Mapping}

Table~\ref{tab:classes} gives the assignment rule of Section~\ref{sec:3.5} with representative commercial models for each class.

\begin{table}[htbp]
\centering
\setlength{\tabcolsep}{4pt}
\small
\begin{tabular}{>{\raggedright\arraybackslash}p{1.3cm} >{\raggedright\arraybackslash}p{5.5cm} >{\raggedright\arraybackslash}p{5.5cm}}
\toprule
Class & Size proxy & Representative commercial models \\
\midrule
\textbf{A --- Small} & Dense $\leq$15B, or MoE $<$50B total; closed: mini/nano/flash/lite/haiku tier; seven measured members & GPT-4o mini, Claude Haiku, Gemini Flash, Mistral 7B, Gemma 12B, Mixtral 8$\times$7B (47B total)* \\
\textbf{B --- Mid} & Dense $>$15B to 70B, or MoE 50B to $<$200B total; closed: standard tier (sonnet, pro, plus); five measured members & GPT-4o, Claude Sonnet, Gemini Pro, Llama 3.3 70B, Perplexity Sonar, Mistral Small 3 (24B), Gemma 27B, Llama 4 Scout (109B total), Mixtral 8$\times$22B (141B total) \\
\textbf{C --- Frontier} & Dense $>$70B, or MoE $\geq$200B total; closed: frontier tier (opus, ultra, max, gpt-5, full-size o-series); three measured members & Claude Opus, GPT-4 / GPT-4 Turbo, GPT-5, Qwen2.5 72B, Llama 3.2 90B, Llama 3.1 405B, Llama 4 Maverick (400B total), DeepSeek V3 / V3.1 (671B total) \\
\bottomrule
\end{tabular}
\caption{Model class taxonomy: the assignment rule and representative commercial models per class.}
\label{tab:classes}
\end{table}

*Mixtral 8$\times$7B (47B total, about 13B active) is Class A under the mixture-of-experts rule. TokenPowerBench~\citep{niu2026} states that it ``consumes roughly the same energy per token as a dense 8B model'', although its Figures 2 and 3 plot it nearer the 70B dense model, so the source does not settle the point. For GPT-4o and Claude Sonnet, Class B is a low-side assignment, because the energy reported by \citet{jegham2025} would place them at or above Class C.

Table~\ref{tab:classmap} is \textbf{an extract of \texttt{model\_\allowbreak class\_\allowbreak mapping.csv}}, which has 187 rows; the 37 printed here are the ones a preparer meets most often, and the deposited file is the authoritative version. The matching rule, the identifier normalisation and the tie-break are in Section~\ref{sec:3.5}, and the fallback is the file's last row rather than prose. The assignment rule uses the disclosed total parameter count where known, marked [size], and the vendor pricing tier where not, marked [tier]. \emph{A disclosed total parameter count always overrides the tier word.} Three 2026 cases turn on that and all three go the opposite way from the tier word: \texttt{deepseek-v4-flash} (mixture-of-experts, 284B total), \texttt{glm-5.3-flash} (320B) and the MiMo V2.x \texttt{-flash} variants (309 to 310B) are efficiency-tier names at or above the 200B cut and are therefore Class C. The tier vocabulary has also been extended beyond the words printed below, with \emph{plus} added to the standard tier and \emph{max} to the frontier tier, and a row whose tier cannot be determined from the provider's own model list or pricing page takes Class C and says so. The file also carries a fourth outcome besides the three classes: image, video, speech, optical-character-recognition, embedding, moderation and reranking identifiers are marked out of scope rather than classed, because no measurement exists for them under the protocol used here (Table~\ref{tab:inclusion}).

Nine of the rows below are flagged in the deposited file as contradicted or weakened by current provider pages, and a preparer should check the file rather than this extract: \texttt{gemini-ultra} (Google's current list has no tier above Pro and no Ultra model, so the row covers historic identifiers only); \texttt{gemini-*-flash} (later Flash generations are priced at the level of an earlier Pro); \texttt{gemini-*-pro} (a text Pro exists only for two generations as of October 2026); \texttt{claude-opus-*} (Opus is no longer the top Anthropic tier, two generations being priced above it); \texttt{sonar} (Perplexity's current model and model-card pages state no base model and no size); \texttt{sonar-reasoning*} (the non-Pro identifier is no longer on Perplexity's list); \texttt{mistral-medium} (superseded for Medium 3.5, whose disclosed 128B dense size puts it in Class C under the size rule); \texttt{deepseek-chat} (retired from DeepSeek's own list); and \texttt{grok-*} (xAI's current list has no named mini or fast text tier, so every current text identifier is Class C). The deposited file also records that \texttt{gemma-*-2b} has no Gemma 3 release at that size.

\textbf{Identifier normalisation and the tie-break.} The reference calculator implements the matching rule of Section~\ref{sec:3.5} once, in \texttt{normalise\_\allowbreak identifier()}. In order it lower-cases the identifier and strips surrounding whitespace; splits on \texttt{/}, drops empty and purely numeric segments and the thirteen structural path words a platform inserts (\texttt{models}, \texttt{model}, \texttt{publishers}, \texttt{providers}, \texttt{registries}, \texttt{versions}, \texttt{version}, \texttt{projects}, \texttt{locations}, \texttt{deployments}, \texttt{inference-profile}, \texttt{endpoints}, \texttt{foundry}, an \texttt{azureml:} scheme and a Bedrock ARN prefix), and keeps the last remaining segment; strips a leading dot-separated cloud-partition prefix (\texttt{us.}, \texttt{eu.}, \texttt{apac.} and the rest) or vendor name (\texttt{anthropic.}, \texttt{amazon.}, \texttt{meta.} and the rest); and also offers a vendor-rejoined form, so that Bedrock's \texttt{deepseek.r1-v1:0} can match \texttt{deepseek-r1}, and a dots-to-hyphens form, so that \texttt{claude-3.5-sonnet} can match \texttt{claude-3-5-sonnet}. Date, version, \texttt{:0}, \texttt{-latest} and \texttt{-preview} suffixes need no handling, because patterns are matched as prefixes. All the candidate strings are tried and the longest matching pattern wins. Where two patterns tie, the order is settled in sequence by the length of the literal prefix, then the total number of literal characters in the pattern, which settles a wildcard pattern against a bare prefix of the same literal length, then the total pattern length, and finally the pattern string itself, so the order is total and reproducible. The prefix rule as stated in the main text does not settle the wildcard case on its own, and this is the one place where the implementation is more specific than the paper. Worked examples, all tested in the end-to-end suite, include \texttt{anthropic.claude-3-haiku-20240307-v1:0} to \texttt{claude-3-haiku*}, \texttt{us.anthropic.claude-sonnet-4-20250514-v1:0} to \texttt{claude-sonnet-*}, \texttt{meta.llama3-3-70b-instruct-v1:0} to \texttt{llama3-3-70b}, \texttt{publishers/google/models/gemini-2.5-pro} to \texttt{gemini-*-pro} and an \texttt{azureml://registries/.../Llama-3.3-70B-Instruct/versions/4} path to \texttt{llama-3.3-70b}.

\small
\setlength{\tabcolsep}{4pt}
\begin{longtable}{>{\raggedright\arraybackslash}p{4.0cm} >{\raggedright\arraybackslash}p{2.1cm} >{\raggedright\arraybackslash}p{0.7cm} >{\raggedright\arraybackslash}p{7.4cm}}
\caption{Commercial model identifier to model class mapping: an extract of 37 of the 187 deposited patterns (longest match wins; an unmapped identifier takes Class C and is flagged, and a non-text modality is marked out of scope).}\label{tab:classmap}\\
\toprule
Pattern & Provider & Class & Rationale \\
\midrule
\endfirsthead
\multicolumn{4}{l}{\emph{Table~\thetable\ (continued)}} \\
\toprule
Pattern & Provider & Class & Rationale \\
\midrule
\endhead
\bottomrule
\endfoot
\bottomrule
\endlastfoot
gpt-4o-mini & OpenAI & A & [tier] Mini $=$ efficiency tier \\
gpt-4.1-mini & OpenAI & A & [tier] Mini tier of 4.1 \\
gpt-4.1-nano & OpenAI & A & [tier] Nano tier \\
gpt-5-mini & OpenAI & A & [tier] Mini tier of GPT-5; the family reasons by default, so a Tier 2a billed output count already includes the hidden tokens \\
gpt-5-nano & OpenAI & A & [tier] Nano tier of GPT-5 \\
o1-mini & OpenAI & A & [tier] Mini tier; reasoning raises tokens per interaction, not the per-token class, as for o3-mini \\
o3-mini;o4-mini & OpenAI & A & [tier] Mini tier; reasoning raises tokens per interaction, not the per-token class \\
gpt-4o;chatgpt-4o-latest & OpenAI & B & [tier] Mid price tier; size undisclosed; energy reported by \citet{jegham2025} would place it at or above Class C, so B is a low-side assignment \\
gpt-4.1 & OpenAI & B & [tier] Successor of 4o at the same tier \\
gpt-4;gpt-4-turbo & OpenAI & C & [tier] Frontier closed model, no disclosed size \\
gpt-5 & OpenAI & C & [tier] Frontier closed model; as a prefix this also covers the later GPT-5 generations and their chat, codex and pro variants \\
o1;o3;o3-pro & OpenAI & C & [tier] Full-size o-series reasoning models \\
claude-haiku-*;claude-3-haiku*;claude-3-5-haiku* & Anthropic & A & [tier] Haiku $=$ small tier \\
claude-sonnet-*;claude-3-sonnet*;claude-3-5-sonnet*;claude-3-7-sonnet* & Anthropic & B & [tier] Sonnet $=$ mid tier; energy reported by \citet{jegham2025} would place it at or above Class C, so B is a low-side assignment \\
claude-opus-*;claude-3-opus* & Anthropic & C & [tier] Opus tier; no longer the top Anthropic tier \\
gemini-*-flash;gemini-*-flash-lite & Google & A & [tier] Flash $=$ efficiency tier; later generations are priced at an earlier Pro level \\
gemini-*-pro & Google & B & [tier] Pro $=$ mid tier \\
gemini-ultra & Google & C & [tier] Ultra $=$ frontier tier; historic identifiers only \\
llama-3.1-8b;llama-3.2-1b;llama-3.2-3b;llama-3.2-11b* & Meta & A & [size] Open 1--11B dense ($\leq$15B); the 8B variant is a measured class member \\
gemma-*-2b;gemma-*-4b;gemma-*-9b;gemma-*-12b & Google & A & [size] Open 2--12B dense ($\leq$15B); the 12B variant is a measured class member \\
llama-3.2-90b*;qwen2.5-72b & Meta/Alibaba & C & [size] 90B dense (vision) and 72B dense, above 70B \\
gemma-*-27b & Google & B & [size] Open 27B dense; a measured class member \\
llama-3.1-70b;llama-3.3-70b & Meta & B & [size] Open 70B dense (upper edge of B); both are measured class members \\
llama-4-scout & Meta & B & [size] MoE 109B total / 17B active; a measured class member \\
llama-3.1-405b;llama-4-maverick & Meta & C & [size] 405B dense; Llama 4 Maverick MoE 400B total / 17B active; both are measured class members \\
deepseek-v3;deepseek-chat & DeepSeek & C & [size] MoE 671B total / 37B active \\
deepseek-r1 & DeepSeek & C & [size] Reasoning variant of V3 \\
qwen3-235b-a22b & Alibaba & C & [size] MoE 235B total / 22B active; a measured class member \\
ministral-*;mixtral-8x7b & Mistral & A & [size] Ministral 3--14B dense; Mixtral 8x7B MoE 47B total / 13B active; energy near a dense 8B model per TokenPowerBench \\
mistral-7b;open-mistral-7b & Mistral & A & [size] 7.3B dense; retired from Mistral's own list and kept for historical exports and third-party hosts \\
mistral-small & Mistral & B & [size] Mistral Small 3 is 24B dense \\
mistral-medium;mixtral-8x22b & Mistral & B & [tier] Medium: standard tier; [size] 8x22B MoE 141B total / 39B active. Superseded for Medium 3.5, whose 128B dense size puts it in C under the size rule \\
mistral-large & Mistral & C & [size] 123B dense (Mistral Large 2); the bare name now resolves to Mistral Large 3, a mixture-of-experts model of 675B total and 41B active parameters, also C under the size rule \\
sonar & Perplexity & B & [size] Built on Llama 3.3 70B per an earlier Perplexity statement; the current pages state no base model or size \\
sonar-pro & Perplexity & B & [tier] Pro tier \\
sonar-reasoning* & Perplexity & C & [tier] Reasoning tier; no longer on Perplexity's list \\
grok-* & xAI & C & [tier] Frontier closed model; no named smaller text tier currently listed \\
\end{longtable}
\normalsize

\subsection*{S7. Register of Assumptions}

Every assumption used in the paper, grouped by the section where it is discussed, with its value, source, the reason that source was chosen, and the expected direction of error with its sensitivity. In the bias column, Low means the factor is more likely an underestimate and High more likely an overestimate. Unknown means the direction is not determined, Linear that the result scales in proportion, and Symmetric that the band has no preferred direction. The label n/a marks a definitional row, a reporting rule or a comparison, not a factor input. Table~\ref{tab:changes} first lists every value changed between dataset versions, and Table~\ref{tab:register} is the register itself.

\footnotesize
\setlength{\tabcolsep}{3pt}
\begin{longtable}{>{\raggedright\arraybackslash}p{2.7cm} >{\raggedright\arraybackslash}p{2.3cm} >{\raggedright\arraybackslash}p{2.6cm} >{\raggedright\arraybackslash}p{3.2cm} >{\raggedright\arraybackslash}p{4.0cm}}
\caption{Values changed between dataset v2 and v3, with the source and the reason for each change.}\label{tab:changes}\\
\toprule
Value & v2 & v3 & Source & Reason \\
\midrule
\endfirsthead
\multicolumn{5}{l}{\emph{Table~\thetable\ (continued)}} \\
\toprule
Value & v2 & v3 & Source & Reason \\
\midrule
\endhead
\bottomrule
\endfoot
\bottomrule
\endlastfoot
Operating-point rule & Mixed batch points and extrapolation & Minimum J/token at median ITL $\leq$ 50 ms, any accelerator count & ML.ENERGY v3~\citep{mlenergy2025} & The unconstrained minimum marks where each sweep ended, not a serving regime \\
Class value rule & One named proxy model per class & Arithmetic mean of the measured members of the class (n $=$ 7 / 5 / 3 on H100) & ML.ENERGY v3~\citep{mlenergy2025}; Table~\ref{tab:classmembers} & One model's energy is not the class's, the previous small-class proxy was the minimum of its class, and for an additive inventory the arithmetic mean is the expected value if a closed model is equally likely to be any measured member, which weights each model equally \\
Class energy, A/B/C, H100 & 0.12 / 0.485 / 0.617 J per token (two extrapolated) & 0.1309 / 0.4285 / 1.5958 J per token (class means of direct measurements, restated to the reference mix; the leaderboard values average 0.1221 / 0.3916 / 1.4666) &~\citep{mlenergy2025} & Direct measurement at a stated operating point replaces extrapolation, and the class mean replaces a single model \\
H100 facility energy, A/B/C & 0.040 / 0.162 / 0.206 Wh/1k & 0.0720 / 0.2357 / 0.8777 Wh/1k & Rows above $\times$ 1.50 host $\times$ 1.10 idle $\times$ PUE 1.20 & Follows the energy rows and the wider serving boundary \\
B200 multiplier & 0.36 (all classes), low scenario & 0.87 / 0.67 / 1.03, hardware sensitivity & Table~\ref{tab:energy} & Same uplifts and operating-point rule on both platforms \\
Energy bands & $\pm$30--50\% (author judgement) & Class range: A $-$40/$+$47\%, B $-$38/$+$24\%, C $-$67/$+$55\% & Table~\ref{tab:classmembers} & Lowest and highest measured member relative to the class mean \\
Reporting range & B200 scenario to central & Central and conservative high ($H$ 1.88, $I$ 1.20 with $U$ 0.60, PUE 1.45) & \citet{schneider2025}; \citet{oviedo2026}; LBNL~\citep{shehabi2024} & The B200 scenario lay below the central value; the high case now moves PUE as well as the two best-evidenced uplifts \\
Embodied term, A/B/C & 0.75 / 3.03 / 3.85 g/M (energy-proportional) & 0.89 / 3.34 / 12.84 g/M (time allocation at $U$ 0.75, manufacturing plus transport, class means) & \citet{dell2025}; measured throughput~\citep{mlenergy2025} & Direct method; power no longer enters; transport stage added \\
Host uplift & Not modelled; the high case divided electricity by 0.60 & $\times$1.50 inside the central factor &~\citep{gamazaychikov2026}; \citet{latif2025}; Google~\citep{elsworth2025} & Non-accelerator server energy moves from a bound in the high case into a measured central value \\
Idle uplift & Excluded & $\times$1.10 inside the central factor & Google~\citep{elsworth2025}; \citet{oviedo2026} & Provisioned idle capacity moves into the central value \\
Hardware transport & Excluded (sized 427--712 kg) & 672 kg per server, inside the embodied term & Dell PCFs~\citep{dell2025}; \citet{schneider2025} & The transport stage is published beside manufacturing \\
Quantified exclusions & Listed, not quantified & Storage 0.11\%, external network 0.17--0.43\%, end-of-life 0.12--0.17\% location-based (0.38\% market-based) of operational serving emissions; 0.71\% together on the location basis &~\citep{gamazaychikov2026} & Bounds are now published with the exclusions, on the basis the rest of the paper uses \\
Construction & Sized at $\sim$1.7 g/M, mid class, not adopted & 1.50 g/M, mid class, a term of the factor & \citet{schneider2025} Table~1 per machine kWh & Adopted as a fourth term on a server (IT) electricity basis \\
Training term, A/B/C & 0.84 / 4.08 / 17.86 g/M & 0.60 / 2.93 / 12.84 g per million \emph{billed} tokens & \citet{meta2026} $\div$ $6.95\times10^{14}$ tokens & Denominator derived from a published service life and a disclosed serving rate \\
Model development & Sensitivity paragraph only & Separate reported line at $d = 0.67 / 1.5 / 2.5$ &~\citep{gamazaychikov2026}; \citet{patterson2022}; Epoch AI~\citep{denain2026} & Final training runs are 9.6--22.6\% of development compute \\
Named footprints & None & Inference footprint; development-inclusive total &~\citep{gamazaychikov2026}; Section~\ref{sec:3.1} & Model and region comparisons need a serving-only figure \\
Factor terms & Three & Four: electricity, embodied, construction, training & Section~\ref{sec:3.4} & Construction adopted; transport folded into embodied \\
Mid class, global average & 0.0848 kg/M (three terms, per output token) & 0.1187 kg/M inference footprint (per M effective) and 0.0388 kg/M at the default mix (per M billed) & Table~\ref{tab:carbon} & Follows the rows above; the two are never summed \\
eGRID basis & CO\textsubscript{2} output rates labelled CO\textsubscript{2}e & CO\textsubscript{2}e output rates $\times$ 1.1232 & eGRID2023~\citep{epa2025}; Ember US~\citep{ember2026} & Lifecycle basis, as for the Ember rows \\
N.\ Virginia grid & RFCE 0.2708 & SRVC 0.3038 & eGRID2023~\citep{epa2025} & Dominion Energy territory; lifecycle scaling \\
Global average grid & 0.400 (author round value) & 0.483 & Ember world 2023~\citep{ember2026} & Sourced value replaces an unsourced one \\
Model mapping & A printed table of 36 patterns, with no machine-readable file & \texttt{model\_\allowbreak class\_\allowbreak mapping.csv}, 187 patterns with the fallback as data, applied by the reference calculator & Provider model and pricing pages & A third party could not resolve an identifier from the deposit; the printed table is now an extract \\
On-site water source & Li et al.\ 2023 Microsoft sites and author estimates & Published region values and derived provider site ratios, under a stated precedence & \citet{aws2026b}; \citet{microsoft2026a}; \citet{meta2025} & No author estimate remains \\
On-site basis & Not stated; totals added withdrawal to consumption & Converted to consumption throughout; the as-published basis becomes the sensitivity & Provider definitions~\citep{li2025,aws2026b,microsoft2026a,meta2025}; AWS's 1.5 relation~\citep{aws2026b} & A total that adds withdrawal to consumption is not a quantity of anything \\
Scarcity weighting & None & AWARE 2.0 factors applied to the on-site term only & AWARE 2.0~\citep{seitfudem2025} & Only the data-centre location is known, so the off-site term and the total are left volumetric \\
Water terminology & ``Water footprint'' & Water-use inventory result for the volumetric figures; water scarcity footprint only for the weighted on-site figure & ISO 14046~\citep{iso2014} & The standard reserves the unqualified term for a comprehensive impact assessment \\
Hydropower convention & Gross evaporation only & Both conventions reported as co-equal, with a high-hydropower sensitivity & \citet{reig2020}; \citet{bakken2017}; \citet{mekonnen2012} & The ranking depends on it, and the allocated convention is not an upper bound \\
Class rule & Size bands overlapping; MoE by active parameters & Dense by total: A $\leq$15B, B $>$15--70B, C $>$70B; MoE by total: A $<$50B, B 50--$<$200B, C $\geq$200B; closed by pricing tier, with size taking precedence & Section~\ref{sec:3.5}; model cards & One explicit rule, thresholds tied to measured models \\
Token basis & Output-token energy applied to all billed tokens & Effective tokens, with input weighted at 0.30 and cached input at 0.05 and a divisor of 1.339 & Section~\ref{sec:3.3};~\citep{chung2025},~\citep{openai2026b,anthropic2026b,google2026b} & The measured figure is per output token and already contains the benchmark's own prefill \\
Tier 2b tokens per exchange & 400 / 3,000 / 2,000 (author assumptions) & 1,210 / 3,676 / 1,952, plus an agentic row at 85,000 per model request & Azure traces~\citep{azure2024}; WildChat~\citep{zhao2024}; routing platform and \citet{anthropic2026c} & Billed input includes re-sent context, which published per-prompt figures do not \\
Tier 2b messages per seat & 110 central, 2,000 high & 30 central (20--65), 325 heavy & UK government pilots~\citep{ukgov2025}; vendor thresholds~\citep{adminreports2026,openai2026a} & The 110 was a consumer figure, and 2,000 exceeded every vendor power-user threshold \\
Worked example, Tiers 2a + 2b & 86.8 kg central (62.1 B200; 138.5 high) & 8.5 kg (6.7 B200; 13.6 high; 39.4 at heavy use) & Tables~\ref{tab:carbon}, \ref{tab:regionindep}; Sections~\ref{sec:3.2} and \ref{sec:3.3} & Follows the rules and factors above, including the Tier 2b reasoning uplift; the chain 86.8 $\to$ 135.3 $\to$ 21.6 $\to$ 8.5 kg is in S8 \\
Worked example, Tier 1 & 944.8 kg, added to the AI total & 769.0 kg deflated, a subscription screening line & Eurostat HICP~\citep{eurostat2026} & Deflation rule; AI share not estimable \\
Worked example, sum of lines & $\sim$975 kg & Removed; the spend-based screening line and the token-based figure are reported separately and not added, because they are on different boundaries & Table~\ref{tab:casestudy} & Review round 1, R4.12 \\
Google / Mistral comparison & Per million tokens, at an assumed 400 tokens per prompt & Per query: Google 2.43 times the mid class and 0.65 times the frontier class on energy, Mistral 21.4 and 5.7 times on carbon, per chat exchange &~\citep{elsworth2025},~\citep{mistral2025} & No source discloses a token count, so a per-token restatement rests on an undisclosed assumption \\
\end{longtable}
\normalsize

\footnotesize
\setlength{\tabcolsep}{3pt}
\begin{longtable}{>{\raggedright\arraybackslash}p{2.6cm} >{\raggedright\arraybackslash}p{2.4cm} >{\raggedright\arraybackslash}p{2.7cm} >{\raggedright\arraybackslash}p{4.4cm} >{\raggedright\arraybackslash}p{2.7cm}}
\caption{Register of assumptions: every assumption used in the paper with its value, source, the reason that source was chosen, and the expected direction of error.}\label{tab:register}\\
\toprule
Assumption & Value & Source & Why this source & Bias (sensitivity) \\
\midrule
\endfirsthead
\multicolumn{5}{l}{\emph{Table~\thetable\ (continued)}} \\
\toprule
Assumption & Value & Source & Why this source & Bias (sensitivity) \\
\midrule
\endhead
\bottomrule
\endfoot
\bottomrule
\endlastfoot
\multicolumn{5}{l}{\textit{Sections~\ref{sec:1} and \ref{sec:2}: scope and background}} \\
\srow{r:cat}Accounting category & Cat.~1 (pay-per-use, subscription); Cat.~2 or 8 (capitalised, dedicated capacity) & GHG Protocol Scope 3 Standard~\citep{wri2011} & Definitional text that ESRS E1 refers to & n/a (definitional) \\
\srow{r:reg}Regulatory scope & CSRD wave 1 from FY2024; Omnibus I scope from FY2027; ESRS E1-6 to FY2026 and E1-8 from FY2027 &~\citep{euparl2022},~\citep{euparl2025},~\citep{euparl2026b},~\citep{eucom2026a} & Official Journal texts are the only binding source & n/a (legal fact). The stop-the-clock directive postponed waves two and three by two years, to FY2027 and FY2028, and wave-one undertakings below the new thresholds still report FY2024, and FY2025 and FY2026 unless their Member State takes the exemption the Omnibus allows \\
\srow{r:gpu}Accelerator energy per output token & Class means, H100 0.1309 / 0.4285 / 1.5958 J; B200 0.1133 / 0.2853 / 1.6473 J (A/B/C), restated to $R_{\mathrm{bench}}$ (raw run files, \texttt{ml-energy/\allowbreak benchmark-v3}) & ML.ENERGY Leaderboard v3~\citep{mlenergy2025}, lm-arena-chat, 16 February 2026 & Only open benchmark that measures the same models on H100 and B200 under one protocol; measured in a window where the batch is saturated at the configured maximum & Low (accelerator only; a saturated batch of 128--512 applied as a fleet average; rows~\ref{r:oppoint}, \ref{r:gpushare}) \\
\srow{r:oppoint}Operating point & Minimum J/token at median ITL $\leq$ 50 ms, any accelerator count & Author choice; data~\citep{mlenergy2025} & Interactive streaming, at least 20 tokens/s per user, a ceiling with no external source; the frontier H100 members sit at 46.6--48.0 ms, so that class is the one exposed to it; an unconstrained minimum marks where each sweep ended, not a serving regime & Low. The mean over all configurations meeting the limit is $\times$2.10 / $\times$2.88 / $\times$2.43 on the class energy, and the minimum at the smallest accelerator count $\times$1.12 / $\times$1.18 / $\times$0.83 \\
\srow{r:classrule}Class value rule & Arithmetic mean of the measured members (n $=$ 7 / 5 / 3 on H100), one entry per distinct model, bfloat16 preferred & Author rule; data~\citep{mlenergy2025} & A single named model is not the class; the previous small-class proxy was the minimum of its class, which made its band one-sided & n/a (definitional). Geometric mean $\times$0.95/0.96/0.83, median $\times$0.98/1.16/1.12, previous named proxies $\times$0.54/1.12/0.93 \\
\srow{r:gpushare}Accelerator share of node energy & ``over 60\%'' of total node energy; about 0.8 read off its Figure 3 for all ten models (total over GPU about 1.2 to 1.3, a digitised reading) & Token\-Power\-Bench~\citep{niu2026} & Only peer-reviewed LLM-inference study that splits node energy by component & n/a (a floor on the accelerator share, so a ceiling of 1.667 on the host uplift; kept as a consistency check that lies inside the band of 1.43 to 1.88 adopted in row~\ref{r:host}, not as its high value) \\
\srow{r:eeiogap}EEIO vs physical gap & 1.7 to 12.3 times on deflated spend, per million billed tokens of the chat mix (2.0 to 15.1 times undeflated) & Author calculation at GPT-4o (USD 2.50/10) and GPT-4o mini (USD 0.15/0.60 per M input/output tokens) list prices~\citep{openai2026b}, global average and US East (Section~\ref{sec:7.1}) & Only like-for-like per-token comparison available; both sides on the 928-in/282-out mix and on the deflated basis the method's own Tier 1 rule produces. GPT-4o deflated gives 0.3614 kg against 0.0440 kg for the mid class at the global average, 8.2 times, and 12.3 times in US East; GPT-4o mini gives 0.0217 kg against 0.0131 kg for the small class, 1.7 and 2.5 times & n/a (Sweden: small class 10.7 times, mid class 46.7 times deflated; deflation itself is $-$18.6\%) \\
\srow{r:fx}Exchange rate & 0.885 EUR/USD & ECB euro reference rate, 2025 annual average~\citep{ecb2026} & Used only to convert list and seat prices; the annual average of the FY2025 reporting year & Linear ($\pm$10\% rate moves EEIO-per-token $\pm$10\%) \\
\multicolumn{5}{l}{\textit{Section~\ref{sec:3}: boundary, activity data, factor terms and rules}} \\
\srow{r:boundary}Boundary exclusions & Provider-side storage, external network, hardware end-of-life and data-centre networking outside with a bound or reported negligible; transmission losses, behind-the-meter generation and the recycling credit as stated conventions; end-user devices, provider overhead, fine-tuning jobs and other modalities out of scope & Scope 3 Standard~\citep{wri2011}; bounds from \citet{gamazaychikov2026} & Category 1 minimum is cradle-to-gate, and every exclusion is disclosed with a bound where one exists and a sensitivity where it does not & Low (row~\ref{r:excl}) \\
\srow{r:excl}Quantified exclusion bounds & Storage about 0.11\%; external network 0.17--0.43\%; hardware end-of-life 0.12--0.17\% location-based (0.38\% market-based) of operational serving emissions &~\citep{gamazaychikov2026} & Only published method that bounds each excluded element on a comparable basis & Low (together 0.71\% on the location basis and 0.92\% market-based in the source's own cut-off test; end-user devices sit in the customer's own inventory; provider overhead is not quantified) \\
\srow{r:tokbasis}Token basis & Effective tokens $=$ (output $+$ 0.30 uncached input $+$ 0.05 cached input) $/$ 1.339; electricity, embodied and construction on effective tokens, training on billed tokens & Section~\ref{sec:3.3}; \citet{mlenergy2025}; \citet{chung2025} & The measured figure is accelerator energy per request over output tokens only, confirmed exactly on all 328 configurations, so prefill is already inside it and the divisor stops it being charged twice & n/a (rule). The effective-token count is 2.9 times below the billed count for a chat exchange and 17 times below it for an agentic request \\
\srow{r:win}Uncached input weight $W_{\mathrm{in}}$ & 0.30 (0.10--0.40) & ML.ENERGY prefill--decode disaggregation~\citep{chung2025}; provider prices~\citep{openai2026b,anthropic2026b,google2026b}; Greenpixie~\citep{horswill2026}; \citet{vellaisamy2026} & One four-row energy table read two ways, one qualitative statement and one price signal, spanning 0.009--0.333; the lowest published values come from low-batch serving, the opposite regime to the batch-128-to-512 operating points used here & Central: the adopted value is a within-cell reading chosen by the author, 0.295--0.304, since the source table states no batch size or hardware, and the band is two-sided about it ($-$24\% to $+$10\% on the worked example's token-based total; $-$33.1\% to $+$14.3\% on the mid-class factor per M billed) \\
\srow{r:wcache}Cached input weight $W_{\mathrm{cache}}$ & 0.05 (0.02--0.10) & Provider cache-read prices~\citep{openai2026b,anthropic2026b,google2026b}; ML.ENERGY~\citep{chung2025} & No published energy measurement of a cache read exists. The price ratio of 0.10 of an uncached input token is a lower bound, because it reflects avoided compute and not the decode-side attention the provider still pays; the upper bound is the decode part of the marginal energy of an input token, 48.5--49.6\% of it and 0.084--0.085 of an output token on the balanced 8B rows, rounded up & Low for cache-heavy workloads: the same share applied to $W_{\mathrm{in}} = 0.30$ gives 0.15 and the decode-attention scaling 0.061--0.076, both above the adopted 0.05, which sits between them and the price-implied 0.03 on the energy basis (0.10 times $W_{\mathrm{in}} = 0.30$) as a named departure from the conservative-end rule (a factor of 2.5 on an agentic request, and on a workload without cache reads only through the restatement of the benchmark runs) \\
\srow{r:rbench}Reference input:output ratio $R_{\mathrm{bench}}$ and the mix of each benchmark run & 1.13 (reference); runs 0.47--2.01, cache-hit share median 62\% & Raw run files of the benchmark~\citep{mlenergy2025}; task definition~\citep{mlenergy2026}; J.-W.~Chung, personal communication & Server-side prompt, generated and cache-hit token counts per run; 300 of 328 configurations with prefix caching on; 21 with a recomputation pattern, for which the hit share of clean runs of the same model is used & None from the reference value; the restatement is $+$7.2\% / $+$9.4\% / $+$8.8\% on the class energy (A/B/C, H100), and the hit share is a whole-run total \\
\srow{r:highcase}Conservative high case & $H = 1.88$, $I = 1.20$ with $U = 0.60$, PUE $= 1.45$, on electricity, construction and embodied; training central & \citet{schneider2025}; \citet{oviedo2026}; LBNL~\citep{shehabi2024}; author rule & Compounds the upper end of the two best-evidenced uplifts with the upper end of the worst-evidenced one, PUE, at the central operating point & n/a (reporting rule; $+$63.7\% on the mid-class global-average inference footprint, $+$58.9\% on the convenience factor) \\
\srow{r:b200mult}B200 multiplier & 0.87 / 0.67 / 1.03 (A/B/C), hardware sensitivity & Table~\ref{tab:energy} & Measured B200 over H100 facility energy at the same uplifts and operating-point rule, with different mid and frontier members on the two platforms & n/a (at the minimum of the smallest accelerator count 0.69 / 0.41 / 0.53, which mixes hardware with latency; DeepSeek V3.1 alone as the frontier class gives 0.95 of that class's B200 mean) \\
\srow{r:b200emb}B200 embodied and training & Held at H100 values & NVIDIA HGX B200 PCF (2,274 kg vs 1,312 kg H100 baseboard)~\citep{nvidia2025a,nvidia2025b} & No B200 server footprint to allocate; training does not depend on serving hardware & Unknown (more embodied carbon per board, more tokens per hour) \\
\srow{r:host}Host uplift $H$ & $\times$1.50 on accelerator energy (1.43 and 1.88 in the sensitivities; the evidence spans 1.41--2.22) & \citet{latif2025}; Google~\citep{elsworth2025}; split from \citet{gamazaychikov2026} & Latif et al.\ meter an eight-accelerator H100 node at 7.9--8.4 kW at the rack power distribution unit and define GPU utilisation as measured over rated accelerator power, which gives 1.52 at 93\% utilisation and 1.51 at 100\% (1.42 for the GPU-only burn test without CPU load), assuming only the 700 W per-accelerator rating behind their utilisation figure, not full-power operation. TokenPowerBench is a ceiling, not a central value (row~\ref{r:gpushare}) & Unknown (serving accelerators draw well below their rated power, which raises the non-accelerator share and points low, but the high value of 1.88 is Google's all-workload TPU machine ratio, a wider boundary than the GPU inference node behind the 1.667 ceiling, which is why the high exceeds the ceiling, and TokenPowerBench's digitised node-to-accelerator reading of about 1.2 to 1.3 sits below the low of 1.43; $-$4.5\% to $+$24.6\% on the mid-class inference footprint). The paper has now been read in the original; the 5.6 kW tray figure behind the lower 1.41 reading is NVIDIA's nameplate, not a Latif measurement, and the 1.10/1.36 component split is from \citet{gamazaychikov2026}, whose 1.10 leg is their own node telemetry and whose 1.36 is derived from Latif et al. \\
\srow{r:idle}Idle uplift $I$ & $\times$1.10 on host-inclusive server energy (1.05--1.20) & Google~\citep{elsworth2025} (0.02 Wh idle against 0.20 Wh of accelerator plus host; 1.11 to 1.13 from Google's printed percentage shares); \citet{oviedo2026} for the high value & Only first-party measurement of provisioned idle energy in a production serving fleet & Low (Google counts only capacity allocated to the same model, so fleet-wide spare capacity is outside it; $-$4.4\% to $+$8.8\% on the mid-class inference footprint). Oviedo et al.\ model no range: Supplemental Note S5 gives 1.08 for diurnal load times 1.10 for redundancy, about 1.19, so the high value is close to their point estimate and the low value of 1.05 is our own sensitivity with no source behind it \\
\srow{r:constr}Data-centre construction $k_c$ & 7.66 g CO\textsubscript{2}e per kWh of \emph{server} (IT) electricity (low 5.3); 1.50 g/M for the mid class & \citet{schneider2025} Table~1 (59, 59, 109, 117 and 218 kg per TPU over six years), divided by the matching machine electricity & Only source reporting construction alongside machine energy, allocated pro rata to data-centre energy and amortised over a 20-year building life. The five machines give 7.58--7.66 g per kWh of machine electricity, and the basis travels with the constant because applying it after PUE would overstate the term by 20\% & Unknown; 7.66 g is the top of a 1\%-wide range, and the low 5.3 g is the Carbon Trust low scenario taken directly on the same basis (117 kg CO\textsubscript{2}e/m\textsuperscript{2}/yr over 60 years, 5 kW/m\textsuperscript{2} of IT load at 50\% utilisation; \citealp{carbontrust2026}), worth $-$0.4\% on the mid-class inference footprint. That report prints no per-kWh figure, its medium and high rack densities give 1.71 and 0.76 g, and two of the three authors of \citet{gamazaychikov2026} sat on its expert working group. Schneider et al.\ call their own estimate of this stage rough, and \citet{mistral2025} puts data-centre construction below 1\% of a model's lifecycle greenhouse-gas emissions \\
\srow{r:server}Server manufacturing footprint & 3,300 kg CO\textsubscript{2}e (range 2,872--4,241) & Dell PCFs~\citep{dell2025}; NVIDIA HGX H100 PCF~\citep{nvidia2025a} & Only retrievable server footprints of this class; weighted to the two XE9680 (eight-accelerator HGX) variants; NVIDIA 1,312 kg baseboard plus host brackets it. Each Dell value is the sum of the nine printed component lines, Dell printing no manufacturing subtotal & Low ($-$13\%/$+$29\% across the Dell range, worth $-$0.3\%/$+$0.7\% on the footprint) \\
\srow{r:transport}Transport stage & 672 kg per server, inside the embodied term (range 427--712) & Dell PCFs~\citep{dell2025}; \citet{schneider2025} & Mean of the two XE9680 variants; Google's TPU v5e machine (Appendix B.1) implies $\times$1.207 against 2,277 kg of manufacturing & Slightly low ($-$0.2\% to $+$0.0\% on the mid-class footprint across the Dell range) \\
\srow{r:accel}Accelerator content of Dell servers & Eight accelerators, inferred & Dell PCFs~\citep{dell2025} & GPU Quantity and GPU Memory are blank on all four reports, so the count rests on the processor count, the mass, the installed memory and the size of the mainboard line; the accelerator model is not declared either & Unknown (if the footprints were for four-accelerator servers, an eight-accelerator equivalent would need two: $\times$2 embodied; the NVIDIA bottom-up cross-check, 2,800--3,800 kg, bounds the eight-accelerator case at $\times$0.85--1.15) \\
\srow{r:throughput}Embodied allocation and throughput & Time allocation; class harmonic-mean throughputs 37,865 / 10,044 / 2,615 output tokens/s per eight-accelerator server &~\citep{mlenergy2025}, the member points of Table~\ref{tab:classmembers} & Direct method with measured throughput; server power does not enter. Each class value is the mean of its members' own terms & n/a (the smallest-accelerator-count point lowers the mid and frontier terms, because larger batches serve more tokens per second) \\
\srow{r:life}Hardware life & 5 years (range 4--6) & Operator Form 10-K filings~\citep{secfilings2026};~\citep{schneider2025};~\citep{dell2025};~\citep{gamazaychikov2026} & Audited depreciation lives: Amazon shortened a subset to five years in 2025 citing AI, Alphabet files six, Microsoft two to six and Meta five to 5.5; Dell assumes 4, Schneider et al.\ 6 and \citet{gamazaychikov2026} 3 & Unknown ($+$25\% at 4 years, $-$17\% at 6 on the embodied term). Six years is Alphabet's general life and the upper end of the Amazon and Microsoft classes, so 5 is a mildly conservative choice; useful life is an accounting estimate \\
\srow{r:util}Utilisation & 0.75, tied to $I$ by $I = 1 + 0.30(1-U)/U$ (0.60 with $I$ 1.20; 0.857 with $I$ 1.05) & Derived from the measured $I$~\citep{elsworth2025}; the 0.30 idle-power fraction is an author assumption & No provider discloses utilisation, and $I$ and $U$ describe one fleet, so the two are tied rather than chosen separately; the pair is asserted in the recompute script & Unknown ($+$25\% on the embodied term at $U$ 0.60); the idle hours are charged electricity through row~\ref{r:idle} and hardware through this row \\
\srow{r:embband}Embodied band & $-$39\%/$+$95\%; upper tail open & Author calculation; Google~\citep{elsworth2025} & Footprint plus transport, life and the tied utilisation pairs combined at their extremes & Skewed high (Google's restated Scope 1 and embodied Scope 3 emissions, 24 g per million effective tokens, are 7.1 times the mid-class term, above the band) \\
\srow{r:training}Training emissions & 420 / 2,040 / 8,930 tCO\textsubscript{2}e & Meta Llama 3.1 model card~\citep{meta2026} & Only verified per-model location-based disclosure spanning all three classes & Low (final runs only) \\
\srow{r:trainc}Frontier-class training proxy & Llama 3.1 405B (12.84 g/M) &~\citep{meta2026} & Only disclosed frontier-scale training footprint & Low for closed frontier models \\
\srow{r:lifetok}Lifetime tokens served & $6.95\times10^{14}$ per model, $(S/12) \times 365 \times V$ &~\citep{anthropic2026a},~\citep{aws2026a},~\citep{microsoft2026b}; GPT-5-Codex~\citep{openai2025a} & Replaces an author assumption with two published quantities, a service life and a serving rate; the provider totals of \citet{openai2025a} and \citet{pichai2026} remain as scale checks (S1.3) & High for the small class, low for the frontier class (class volumes 10/2/1 times the denominator: 0.06/1.47/12.84 g) \\
\srow{r:slife}Model service life $S$ & 12 months (6--18) & \citet{anthropic2026a}; AWS Bedrock~\citep{aws2026a}; Microsoft Foundry~\citep{microsoft2026b} & Minimum published by two hosting platforms: Bedrock committed to at least 12 months for models launched before 7 September 2026, and Microsoft Foundry sets a retirement date 18 months after launch for generally available models on its standard lifecycle, which closes them to new customers at 12 months, and a 12-month lifecycle for models from Anthropic, DeepSeek, Fireworks and Mistral AI. As of 2 October 2026 all 14 active Anthropic models also sit at release plus 12 months, but that is a pattern inferred from the dates Anthropic lists rather than a commitment it states, no published date has been brought forward, most completed retirements ran longer, and Claude Sonnet 3.7 ran five days short of twelve months against its announcement date, the deposited series anchoring on the snapshot date and recording 12.0 months & Unknown ($+$7.6\% at 6 months, $-$2.5\% at 18 on the mid-class factor per million billed tokens); the 80-row series is 45 completed retirements, 21 announced dates and 14 floors at release plus 12.0 months by construction, with 10 censored, so its 12.8-month median is not independent of the 12-month floor it supports \\
\srow{r:vday}Per-model serving rate $V$ & $1.905 \times 10^{12}$ tokens/day & GPT-5-Codex, over 40 trillion tokens in its first 21 days~\citep{openai2025a} & The only per-model serving volume any provider has disclosed & Low on the factor, the rate itself being high (a first-21-day rate for a model the provider described as one of its fastest growing ever, and a coding model rather than a general assistant). The source does not state whether the count is input plus output or output only, or whether cache reads are inside it; any other reading makes the denominator too small and the term too high \\
\srow{r:shape}Usage shape factor & 1.0 central; 0.4 as a sensitivity & Per-author weekly token series on a public routing platform & Lifetime usage ramps, peaks and decays, so a lifetime-average rate is below a peak rate; the median ratio of the mean active week to the peak week is 0.40 across 14 provider accounts & Not adopted as a constant, because the series is per provider rather than per model and is confounded by 21-fold growth of the platform over the window (0.4 raises the mid-class factor per million billed tokens by 11.3\%) \\
\srow{r:dev}Development-to-serving ratio $d$, all model development including the final training runs & 1.5 (0.67--2.5), separate line only &~\citep{gamazaychikov2026}; \citet{patterson2022}; Epoch AI~\citep{denain2026} & 1.5 is the Sustainable AI Group value, from Epoch AI's split of Anthropic's 2025 compute spending, whose USD 2.70 billion serving leg is Epoch's inference from a reported 40\% gross margin rather than a disclosed figure, and an industry capacity allocation of 50/10/40; 0.67 is Google's measured two fifths training against three fifths inference, over one week of April in each of three years; 2.5 is Epoch AI's reconstruction of OpenAI's 2024 spending at about USD 5 billion against about USD 2 billion & n/a (never added to the factor; for the mid class at the global average the line is 0.0694--0.2588 kg per M effective tokens and a development-inclusive total is 1.6 to 3.2 times the inference footprint). A spend or capacity ratio used as an energy ratio; the agreement of the two reconstructions to 1.2\% is an artefact of round numbers. The line carries no embodied or construction term, so it is about 4\% short of the footprint's boundary \\
\srow{r:trainband}Training band & Service life 6 to 18 months, giving $\times$2.0 to $\times$0.67 on the term &~\citep{anthropic2026a},~\citep{aws2026a},~\citep{microsoft2026b} & Spans the published provider minima rather than an author-chosen order of magnitude & Symmetric in $1/S$ (row~\ref{r:slife}) \\
\srow{r:tokens}Tokens per exchange & Short chat 928 in / 282 out; document-heavy 3,293 / 383; inline completion 1,944 / 8 (input $=$ median billed total minus median output; the raw input medians are 3,130 and 1,930) & Azure production inference traces~\citep{azure2024}; WildChat-1M~\citep{zhao2024} & Provider-counted context tokens from all 44.1 M requests of the two public one-week 2024 traces, and output lengths from a conversation corpus; these are billed inputs including re-sent history & Replaces 400 / 3,000 / 2,000. Short chat was 3.0 times too low and the inline-completion output 62 times too high. Low for an annual total: these are per-message medians, and the deposited means are 1.30 to 1.83 times them on billed tokens (S9.7) \\
\srow{r:agentic}Agentic coding request & 85,000 tokens per model request (71,000--119,000), 91\% of input from cache (0.91--0.95), 400 output & Routing-platform application rankings~\citep{openrouter2026}; Claude Code cost documentation~\citep{anthropic2026c} & Two unrelated routes, 258.7 M routed requests and two vendor-documented examples on one page, combined here as if one session, land within 27\% of each other (71,179 against 96,884 tokens per request) & New category. One human instruction triggers many model requests, so the unit is the model request. The cache share is illustrative, not measured \\
\srow{r:t2aband}Tier 2a range & The class range on the energy term (Table~\ref{tab:energy}) & ML.ENERGY v3 class members~\citep{mlenergy2025} & Spread of the measured models of the class around the class mean & Data-driven; misses one-sided biases; removed entirely where the model has its own factor \\
\srow{r:t2bband}Tier 2b band & The class range plus $-$50\%/$+$130\% on estimated tokens & Interquartile ranges of the measured distributions~\citep{azure2024,zhao2024} & The measured spreads are right-skewed: $-$50\%/$+$115\% short chat, $-$38\%/$+$132\% document-heavy, $-$51\%/$+$73\% inline completion & Replaces a symmetric $\pm$50\%, which understated the upper tail (worked example Tiers 2a $+$ 2b 7.0 to 12.6 kg). It is a per-message spread carried as a scenario, not an interval on an annual total and not an extreme \\
\srow{r:eeio}EEIO factor & Producing-country factor; example 0.1181 kg CO\textsubscript{2}e/EUR (Austria, 2019) & EXIOBASE 3.8.2~\citep{stadler2018}, via \citet{climatiq2026} & A verified United States factor for the example's United States vendor could not be retrieved & Unknown (Germany 0.1333, $+$13\%) \\
\srow{r:eeioprice}EEIO price basis & Spend deflated to 2019 EUR with the euro-area HICP, the EEIO model's own price basis being basic prices &~\citep{stadler2018}; Eurostat HICP~\citep{eurostat2026} & Model base year; euro-area HICP 104.80 (2019) to 128.75 (2025), $+$22.9\% & n/a in the method (undeflated 944.8 kg vs 769.0 kg on the worked example) \\
\srow{r:eeiover}EEIO version & 3.8.2, reference year 2019 (a now-cast beyond the 2011 core series) &~\citep{stadler2018} & Version with a retrievable Climatiq entry. Later releases exist and are not adopted: from version 3.9 the licence is non-commercial and does not permit publishing the coefficients inside a derived dataset & Unknown \\
\srow{r:aishare}AI share of SaaS spend & Not estimated; whole subscription kept as a screening line & Author rule & Share not observable without a separately priced AI add-on & n/a \\
\multicolumn{5}{l}{\textit{Section S9: parameterisation}} \\
\srow{r:class}Class of closed models & Provider pricing tier, with disclosed size taking precedence; unmapped $=$ C & Author mapping, Table~\ref{tab:classmap} & No provider discloses size; tier is on every billing record & Low for the mid and frontier classes (one class of misclassification is 3.3 to 3.7 times on the electricity term, two classes 12.2 times) \\
\srow{r:thresh}Class size thresholds & Dense: A $\leq$15B, B $>$15--70B, C $>$70B; MoE by total parameters: A $<$50B, B 50--$<$200B, C $\geq$200B & Author rule; ML.ENERGY v3~\citep{mlenergy2025}; model cards & Each cut falls between measured models of different energy; all experts must be resident in accelerator memory, which sets the accelerator count & Unknown (Llama 4 Maverick, 400B MoE, draws 0.33 times the frontier-class mean) \\
\srow{r:precision}Precision & bfloat16 where a measurement exists, FP8 only where none does &~\citep{mlenergy2025} & Two of the three frontier-class H100 members have no bfloat16 point at the operating point & High for the frontier class (using the FP8 measurement of every member that has one lowers that class's footprint by 10\%) \\
\srow{r:ebands}Class range & A $-$40/$+$47\%, B $-$38/$+$24\%, C $-$67/$+$55\% (H100) &~\citep{mlenergy2025}, Table~\ref{tab:classmembers} & Lowest and highest measured member relative to the class mean at the 50 ms rule & Data-driven; excludes one-sided biases \\
\srow{r:pue}PUE & 1.20 on both platforms & Microsoft 1.17~\citep{microsoft2026a}; AWS 1.14~\citep{amazon2024}; Google $\approx$1.09~\citep{google2026a}; Uptime 1.56~\citep{uptime2024} & APIs are served from hyperscale facilities; a facility property, not tied to the chip & Unknown; low if any serving is outside hyperscale facilities. 1.20 is an author choice above the three disclosed hyperscaler values and well below the industry average; one low--high pair, 1.09 and 1.45, is used throughout, worth $-$8.8\% and $+$20.0\% on the mid-class inference footprint \\
\srow{r:usgrid}US grid intensity & eGRID2023 subregion CO\textsubscript{2}e total-output rates $\times$ 1.1232 &~\citep{epa2025}; Ember US 2023~\citep{ember2026}; Scope 2 Guidance~\citep{wri2015} & Subregion approximates the power pool; the scaling puts United States rows on Ember's lifecycle basis & Unknown (unscaled: US East mid-class footprint 0.0686 against 0.0765, $-$10\%) \\
\srow{r:vasub}Cloud region to grid & us-east-1 to SRVC (0.3038) & eGRID2023~\citep{epa2025} & Ashburn--Loudoun is Dominion Energy territory; same subregion as the Virginia water factor & Negligible (RFCE 0.2718 vs SRVC 0.2705, CO\textsubscript{2}e, unscaled) \\
\srow{r:nonus}Non-US grid intensity & Ember national, 2023, lifecycle CO\textsubscript{2}e &~\citep{ember2026} & Only open annual series covering all ten non-United States countries, the EU-27 and the world & Unknown; generation-based, imports not reflected \\
\srow{r:eu27}EU-27 grid & 0.236 kg/kWh & Ember EU 2023~\citep{ember2026} & Rule for EU data zones & Unknown \\
\srow{r:sgp}Singapore grid & 0.501 kg/kWh & Ember 2023~\citep{ember2026}; EMA 0.412 kg CO\textsubscript{2}~\citep{ema2025} & Cross-country consistency; EMA is combustion CO\textsubscript{2} only & Unknown ($-$18\% on EMA) \\
\srow{r:global}Global average grid & 0.483 kg/kWh & Ember world 2023~\citep{ember2026} & Same basis as the other Ember rows & Unknown \\
\srow{r:gridyear}Grid data year & 2023 for FY2025 &~\citep{epa2025},~\citep{ember2026} & eGRID2023 is the latest eGRID year; Ember rows are held at the same year for consistency (Ember 2024 and 2025 are published) & High for EU rows (Ember 2024: Germany $-$7\%, France $-$24\%) \\
\srow{r:wue}On-site WUE & 0.020--1.570 L/kWh across the published and derived rows & AWS regional series~\citep{aws2026b}; \citet{microsoft2026a} and \citet{meta2025} site disclosures; \citet{li2025} & Precedence: published named-region value, then a site ratio derived from published site water and site electricity, then the 2023 Microsoft series, then a continental aggregate. No author estimate remains & Operator-specific, not national: Microsoft's northern-Virginia sites are 4.6 times the AWS region value used here, so that row is a low end \\
\srow{r:wbasis}On-site basis & As published, withdrawal for fifteen rows and consumption for two; converted to consumption for all seventeen & AWS and Meta publish withdrawal over IT load; Microsoft's site table is withdrawal, informed by GRI 303-3, while its fleet WUE page names no basis; Li et al.\ is consumption, and Li et al.\ treat Microsoft's reported WUE as consumption~\citep{li2025,aws2026b,microsoft2026a,meta2025} & Mixing bases is unavoidable while providers disagree, so the basis travels with each row & Resolved: every row is reported on a consumption basis, the withdrawal rows divided by the 1.5 implied by the AWS page's own industry-average footnote (0.84 litres withdrawn per 0.56 consumed at three cycles of concentration), a derived relation rather than a statement about the AWS fleet. Leaving them as published raises totals by 0.1--11.3\% allocated and 0.4--15.6\% excluded, and swaps the two lowest regions on the allocated ranking \\
\srow{r:wdenom}On-site denominator & Server energy for published values, facility energy for derived site ratios & Provider definitions~\citep{aws2026b,microsoft2026a,meta2025} & A published value is per kWh of IT load; a derived site ratio has total site electricity in the denominator and all site water in the numerator & Unknown; applying a derived ratio to IT load instead would lower those four rows by a factor of 1.20 \\
\srow{r:ewif}Off-site water intensity & Consumption factors, for example Sweden 6.019, United States average 3.142 L/kWh & \citet{reig2020}, verified against the source rather than its reproduction & One method covering eGRID subregions and 47 countries; no alternative covers both & Low for the United States rows: the two published United States averages are 4.35 and 4.52 L/kWh~\citep{shehabi2024} and Meta's own disclosure implies 4.00~\citep{meta2025}. Generation mixes date from 2018 \\
\srow{r:hydro}Hydropower convention & Both conventions reported as co-equal, neither preferred & \citet{reig2020}; \citet{bakken2017}; \citet{siddik2021} & The literature settles neither gross against net nor how to allocate a multipurpose reservoir. About 70\% of a United States factor of this family is allocated reservoir evaporation & n/a (reporting rule). Allocating is not an upper bound: the factors applied are 11--23 times below the most-cited global average, which the high case uses \\
\srow{r:hydroex}Hydropower-excluded factor & Published factor $\times$ (1 $-$ hydro share of a reconstruction of it) & \citet{reig2020} App.~4--6 & The reconstruction reproduces the published factor within a factor of 0.60--1.20, except the Netherlands, whose hydropower share is zero & Low for hydro-heavy grids. A directly published hydropower-excluded intensity exists for United States balancing authorities~\citep{siddik2024} and should replace the reconstruction for the United States rows in a later dataset version \\
\srow{r:wueest}Continental on-site values & France, Belgium and the EU-27 at 0.05 L/kWh & AWS Europe aggregate~\citep{aws2026b} & AWS publishes a PUE but no water figure for its Paris region, and neither Microsoft nor Meta operates a publishing site in either country & Replaces author estimates. Three previous estimates were wrong in the same direction (UK 0.05 against 0.14, Germany 0.08 against 0.17, Japan 0.20 against 1.22) \\
\srow{r:wueproj}Projected-site value & Finland 0.010 L/kWh & \citet{li2025} & The weakest row in the table: a Microsoft projection for a site under construction as of July 2023 & Low: comparable operating Nordic sites give 0.020 to 0.513 L/kWh \\
\srow{r:oregon}Oregon WUE & 0.120 L/kWh & AWS \texttt{us-west-2} published value~\citep{aws2026b} & Replaces a neighbouring-site proxy of 0.950 that was eight times too high; Meta's own Oregon site gives 0.190 & Nearest-neighbour substitution failed here by 5--8 times, which is a caution for any proxy rule \\
\srow{r:sgpw}Singapore and EU-27 off-site water & No value & None exists & Singapore is confirmed absent from the off-site source and from Li et al., who exclude it for the same reason & n/a. Both rows carry an on-site value (Singapore 0.206 mL per 1,000 effective tokens on a consumption basis, 0.308 as published, which alone is comparable with Ireland's whole hydropower-excluded total) \\
\srow{r:globw}Global average water & 0.550 L/kWh on-site, 3.142 L/kWh off-site & \citet{li2025} United States average & A consumption-basis value matching the basis of the off-site term, chosen by judgement. Google's fleet value of 1.15 L/kWh, also consumption-based, is the other candidate (S1.5) & Unknown; it is a United States average standing in for the world \\
\srow{r:aware}Scarcity weighting & AWARE 2.0 non-agricultural annual factors, 0.64 (Japan) to 13.56 (Texas) & AWARE 2.0~\citep{seitfudem2025} & Country factors outside the United States and state factors inside it. The method is defined at watershed and monthly resolution, and its developers supply country and annual aggregates only for cases where the watershed and the month are unknown, recording that these carry very large uncertainty; classing a data centre in the non-agricultural aggregation, which is the one built for domestic and industrial consumption, is the inference, since they nowhere name the sector & n/a (a second reporting convention), applied to the on-site term only because the basin of the off-site water is unknown; it therefore characterises 0.2\% to 31\% of the volume \\
\srow{r:wpue}Water PUE & 1.20 uniform & Author choice & Consistency with the carbon chain & $-$5\% to $+$7\% (site PUEs 1.14--1.28) \\
\multicolumn{5}{l}{\textit{Sections~\ref{sec:results} and \ref{sec:7}: comparison, application and discussion}} \\
\srow{r:firm}Firm size & 200 staff & Author assumption & Outside CSRD scope under Omnibus I~\citep{euparl2026b}; represents supplier firms that report voluntarily or answer a customer data request that no statute obliges them to meet for FY2025 and that, from FY2027, they have a statutory right to decline, Scope 3 being outside the value-chain cap's closed list of datapoints~\citep{euparl2026b,eucom2026b} & None on the factors; with seats and tokens scaled in proportion, a 1,000-person firm gives about five times 8.5 kg of token-based AI emissions \\
\srow{r:apitok}API token volume & 120 M billed tokens/yr, 75\% input & Author assumption & About USD 525/yr at gpt-4o list prices, one internal integration; the input share is the one the billing export reports & Linear (Tier 2a 5.4 kg per 120 M billed, which is 42.6 M effective) \\
\srow{r:apiclass}API model class & gpt-4o $=$ mid class & Vendor pricing tier & Size undisclosed; the energy reported by \citet{jegham2025} would place it at or above the frontier class & Low (as the frontier class: $42.6 \times 0.4425 + 120 \times 0.01284 = 20.4$ kg against 5.4 kg) \\
\srow{r:region}Serving region & Global average (0.4832) & Ember world 2023~\citep{ember2026}; Section~\ref{sec:3.5} & Region not disclosed by OpenAI & Unknown (US East: Tiers 2a $+$ 2b 5.7 kg vs 8.5 kg) \\
\srow{r:seats}Seat users & 50 of 200 staff & Author assumption & Partial roll-out to client-facing roles & Linear (25 or 200 seats: Tier 2b 1.6 or 12.6 kg vs 3.1 kg) \\
\srow{r:msgs}Messages per active seat & 30 per month (20--65); heavy use 325 & UK Department for Business and Trade telemetry on 1,000 licences~\citep{ukgov2025}; Microsoft's and OpenAI's own power-user thresholds~\citep{adminreports2026,openai2026a} & The only public absolute per-seat activity counts come from government pilot evaluations, and two readings, from one department and from the cross-government experiment it formed part of, land within 5\% of 30 per 30-day-active seat, although they are not independent: the 80\% active rate is the whole experiment's and the action counts the department's, so the conversion applies the experiment's rate to one participant's counts & Replaces 110 central and 2,000 high (worked example Tier 2b 3.1 kg, or 34.0 kg at 325). The 110 was a consumer figure obtained as arithmetic on two aggregates reported by \citet{chatterji2025}, who do not state it, and 2,000 was six times OpenAI's own power-user threshold. Measured on Microsoft 365 Copilot and counting \emph{actions}, while that product's own administrative report counts prompts submitted and active days, so two units are bridged and the default is likely low-side for a chat assistant \\
\srow{r:seatmodel}Seat-product model & GPT-4o 223 days, GPT-5 142 days of FY2025 & ``Introducing GPT-5''~\citep{openai2025b}; ChatGPT Enterprise and Edu release notes, 12 August 2025~\citep{openai2025d} & Dates of the default-model change in ChatGPT Enterprise & n/a (the measured 4.98\% share and the central multiplier of 25 are applied, which scales the Tier 2b output tokens by 2.195 and its billed count by 1.279; row~\ref{r:reasonshare}) \\
\srow{r:notion}Notion spend & EUR 8,000/yr & Author assumption; Notion Business list price~\citep{notion2026} & About 34 Business seats at EUR 19.50/member/month billed annually & Linear (EUR 4,000 or EUR 16,000: 384.5 or 1,538.1 kg deflated vs 769.0 kg) \\
\srow{r:google}Google calibration & 0.24 Wh per median Gemini Apps text prompt, compared per query & \citet{elsworth2025} & Only first-party per-prompt measurement of a full serving stack. The source discloses no token count, so the comparison is made per query & n/a (2.43 times the mid-class facility energy per chat exchange and 0.65 times the frontier-class value; its disclosed Scope 1 and embodied Scope 3 emissions, 0.010 g per prompt, restated on the effective-token basis, are 24 g per million effective tokens, 7.1 times the mid-class embodied term) \\
\srow{r:mistral}Mistral calibration & 1.14 g CO\textsubscript{2}e per 400-token response, compared per query (excludes training) & Mistral LCA~\citep{mistral2025} & Only externally peer-reviewed per-response lifecycle assessment. Its 400 tokens are output tokens only, so it is not a 400-token exchange & n/a (21.4 times the mid-class and 5.7 times the frontier-class grams per chat exchange, and 22.9 and 6.2 times against the inference footprint alone) \\
\srow{r:reason}Reasoning multiplier & 10--50 times, mid 25 & Author assumption; AI Energy Score v2~\citep{luccioni2025} & No provider publishes hidden-to-visible ratios; AI Energy Score reports about 30 times energy on average & Skewed high; the same source reports 150--700 times for one model with reasoning on. AI Energy Score reports an \emph{energy} ratio, not a hidden-to-visible \emph{token} ratio, so it corroborates the order of magnitude and not the quantity \\
\srow{r:reasonshare}Reasoning share & 4.98\% of conversations (unweighted interquartile range across the 601 clusters, 2.92--6.44\%) & Anthropic Economic Index open data~\citep{anthropic2025}, 601 clusters covering 96\% of records & The only public per-conversation measurement of how often a reasoning mode is used; the 152 clusters whose name contains a coding keyword (24\% of records) are 6.77\% & Replaces an illustrative 10\%. At the central multiplier the output token count rises $0.9502 + 0.0498 \times 25 = 2.195$ times, and the billed count 1.279 times at the short-chat mix; input tokens are not multiplied. It is a share of conversations, not a token share \\
\srow{r:locbasis}Location-based basis & Annual location-based averages & Scope 2 Guidance~\citep{wri2015};~\citep{wri2011} & Customer energy attribute certificates do not apply to supplier emissions, a reading of the Scope 2 Quality Criteria rather than a stated rule & High vs a provider market-based figure. A customer's own instruments may not be applied to a supplier's emissions, a provider's public renewable claim may not reduce a Category 1 figure without Tier 3 data, and the rule that a location-based figure accompanies any market-based one is this paper's convention rather than a standard's \\
\end{longtable}
\normalsize

\subsection*{S8. Version History}

\textbf{v3 against v2.} A source audit stated every assumption with its value, source, reason and direction of bias (S7) and led to the method changes below. Table~\ref{tab:changes} carries the full list of changed values.

\begin{itemize}[leftmargin=1.4em,topsep=3pt,itemsep=3pt,parsep=0pt]
\item \emph{Operating point and class value.} Energy values are direct ML.ENERGY v3 measurements at a latency-constrained operating point, which replaces v2's extrapolated values; the minimum at the smallest accelerator count with no latency limit is kept as a sensitivity. The class value is now the arithmetic mean of the measured members of each class, replacing one named proxy model per class. That change lowers the mid-class energy by 11\% and raises the small-class energy by 86\%, and it removes the one-sided band the previous small-class proxy produced, because that proxy was the lowest-energy member of its own class.
\item \emph{Serving boundary.} The boundary widens from accelerator energy times PUE to accelerator energy times a host uplift, an idle uplift and PUE, each varied separately. v2 carried non-accelerator server energy only in a conservative high case that divided the electricity term by 0.60, a factor of 1.667 taken from the TokenPowerBench bound; the new central chain of $1.50 \times 1.10 = 1.65$ reproduces it to within 1\%, so replacing a bound with the host and idle uplifts barely moves the number but makes it auditable. The high case is rebuilt on $H = 1.88$, the tied $I = 1.20$ / $U = 0.60$ pair and PUE 1.45, so that it moves the weakest-evidenced link in the chain as well as the two strongest. The B200 scenario becomes a hardware sensitivity instead of a low scenario.
\item \emph{Utilisation tied to the idle uplift.} Utilisation and the idle uplift describe one fleet and are no longer chosen independently. At an assumed idle power of 0.30 of serving power the measured $I = 1.10$ implies $U = 0.75$, which replaces the earlier independent assumption of 0.60 and lowers the embodied term by a fifth.
\item \emph{Grid.} Intensities move to a lifecycle basis, eGRID CO\textsubscript{2}e subregion rates are rescaled to Ember's United States total, N.\ Virginia moves to the SRVC subregion, and an EU-27 row is added.
\item \emph{Non-electricity terms.} Embodied carbon is allocated by service time with measured throughput and now includes the transport stage. Data-centre construction becomes a term of its own, applied to \emph{server} electricity at 7.66 g per kWh, which is the basis on which the source's own figures are expressed; an intermediate version applied the facility-basis restatement, 6.9 g, to server electricity, which understated the term by 10\%. The training denominator moves from an assumed lifetime token count to a published service life times the one disclosed per-model serving rate.
\item \emph{Model development.} All model development, including the final training runs, is reported as a separate line and never added to the factor, and two named footprints are defined, an inference footprint and a development-inclusive total.
\item \emph{Token basis.} Tokens are weighted before the factor is applied, and the electricity, embodied and construction terms are charged on effective tokens while training is charged on billed tokens. v2 applied the output rate to every billed token; against that rule the effective-token count is 2.9 times lower for a chat exchange and 17 times lower for an agentic request. The cached-input band was tightened from 0.01--0.20 to the two derived bounds, 0.02--0.10, after an independent review of an intermediate draft, and each benchmark measurement is now restated from the token mix and cache use of its own run, read from the benchmark's raw run files, to a reference mix; this replaces the bounding $D = 1$ sensitivity of an intermediate draft, which was needed while the benchmark's prefix-caching state could not be established.
\item \emph{Activity defaults.} The Tier 2b defaults are replaced by measured ones, a new agentic-coding category is added, and the symmetric band is replaced by an asymmetric one.
\item \emph{Structural range (8 October 2026).} The uncached-input weight moves from 0.20 to 0.30, the reading matched to the benchmark's batch sizes, with the band of 0.10 to 0.40 kept; the divisor becomes 1.339 and the default multiplier 0.302, which raises every per-billed figure by about 19\%. Every result now carries a parameter interval and a structural range on two typed axes, printed side by side (Table~\ref{tab:structrange}). An earlier statement that the independent per-query estimates lie between the two operating-point rules was wrong for the mid class, whose estimates sit at or above the mean-rule value, and is corrected. A training-high sensitivity, the development-inclusive line of the worked example and two reasoning-share sensitivities are added.
\item \emph{Resolution and packaging.} The commercial-model mapping ships as a 187-row CSV applied by a reference calculator, with the identifier normalisation and tie-break stated, instead of a printed table alone; the token-basis constants, a region crosswalk and a per-region water file ship as data; and the deposit carries an end-to-end test suite.
\item \emph{Worked example.} The example now follows the paper's own rules and the corrected defaults, with Tier 1 shown deflated as a subscription screening line. The manuscript submitted on 9 September 2026 printed 5.3 kg for the OpenAI line and 21.0 kg for the ChatGPT Enterprise line, 26.3 kg token-based, beside a 944.8 kg Notion line and a 971 kg sum, all on the v1 electricity-only factors. The 30 September recompute on the v2 dataset, which added the embodied and training terms, gave \textbf{86.8 kg} for the same activity, not 138.5 kg, which was that version's \emph{conservative high case}; an intermediate draft conflated the two, and its sum with the Notion line was about 975 kg. Three steps separate 86.8 from the present 8.5 kg, and they do not all pull the same way:
\begin{enumerate}[leftmargin=1.6em,topsep=2pt,itemsep=1pt,parsep=0pt]
\item 86.8 kg --- the 30 September version: three-term factor, 2,000 messages per seat per month at 400 tokens per exchange (the activity the submitted manuscript also used), every billed token charged at the output rate.
\item 135.3 kg --- the same activity and the same token rule, repriced at the present wider four-term boundary and the present class rule.
\item 21.6 kg --- the present activity defaults (30 messages per seat per month, 1,210 billed tokens per exchange, 1,547 after the Tier 2b reasoning uplift on output tokens) on the previous token rule.
\item 8.5 kg --- the present rule, charging effective tokens for the compute-proportional terms and billed tokens for training.
\end{enumerate}
Step 2 raises the figure by 56\% because the boundary widened; steps 3 and 4 lower it by 84\% and 60\% because the previous seat message rate, six times a vendor power-user threshold, falls to 30 a month, which outweighs the near fourfold rise in tokens per exchange, and because a billed token is not an output token.
\item \emph{Reasoning uplift and the Notion price.} Two activity corrections were made after the deep review of 3 October 2026. The worked example now applies the method's own Tier 2b reasoning default, the measured 4.98\% share at the central multiplier of 25, which scales the Tier 2b output tokens by 2.195 and its billed count by 1.279, and raises the token-based total from 7.4 to 8.5 kg; the previous no-uplift values are kept beside it. And the Notion seat count is read from the euro list price, 34 seats at EUR 19.50 rather than 38 at USD 20, which removes an unstated exchange rate from the example.
\item \emph{Comparison with published estimates.} The comparison is made per query at the token counts each source discloses, rather than per 1,000 tokens. The EEIO comparison is stated per million billed tokens of a named mix.
\item \emph{Water.} On-site values follow a stated precedence and no author estimate remains; the confirmed set grows from five regions to eleven; the two hydropower conventions are reported as co-equal with a high-hydropower case added; and the terminology follows ISO 14046. On-site water is reported on a \emph{consumption} basis throughout and scarcity weighting is applied to the \emph{on-site term only}, so the earlier scarcity-weighted totals are withdrawn and there are three rankings rather than four. Under the corrected definitions the Netherlands is no longer among the highest on every measure, and Ireland remains the only region among the lowest on all of them.
\item \emph{Corrected statements.} Four statements of the previous version no longer hold. Microsoft is not the only hyperscaler publishing site-level water intensity: AWS publishes a 2025 value for seventeen of its twenty-seven named regions~\citep{aws2026b}, and Microsoft and Meta publish site water alongside site electricity~\citep{microsoft2026a,meta2025}. The Li et al.\ values were described as being ``as of July 2023''; that date in the source qualifies the under-construction status of the seven projected sites, and the other rows are the most recent annualised values reported by Microsoft in 2023~\citep{li2025}. The allocated-hydropower convention was described as an upper bound; it is not. And a 400-token exchange was attributed to Mistral, whose figure is an output count only.
\item \emph{Attributions corrected against the primary sources.} The four papers previously known only through \citet{gamazaychikov2026} have now been read in the original, and four attributions change with them. The host uplift is cited to \citet{latif2025} directly, at the 1.51 to 1.52 their own utilisation definition gives rather than the 1.41 that assumes every accelerator at its full 700 W rating; the 1.10 and 1.36 component split stays with \citet{gamazaychikov2026}. The idle range of 1.05 to 1.20 was attributed to \citet{oviedo2026}, who model no such range but a point estimate of about 1.19; the central 1.10 is Google's measured value and the low 1.05 is ours. The high PUE of 1.45 was described as a United States average; it is the top of the Lawrence Berkeley National Laboratory 2024 projection band, and that assessment puts AI-specialised facilities at 1.14. The low construction value is restated on the Carbon Trust's own basis, at 5.3 g per kWh of server electricity taken directly rather than 5.8 g restated at a PUE, with its scenario and its authorship overlap with \citet{gamazaychikov2026} stated.

\item \emph{Standards statements corrected.} The ESRS requirement is date-stamped to E1-6 and E1-8 and the cloud sub-category is cited to Application Requirement 24(g) and AR 51; the GHG Protocol revision is restated as the consolidated ISO co-branded Corporate Standard; the data-quality scale is labelled as our own numbering of the Standard's four qualitative levels; the claim that no standard exists is narrowed against ITU-T L.1801; and the revised ESRS E3 water metrics are identified as own-operations metrics, so a purchased-inference water figure has no ESRS E3 datapoint.
\item \emph{Structure.} The paper is restructured as a methodology paper: the method (Section~\ref{sec:3}) is stated without values, the factor tables become parameterisation (Section S9), and the evidence that the method works is gathered into one section (Section~\ref{sec:5}). The CSRD scope is updated for the Omnibus I Directive (EU) 2026/470, the AI Act description extended with the Article 53(2) exemption and the Article 111(3) transition, and the host uplift re-sourced from TokenPowerBench's bound to metered-node measurements.
\end{itemize}

\textbf{v2 (September 2026 manuscript).} The v2 manuscript extended the v1 preprint (arXiv:2606.10660) and the v1 dataset. Its substantive changes, with the v2 values that v3 supersedes, were: two region-independent terms (embodied hardware manufacturing and training amortisation) added per model class and summed with electricity into the reporting formula; a B200 multiplier of 0.36 applied to the electricity term only; the Singapore-to-Sweden spread compressed from about 12 to about 6 times by the lifecycle terms, the v1 figure of 13 times having been a rounding artefact of a three-decimal display; a case-study total of about 975 kg CO\textsubscript{2}e; a revised disclosure text naming all three terms; a new limitation documenting the calibration against Google's and Mistral's per-prompt disclosures; data-centre construction sized at about 1.7 g per million tokens for the mid class but not adopted as a fourth constant; the regional table extended from nine cloud regions to fifteen; two v1 dataset rows corrected; and a new model class mapping, corrected reference and caption numbering and a new Data Availability section.

\subsection*{S9. Parameterisation of the Inputs}

This section gives the full derivation of every input summarised in Section~\ref{sec:4} of the main text: the value, its range, the evidence behind it and the alternatives considered.

\subsubsection*{S9.1 Accelerator Energy per Token}

The energy anchor is the ML.ENERGY Leaderboard v3~\citep{mlenergy2025}, which measures accelerator energy in joules per output token for open-weight models under one protocol (the \texttt{lm-arena-chat} task, vLLM, at each model's published precision, bfloat16 where available, data of 16 February 2026) across sweeps of accelerator count and serving batch size. It is the only open, maintained benchmark that measures the same models on H100 and B200 under one protocol, and every value used here is a leaderboard row. The leaderboard publishes one energy value per model and configuration, computed over the steady state of a run through its request dataset, where the batch is saturated~\citep{chung2025}, and reports no run-to-run spread, so the class range below is a spread between models, and the repeatability of a single configuration is inherited from the benchmark rather than tested here.

\textbf{Operating-point rule.} For each model and accelerator type, keep the configurations whose median inter-token latency is at most 50 ms, that is at least 20 tokens per second streamed to each user, and take the minimum joules per output token among them. The 50 ms ceiling is the author's choice and rests on no external source. The frontier class is the one exposed to it, because its H100 members sit at 46.6 to 48.0 ms, just inside the limit. Two reasons justify taking the minimum rather than an average. The sweep covers many accelerator counts and batch sizes, most of which no operator would run, so an average over the sweep measures which points happened to be tested rather than how a model is served. And the lowest energy among the points that still stream at interactive speed is the configuration a cost-minimising provider would choose. The rule has a direction. It is a best-case serving point, so it biases the electricity term low, the idle uplift covers part of that, and the residual is not bounded. ML.ENERGY also measures in a window where the batch is saturated at the configured maximum~\citep{chung2025}, and the operating points used here are at configured batches of 128 to 512, so the figure is a best-case efficiency at high concurrency applied as a fleet average.

One sensitivity bounds the rule from above and none bounds it from below. Taking the \emph{mean over all configurations that meet the latency limit} instead of the minimum raises the class energy by 2.10, 2.88 and 2.43 times for the small, mid and frontier classes on H100 and, with the embodied term recomputed at the throughput of the same configurations, the mid-class inference footprint at the global-average grid from 0.1187 to 0.3446 kg CO\textsubscript{2}e per million effective tokens, which is \emph{a rise of 190\%}: the largest single sensitivity in this paper, and the upper end of the serving-regime range of Section~\ref{sec:3.8} (Table~\ref{tab:structrange}). The other configuration sensitivity is not a lower bracket on it. Taking each member at the lowest-energy point of its smallest accelerator count, with no latency constraint, moves the mid-class footprint by $+$18\% and the frontier-class one by $-$17\%, because that point is an orthogonal configuration choice rather than a looser version of the same rule. The frontier-class members all sit within a few milliseconds of the latency ceiling, so that class is the one most exposed to the choice of ceiling.

The comparison of Section~\ref{sec:5.1} shows what the published per-query estimates say about the choice. No published figure describes a small-class model, so Class A is untested. For the mid class the three energy estimates that can describe it, Jegham et al.'s, Google's and the ChatGPT average, do \emph{not} lie between the two rules. At the token counts behind them they are 2.4 to 7.2 times the central value and 0.84 to 2.51 times the mean-rule value, above the mean-rule value in four of the five rows and just below it in the fifth, Google's figure at the chat exchange; they sit at or above the top of the serving-regime range, and all are closer to the mean rule. For the frontier class the three that can describe it, Oviedo et al.'s, Google's and the ChatGPT average, bracket the central value, at 0.65 to 1.57 times it, and are 0.27 to 0.65 times the mean-rule value: three rows lie between the two rules and two, both at the chat exchange, below the central value. Mistral's lifecycle figure is above both rules for every class. The central value is therefore more likely low than high for the mid class, which is why the mean-over-configurations rule is printed as the upper end of the serving-regime range (Table~\ref{tab:structrange}) rather than discarded.

\textbf{The class value is the mean of the measured models.} For each class and accelerator type it is the arithmetic mean of the members' joules per output token at the operating point, one entry per distinct model, taking the bfloat16 measurement where one exists at that point on that accelerator type and the eight-bit floating-point (FP8) measurement only where none does; membership follows the size rule of Section~\ref{sec:3.5}. The class embodied term is the mean of the members' own time-allocated embodied terms, and construction follows from the class energy. On H100 the classes have seven, five and three members, listed with their operating points in Table~\ref{tab:classmembers}; on B200 they have seven, four and three. The class means are 0.1309, 0.4285 and 1.5958 J per output token on H100, and the \emph{class range}, the lowest and highest member relative to the mean, is $-$40\%/$+$47\%, $-$38\%/$+$24\% and $-$67\%/$+$55\%. Two properties of the frontier class travel with any use of its value: it has three members, and two of them are FP8 measurements, so part of its spread is a precision spread. Three alternative central statistics are carried as sensitivities, the geometric mean, the median and the named proxy model of an earlier draft of this paper; the last would lower the small-class footprint by 46\%, raise the mid class by 12\% and widen the small-to-frontier ratio on the inference footprint from 12.3 to 21.4 (S1.1).

\textit{Values: class means 0.1309 / 0.4285 / 1.5958 J per output token on H100 (n $=$ 7, 5, 3) and 0.1133 / 0.2853 / 1.6473 on B200 (n $=$ 7, 4, 3); class range $-$40\%/$+$47\%, $-$38\%/$+$24\%, $-$67\%/$+$55\% (H100). Bias: low, because the operating point is a best-case serving configuration applied as a fleet average. Sources, evidence and bias for every symbol are in Table~\ref{tab:master}.}

\subsubsection*{S9.2 Server and Facility Overheads}

Three uplifts convert measured accelerator energy to the energy a facility buys. S1.1 gives every candidate value and the reason each central was chosen.

The \textbf{host uplift}, $H = 1.50$ (1.43 to 1.88), charges the rest of the server, meaning central processing unit (CPU), memory, fans, power supplies, network interface and local storage, as a multiple of accelerator energy. The central value is supported directly by \citet{latif2025}, who meter an eight-accelerator H100 node at the rack power distribution unit and define their graphics processing unit (GPU) utilisation as measured accelerator power over rated accelerator power, so the measured accelerator power is recoverable and the ratio can be formed at one utilisation instead of dividing a part-load node measurement by a full-load nameplate: 7.92 kW during Llama2-13b training at 93\% utilisation gives 1.52, and 8.43 kW in the GPU-plus-CPU burn stress test at 100\% gives 1.51; the GPU burn without CPU load, 7.97 kW at 100\%, gives 1.42. The lower reading of 1.41 taken from that paper in \citet{gamazaychikov2026} divides a 93\%-utilisation node measurement by a 5.6 kW tray nameplate, which is NVIDIA's rated power rather than a Latif measurement, and biases the ratio down. The low value is Google's measured Gemini serving split~\citep{elsworth2025} and the high value the mean of the machine-to-accelerator power ratios of Google's five tensor processing unit (TPU) generations~\citep{schneider2025}. The provenance of the 1.10 and 1.36 component split, and the TokenPowerBench ceiling of 1.667 that is kept as a consistency check rather than a reporting bound, are in S1.1. Serving accelerators draw below their rated power in the mid and frontier classes, 80\% and 71\% of 700 W on average, which raises the non-accelerator share there and points to 1.50 understating. Against that, the high value of 1.88 is Google's all-workload TPU machine ratio, a wider boundary than the GPU inference node behind the 1.667 ceiling, which is why the high exceeds the ceiling, and TokenPowerBench's digitised node-to-accelerator reading of about 1.2 to 1.3 sits below the low of 1.43 (S1.1), so the direction of the bias in $H$ is unknown.

The \textbf{idle uplift}, $I = 1.10$ (1.05 to 1.20), charges capacity held ready to meet latency targets and absorb spikes and failover. The central value is Google's measurement of provisioned idle machines at 0.02 Wh against 0.20 Wh of active accelerator and host energy per median prompt (1.11 to 1.13 from Google's printed percentage shares)~\citep{elsworth2025}, and Google's figure counts only machines allocated to the same model, so it excludes fleet-wide spare capacity. Oviedo et al.\ model no band for this: their Supplemental Note S5 gives a point decomposition of 1.08 for a diurnal load between 50\% and 100\% of peak and 1.10 for provisioned redundancy, about 1.19 together~\citep{oviedo2026}. The high value of 1.20 is therefore close to that point estimate, while the low value of 1.05 is our own sensitivity and rests on no source. The idle uplift and the utilisation $U$ of the embodied term describe one fleet and are therefore tied: writing $p$ for the power an idle accelerator draws as a fraction of its serving power, $I = 1 + p\,(1-U)/U$, because for every serving hour there are $(1-U)/U$ idle hours at $p$ of the serving draw. No provider discloses $p$; this paper takes $p = 0.30$, so the measured $I = 1.10$ implies $U = 0.75$ and the low-utilisation pair is $U = 0.60$ with $I = 1.20$. The identity is asserted in the recompute script for both pairs, and because $p$ is an assumption the sensitivity table moves $I$ alone and the pair separately.

\textbf{PUE} is 1.20 (1.09 to 1.45) on both accelerator platforms. The value is an author choice, not a sourced one. It sits above the three disclosed hyperscaler fleet values (Google about 1.09, \citealp{google2026a}; AWS 1.14, \citealp{amazon2024}; Microsoft 1.17, \citealp{microsoft2026a}), with Google's about 9\% below it, but close to the LBNL 2023 hyperscale aggregate of about 1.22 and Oviedo et al.'s centre of 1.21 (S1.1). It sits below the industry average of 1.56 reported by the 2024 Uptime Institute survey, a mean of operators' reported annual PUE for their largest data centre (526 respondents)~\citep{uptime2024}. PUE is a property of the facility rather than of the chip, and the margin covers serving that is not in a hyperscale facility. The low value is Google's disclosed fleet value and the high value of 1.45 the top of the Lawrence Berkeley National Laboratory 2024 projection band for that year, not a United States average; that series and Oviedo et al.'s log-normal model, whose geometric centre of 1.21 is about 1 per cent above the value used here, are in S1.1~\citep{oviedo2026,shehabi2024}. The low and high values scale the electricity term by $-$9\% and $+$21\%; construction is charged on server electricity and does not move with PUE.

\textit{Values: $H$ 1.50 (1.43 to 1.88), $I$ 1.10 with the tied $U$ 0.75 (1.05 with $U$ 0.857; 1.20 with $U$ 0.60), PUE 1.20 (1.09 to 1.45). Bias: $I$ low; $H$ unknown, its readings lying on both sides of the band (S1.1); PUE unknown, and low if any serving is outside hyperscale facilities.}

\subsubsection*{S9.3 Grid Carbon Intensity}

Electricity emissions are facility kWh times $g$, an annual location-based grid intensity for calendar year 2023 on a lifecycle basis, so that upstream fuel-cycle emissions are already inside the factor and need no separate term. Non-United States rows, the EU-27 and the world average use \citet{ember2026} national averages (lifecycle CO\textsubscript{2}e over a 100-year horizon, with lifecycle factors from the Intergovernmental Panel on Climate Change (IPCC) Fifth Assessment Report (AR5) Working Group III Annex III and, for coal and gas, UNECE and Jordaan et al.), the only open annual series covering all ten non-United States countries and both averages under one method. United States rows use the Emissions \& Generation Resource Integrated Database (eGRID2023)~\citep{epa2025} subregion CO\textsubscript{2}e total-output rates, which count stack emissions only, scaled by 1.1232 to the Ember basis; that factor is the ratio of Ember's United States 2023 intensity to eGRID's national rate of 349.67 g/kWh, 1.1232 at the Ember vintage the constant was set from, the current release giving 392.9 g/kWh and 1.1236. A uniform factor keeps the order of the United States rows. Because the uplift is multiplicative, it understates the lifecycle addition for low-stack-intensity subregions and overstates it for coal-heavy ones; the direction of that error follows the subregion's clean share and is not bounded here. Subregions approximate the power pool a data centre draws from, which state boundaries do not. All values are location-based, so the figures do not reflect a provider's renewable procurement, and all are per kWh generated and therefore exclude transmission and distribution losses (Section~\ref{sec:5.5}).

\textit{Values in Table~\ref{tab:carbon}; global average 0.4832 kg CO\textsubscript{2}e per kWh. The per-subregion derivation and the unscaled sensitivity are in S1.2. Bias: high on the grid year for every row and most for the European ones, because the 2024 Ember values are lower throughout, by 24\% for France and 10\% for the EU-27 against 2\% for the world average. The excluded transmission and distribution losses, about 4\% on the United States rows (Section~\ref{sec:3.1}), pull the other way, so the net bias is high for the EU rows and unknown for the others.}

\subsubsection*{S9.4 Embodied Hardware, Transport and Construction}

The central manufacturing value of 3,300 kg CO\textsubscript{2}e for an eight-accelerator H100-class server is an author-chosen round value inside the 2,872 to 4,241 kg range of Dell's four published PowerEdge product carbon footprints~\citep{dell2025}, weighted towards the two XE9680 variants, the air-cooled XE9680 (offered with H100 or H200 SXM accelerators) and the liquid-cooled XE9680L (offered with B200 or H200). The reports do not name the accelerator fitted. Those four values are sums of the component lines printed on page 2 of each report, because Dell prints no manufacturing subtotal, and the accelerator count is the inference, since GPU Quantity and GPU Memory are blank on all four reports; the per-stock-keeping-unit values and the NVIDIA baseboard cross-check are in S1.3. Dell reports transport separately across a 427 to 712 kg span, and the central 672 kg is the mean of the two XE9680 variants. Transport sits inside the embodied term, so $M_{\mathrm{mfg}} + M_{\mathrm{tr}} = 3{,}972$ kg central with a band of 3,299 to 4,953 kg, the top of which pairs one platform's manufacturing value with another's transport value rather than describing any one server; because the Dell footprints are cradle-to-gate, raw material acquisition and component fabrication are inside them and no end-of-life stage or recovery credit is (Table~\ref{tab:inclusion}).

The footprint is allocated by time over the service life $L$ at utilisation $U$ and divided by $T_8$, the measured output throughput of each member at the class operating point, restated to the reference mix like the energy (Section S9.6) and scaled to a full eight-accelerator server, that is its measured rate times eight divided by its accelerator count, which assumes independent replicas on one server; with $L = 5$ years and $U = 0.75$, $L \times 8{,}760 \times 3{,}600 \times U = 118{,}260{,}000$ s. The class value is the mean of the members' own terms, and the harmonic means of the scaled throughputs, 37,865, 10,044 and 2,615 output tokens per second, give the same values; for the mid class, $3{,}972{,}000 / (118{,}260{,}000 \times 10{,}044) = 3.34 \times 10^{-6}$ g per token, that is 3.34 g per million effective tokens, against 0.89 and 12.84 g. $L = 5$ years is the useful life Amazon has applied since January 2025 to a subset of its servers, shortened from six years because of the pace of AI hardware development~\citep{secfilings2026}; six years is Alphabet's general life and the upper end of the Amazon and Microsoft classes, and the life used by the assessment the construction constant comes from~\citep{schneider2025}, so five is a mildly conservative choice rather than the central reading of the evidence, and four years raises the term by 25\% while six lowers it by 17\%. The embodied band, $-$39\%/$+$95\%, and the comparison with Google's own disclosed embodied emissions per prompt, which sit above it, are in S1.3, so the upper tail is open.

Data-centre construction is a term of its own, independent of the grid because building emissions do not vary with carbon intensity, and it applies to \emph{server} electricity, which is the basis on which the source's own figures are expressed. Dividing each of the five per-TPU construction values of \citet{schneider2025} by the matching machine electricity gives 7.58 to 7.66 g CO\textsubscript{2}e per kWh of server electricity, and the constant adopted is the upper end, $k_c = 7.66$ g, which is the conservative choice inside a range only 1\% wide; carrying the basis in the constant's name matters, because the same number applied after PUE would overstate construction by 20\% at a PUE of 1.20. The low value is 5.3 g, read from the Carbon Trust low-density scenario~\citep{carbontrust2026}, whose denser scenarios give lower values, and no high value is defined. That report prints no per-kWh figure: its materiality table gives 117 kg CO\textsubscript{2}e per square metre per year for building and services embodied carbon over a 60-year life, and dividing by the information technology electricity of its low scenario, 5 kW per square metre at 50\% utilisation, gives 5.3 g per kWh of \emph{server} electricity, which is already the basis $k_c$ uses, so no PUE conversion applies. The three Carbon Trust scenarios and the working-group overlap with~\citep{gamazaychikov2026} are in S1.3. At the low value construction is about 0.9\% of the mid-class inference footprint, below the 1\% inclusion test of Section~\ref{sec:3.1}, which is applied at central values, so the inclusion of the term rests on the central constant. For the mid class, $0.00766 \times 0.1964 = 0.00150$ kg, that is 1.50 g per million effective tokens.

\textit{Values: embodied 0.89 / 3.34 / 12.84 g and construction 0.46 / 1.50 / 5.60 g CO\textsubscript{2}e per million effective tokens (Table~\ref{tab:regionindep}); embodied band $-$39\%/$+$95\%; construction low 5.3 g per server kWh with no high value. The filed lives and the construction arithmetic are in S1.3. Bias: embodied low, with the upper tail open; the Dell manufacturing values are sums of printed component lines and the accelerator content of those servers is inferred, not declared.}

\subsubsection*{S9.5 Training, Service Life and the Development Ratio}

Meta's Llama 3.1 model card gives location-based training emissions of 420, 2,040 and 8,930 tCO\textsubscript{2}e for the 8B, 70B and 405B models~\citep{meta2026}. It is the only verified per-model training disclosure for a family spanning all three classes, its location-based column matches the location basis of the grid factors, although the disclosure does not state whether its factors are combustion-only or life-cycle, and it covers the final training runs only.

The denominator is the tokens a model serves over its service life, $N = (S/12) \times 365 \times V$. For $S$ the central value is 12 months, the minimum published by two large hosting platforms: AWS Bedrock committed to at least 12 months for models launched before 7 September 2026~\citep{aws2026a}, and Microsoft Foundry sets a retirement date 18 months after launch for generally available models on its standard lifecycle, which closes them to new customers at 12 months, and a 12-month lifecycle for models from Anthropic, DeepSeek, Fireworks and Mistral AI~\citep{microsoft2026b}. Anthropic's published retirement dates, and a compiled series of 80 closed text models whose 12.8-month median is not independent of the 12-month floor it would support, are in S1.3~\citep{anthropic2026a}. For $V$ there is exactly one per-model serving volume ever disclosed: OpenAI stated on 6 October 2025 that GPT-5-Codex, released on 15 September 2025, was ``serving over 40 trillion tokens in the three weeks since launch''~\citep{openai2025a}, which is 1.905 trillion tokens per day over 21 days. With $S = 12$ months that gives $N = 6.95 \times 10^{14}$ tokens, and dividing the Meta disclosures by it gives 0.60, 2.93 and 12.84 g CO\textsubscript{2}e per million \emph{billed} tokens.

Five weaknesses in $V$ are stated rather than smoothed. It rests on one datapoint; that datapoint is a first-21-day rate for a model the provider described as one of its fastest growing ever, so it is an early-life rate that biases the denominator high and the training term low; it is a coding model, not a general assistant; \emph{the disclosure does not say what it counts}, since whether the 40 trillion are input plus output or output only, and whether cache reads are inside them, is not stated, while the training term here is charged on billed tokens including cached ones; and the numerator and the denominator come from different companies, so the frontier-class term charges Meta's 405B training over a serving volume that an open-weight 405B model demonstrably does not reach. A usage shape factor of 0.4 and class-specific volumes are carried as named sensitivities rather than as central constants.

A disclosed training figure is the compute of the run that produced the released model, and Epoch AI's reconstructions put final training runs at 9.6\% of research and development compute at OpenAI, 22.6\% at MiniMax and 12.3\% at Z.ai~\citep{denain2026}, so a method anchored on published training-run figures captures between a tenth and a quarter of the development activity it is meant to represent. Three estimates of the ratio $d$ of all development, including the final runs, to serving energy are available, they do not agree, and none is an energy measurement of a development fleet: $d = 0.67$ from \citet{patterson2022}, the only first-party energy split in the set; $d = 1.5$, the value the Sustainable AI Group adopts~\citep{gamazaychikov2026}; and $d = 2.5$, Epoch AI's reconstruction of OpenAI's 2024 spending~\citep{denain2026}. The three reconstructions, and why the agreement of two of them to 1.2\% is an artefact of round numbers rather than corroboration, are in S1.3. Two caveats travel with the number: a spend or capacity ratio is used here as an energy ratio, and the source allocation is a same-period flow between research and serving rather than an amortisation of one model's development over its own service life. Two properties travel with the line wherever it is reported. It carries no embodied or construction term for the development fleet, so it is about 4\% short of the boundary the inference footprint uses and the two are not boundary-comparable; and pro-rating one company-wide ratio across the classes assumes development is distributed in proportion to serving energy, whereas development compute is concentrated in frontier models and small models are commonly distilled from them, so the per-class split understates the frontier class and overstates the small one.

\textit{Values: training 0.60 / 2.93 / 12.84 g CO\textsubscript{2}e per million billed tokens; $S$ 12 months (6 to 18), $V$ $1.905 \times 10^{12}$ tokens per day, $N = 6.95 \times 10^{14}$ tokens; $d = 1.5$ (0.67 to 2.5), reported as a separate line only. Bias: the training term low, because it counts final runs only, over a denominator that is itself biased high; $d$ rests on spend reconstructions used as energy ratios.}

The three region-independent terms by class, with their ranges and the high-case embodied term, are in Table~\ref{tab:regionindep} in Section~\ref{sec:4.9}.

\subsubsection*{S9.6 Token Weights}

ML.ENERGY defines its published quantity as the total accelerator energy of a request divided by that request's output tokens, confirmed exactly on the export used here for all 328 configurations, so prefill energy is already inside it. How much is inside is measured run by run. The benchmark's raw run files record, for every configuration, the server-side counts of prompt tokens, generated tokens and prompt tokens served from the serving engine's prefix cache~\citep{mlenergy2025}. The \texttt{lm-arena-chat} task sends each conversation turn as its own request with the full history~\citep{mlenergy2026}, so the workload is not input-starved: the runs carry 0.47 to 2.01 input tokens per output token across the 328 runs, with a median of 1.06. Prefix caching was on in every run except those of the two NVIDIA Nemotron models, and where it was on the cache served a median 62\% of the prompt tokens (5th to 95th percentile 54\% to 84\%). Each measurement is therefore restated from the mix of its own run to one reference mix, $R_{\mathrm{bench}} = 1.13$ uncached input tokens per output token, by the factor $D/D_{\mathrm{run}}$, where $D_{\mathrm{run}} = 1 + R_{\mathrm{run}}\,(W_{\mathrm{in}}(1 - h_{\mathrm{run}}) + W_{\mathrm{cache}} h_{\mathrm{run}})$ and $h_{\mathrm{run}}$ is the cache-hit share of the run. At the 29 operating points behind the class values $D_{\mathrm{run}}$ runs from 1.08 to 1.45 against $D = 1.339$, and the restatement raises the class energy by 7.2\%, 9.4\% and 8.8\% for the small, mid and frontier classes on H100 (Table~\ref{tab:classmembers}). The reference mix is a convention. Any other choice gives the same result for every quantity computed through effective tokens, and 1.13 is kept from the earlier reconstruction of the task's mix. $D$ cancels in every quantity per billed token, and the prefix caching measured in each run enters through its own $D_{\mathrm{run}}$, so the caching of the benchmark is already measured per run; applying a single all-run divisor to the class energy as well would count it twice. In 21 of the 328 configurations the cache counter exceeds the prompt tokens, which the benchmark's authors attribute to preemption and recomputation at large batch sizes (J.-W.~Chung, personal communication, 5 October 2026). Five of the 29 operating points are among them, and for these the hit share is the median of the clean runs of the same model and accelerator type (S1.4). At the reference mix the numerator and the divisor cancel, so effective tokens equal output tokens there; that is what lets a figure measured per output token be carried as a figure per effective token.

The evidence behind $W_{\mathrm{in}} = 0.30$ is one four-row energy table read two ways, one qualitative statement and one commercial signal, and S1.4 gives every reading. The table is ML.ENERGY's own prefill--decode disaggregation experiment, which runs the two phases on separate accelerators and so measures the split rather than modelling it; differencing its published energies gives 0.17 to 0.27 and reading its balanced 512-in/512-out column directly gives 0.295 to 0.304. The spread is batch size rather than disagreement, because prefill is compute-bound while decode amortises its weight reads across the batch, so $W_{\mathrm{in}}$ rises with concurrency, and the low-batch fits of \citet{vellaisamy2026} give a lower bound for production serving. The factors here are measured at configured batches of 128 to 512, but Table 4 of \citet{chung2025} does not state the batch size or the hardware of its runs, so the adopted 0.30 is a within-cell reading chosen by the author, the 0.295 to 0.304 of the balanced column, and not a reading matched to this paper's batches. The table's cells have mean lengths of 512/512, 512/4096 and 4096/512. On the 8B model the prefill energy per input token is nearly constant across them, at about 0.0165 to 0.017 J, so the ratio moves with the decode energy per output token. The 8B 512/512 cells give about 0.30 and the 70B 512/512 cell about 0.54, above the band; the 70B row is not used, because it is read as a deployment artefact rather than an energy ratio (S1.4). The band of 0.10 to 0.40 is kept as the author's sensitivity, since no reading above reaches 0.40, so it now reaches further below the central value than above it, and the token weight is the input whose band moves the factor per million billed tokens furthest below the central value (Section~\ref{sec:5.5}).

On a cache hit the stored key and value states of the cached prefix are reused and its prefill compute is skipped, but each decode step still attends over that context and still reads those states from memory. \emph{No published energy measurement of a cache read exists}, so $W_{\mathrm{cache}} = 0.05$ sits between two bounds of different kinds and the band is exactly those two: a commercial lower bound of about 0.02 of an output token, from cache-read prices of 0.10 times uncached input on current models, and a physical upper bound of 0.10, the decode part of the marginal energy of an input token in the same disaggregation table, 0.084 to 0.085 of an output token on its two balanced Llama 3.1 8B deployments, rounded up (S1.4). The physical readings, 0.061 to 0.15, put the central value low, and on the energy basis the price signal gives 0.03, the cache-read price ratio of 0.10 times $W_{\mathrm{in}} = 0.30$, so 0.05 sits between the price-implied value and the readings. That is a named departure from the rule that sets a central value at or towards the conservative end of its evidence, and the weight is therefore biased low for cache-heavy workloads, which in practice means agentic coding. The band of 0.02 to 0.10 spans the two derived bounds. On a workload with no cache reads it still moves the class energy through the restatement of each benchmark run.

The divisor $D = 1 + W_{\mathrm{in}} R_{\mathrm{bench}} = 1.339$ is therefore the divisor of the reference mix and no longer an assumption about the benchmark. Because the restatement uses the same weights, a change in $W_{\mathrm{in}}$ or $W_{\mathrm{cache}}$ now moves the class energy as well as the token count. With an input share $s_{\mathrm{in}} = 0.85$ and no cache credit, $m = (0.15 + 0.30 \times 0.85)/1.339 = 0.302$ (0.211 to 0.337); the 0.85 is a round value chosen inside the spread of the workload defaults, which runs from 76.7\% for the composite short-chat default to 99.5\% for an agentic request. The Azure conversation trace itself is 93.9\% input by tokens, so 0.85 is not above every chat observation; for a chat-dominated account billed at the short-chat default the matching share is 76.7\% and the matching multiplier 0.346, which is also the effective-to-billed ratio of the short-chat workload in Table~\ref{tab:workload}.

\textit{Values: $W_{\mathrm{in}}$ 0.30 (0.10 to 0.40), a within-cell reading of the ML.ENERGY prefill--decode disaggregation experiment~\citep{chung2025}, read against the Greenpixie figures~\citep{horswill2026}, with the low-batch fit of \citet{vellaisamy2026} as a lower bound outside the band (S1.4). Provider list prices for uncached input are a commercial cross-check~\citep{openai2026b,anthropic2026b,google2026b}. $W_{\mathrm{cache}}$ 0.05 (0.02 to 0.10), with the lower bound from the same providers' cache-read prices~\citep{openai2026b,anthropic2026b,google2026b}. Its upper bound is the decode part of the marginal input-token energy in the same disaggregation table, 0.084 to 0.085 of an output token on the balanced 8B rows~\citep{chung2025}. $R_{\mathrm{bench}}$ 1.13 (reference mix; measured run values 0.47 to 2.01) and $D$ 1.339 (run values 1.08 to 1.45 at the operating points), from the ML.ENERGY raw run files~\citep{mlenergy2025} and task definition~\citep{mlenergy2026}. $s_{\mathrm{in}}$ 0.85 and $m$ 0.302 (0.211 to 0.337) complete the set, and all six are deposited as \texttt{token\_\allowbreak basis\_\allowbreak constants.csv}. Bias: $W_{\mathrm{in}}$ central, at a within-cell reading with a band reaching below it; $W_{\mathrm{cache}}$ low, and no measurement exists for the cached weight, which now also enters the restatement of each benchmark run.}

\subsubsection*{S9.7 Tier 2b Activity Defaults}

Where a seat product reports messages or seats but no tokens, the tokens are estimated from measured workload distributions wherever one exists. A short chat exchange is 928 input and 282 output tokens, a document-heavy exchange 3,293 and 383, an inline code completion 1,944 and 8, and an agentic coding request 84,600 input, of which 91\% are cache reads, and 400 output. The chat input and output medians come from different corpora, an Azure production trace and WildChat-1M, so the 1,210-token exchange is a composite rather than a measured pair. The Azure medians are taken over every request in the public one-week 2024 traces, 27.3 M conversation requests from 12 to 18 May 2024 and 16.8 M coding requests from 10 to 16 May 2024~\citep{azure2024}. Input is the billed input, so it includes re-sent context, and it is the median billed total minus the median output for the document-heavy and inline rows, because medians of input, output and total do not sum. An active enterprise seat sends 30 messages a month, and a heavy-use seat 325. The 30 restates United Kingdom (UK) government telemetry on licensed seats onto a thirty-day-active basis using the activity rate of a second, larger pilot, which is an assumption rather than one measurement on one population; the 325 is OpenAI's own power-user threshold for ChatGPT Enterprise, 75 or more messages a week, restated per month. Central values are per-message medians and the bands are interquartile ranges of single messages, so each band is a per-message spread carried as a scenario, not an interval on an annual total and not an extreme; the seat band is neither, being a derived diary low and a vendor threshold. The medians bias an annual total low. A sum of $N$ messages equals $N$ times the mean, and the means in the deposited evidence (\texttt{conversation\_\allowbreak token\_\allowbreak stats.csv} and \texttt{tier2b\_\allowbreak defaults.csv}) are higher than the medians. For the short chat they are 1,631.6 input and 330.2 output tokens against the medians of 928 and 282, a mean-to-median ratio of 1.62 on billed tokens and 1.46 on effective tokens. The document-heavy mean is 6,727 billed tokens against the median 3,676 (1.83), and the inline mean 2,534 against 1,952 (1.30). On means the case-study Tier 2b line would be about 4.3 kg instead of 3.1 kg and the token-based total about 9.7 kg instead of 8.5 kg, within the printed activity band. The median is kept because the short-chat exchange is a composite of two corpora, input from one and output from the other, so its mean is no more a measured pair than its median. The agentic request size of 85,000 tokens is the author's choice between the applications listed in Table~\ref{tab:tier2b}, close to Codex's 85,269 tokens per request; Claude Code's 96,884 carries more request volume and would raise the row by about 14\%. Agentic coding is a new category: one human instruction triggers many model requests, each re-sending the whole session, so the preparer counts model requests, not human turns, and the cache share rests on a single documented session and is illustrative rather than measured. The Tier 2b band of $-$50\%/$+$130\% on estimated tokens is asymmetric because the measured distributions are right-skewed, their interquartile ranges being $-$50\%/$+$115\%, $-$38\%/$+$132\% and $-$51\%/$+$73\% for the first three workloads. No public source gives a reasoning-\emph{token} multiplier, so the 10 to 50 times range used for Tier 2b message-based estimates, with a round mid value of 25, stays an author assumption; the measured share of conversations that use a reasoning mode is 4.98\%, so at the central multiplier a product's estimated output tokens rise by 2.2 times, and its billed tokens by 1.28 times at the short-chat mix. Coding clusters, here the 152 clusters whose name contains code, coding, program, debug, software, algorithm, api, script, function or bug as a case-insensitive substring (24\% of records, including a few non-coding clusters the substring rule picks up), use a reasoning mode in 6.77\% of conversations~\citep{anthropic2025}. That share has a vintage: it is measured on Claude.ai in February and March 2025, before reasoning models became the default route on major chat products, and it is applied here to products of other vendors. The one later public figure is a statement by OpenAI's chief executive on 10 August 2025 that, after GPT-5, the share of daily users on reasoning models had risen to 24\% on the Plus tier and 7\% on the free tier~\citep{altman2025}. It is a share of users rather than of messages, and it does not cover Enterprise, so it cannot replace the measured share; it is carried as two named sensitivities on the worked example, which at the central multiplier raise the token-based total by 51\% and 5\% (Section~\ref{sec:6.1}). Table~\ref{tab:tier2b} in S1.4 gives every default with its band and the evidence behind it, and Table~\ref{tab:workload} the effective-token count each workload produces.

\textit{Values as above, with every band and source in Table~\ref{tab:tier2b}. The chat, document-heavy and inline token counts come from Azure production inference traces~\citep{azure2024} and WildChat-1M~\citep{zhao2024}. The agentic request size and cache share come from OpenRouter application rankings~\citep{openrouter2026} and Claude Code cost documentation~\citep{anthropic2026c}. The active-seat default rests on UK government pilot evaluations~\citep{ukgov2025}. The heavy-use seat rests on the ChatGPT Enterprise and Microsoft 365 Copilot power-user definitions~\citep{adminreports2026,openai2026a}, and the reasoning share on the Anthropic Economic Index~\citep{anthropic2025}. Bias: the token defaults are per-message medians and bias an annual total low, the deposited means being 1.30 to 1.83 times them on billed tokens; the seat default counts actions on one product and is likely low for a chat assistant; the reasoning share predates default reasoning routing and is likely low for 2025 and later; the reasoning multiplier is an author assumption and the agentic cache share is illustrative rather than measured.}

\subsubsection*{S9.8 Water Inputs}

The on-site values follow a stated precedence and every one is a published figure or an operator-derived ratio rather than an author estimate, although two rows substitute a value from another place. Eight rows take a water use effectiveness that the operator publishes for the named cloud region, from the AWS 2025 regional series, in which seventeen of twenty-seven named regions carry a 2025 value, defined as water withdrawn per kWh of information technology load~\citep{aws2026b}; the eight run from 0.02 L/kWh in Stockholm and Dublin to 1.22 in Tokyo and 1.57 in Singapore. Four take a site ratio derived from a provider's own published site water and site electricity, three Microsoft sites~\citep{microsoft2026a} and one Meta site~\citep{meta2025}. Finland and the global row keep the 2023 Microsoft site series of \citet{li2025}, Finland being a site under construction in July 2023 and the global row a United States average standing in for the world, and France, Belgium and the EU-27 take the AWS Europe continental aggregate. Only the first two rungs count as confirmed, which gives twelve confirmed on-site values and eleven confirmed totals, Singapore having no off-site source. The caveats on the derived site ratios, including their partial circularity and the several-fold spread between operator values in one metropolitan area, are in S1.5. A region's on-site value is applied whichever provider serves the model from it, which is a proxy whenever the serving provider is not the operator that published the value; the spread between operators in S1.5 is the size of that error.

Providers do not share a basis, so the withdrawal rows are converted to consumption using the only documented conversion. The page accompanying the AWS series derives an industry-average withdrawal figure of 0.84 litres per kWh from a consumption figure of 0.56 at three cycles of concentration, which implies withdrawal equal to 1.5 times consumption~\citep{aws2026b}; AWS does not state that ratio, and it is a derived relation from a benchmark calculation rather than a property of its own fleet. The 0.62, 0.72 and 0.78 ratios disclosed by \citet{microsoft2026a}, \citet{meta2025} and \citet{google2026c} bracket it, and the as-published figures become a sensitivity (S3). The energy water intensity factors are the consumption factors of \citet{reig2020}, verified against the source document rather than its reproduction elsewhere, and Singapore is absent from that source and therefore carries no off-site value. The hydropower-excluded variant multiplies each factor by one minus the hydropower share of a reconstruction of it from the source's own per-technology factors and generation mixes; that reconstruction reproduces the published factor within a factor of 0.60 to 1.20 for fourteen of the fifteen rows and only 0.27 for the Netherlands, whose reconstructed hydropower share is zero and whose hydropower-excluded factor is therefore unchanged (S1.5). A directly published hydropower-excluded intensity exists for United States balancing authorities~\citep{siddik2024} and is the natural replacement for the United States rows in a later dataset version. Every generation mix behind the off-site term dates from 2018 while the grid carbon intensities are 2023. The on-site consumption is multiplied by the Available Water Remaining (AWARE) 2.0 characterisation factor of the region, non-agricultural sector, annual value~\citep{seitfudem2025}, country-level outside the United States and state-level inside it, and none is applied to the global-average or EU-27 rows, which are not locations; the resolution caveat the method's developers attach to those aggregates, and the sector choice, are in S1.5.

\textit{Values in Table~\ref{tab:water}, 17 regions, of which 12 have a confirmed on-site value and 11 a confirmed total (Singapore has no off-site source). On-site water comes from the AWS regional series~\citep{aws2026b} and from site ratios derived from \citet{microsoft2026a} and \citet{meta2025}. Where nothing more specific is published, the 2023 Microsoft site series of \citet{li2025} is used. Off-site water uses the consumption factors of \citet{reig2020}, and the scarcity weighting uses AWARE 2.0~\citep{seitfudem2025}; the derivation and the cross-checks are in S1.5. Bias: the off-site factors used are about 30\% below the two published United States averages, so the United States rows are low.}

\subsection*{S10. Further Consistency Checks}

Three further checks complement those of Section~\ref{sec:5} of the main text.

\subsubsection*{S10.1 The Activity Defaults, Checked without Customer Data}

The Tier 2b defaults can be tested against published material alone.

\textbf{Cost inverted to tokens.} Anthropic documents that Claude Code charges by API token consumption and by default computes its displayed cost locally from token counts at list price, that across enterprise deployments the average is about USD 13 per developer per active day and USD 150 to 250 per developer per month, and gives a worked \texttt{/usage} example: a Claude Sonnet 4.6 session of 1.2k input, 5.3k output, 940.0k cache-read and 50.0k cache-write tokens at USD 0.55~\citep{anthropic2026c}. Pricing that session at the mid-tier list rates reproduces the documented cost to 0.5\%, so the sample's token mix and the published list prices are mutually consistent. The page does not say which model the aggregate cost figures were incurred on, so inverting them assumes the same mid-tier price and the same cache-heavy mix: at that tier the monthly figure gives 270 to 451 million billed tokens per developer-month, and at the other two current tiers the same spend implies 162 to 676 million. At the 85,000 tokens per model request of Table~\ref{tab:tier2b} the mid-tier range is about 3,200 to 5,300 model requests a month, and the separate USD 13 per active day anchor gives 23.4 million tokens a day, about 276 requests per active day, which is within the range an agentic loop issues. At the inline-completion default of 1,952 tokens the same monthly token count would be 139,000 to 231,000 exchanges a month. That count is computed under the agentic cache-heavy mix, so the test shows that the agentic row is consistent with the vendor's cost anchor under its own mix and does not by itself reject inline completion. Two caveats. Cache reads are 94.8\% of the input tokens of the documented breakdown, and a separate prompt-cache example on the same page reports 91\% over fourteen requests, which the page does not state describes the same session, so the 0.91 used here rests on an illustration. And an independent platform-wide figure for a different route to the same models, OpenRouter at more than 8 million developers over a run rate of about 1.5 quadrillion tokens a year~\citep{menlo2026}, implies at most about 15.6 million tokens per developer-month, roughly seventeen times below the vendor's own cost anchor. The two are not in conflict: one counts a platform-wide developer base that is likely to include many dormant accounts and the other actively deployed enterprise developers, and together they bracket the range a preparer should expect.

\textbf{Billed input against ``tokens per prompt''.} The Tier 2b input column is the billed input, including re-sent context. On the WildChat-1M sample the token-weighted ratio of billed input to user-message tokens is 6.4 (1.0 for single-turn conversations, 10.8 for multi-turn), which is why a default built from published per-prompt figures understates Tier 2b activity: those figures are 94.9 tokens per prompt for Chatbot Arena~\citep{chiang2024} and 295.6 for WildChat~\citep{zhao2024}, against a billed input of 928 from the Azure trace. The three are different corpora, so the amplification factor explains the direction rather than the exact gap. Output length is the quantity the literature does get right, with published assumptions of 300 output tokens for a short query~\citep{jegham2025} and 500 tokens per response~\citep{you2025} against a measured median of 282.

\textbf{The seat rate.} Only one of the readings behind 30 messages per 30-day-active seat per month is a telemetry count. The 1,000-licence study measures 24 actions per licensed seat per month, which becomes 30 per 30-day-active seat at the 80\% thirty-day-active rate reported by the 20,000-seat cross-government experiment in which the department took part~\citep{ukgov2025}. Mapping that experiment's own survey frequency bands gives 31, so two readings at two scales land within 5\% of each other by two different methods. The first study's 300-diarist sub-study gives 20.6, which is the low end of the band of Table~\ref{tab:tier2b} rather than a third convergent reading. The readings are not independent: the rate used to convert the first comes from the second, the rate is measured on the whole experiment and the counts on one department within it, and the band mapping rests on self-reported frequency rather than telemetry. The number also counts \emph{actions} on Microsoft 365 Copilot, whose own administrative report counts prompts submitted and active days~\citep{adminreports2026}, so two units are being bridged as well, and it is a low-side default for a chat assistant; where a product reports its own count the count is used instead.

\subsubsection*{S10.2 A Common Workload under Another Method's Defaults}

Methods can be compared only on the same workload. Table~\ref{tab:bistline} runs two of the three Tier 2b text workloads through the published hyperscaler defaults of \citet{bistline2026} and through this method, at their unknown-provider default grid of 341 g (U.S. national average) and at two of this paper's; the document-heavy workload behaves like the chat one and is in the deposit.

\begin{table}[htbp]
\centering
\setlength{\tabcolsep}{3pt}
\footnotesize
\begin{tabular}{>{\raggedright\arraybackslash}p{5.0cm} >{\raggedleft\arraybackslash}p{1.7cm} >{\raggedleft\arraybackslash}p{1.9cm} >{\raggedleft\arraybackslash}p{1.9cm} >{\raggedleft\arraybackslash}p{1.6cm} >{\raggedleft\arraybackslash}p{1.5cm}}
\toprule
Method & Wh per exchange & g at their grid 0.341 & g at the global average & g at US East & Wh ratio to Bistline \\
\midrule
\multicolumn{6}{l}{\textit{Short chat: 928 input, 282 output tokens}} \\
Bistline et al.\ defaults (training added to match the four-term boundary) & 0.6244 & 0.3545 & 0.4433 & 0.3313 & 1.00 \\
This method, Class A (central) & 0.0301 & 0.0116 & 0.0158 & 0.0104 & 0.05 \\
This method, Class B (central) & 0.0986 & 0.0392 & 0.0532 & 0.0355 & 0.16 \\
This method, Class C (central) & 0.3673 & 0.1485 & 0.2007 & 0.1349 & 0.59 \\
This method, Class C (high case) & 0.6069 & 0.2324 & 0.3187 & 0.2098 & 0.97 \\
\midrule
\multicolumn{6}{l}{\textit{Agentic coding, per model request: 84,600 input, 400 output tokens}} \\
Bistline et al.\ defaults (training added to match the four-term boundary) & 30.2016 & 20.2437 & 24.5381 & 19.1203 & 1.00 \\
This method, Class A (central) & 0.3512 & 0.1777 & 0.2276 & 0.1646 & 0.01 \\
This method, Class B (central) & 1.1500 & 0.6652 & 0.8288 & 0.6225 & 0.04 \\
This method, Class C (central) & 4.2825 & 2.6421 & 3.2510 & 2.4828 & 0.14 \\
This method, Class C (high case) & 7.0752 & 3.6201 & 4.6262 & 3.3569 & 0.23 \\
\bottomrule
\end{tabular}
\caption{The same two workloads under Bistline et al.'s published hyperscaler defaults and under this method, in Wh and g CO\textsubscript{2}e per exchange (per model request for the agentic rows).}
\label{tab:bistline}

\begin{minipage}{\textwidth}\footnotesize \textit{Bistline et al.\ defaults: energy $=$ (input $\times$ 0.32 $+$ output $\times$ 0.96 Wh per 1,000 tokens) $\times$ PUE 1.10; operational emissions at their unknown-provider default grid of 0.341 kg/kWh (U.S. national average) and, for comparison, at this dataset's global-average (0.4832) and US East (0.3038) grids; plus 0.017 kg embodied and 0.1 kg training per million tokens on their own unit, all tokens, which gives their chat row 0.0206 g of embodied and 0.1210 g of training and their agentic row 1.4450 g and 8.5000 g. They report that training default as a separate line, as this paper reports its development line, and it is added into their total here so that both columns carry the same four-term boundary; the like-for-like comparison of the two separate lines is in Table~\ref{tab:decomp}. This method: facility Wh $=$ effective tokens $\times$ facility Wh per 1,000 effective tokens at the class mean on H100, and emissions $=$ effective tokens $\times$ inference footprint $+$ billed tokens $\times$ the training term. Cached input is charged as ordinary input in the Bistline column, because their defaults have no cache rate.}\end{minipage}
\end{table}

\textbf{Why the results differ.} Three differences account for almost all of it, and Table~\ref{tab:decomp} sizes them at the default token mix. First, their input intensity is not a per-token measurement: it is back-solved from Google's 0.24 Wh per prompt at an assumed prompt length of about 500 tokens and a 3-to-1 decode-to-prefill electricity ratio they take from Splitwise~\citep{patel2024}, so their per-token energy inherits a per-prompt figure that already contains a whole serving stack, and the PUE of 1.10 they then apply charges facility overhead a second time. Part of the electricity gap is also this method's own best-case operating point, because the mid class at the chat exchange is 2.43 times below the same Google figure (Table~\ref{tab:comparison}), which also supplies this method's idle uplift. Second, their training default of 0.1 kg per million tokens is 34 times this method's mid-class training term, because it is not derived from a disclosed training footprint over a disclosed serving volume; substituting it raises this method's mid-class factor per million billed tokens from 0.0388 to 0.1359, a rise of 250\%. That is the largest ratio between the two methods but not the largest contribution: at the default mix their training default is 32\% of the gap in kg per million billed tokens, their energy intensity 62\% and their embodied term 5\%. Their default is also larger than this method's whole development line: it is 1.3 to 4.8 times that line per million billed tokens, 0.0210 to 0.0783 kg at $d = 0.67$ to 2.5 (Table~\ref{tab:decomp}), so the development-inclusive figure of Table~\ref{tab:structrange} is not a concession towards their default but remains well below it. Third, their defaults have no cache rate, so the 76,986 cache-read tokens of an agentic request are charged at the same intensity as fresh input, which is why, on energy for the mid class, the gap widens from 6.3 times on a chat exchange to 26.3 times on an agentic one (Table~\ref{tab:bistline}). The two methods are closest where the workload is output-heavy and the model is frontier-class, which is the regime Google's per-prompt figure describes.

\begin{table}[htbp]
\centering
\setlength{\tabcolsep}{4pt}
\footnotesize
\begin{tabular}{>{\raggedright\arraybackslash}p{4.2cm} >{\raggedleft\arraybackslash}p{2.3cm} >{\raggedleft\arraybackslash}p{2.3cm} >{\raggedleft\arraybackslash}p{1.8cm} >{\raggedleft\arraybackslash}p{1.8cm} >{\raggedleft\arraybackslash}p{1.3cm}}
\toprule
Term & Bistline et al. & This method & Difference & Share of the gap & Ratio \\
\midrule
Electricity & 0.22111 & 0.03445 & 0.18666 & 62.4\% & 6.4 \\
Training & 0.10000 & 0.00293 & 0.09707 & 32.4\% & 34.1 \\
Embodied plus construction & 0.01700 & 0.00147 & 0.01553 & 5.2\% & 11.6 \\
\textbf{Total} & \textbf{0.33811} & \textbf{0.03885} & \textbf{0.29926} & \textbf{100\%} & \textbf{8.7} \\
\midrule
\multicolumn{6}{l}{\textit{Separately reported lines, kg CO\textsubscript{2}e per million tokens}} \\
Bistline et al.\ training default & 0.10000 & --- & --- & --- & 1.0 \\
This method, training term & --- & 0.00293 & --- & --- & 34.1 \\
This method, development line, $d = 0.67$ & --- & 0.02098 & --- & --- & 4.77 \\
This method, development line, $d = 1.5$ & --- & 0.04697 & --- & --- & 2.13 \\
This method, development line, $d = 2.5$ & --- & 0.07829 & --- & --- & 1.28 \\
\bottomrule
\end{tabular}
\caption{Decomposition of the gap to the defaults of \citet{bistline2026} at the default token mix, mid class, global-average grid, in kg CO\textsubscript{2}e per million \emph{billed} tokens, and the like-for-like comparison of the lines each method reports separately. Ratio is their value over this method's. Their total without their training line is 0.23811.}
\label{tab:decomp}
\end{table}

\subsubsection*{S10.3 Reproducibility}

Every factor, table and headline figure in this paper is generated, and every generated number is checked. No accelerator workload was run for this paper: every energy measurement is taken from the published sources named in S9, and the generators, the calculator and the verifier below run on a laptop processor, so the paper's own computation adds nothing material to the footprint it reports. \texttt{recompute\_\allowbreak v3\_\allowbreak lifecycle.py} is the single definition of every factor term and writes the factor file, the printed tables, the headline figures, the water results and a JavaScript Object Notation (JSON) file carrying every number printed here with its superseded value alongside; on every run it asserts its own sanity values to three significant figures, the idle-and-utilisation identity for both the central and the high-case pair, a 408-cell back-compatibility check against the superseded definitions, the default effective-to-billed multiplier and the effective-token count of the chat exchange.

\texttt{build\_\allowbreak detailed.py} writes the detailed and lookup tables by importing that script, so no constant or formula is duplicated, and runs 558 cross-checks against the factor file. \texttt{ai\_\allowbreak inference\_\allowbreak emissions.py}, the reference calculator, reads only the published comma-separated values (CSV) files and carries no factor value of its own; \texttt{-{}-selftest} runs 447 internal checks. \texttt{tests\_\allowbreak end\_\allowbreak to\_\allowbreak end.py} runs 589 end-to-end checks: eight third-party scenarios that use only files a third party would receive, AWS Bedrock, Azure and Vertex AI style identifiers resolved through the mapping, the three class values for five regions against the factor file, the worked-example and fifty-developer figures against the JSON, and the mapping tested against the deposited service-life list and the OpenRouter top 150 by tokens. \texttt{verify\_\allowbreak printed\_\allowbreak numbers.py} runs 2,019 checks: every number printed in this paper, table cell by table cell, against the generated dataset, and the figures restated in the abstract, introduction, discussion, conclusion and suggested disclosure text equal the table values they come from. All five are deposited, and all five run on Python 3.8 or later with the standard library only.

\subsubsection*{S10.4 The Main-Text Checks in Full}

The paragraphs of Section~\ref{sec:5} as they stood before condensation, with the commentary that repeats Tables~\ref{tab:comparison}, \ref{tab:chains} and \ref{tab:sens}.

Three checks are reported here: a comparison with published per-query estimates, the consistency of the serving chain with a first-party energy decomposition, and a sensitivity analysis. Three further checks are in S10: the Tier 2b defaults tested against published material alone, which find the agentic-coding row consistent with a documented cost under its own cache-heavy mix (S10.1); another method's published defaults on a common workload (S10.2); and the generation and checking of every printed number by deposited scripts (S10.3).

What this paper does not do should be stated before the checks. It does not compare predicted with metered energy for any commercial or self-hosted service; it does not test the method on a real organisation's billing data, since both worked examples use invented or derived activity; and its propagated interval is not calibrated against any observed error. The comparisons below are therefore consistency and plausibility checks, and each reaches only part of the chain.

The comparison is made per query, because every external figure is published per prompt, per query or per response, and each is restated at the token counts it discloses (Table~\ref{tab:comparison}); only Jegham et al.\ and Oviedo et al.\ state both an input and an output count. One chat exchange at the Tier 2b short-chat default is 928 input and 282 output tokens, that is 418.5 effective tokens; at it the three classes draw 0.0301, 0.0986 and 0.3673 Wh of facility electricity, 0.0498, 0.1630 and 0.6069 Wh in the conservative high case, and emit 0.0158, 0.0532 and 0.2007 g CO\textsubscript{2}e at the global-average grid.

\textbf{Where the method sits, plainly.} No published per-query figure describes a small-class model; for the mid class the per-query values lie \emph{below} every figure that can describe it, by 2.4 to 7.2 times on energy against the three figures that can describe it and 18 to 31 times on carbon. For the frontier class the three energy figures that can describe it, Oviedo et al.'s, Google's and the ChatGPT average, the last two at two token counts, \emph{bracket} the central value, from 0.65 to 1.57 times, and Mistral's figure is 4.7 to 8.3 times above it; under the conservative high case the energy figures are 0.40 to 0.95 times the frontier-class value, so none remains above it. The comparison is class-matched. Jegham et al.'s GPT-4o figure is read against the mid class only, because the mapping assigns gpt-4o to it (S6), and Oviedo et al.'s against the frontier class only, because it describes models above 200B parameters.

The operating-point rule of Section S9.1 is what that spread is mostly about, but the estimates do not lie between the two rules. For the mid class they sit at or above the top of the serving-regime range, at 0.84 to 2.51 times the mean-rule value and above it in four of five rows, which is why the central value is more likely low than high there; for the frontier class they bracket the central value and lie at 0.27 to 0.65 times the mean-rule value (S9.1). No published figure describes a small-class model, so Class A is untested here and its 8-to-24-times gap is not evidence about the method. The frontier-class bracketing is conditional on the token counts rather than on the factor, whereas the mid-class conclusion survives any exchange length up to about 2.4 times the one assumed here and the small-class one up to about eight times. No check in this paper tests the accelerator-energy term against an independent measurement of the same models at a production operating point, which remains the method's largest untested link. Figure~\ref{fig:cmpenergy} in S2 plots the comparison on a per-1,000-token basis.

Google reports the median Gemini Apps text prompt in May 2025 as 0.14 Wh of active accelerators, 0.06 Wh of host CPU and memory, 0.02 Wh of idle machines and 0.02 Wh of facility overhead, totalling 0.24 Wh, and states that a scaling of 1.72 on accelerator energy would be needed to reach a production serving boundary~\citep{elsworth2025}. That 1.72 is 0.24/0.14, so rebuilding it from the same four components as $1.43 \times 1.10 \times 1.09 = 1.71$ is an identity rather than a test. What it establishes is that this method's chain has the same four links as the only serving stack disclosed component by component.

The test is in the values, and two of the three links are inputs rather than results. The idle uplift of 1.10 is taken from this disclosure (Section S9.2) and is therefore not tested by it, and Google's implied host ratio of 1.43 is this method's low bound for the host uplift. The measurements of \citet{latif2025}, from which the central host uplift is taken, imply 1.51 to 1.52 at the rack power distribution unit, so this compares the two sources of the host band rather than testing the chain. Google's own table is printed to two decimal places, so with every entry free to sit anywhere inside its rounding it pins the host uplift only to 1.38 to 1.48 and the idle uplift to 1.07 to 1.13, and Latif's ratio lies 2 to 3\% above the top of that host band rather than inside it. The chain used here is $1.50 \times 1.10 \times 1.20 = 1.98$, 15\% above Google's, of which the PUE accounts for 10\% and the host uplift for 5\%. Table~\ref{tab:chains} sets the three published serving chains side by side.

Five things follow. The choice of operating point moves the result by more than any input, by nearly a factor of three, and Section~\ref{sec:5.1} shows that the published per-query estimates for the mid class sit at or above the top of the range it spans. The training denominator is the second lever, and it accounts for a third of the gap to \citet{bistline2026}; their input energy intensity accounts for most of the rest. The compounded high case is worth $+$64\% on the footprint, within which PUE and the host uplift carry roughly equal weight and the idle uplift, moved with its tied utilisation, 9.5\%. The hardware block moves almost nothing: each of its four inputs changes the result by at most 0.7\% on its own and all four together by under 2\%. And on the per-billed denominator the token basis matters more than any input of the embodied or construction terms, with the band of $W_{\mathrm{in}}$ around its within-cell 0.30 worth $-$33\% to $+$14\%. Restating each benchmark run to the reference mix, rather than taking the measured values as they are, is worth $+$9\% on the mid-class footprint. Figure~\ref{fig:evidence} in S2 grades every input of Table~\ref{tab:master} by the kind of evidence it rests on; for a closed commercial model only the accelerator energy and the training emissions change class, and both changes follow from the resolution rule.

\subsection*{S11. Second Worked Example and Suggested Disclosure Text}

The second worked example and the disclosure text apply the method of the main text at a different scale and show what a preparer would write.

\subsubsection*{S11.1 Fifty Developers Using Agentic Coding Tools}

The same rules, applied to a workload at the other end of the volume range. Fifty developers each use an agentic coding assistant. A preparer with this product reads the tokens from the vendor's analytics export and is at Tier 2a (Section~\ref{sec:3.2}); the inversion from documented cost used here stands in for that export, because no such export is public, and it therefore carries the Tier 2b activity data-quality score of 4 rather than 2. Tokens per developer-month are inverted from the vendor's own documented cost per developer, USD 150 to 250 per month at the vendor's documented standard-tier list price, giving 270.5 to 450.8 million billed tokens a month with a midpoint of 360.7 million (Section S10.1); across the vendor's three current price tiers the same spend implies 162.3 to 676.2 million, which is the price-tier assumption this example rests on. This is the vendor's documented cost anchor, not a platform average. The platform-wide figure of Section S10.1 is about seventeen times lower, so these results describe an actively deployed enterprise fleet rather than a median one.

The workload mix is the agentic row of Table~\ref{tab:tier2b}: 85,000 billed tokens per model request, 400 output and 91\% of input from cache, so the effective-token share is 0.0574 and the fleet's 216.4 billion billed tokens a year become 12.42 billion effective tokens. Pricing the tokens at the standard tier means the documented default model maps to the mid class under Section~\ref{sec:3.5}; the frontier row is shown as well, because a preparer who cannot evidence which model served a request takes the highest-intensity class under the rule of Section~\ref{sec:3.5}, and because the assistant can route individual requests to a frontier model. The frontier row applies the frontier-class intensity to a token volume inverted at the mid-tier price. At the top-tier price the same spend buys about 0.6 of that volume, 162.3 against 270.5 million billed tokens per developer-month at USD 150, so the class-consistent frontier figure is about $0.600 \times 8{,}277 = 4{,}966$ kg, about 5.0 tonnes, against the 8.3 tonnes of the table. Table~\ref{tab:illus2} gives the result.

\begin{table}[htbp]
\centering
\setlength{\tabcolsep}{4pt}
\footnotesize
\begin{tabular}{>{\raggedright\arraybackslash}p{0.85cm} >{\raggedright\arraybackslash}p{2.0cm} >{\raggedleft\arraybackslash}p{2.0cm} >{\raggedleft\arraybackslash}p{2.1cm} >{\raggedleft\arraybackslash}p{1.4cm} >{\raggedleft\arraybackslash}p{1.5cm} >{\raggedleft\arraybackslash}p{1.4cm} >{\raggedleft\arraybackslash}p{1.5cm}}
\toprule
 & & \multicolumn{2}{c}{Central} & \multicolumn{2}{c}{B200 sensitivity} & \multicolumn{2}{c}{Conservative high} \\
\cmidrule(lr){3-4}\cmidrule(lr){5-6}\cmidrule(lr){7-8}
Class & Region & without dev. & with dev. & without dev. & with dev. & without dev. & with dev. \\
\midrule
B & Global average & \textbf{2,110} & 3,404 & 1,631 & 2,280 & 3,050 & 5,336 \\
B & US East & 1,585 & 2,163 & 1,281 & 1,454 & 2,182 & 3,384 \\
C & Global average & \textbf{8,277} & 12,681 & 8,449 & 13,085 & 11,777 & 19,877 \\
C & US East & 6,321 & 8,058 & 6,430 & 8,313 & 8,546 & 12,607 \\
\midrule
\multicolumn{2}{l}{B, global, token volume low--high} & 1,582--2,637 & 2,553--4,255 & --- & --- & --- & --- \\
\multicolumn{2}{l}{C, global, token volume low--high} & 6,207--10,346 & 9,510--15,851 & --- & --- & --- & --- \\
\multicolumn{2}{l}{B, global, token weights low--high} & 1,369--2,955 & 1,693--5,354 & --- & --- & --- & --- \\
\multicolumn{2}{l}{C, global, token weights low--high} & 5,514--11,426 & 6,307--19,946 & --- & --- & --- & --- \\
\bottomrule
\end{tabular}
\caption{Fifty developers using agentic coding tools, in kg CO\textsubscript{2}e per year. ``Without dev.'' is the inference footprint on effective tokens plus training on billed tokens; ``with dev.'' replaces the training term with the development line at $d = 1.5$ on effective tokens (Section~\ref{sec:3.4}).}
\label{tab:illus2}

\begin{minipage}{\textwidth}\footnotesize \textit{Central token volume is 360.7 million billed tokens per developer-month, 216.4 billion for the fleet per year and 12.42 billion effective tokens; the low and high volume rows are 270.5 and 450.8 million per developer-month, which are the ends of the standard-tier range. Across the three price tiers the range widens to 162.3 to 676.2 million, which gives 949 to 3,956 kg for the mid class and 3,725 to 15,519 kg for the frontier class. The frontier rows are inverted at the mid-tier price, and at the top-tier price the central frontier figure is about 5.0 tonnes (S11.1 text). The token-weight rows hold the volume at the central value and move $W_{\mathrm{in}}$ and $W_{\mathrm{cache}}$ together to their low and high values, which changes effective tokens from 12.42 billion to 6.18 and 19.54 billion. The B200 ``with dev.'' column falls further below the central one than the ``without dev.'' column does, because the development line is proportional to accelerator energy, which the B200 lowers more than it lowers the footprint.}\end{minipage}
\end{table}

Three things follow. The spread that matters is not the uncertainty in the factor but the class assignment: the frontier class is about four times the mid class on the same activity, which is 6.2 tonnes a year for fifty people. The development line is the second lever, adding 61\% for the mid class and 53\% for the frontier class at the central ratio, and it is the quantity the method deliberately keeps outside the factor. And the token-weight band is a factor of 2.1 to 2.2 here, against $-$24\% to $+$10\% on the chat-dominated firm above, because 91\% of the billed tokens are cache reads and the weight on a cache read has no energy measurement behind it. A preparer in this position reports the range, names the class assumption, and presses the vendor for the model actually served.

\subsubsection*{S11.2 Suggested Disclosure Text}

The following applies the reporting rules of Section S12.3, which are this method's proposal, and is written to fit ESRS E1-8 (E1-6 for financial years before 2027, unless the undertaking opts for the revised standards for 2026), for the Scope 3 figure that paragraphs 49 to 52 of the standards for voluntary use established by Commission Delegated Regulation (EU) 2026/1560 attach to module B3, where it is discretionary and sits outside the value-chain cap~\citep{eucom2026b}, or for a customer data request. The AI figure is aggregated into the Category 1 total rather than disclosed as a category of its own; the one place it appears in its own right is the cloud computing and data-centre services sub-category of Application Requirement 24(g) under the revised standards, and of AR 51 under the current ones, where it is required if material, although the table that AR 48 requires labels the row optional~\citep{euparl2022,eucom2026a}.

\begin{quote}\small
\textit{\textbf{Boundary and basis.} Scope 3 Category 1 includes purchased AI inference. The factor covers serving electricity at a full-server and facility boundary on location-based 2023 grid intensities on a life-cycle basis, embodied server manufacturing and transport, data-centre construction and amortised final-run training, at the 100-year global warming potential (GWP-100) on IPCC AR5, the set on which the grid sources publish. Excluded: model development other than the final training runs, reported as a development line that covers all development including those runs and is worth 1.3 times the inference footprint at the central sector ratio and 0.6 to 2.2 times across its range; provider overhead outside compute, unquantified; storage, network transmission and hardware end-of-life, together 0.71\% of operational serving emissions; and end-user devices, which are in our own inventory and are not bounded. The figure is gross.}

\textit{\textbf{Activity data and factors.} Factors come from the published dataset~\citep{llopis2026} at the version named in our file; this is the average-data method; the activity data of both token-based lines come from the suppliers' own exports, so under the methodology's convention 100\% of the figure is calculated using data obtained from suppliers, or about 63\% (5.4 of 8.5 kg, Table~\ref{tab:casestudy}) counting the API line alone, because the seat line's tokens per message, reasoning share and reasoning multiplier are defaults, and 0\% uses supplier-specific emission factors. The API export gives 120 million billed tokens, 75\% input, no cache reads, so no cache credit is claimed. The seat export gives 18,000 messages, converted at the default 928 input plus 282 output tokens each, the output tokens alone raised 2.195 times by the measured 4.98\% reasoning share at a multiplier of 25 (619.0 output and 1,547 billed tokens per message), GPT-4o (mid class) and GPT-5 (frontier class) weighted by the days each was the documented default, both classed by pricing tier. Input tokens are weighted, uncached at 0.30 and cached at 0.05 of an output token, and divided by 1.339. The region is undisclosed, so the global-average grid applies.}

\textit{\textbf{Uncertainty.} FY2025 token-based AI inference emissions are 8.5 kg CO\textsubscript{2}e, 13.6 kg in a conservative high case and 7.0 to 12.6 kg across the band on the token conversion, 1.5 to 8.8 kg across the seventeen printed regions, with a class range on the energy term of $-$38\%/$+$24\% (mid) and $-$67\%/$+$55\% (frontier). The figure is the low end of a stated structural range. Energy per token is taken at the most efficient configuration that meets interactive latency; the mean over all such configurations, the upper end of the serving-regime range, would raise the figure by about 170\% on this firm's model and token mix. On a different boundary, with all model development, including the final training runs, in place of the training term, the figure would be 18.5 kg (12.7 to 25.5 kg); that figure is reported beside the result and not added to it. No low case is defined. Data quality is 2 and 3 at the API and 4 and 3 at the seat product, on the methodology's one-to-five scale, whose levels 1 to 4 are the Scope 3 Standard's very good, good, fair and poor.}

\textit{\textbf{Exclusions and double counting.} The corresponding supplier spend has been removed from our spend-based Category 1 line, and the AI figure is aggregated into the Category 1 total. AI embedded in other subscriptions stays in the spend-based line and is not separately estimated (Notion, 769.0 kg CO\textsubscript{2}e for the whole subscription, on spend deflated to 2019 prices).}
\end{quote}

\textbf{What to retain.} Six items make the figure auditable and none of them is in the disclosure: the billing and analytics exports with their date ranges, reconciled to the invoices; the list of model identifiers and which pattern each matched, with the share of activity that fell back; the dataset version and the grid year; the region rule applied and the evidence for it; the cache treatment and the input share used; and the class weighting by default-model dates with the source for those dates.

\subsection*{S12. Boundary Justification, Water Method and Reporting Rules}

These subsections were part of the Methods section in the full version of this paper and are kept here in full.

\subsubsection*{S12.1 Why These Terms Belong in Category 1}

Three propositions place that boundary, each resting on a rule of the standards cited rather than on convenience.

\emph{The minimum boundary is cradle-to-gate.} The Scope 3 Standard requires a reporting company to account for each category according to the minimum boundaries of its Table 5.4, and for Category 1 that boundary is all upstream, cradle-to-gate emissions of the purchased good or service, defined as every emission in the product's life cycle up to the point of receipt~\citep{wri2011}. Because the boundary runs over the product's life cycle rather than over the supplier's reported inventory, a serving-electricity-only factor falls short of that minimum rather than merely being less complete than it could be, the server and the building in which it runs being the provider's capital goods and therefore upstream of the service it sells. Development other than the final runs is not treated as part of that life cycle, because it is a same-period flow of sector compute that no single model owns (S1.3). It enters only the development line, which covers all development, including the final runs, and is reported beside the factor. The revised ESRS point the same way, placing purchased cloud computing and data-centre services as a sub-category of Category 1 rather than as a category of their own~\citep{eucom2026a}. How a preparer assembles that boundary in practice, and why the allocation rules of the Standard's Chapter 8 then apply, is in S5.

\emph{Allocating capital goods over a service life is not depreciation.} The Standard's recommendation that capital goods be reported in full in the year of acquisition governs a reporting company's own Category 2 inventory~\citep{wri2011}, not the way a supplier spreads capital-goods emissions across the units of service it sells.

\emph{There is no double counting with the preparer's own Category 2 or Category 8.} The accelerators and the data centre belong to the provider, are never on the preparer's balance sheet and are never leased to it, so neither the preparer's own capital-goods category nor its upstream-leased-assets category applies~\citep{wri2011}; supplier Scope 1 and Scope 2 emissions appearing inside a customer's Category 1 are the intended behaviour of value-chain accounting.

A fourth statement is a limitation rather than a proposition: \emph{the factor is not reconciled to any provider's inventory}. No provider publishes an allocation of its reported emissions to customer usage, so the factor is built bottom-up and the sum of every customer's allocated share is not guaranteed to equal any provider total. A preparer should not represent a figure built this way as the provider's own allocation.

\subsubsection*{S12.2 Water as an Extension of the Same Chain}

The same activity data and the same energy chain support a water-use inventory, with two terms and no change to the carbon factor; neither water term is ever added to a carbon figure. The \emph{on-site} term is the water used for cooling and humidification at the data centre, a published water use effectiveness in litres per kWh multiplied by the energy of the denominator that the published value itself uses, which is server energy where that denominator is IT load and facility energy where the value is a site ratio derived from a provider's own published site water and site electricity. The \emph{off-site} term is the water consumed in generating the electricity, an energy water intensity factor applied to facility energy. Both are reported in millilitres per 1,000 effective tokens.

Three conventions must be declared by the preparer, because each changes the answer and none is settled by the literature. \emph{Basis}: providers publish withdrawal or consumption and often do not say which, and a total that adds withdrawal to consumption is not a quantity of anything, so every on-site value is converted to a consumption basis with the one documented conversion available (Section S9.8), and the as-published figures become a sensitivity. \emph{Hydropower}: reservoir evaporation at hydroelectric dams may be allocated to the electricity or excluded from it, and the two conventions are reported side by side with neither preferred, because the allocated share dominates the off-site factor and the published evidence settles neither gross against net evaporation nor how to allocate a multipurpose reservoir. \emph{Scarcity weighting}: a water scarcity footprint multiplies water consumed in a basin by that basin's characterisation factor. The only location the method knows is the data centre, and the factor set is applied at country or state resolution, which its developers provide only where the basin and the month are unknown and record as carrying very large uncertainty (Section S9.8), so the weighting is applied to the on-site term alone and never to the total. Three results follow and are never combined; they are named per International Organization for Standardization (ISO) 14046~\citep{iso2014}, as S3 sets out. A preparer who wants a single volumetric figure still reports the total under both hydropower conventions, as S3 sets out. Water has no unknown-region fallback: where no value exists for a region, the result is ``not available'' rather than a number.

\subsubsection*{S12.3 Reporting Rules and Maintenance}

\textbf{What to disclose.} These reporting rules are a proposal of this method and go beyond what ESRS E1 or the GHG Protocol prescribe. State the boundary of Section~\ref{sec:3.1} with its exclusions and their size, the location basis and the grid year, the model class or own-model factor assigned to each service and why, the serving region or the rule used in its absence, the token basis with the weights used and any cache credit claimed, the operating-point rule with its named scenario, and the global warming potential set. Report the central value with the class range and the conservative high case, and for a closed model whose class rests on a vendor tier and not on a disclosed size add the adjacent-class values as a scenario, and add the conservative high cases across services, which assumes they move together and is therefore conservative. Report the figure as gross, before any offset or avoided-emissions claim. Report the development-inclusive total beside it, labelled as a different boundary, with the ratio used. Section S11.2 gives a worked disclosure that meets these rules.

\textbf{Further items to disclose.} For each AI service the preparer also states the tier and its GHG Protocol method, the dataset version and the share of activity that fell back to the unmapped-identifier rule. It states the allocation conventions, which are embodied hardware by server time, construction by server electricity and training over lifetime billed tokens. It reports the five uncertainty parts of Section~\ref{sec:3.8}, with the serving-regime end and the development-inclusive total each labelled by its axis, and the data-quality scores with both supplier-data shares. The global warming potential set is IPCC AR5, for the reason given in S5. As ESRS 2 asks of value-chain metrics (general disclosure requirements on metrics, GDR-M, paragraph 49(b)), the preparer discloses its reliance on proxies, namely the class assignment, the Tier 2b defaults and the global-average grid, with the actions planned to obtain supplier data~\citep{eucom2026a}.

\textbf{What to retain for assurance.} The preparer retains the billing and analytics exports with their date ranges, reconciled to the invoices, the identifier-to-pattern matches with the fallback share, the dataset version and grid year, the region rule and its evidence, the cache treatment and input share, and the class weighting by default-model dates.

\textbf{Rules the tier logic does not settle} are in Table~\ref{tab:rules}, with the two a preparer meets most often repeated in Table~\ref{tab:overview}; four further reporting rules, on market-based claims, global warming potentials, factor versions and base-year recalculation, and completeness and integration with the spend-based line, are in S5. Four of them govern the next reporting year rather than this one, so they are stated here as well: pin one dataset version per reporting year; treat a change of dataset version or of grid year, or the discovery of a material error, as a methodology change and disclose it; recalculate the base year only where the resulting change in the AI line exceeds the company's own disclosed significance threshold applied to total Scope 3; and back-cast on spend only after adjusting for the fall in the vendor's price per million tokens.

\textbf{Maintenance protocol.} The factors and the activity defaults rest on fifteen input families, each refreshed from a named source on a stated cadence, each with a named event that triggers an out-of-cycle revision. Table~\ref{tab:refresh} gives all three for every family: accelerator energy on each benchmark release; the model class mapping, model service life and per-model token volume quarterly; the cloud region list semi-annually; grid intensity, the uplifts, PUE, the hardware footprints, hardware life, the development ratio, the token weights, water, the vendor prices and rates, and the activity defaults annually; and the off-site water and scarcity factors on release of a replacement set. The input register, the procedure and a script that re-downloads the machine-readable inputs and reports what has changed will be deposited with the dataset as \texttt{input\_\allowbreak register.csv}, \texttt{REFRESH.md} and \texttt{refresh\_\allowbreak check.py}~\citep{llopis2026}. Two triggers apply to the whole dataset rather than to one input. Any provider disclosing energy per token for a model it serves supersedes the class-based chain for that model and moves the service to the hybrid method, or to Tier 3 if the provider publishes a CO\textsubscript{2}e figure. And where a refresh changes any class factor by more than 5\%, the dataset version number increments and preparers disclose the change.

\subsubsection*{S12.4 Resolution Mechanics}

The resolution rule of Section~\ref{sec:3.5} with its mechanics and examples, as stated before condensation.

A preparer has two strings, a model identifier from a billing export and a region, and needs a factor.

\textbf{Model.} Where the model served is one for which an energy measurement exists under the benchmark protocol, its own factor is used; that is the case for the open-weight models the benchmark measures, each of which carries its own energy, embodied and construction terms and its own development line in the dataset, with the class training term, since no per-model training disclosure exists outside the Llama 3.1 family (Section S9.5). Where no measurement exists, which is every closed commercial model, the model is assigned to a size class and the class value is used. \emph{The class value is the arithmetic mean of the measured models of that class at a common operating point}, defined in Section S9.1 as the lowest-energy configuration of the measured sweep that still streams at interactive speed.

\textbf{Class assignment.} Class is assigned by disclosed total parameter count where the provider discloses it, and by the vendor's pricing tier where it does not. For dense models, the small class is up to 15B parameters, the mid class above 15B up to and including 70B, and the frontier class above 70B. For mixture-of-experts models, also by total parameters, the small class is below 50B, the mid class from 50B to below 200B, and the frontier class at 200B and above; total rather than active parameters, because all experts must be resident in accelerator memory, which sets the accelerator count. Where the size is not disclosed, the efficiency tier (\emph{mini}, \emph{nano}, \emph{flash}, \emph{lite}, \emph{haiku} and similar) is the small class, the standard tier (\emph{sonnet}, \emph{pro}, \emph{plus}, an unsuffixed mid-tier model) the mid class, and the frontier tier (\emph{opus}, \emph{ultra}, \emph{max}, the top reasoning or flagship model of a generation) the frontier class. \emph{A disclosed total parameter count always overrides the tier word}: three 2026 efficiency-tier models disclose sizes at or above the mixture-of-experts frontier cut and are therefore Class C (S6). Reasoning does not change the class; it changes the tokens per interaction. The tier is read from the vendor's price list for the generation, so gpt-4o, priced at the standard tier below GPT-4 Turbo and the full-size o-series of its time, is the mid class, and GPT-5, the flagship of its generation, is the frontier class (Table~\ref{tab:classmap} in S6).

\textbf{The mapping is data, not prose.} The assignment is deposited as \texttt{model\_\allowbreak class\_\allowbreak mapping.csv}, 187 identifier patterns with the basis of each (disclosed size or vendor tier), the own-factor model where there is one, the provider page the assignment was read from and the month it was assigned. It is applied by the reference calculator \texttt{ai\_\allowbreak inference\_\allowbreak emissions.py}, which is deposited with it; no factor-generation script applies it. Identifiers are normalised and matched by a total, reproducible rule, with the longest match winning and the fallback of Section~\ref{sec:3.5} as the file's last row, and the normalisation, the wildcard and the tie-break are stated with worked examples in S6.

\textbf{Region.} The electricity term needs one number, the grid intensity $g$, location-based and on a life-cycle basis, which is not the combustion-only factor a provider uses for its own Scope 2. A customer-pinned region takes its own row; an EU data zone or a contractual EU-only processing commitment, such as the Microsoft EU Data Boundary or an AWS or Google EU data-zone commitment, takes the EU-27 row, while a marketing statement about data residency for storage is not one; an unknown region, global routing or a cross-region profile takes the global-average row. The dataset resolves a region at three levels: a country or economy, a United States grid subregion, or a named public cloud region mapped by hand to one of those with a stated confidence. Outside the United States each cloud region takes its national average, which hides large differences within large countries, and the mappings that rest on a location the provider states only as a state or country are marked as such. For the ten United States regions in that case, the campus locations listed by an industry directory~\citep{datacentermap2026} were tested against the eGRID subregion boundaries on 14 October 2026 and agree with the assumed sites in every case, which the dataset records row by row; the directory is secondary evidence for a site's location and is never used for energy, PUE or water. The same listing places Canada (Central) at Varennes, Quebec, whose grid is far below the Canadian national average, so that region now takes a provincial row, 0.0265 kg CO\textsubscript{2}e per kWh against the national 0.1744, built on the same life-cycle basis as the country rows: Quebec's 2023 generation shares (94.0\% hydro, 5.0\% wind, 0.7\% biomass, the 0.4\% remainder charged as gas; \citealp{cer2026}) multiplied by the Canadian life-cycle factor per fuel implied by Ember's own emissions and generation by fuel for the same year. It is the only subnational row outside the United States and serves the three cloud regions sited in Quebec (AWS Canada (Central), Azure Canada East and Google Cloud Montréal).

\textbf{What the resolution does not do.} It does not distinguish self-hosting from a provider API: the same host, idle and PUE uplifts are applied, because no on-premises variant of those uplifts is available. Self-hosted inference is in any case the preparer's own Scope 2 and Category 2 rather than Category 1, and only the energy chain transfers. And it does not produce a model-specific factor for a closed model (Section~\ref{sec:7.2}).

\subsubsection*{S12.5 Activity Tiers: Product Detail}

The tier discussion of Section~\ref{sec:3.2} with the product-by-product detail, as stated before condensation.

The GHG Protocol does not set a strict hierarchy for purchased goods and services. Its Technical Guidance ranks the four Category 1 methods by how specific each is to the individual supplier, but states that companies need not always use the most specific method as a first preference, and warns that a more specific method is not necessarily a more accurate one; the choice is driven by the size of the activity, business goals, data availability, data quality and cost, falling back to the spend-based method where physical and supplier data of sufficient quality are unavailable~\citep{wri2013}. Applied to AI services, this gives four tiers ordered by the quality of data an organisation can realistically obtain, numbered as the Intergovernmental Panel on Climate Change numbers tiers, with Tier 1 on default factors and Tier 3 the most specific. In GHG Protocol terms Tier 3 is the supplier-specific method, Tiers 2a and 2b the average-data method, the Tier 2a hybrid case the hybrid method and Tier 1 the spend-based method, all four being named in the Technical Guidance rather than in the Standard~\citep{wri2013}; the split of Tier 2 into 2a (exact tokens) and 2b (estimated tokens) is introduced here, and two variants sit beside Tier 2a, the hybrid method, where a supplier gives an energy quantity rather than a CO\textsubscript{2}e figure (S5), and reserved capacity, which is charged on accelerator-hours (Section~\ref{sec:3.4}). A tier is assigned to each AI service individually, not to the company as a whole: a typical organisation has a direct API connection at Tier 2a, an enterprise seat product at Tier 2b and an AI-bundled software subscription at Tier 1 (Figure~\ref{fig:flowchart}).

\textbf{Which products sit where.} No major AI provider yet offers customer-specific, independently assured carbon reporting at model or service level, so Tier 3 is at present empty and the product-level disclosures by Google and Mistral serve as calibration points instead (Section~\ref{sec:5.1}). Tier 2a covers API access with token billing, and Claude Enterprise on a usage-based plan, whose Analytics API returns uncached input, cached input and output tokens by user and by model; on a seat-based plan the same endpoints report usage credits only and the service is Tier 2b~\citep{adminreports2026}. Tier 2b covers seat-licensed products that report activity but no tokens, and Tier 1 covers subscriptions in which the vendor, not the customer, holds the AI API relationship, so token-level data never surfaces in the customer's billing portal. Which admin export each product offers today is in S5; the rule is to \emph{read the product's own message count first, and use the per-seat default only where it cannot be retrieved}. Token counts at every tier are billed tokens, so the input column includes the system prompt, any retrieval or tool context and the conversation history re-sent with every turn; that is several times the ``tokens per prompt'' that conversation corpora report, because those count only the user's own words (Section S10.1).

Microsoft 365 Copilot is the case that rule exists for: Microsoft's own administrative reports give prompts submitted and active days and name no model~\citep{adminreports2026}, so absent a tenant-level statement the service takes the highest-intensity class, and the preparer records the fallback and the share of activity it covers.

\section*{References}
\small
\sloppy
\raggedright

\input{references_supp}
\fussy

\end{document}

%% file: nomenclature.tex
\section*{Nomenclature}
\begin{small}
\noindent\textbf{Symbols}\par\nobreak
\begin{tabular}{@{}p{3.1cm}p{12.7cm}@{}}
$D$ & Prefill divisor, $1 + W_{\mathrm{in}} R_{\mathrm{bench}}$ (1.339 at the reference mix) \\
$D_{\mathrm{run}}$ & The same divisor evaluated at the token mix and cache share of one benchmark run \\
$d$ & Development ratio: energy of all model development, including the final training runs, as a multiple of serving energy; reported as a separate line only \\
$E_{\mathrm{fac}}$, $E_{\mathrm{srv}}$, $E_{\mathrm{GPU}}$ & Facility, server and accelerator electricity per million effective tokens, kWh \\
$E_{\mathrm{train}}$ & Disclosed emissions of the final training runs, kg CO\textsubscript{2}e \\
$e_d$ & Decode energy per output token \\
$\mathrm{EF_{el}}$, $\mathrm{EF_{emb}}$, $\mathrm{EF_{con}}$ & Electricity, embodied-hardware and data-centre-construction terms, kg CO\textsubscript{2}e per million effective tokens \\
$\mathrm{EF_{tr}}$ & Training term, kg CO\textsubscript{2}e per million billed tokens \\
$\mathrm{EF_{dev}}$ & Model-development line, reported separately from the factor \\
$g$ & Grid carbon intensity of the serving region on a life-cycle basis (combustion plus upstream fuel cycle), kg CO\textsubscript{2}e per kWh \\
$H$ & Host uplift: node power over accelerator power \\
$I$ & Idle uplift: fleet power over power at the serving operating point \\
$k$ & Reasoning multiplier: output tokens of a reasoning-mode message relative to an ordinary one, Tier 2b \\
$k_c$ & Data-centre construction emissions per kWh of server electricity, kg CO\textsubscript{2}e per kWh \\
$L$ & Hardware service life, years \\
$M_{\mathrm{mfg}}$, $M_{\mathrm{tr}}$ & Manufacturing and transport footprint of an eight-accelerator server, kg CO\textsubscript{2}e \\
$m$ & Total-token multiplier converting billed tokens to effective tokens at the default input share \\
$N$ & Lifetime tokens a model serves, $(S/12) \times 365 \times V$ \\
$p$ & Idle power of an accelerator as a fraction of its serving power \\
$R_{\mathrm{bench}}$ & Input-to-output token ratio of the reference benchmark mix \\
$S$ & Model service life, months \\
$s_{\mathrm{in}}$ & Assumed input share of billed tokens \\
$s_r$ & Share of conversations that use a reasoning mode, applied per message, Tier 2b \\
$T_8$ & Measured output throughput of an eight-accelerator server at the operating point, tokens per second \\
$T_{\mathrm{billed}}$, $T_{\mathrm{eff}}$ & Billed tokens ($T_{\mathrm{in}} + T_{\mathrm{out}}$) and effective tokens \\
$T_{\mathrm{in}}$, $T_{\mathrm{out}}$ & Input and output tokens; input is split into uncached and cached \\
$U$ & Utilisation of the serving fleet \\
$V$ & Tokens a model serves per day \\
$W_{\mathrm{in}}$, $W_{\mathrm{cache}}$ & Energy of an uncached and of a cached input token relative to an output token \\
\end{tabular}

\medskip
\noindent\textbf{Abbreviations}\par\nobreak
\begin{tabular}{@{}p{3.1cm}p{12.7cm}@{}}
AI & Artificial intelligence \\
API & Application programming interface \\
AWARE & Available Water Remaining, the water-scarcity characterisation method \\
CO\textsubscript{2}e & Carbon dioxide equivalent \\
CPI & Consumer price index \\
CSRD & Corporate Sustainability Reporting Directive \\
EEIO & Environmentally extended input-output \\
ESRS & European Sustainability Reporting Standards \\
EU & European Union \\
FP8 & Eight-bit floating-point precision \\
GHG & Greenhouse gas \\
GPU & Graphics processing unit \\
HICP & Harmonised index of consumer prices \\
ICT & Information and communications technology \\
IPCC & Intergovernmental Panel on Climate Change \\
ISO & International Organization for Standardization \\
PUE & Power usage effectiveness \\
SCI & Software Carbon Intensity \\
TPU & Tensor processing unit \\
WUE & Water use effectiveness \\
\end{tabular}
\end{small}
\medskip